\documentclass[11pt, a4paper, oneside]{Thesis} 

\graphicspath{{./Pictures/}} 

\usepackage[square, numbers, comma, sort&compress]{natbib} 
\hypersetup{urlcolor=blue, colorlinks=true} 
\title{\ttitle} 

\begin{document}

\frontmatter 

\setstretch{1.3} 

\fancyhead{} 
\rhead{\thepage} 
\lhead{} 

\pagestyle{fancy} 

\newcommand{\HRule}{\rule{\linewidth}{0.5mm}} 

\hypersetup{pdftitle={\ttitle}}
\hypersetup{pdfsubject=\subjectname}
\hypersetup{pdfauthor=\authornames}
\hypersetup{pdfkeywords=\keywordnames}


\begin{titlepage}
\begin{center}

\includegraphics[width=1in]{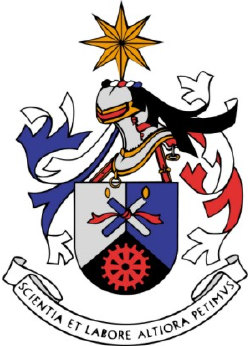} 

\textsc{\Large \univname}\\[1.5cm] 
\textsc{\Large Doctoral Thesis}\\[0.5cm] 

\HRule \\[0.4cm] 
{\huge \bfseries \ttitle}\\[0.4cm] 
\HRule \\[1.5cm] 
 
\begin{minipage}{0.4\textwidth}
\begin{flushleft} \large
\emph{Author:}\\
\authornames 
\end{flushleft}
\end{minipage}
\begin{minipage}{0.5\textwidth}
\begin{flushright} \large
\emph{Supervisor:} \\
\supname \\ 
\emph{Co-supervisor:} \\
\cosupname
\end{flushright}
\end{minipage}\\[3cm]
 
\large \textit{A thesis submitted in fulfilment of the requirements\\ for the degree of \degreename}\\[0.3cm] 
\textit{in the}\\[0.4cm]
\deptname\\[2cm] 
 
{\large December 2013}\\[4cm] 

\vfill
\end{center}

\end{titlepage}





 
%

\clearpage 


\pagestyle{empty} 

\null\vfill 

The work on this dissertation was supported by the Portuguese Agency Funda\c{c}\~ao para a Ci\^encia e Tecnologia (FCT) through
the fellowship SFRH/BD/43709/2008 and a  grant under the Quantum Geometry and Quantum Gravity program of 
the European Science Foundation (ESF).

\vfill\vfill\vfill\vfill\vfill\vfill\null 

\begin{center}
\includegraphics[width=1.9in]{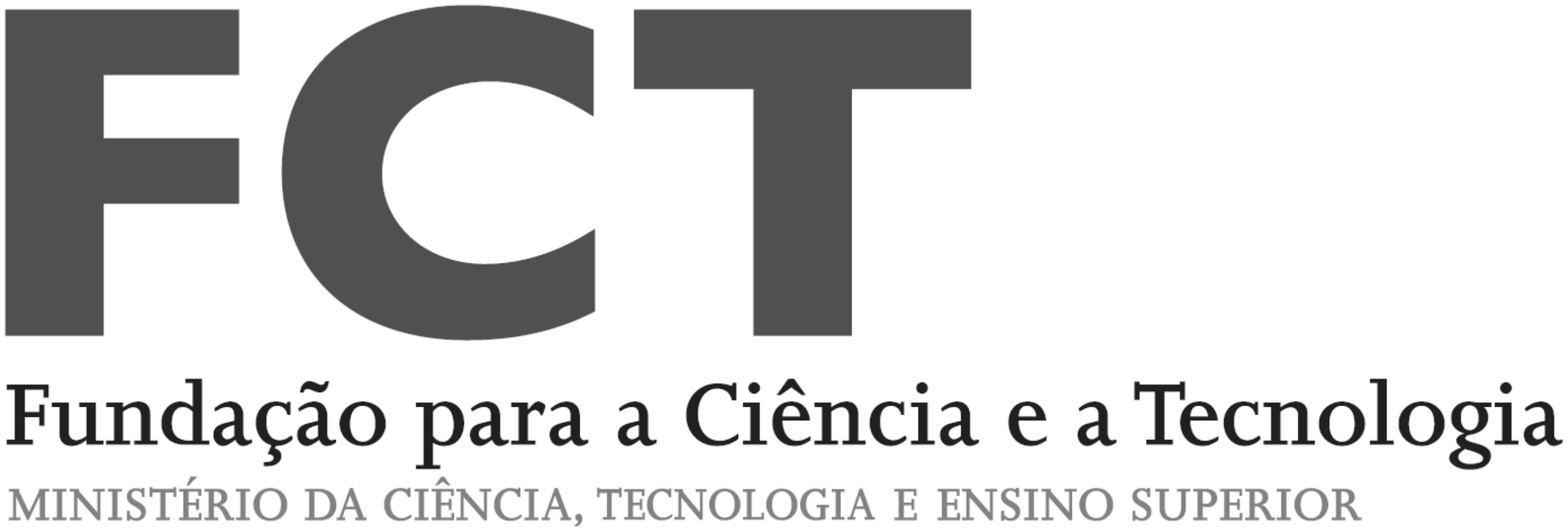} \quad{}  \quad{}  \quad{} \includegraphics[width=1.2in]{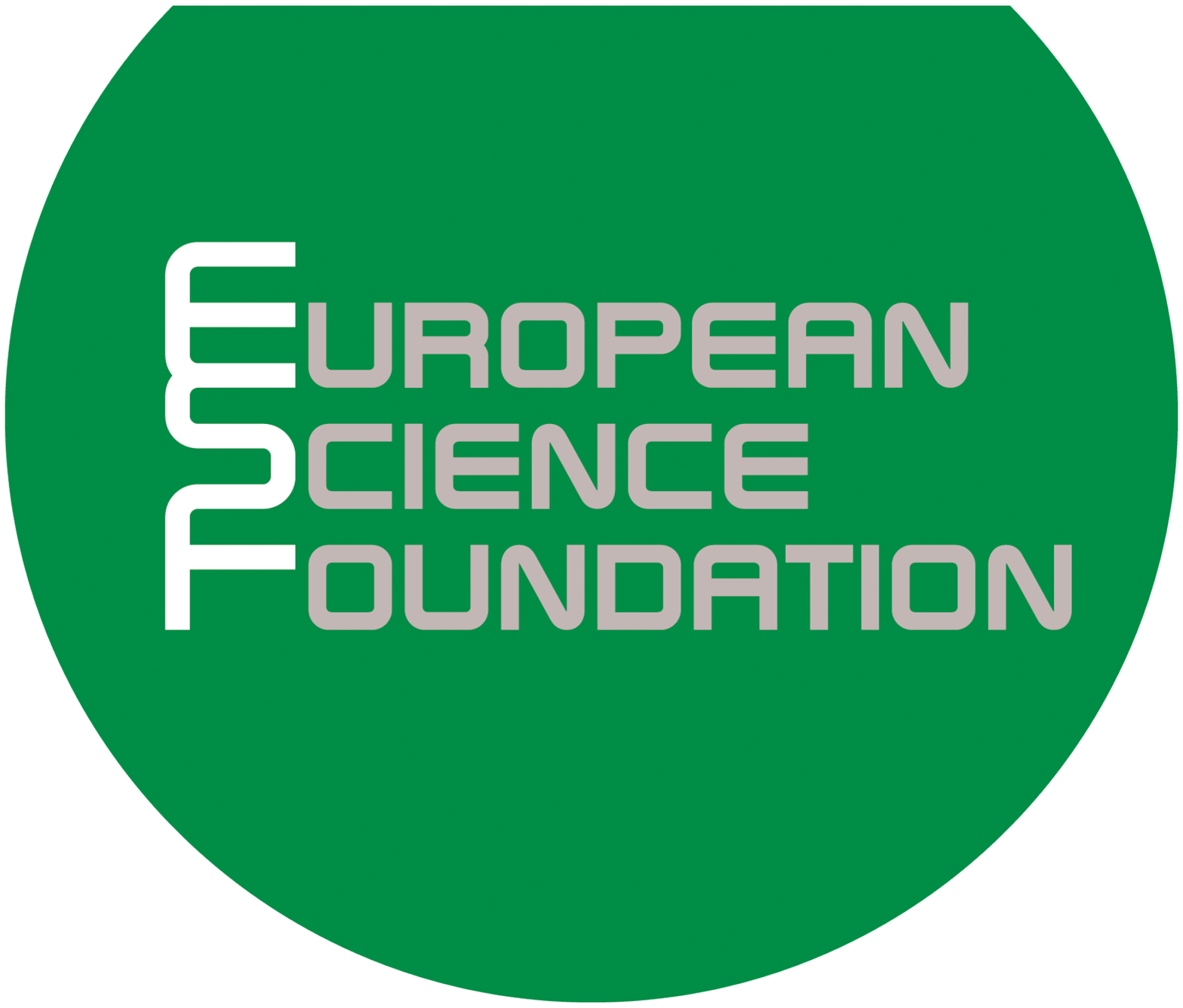}
\par
\end{center}%


\clearpage
\pagestyle{empty}


\abstract{\addtocontents{toc}{\vspace{1em}}  

In this dissertation we study two well known gravitational scenarios in which singularities may appear. 
The first scenario can be brought in by the final state of gravitational collapse (the singularity that is found, e.g., inside the event horizon of every black hole or,  if  no trapped surface is formed, as  a naked singularity instead).
The second scenario is the one corresponding to singularities that may appear at the late time evolution of the universe.

In the context of gravitational collapse, we study a homogeneous spherically symmetric space-time whose matter content includes a scalar field. 
We  investigate a particular class of such space-time, with a tachyon field and a barotropic fluid being present.
By making use of the specific kinematical features of the tachyon, which are rather different from a standard scalar field, we establish several types of asymptotic behavior that our matter content induces. Employing a dynamical system analysis, complemented by a thorough numerical study, we find classical solutions corresponding to a naked singularity or a black hole formation. Furthermore, we find a subset where the fluid and tachyon participate in an interesting tracking behaviour, depending on the initial conditions for the energy densities of the tachyon field and barotropic fluid.

It seems  reasonable that the singularity indicates that the classical theory we use for gravitational collapse (general relativity)  cannot be trusted when the space-time curvature becomes very large; quantum effects cannot be ignored.
We therefore investigate, in a semiclassical manner,  loop quantum gravity (LQG) induced effects on the fate of the classical singularities that arise at the final state of the gravitational collapse. 
We study  the semiclassical interior of a spherically symmetric space-time with 
a tachyon field and barotropic fluid  as matter content. 
We show how, due to two different types of corrections, namely ``inverse triad" and ``holonomy",  classical singularities can be removed.
By employing an \emph{inverse triad} correction, we obtain, for a semiclassical description, several classes of analytical as well as numerical solutions. 
We identify a subset whose behavior corresponds to an outward flux of energy, thus avoiding either a naked singularity or a black hole formation. 
Within  a holonomy correction, we obtain the semiclassical counterpart of our classical solutions for the general relativistic collapse.
We show that  classical singularity  is resolved and replaced by a  bounce.
By employing a phase space analysis in the semiclassical regime, we find that there is no stable fixed point solution, hence no  singular back hole neither naked singularities form.
We further discuss the possible predictions for the exterior geometry within this model.

In the context of the dark energy cosmology, we study the status of dark energy late time singularities. 
We employ several models of dark energy to investigate whether they remove or appease the classical singularities.
In the first model we consider  a Dvali-Gabadadze-Porrati (DGP) brane-world model that has   infra-red (IR)  modifications through an induced gravity term.
It models our 4-dimensional world as a FLRW  brane embedded in a Minkowski bulk. 
A Gauss-Bonnet (GB) term is provided for the bulk action whose higher order curvature terms modify gravity at high energy (with ultra-violet (UV) effects). 
Furthermore, a phantom matter is present on the brane which constitutes the dark energy component of our universe. 
It is shown that a combination of IR and UV modifications to general relativity replaces a big rip singularity by a sudden singularity at late times.

Another model we consider herein to describe the origin of dark energy is the  generalised running vacuum energy (GRVE) model.
The Friedmann equation of the GRVE model looks  much similar to that of a homogeneous and isotropic universe filled with an   holographic Ricci dark energy (HRDE)  component. 
We study the late time behaviour of the universe in the presence of these two models for dark energy. 
Despite the analogy between these two models, it turns out that one of them, a GRVE, is singularity-free in the future while the other, the HRDE, is not. Indeed, a universe filled with an HRDE component can hit, for example, a big rip singularity. We clarify this issue by solving analytically the Friedmann equation for both models and analyzing the role played by the local conservation of the energy density of the different components  filling the universe. In addition, we point out that in some particular cases the HRDE, when endowed with a negative cosmological constant and in the absence of an explicit dark matter component, can mimic dark matter and explain the late time cosmic acceleration of the universe through an asymptotically de Sitter universe.

\paragraph*{Keywords.}


General relativity, singularities, gravitational collapse, quantum gravity,
late time cosmology, dark energy, brane-worlds.

}

\clearpage


\pagestyle{empty} 

\section*{List of publications}

\noindent Much of this dissertation is based on the following publications:
\begin{itemize} 
\item[\tiny{$\blacksquare$}] Mariam Bouhmadi-L\'opez, Yaser Tavakoli, and Paulo Vargas Moniz, \emph{Appeasing the Phantom Menace?}, JCAP {\bf 1004}: 016 (2010).
\item[\tiny{$\blacksquare$}] Yaser Tavakoli, Jo\~ao Marto, A. H. Ziaie and Paulo V. Moniz, \emph{Loop quantum effect and the fate of tachyon field collapse},  J. Phys.: Conf.  Ser.  {\bf 360},  012016 (2012).
\item[\tiny{$\blacksquare$}]  Yaser Tavakoli, Jo\~ao Marto, A. Hadi Ziaie and Paulo Vargas Moniz, \emph{Gravitational collapse with  tachyon field and barotropic fluid}, Gen. Rel. Grav. {\bf 45}, 819 (2013). 
\item[\tiny{$\blacksquare$}] Yaser Tavakoli, Jo\~ao Marto, A. Hadi Ziaie and Paulo Vargas Moniz, \emph{Semiclassical collapse with  tachyon field and barotropic fluid}, Phys. Rev. D \textbf{87}, 024042 (2013). 
\item[\tiny{$\blacksquare$}] Mariam Bouhmadi-L\'opez, Yaser Tavakoli, \emph{Why is the running vacuum energy more bennign than the   holographic Ricci dark energy?}   Phys.~Rev.~D~\textbf{87}, 023515 (2013).  
\item[\tiny{$\blacksquare$}] Yaser Tavakoli, Jo\~ao Marto and Andrea Dapor, \emph{Semiclassical dynamics of horizons in spherically}  {\em symmetric collapse}, Int. J. Mod. Phys. D {\bf 23}, 1450061 (2014).
\item[\tiny{$\blacksquare$}] Jo\~ao Marto, Yaser Tavakoli, Paulo  Moniz,  \emph{Improved dynamics and gravitational collapse  of tachyon field coupled with  a barotropic fluid}, arXiv:1308.4953 [gr-qc].
\item[\tiny{$\blacksquare$}] Yaser Tavakoli, Jo\~ao Marto and Andrea Dapor, \emph{Dynamics of apparent horizons in quantum gravitational collapse}  (proceeding of ERE2012), Springer Proc. Math. Stat. {\bf 60} (2014), arXiv:1306.3458 [gr-qc]. 
\end{itemize}

\noindent Other publications during this doctoral studies:
\begin{itemize} 
\item[\tiny{$\blacksquare$}]  A. H. Ziaie, K. Atazadeh and Y. Tavakoli, \emph{Naked singularity formation in Brans-Dicke theories:}  Class. Quantum Grav. \textbf{27}, 075016  (2010).
\item[\tiny{$\blacksquare$}] Andrea Dapor, Jerzy Lewandowski and Yaser Tavakoli,  \emph{Lorentz symmetry in QFT on quantum  Bianchi~I space-time}: Phys. Rev. D \textbf{86}, 064013 (2012).
\item[\tiny{$\blacksquare$}]  Andrea Dapor, Jerzy Lewandowski and Yaser Tavakoli,  \emph{Quantum field theory on LQC Bianchi   space-times}:  preceeding of MG13 (2013) arXiv:1305.4513 [gr-qc].
\end{itemize}


\pagestyle{empty} 

\section*{Resumo}

Nesta disserta\c{c}\~ao apresenta-se o estudo de dois cen\'arios, bem conhecidos
em relatividade geral, em que  podem surgir singularidades. O primeiro \'e caracter\'istico dos est\'agios finais de colapsos gravitacionais (a
singularidade presente no interior do horizonte de acontecimentos
do \emph{buraco negro}, ou a \emph{singularidade nua}). O segundo
consiste no aparecimento de singularidades nos est\'agios finais da
evolu\c{c}\~ao de cen\'arios particulares da universo.

No contexto do colapso gravitacional iremos considerar o estudo de
um espa\c{c}o tempo homog\'eneo com simetria esf\'erica, em que o conte\'udo
material \'e composto por um campo escalar.
Come\c{c}amos por investigar um tipo particular desse espa\c{c}o
tempo, em que a mat\'eria \'e constitu\'ida por um campo escalar taqui\'onico
e um flu\'ido barotr\'opico, como modelo para o colapso gravitacional.
Fazendo uso das carater\'isticas cinem\'aticas espec\'ificas do campo taqui\'onico,
que s\~ao bastantes diferentes das do campo escalar usual, iremos descrever
os v\'arios comportamentos assimpt\'oticos do modelo anteriormente referido.
Usaremos ferramentas de an\'alise de sistemas din\^amicos, 
apoiadas em m\'etodos num\'ericos, para encontrar as solu\c{c}\~oes cl\'assicas correspondentes
\`{a} forma\c{c}\~ao de singularidades nuas e de buracos negros. Para al\'em destas
solu\c{c}\~oes, ser\~ao tamb\'em apresentadas outras em que a 
presen\c{c}a dominante na  
densidade de energia ser\'a disputada pelo flu\'ido barotr\'opico
e o campo escalar taqui\'onico, dependendo das condi\c{c}\~oes iniciais do
colapso.

\'E razo\'avel assumir que o aparecimento de singularidades parece
indicar que a teoria da relatividade geral \'e incompleta na  descri\c{c}\~ao do espa\c{c}o
tempo quando a curvatura se torna extrema, e que os efeitos qu\^anticos
n\~ao podem ser ignorados. Iremos investigar efeitos, 
que podem ser extraidos da teoria de gravita\c{c}\~ao qu\^antica com lacetes, 
no destino das singularidades cl\'assicas que surgem
nos est\'agios finais do colapso gravitacional. Para ser mais concretos,
aqui tamb\'em iremos considerar um espa\c{c}o tempo homog\'eneo com simetria
esf\'erica, em que o conte\'udo material \'e composto por um campo 
escalar, mas agora no contexto da gravidade semicl\'assica (fornecida a partir loop quantum gravity). Iremos ilustrar como os efeitos
da gravidade qu\^antica removem as singularidades cl\'assicas, quando
differentes tipos de corre\c{c}\~oes derivados da gravidade qu\^antica em
la\c{c}os s\~ao implementadas. V\'arios tipos de solu\c{c}\~oes anal\'iticas e num\'ericas
ser\~ao apresentadas, num contexto semicl\'assico, e decorrentes de corre\c{c}\~oes
do tipo \emph{inverse triad} ao modelo de colapso do taqui\~ao. Iremos
identificar um subconjunto de solu\c{c}\~oes em que \'e descrito um fluxo
de energia emergente, sendo assim evitados a forma\c{c}\~ao de uma singularidade
nua ou de um buraco negro. 
Considerando uma corre\c{c}\~ao holon\'omica, obtemos a extens\~ao semicl\'assica para nossas solu\c{c}\~oes cl\'assicas 
do colapso gravitacional de um campo taqui\'onico com um fluido barotr\'opico. 
Neste caso, a singularidade cl\'assica que surgiria em consequ\^encia do colapso \'e evitada e substitu\'ida por um ressalto (bounce). 
Ao empregar uma an\'alise do espa\c{c}o de fase no regime semicl\'assico, descobrimos que n\~ao h\'a um ponto fixo est\'avel 
correspondente \`as solu\c{c}\~oes cl\'assicas, consequentemente, nenhum buraco negro ou singularidade nua se formam.

No contexto da cosmologia com energia escura,
usamos  v\'arios modelos  para investigar se eles resolvem ou substituem a singularidade cl\'assica por um fen\'omeno mais suave. 
Na  primeira abordagem,  consideramos um modelo  Dvali-Gabadadze-Porrati (DGP)  brana-mundo (\emph{brane-world}) que tem modifica\c{c}\~oes infra-vermelhas (infra-red (IR)) por meio de um termo de gravidade induzida. 
Ele modela o nosso mundo quadrimensional como uma  brana FLRW incorporada num  espa\c{c}o de  imers\~ao (\emph{bulk}) Minkowski. 
Um termo de Gauss-Bonnet (GB)  est\'a previsto para o espa\c{c}o de  imers\~ao, cujos termos de ordem superior da curvatura modificam a gravidade em alta energia (com efeitos ultra-violeta (ultra-violet (UV)).
Al\'em disso, uma mat\'eria fantasma (\emph{phantom}) est\'a presente na brana e constitui a componente da energia escura do universo. \'E mostrado que uma combina\c{c}\~ao de modifica\c{c}\~oes de IR e UV em relatividade geral substitui uma singularidade de tipo \emph{big rip} por uma singularidade s\'ubita (\emph{sudden}), no est\'agio final do universo.

Outro modelo que consideramos aqui 
\'e o modelo generalised running vacuum energy (GRVE). 
A equa\c{c}\~ao de Friedmann do modelo GRVE parece muito similar \`a que descreve um universo homog\'eneo e isotr\'opico preenchido  com uma componente hologr\'afica de Ricci da energia escura (holographic Ricci dark energy (HRDE)). 
Estudamos o comportamento do est\'agio final do universo, na presen\c{c}a destes  dois modelos da energia escura. 
Apesar da analogia entre esses  dois modelos, verifica-se que um deles, o modelo GRVE, \'e  livre de singularidade no futuro, enquanto  o outro n\~ao o \'e. 
Com efeito, um universo preenchido com um componente HRDE pode atingir, por exemplo, uma  singularidade \emph{big rip}. 
N\'os esclarecemos esta quest\~ao ao resolver analiticamente a equa\c{c}\~ao de Friedmann para ambos os modelos, e ao analisar o papel desempenhado pela conserva\c{c}\~ao local da densidade de energia dos diferentes componentes que preenchem o universo. 
Al\'em disso, cabe real\c{c}ar que, em alguns casos particulares, o  HRDE, quando dotado de uma constante cosmol\'ogica negativa e na aus\^encia de uma componente de mat\'eria escura expl\'icita, 
pode imitar a mat\'eria escura e explicar  a acelera\c{c}\~ao que surge nos est\'agios finais do universo atrav\'es de um  universo assimpt\'otico de Sitter.

\paragraph*{Palavras-chave.}

Relatividade geral, singularidades, colapso gravitacional, gravidade qu\^antica, acelera\c{c}\~ao
nos est\'agios finais em cosmologia, energia escura, brana-mundos.

\clearpage


\setstretch{1.1} 

\acknowledgements{\addtocontents{toc}{\vspace{1em}} 

First of all I would dedicate this work 
to my wife and to my family. Thank you for all your support in my life. \par

During the completion of this dissertation I have benefited  from valuable discussions with scientists
from several parts of the world. I am delighted to express my sincerest gratitude to people who helped me during these years of my doctoral studies.\par

I would like to thank first my PhD supervisor, Prof. Paulo Vargas Moniz, for the guidance, patience,
moral support and for all efforts he has done for introducing me to the art of physics during my doctoral program. \par
I wish to thank Dr. Mariam Bouhmadi-L\'opez, my PhD co-supervisor, for being an incessant source of ideas that
gave ground for an interesting part of my work. Thank you for your help, motivation and
fruitful discussions. \par

I would also thank Dr. Jo\~ao Marto, who was abundantly helpful on numerical studies of our works, in particular, for  his offered invaluable assistance during the writing of this dissertation.
Thank you for all fruitful discussions and collaboration during these years. 

I am very grateful to Prof. Hamid Reza Sepangi and Dr. Shahram Jalalzadeh, for their personal and scientific supports and attentions since my Masters program.
I would like also to thank to Prof. Jerzy Lewandowski for giving me the opportunity to work with him and the members of his group
in a very interesting field of research, for their passion in physics and for the support during my stay at the University of Warsaw. 
I would also thank Andrea Dapor for the nice time we spent together in Warsaw and for all the fruitful discussions and collaboration we had. \par

I also thank the following people for one or more of: discussion, collaboration, reading final drafts
of my papers, providing encouragement. Prof.~Martin~Bojowald, Prof.~Jack~Carr, Dr.~R.~Goswami,  Prof.~P.~Joshi, Sravan Kumar, Dr. Filipe Mena, Prof.~Guillermo Mena Marug\'an, Dr. Parampreet Singh, Dr. Jose Velhinho, Dr. John Ward and  A. Hadi Ziaie. \par

I would like to thank my dear friends Ahad Khaleghi and Hamid Reza Shojaie, for all nice time we had together and for all discussions. 
In particular, I wish to express my sincere gratitude to Dr. Ghader Najarbashi, for his
encouragement and cooperation in carrying out the project work during my bachelor studies, and introducing me to scientific research; thank you for the `Quaternionic Days'.
Of course I can not miss to thank my friends, 
for the incredibly nice time we spent together during these years in Covilh\~a. I am also sincerely thankful to Dr. Mohammad Rostami and his wife  for their hospitality during these years in Covilh\~a.
I will never forget you. 
I would like to thank Dr. Rumen Moraliyski in specific for the support and help with \LaTeX ~ program during the writing of this thesis. \par

}

\clearpage 


\pagestyle{fancy} 

\lhead{\emph{Contents}} 
\tableofcontents 

\lhead{\emph{List of Figures}} 
\listoffigures 

\lhead{\emph{List of Tables}} 
\listoftables 


\clearpage 

\setstretch{1.5} 

\lhead{\emph{Abbreviations}} 
\listofsymbols{ll} 
{
\textbf{$\Lambda$CDM} & $\Lambda$ cold dark matter \\
\textbf{AdS} & anti-de Sitter \\
\textbf{BAO} & baryon acoustic oscillations \\
\textbf{CCC} & cosmic censorship conjecture \\
\textbf{CMB} & cosmic microwave background  \\ 
\textbf{DEC} & dominant energy condition \\
\textbf{DGP} & Dvali-Gabadadze-Porrati \\
\textbf{FLRW} & Friedmann-Lema\^itre-Robertson-Walker \\
\textbf{GB} & Gauss-Bonnet \\
\textbf{GRVE} & generalized running vacuum energy \\
\textbf{HRDE} & holographic Ricci dark energy \\
\textbf{IR} & infra-red \\
\textbf{LQC} & loop quantum cosmology \\
\textbf{LQG} & loop quantum gravity \\
\textbf{LSS} & Large Scale Structure  \\
\textbf{NEC} & null energy condition  \\
\textbf{QFT} & quantum field theory  \\
\textbf{SEC} & strong energy condition  \\
\textbf{SN-Ia} & the type Ia supernovea \\
\textbf{UV} & ultra violet  \\
\textbf{WDW} & Weeller-DeWitt \\
\textbf{WEC} & weak energy condition  \\
\textbf{WMAP} & Wilkinson Microwave Anisotropic Probe \\
}



%


\clearpage 


%
%


\setstretch{1.3} 

\pagestyle{empty} 

\dedicatory{To my loving Parents} 

\addtocontents{toc}{\vspace{2em}} 

\clearpage

\mainmatter 

\pagestyle{fancy} 



\chapter{Introduction} 
\label{Intro} 
\setstretch{1.3}
\lhead{Chapter 1. \emph{Introduction}} 


The theory of general relativity is one of the cornerstones of classical physics and has a central role in our
understanding of many areas of astrophysics and cosmology. However, it is expected to be an incomplete theory due to the fact that solutions
to the Einstein's equations can contain singularities \cite{Hawking1974,Wald1984,Canarutto:1988}. 
These are regions where the curvature of space-time, which encodes the strength of the gravitational field, diverges and becomes infinite, so that, the classical framework of the theory breaks down and
has to be replaced by a more fundamental theory of gravity.
In other words, the solutions of Einsten's field equations
predict  models of the physical world, in which an observer reaches, in a finite
interval of proper time, to the `edge of the universe'. Moreover, this edge
cannot be included in an extension of the space-time, because the
observer experiences unbounded curvature; the best-known examples are the singularities present in the
Schwarzschild and FLRW solutions.
Important theorems by Penrose \cite{Penrose1965,Penrose1969}, Hawking \cite{Hawking1967} and Geroch \cite{Geroch1968} indicate that singularities are quite a general feature  of Einstein's field equations and not just a consequence of symmetry assumptions (see also Ref. \cite{Hawking-Penrose1996}). Thus, relativistic singularities seem to be generic and  physically very important. 

From the singularity theorems indicated above, singularities are classified as past-spacelike (including the big bang), timelike (naked singularities), and future-spacelike (including black holes or the final recollapse, the big rip \cite{Caldwell:2003vq, O1,O2},  the sudden singularities \cite{Barrow:2004xh,Nojiri:2004ip}, and the big freeze \cite{BigFreeze5}).
There are three well known scenarios in which these singularities may appear. One is the initial singularity at the creation of the universe, namely, the `big bang' singularity.
The second  are  singularities that could appear at the late time evolution of the universe (such as the  big rip, sudden singularities and the big freeze).
The third  can be brought in by the final state of gravitational collapse (the singularity that is found, e.g., inside the event horizon of every black hole or instead a naked singularity).
Sometimes it is said that the big bang singularity represents the beginning of time and space, however, it seems more reasonable that the singularity indicates that the theory of general relativity cannot be trusted when the space-time curvature becomes very large and that quantum effects cannot be ignored: It is necessary to use a theory of quantum gravity in order to study the very early universe 
and the final state of the  collapsing stars as well as at the very late time where the universe might hit a doomsday.

In this chapter, we will present our motivation and work plan for this dissertation. Before we proceed,
it is worthy  to have a general idea of singularities in general relativity from a geometrical point of view.
This provides an introduction to some of the most well known definitions of  space-time singularities, from which singularity theorems have been extracted.

\section{Singularities in general relativity}

In order to define a space-time singularity, 
one must find a
suitable criterion of completeness analogous to that of metric completeness (in
other terms; how can one decide whether a given space-time has `cuts' or
`holes') and then examine the problem of extendibility.
Therefore, a space-time is said to be singular if a geodesic, i.e. a word-line of a freely falling test object, exists which is not complete and not extendible \cite{Hawking1974}.

There are typical assumptions and properties, so called the \emph{global techniques}, in order to prove the singularity theorems under certain assumptions.
These theorems are now generally accepted as proving that singularities are indeed generic and not artifacts of symmetry. We will give a very brief summary of these global techniques,  involving concrete assumptions:

\noindent $\bullet$~ A stress tensor $T_{\mu\nu}$ which satisfies an energy condition. Some restriction on the stress tensor is usually essential, because every space-time is a solution of Einstein's equations 
\begin{equation}
{\cal R}_{\mu\nu}-\frac{1}{2} g_{\mu\nu} {\cal R} =8\pi G ~T_{\mu\nu}[g],
\label{EinsteinEqua}
\end{equation}
with some stress tensor $T_{\mu\nu}$. Moreover, 
$G$ is the gravitational constant, ${\cal R}_{\mu\nu}$ is the Ricci tensor with ${\cal R}={\cal R}_{\mu}^\mu$ being  the scalar curvature, and $g_{\mu\nu}$ is the space-time metric tensor.

\noindent $\bullet$~ Some assumptions, such as the existence of a ``trapped surface", which specify the type of physical situation being discussed. 
These assumptions are also essential, as there are many non-singular exact solutions of the Einstein's equations, such as those which describe static stars.

\noindent $\bullet$~ An assumption concerning global behavior, such as a universe which is globally hyperbolic.

Let us introduce in the following,  the energy conditions on the stress tensor $T_{\mu\nu}$.

\vspace*{3mm}

\noindent \textbf{Definition.} The ``strong energy condition" (SEC)   implies that  for all timelike vectors $u^\mu$:
\begin{equation}
\left(T^{\mu\nu}-\frac{1}{2}g^{\mu\nu}T\right)u_\mu u_\nu \geq 0,
\end{equation}
where  $T=T^{\mu}_\mu$. In the frame in which $T^{\mu\nu}$ is diagonal, this condition implies that the local energy density $\rho$ plus the sum of the local pressures $p^i$ is
non-negative: $\rho+\Sigma_j p^j\geq 0$, and that $\rho+p^i\geq 0$ for each $p^i$. This condition certainly holds for ordinary forms of matter, although it can be violated by a classical massive scalar field like the inflaton field.

\vspace*{3mm}

\noindent \textbf{Definition.} The ``weak energy condition" (WEC) implies that  for all timelike vectors $u^\mu$:
 \begin{equation}
T^{\mu\nu} u_\mu u_\nu \geq 0.
\end{equation}
This condition requires that the local energy density must be non-negative in every observer's rest frame. Again this seems to be a very reasonable condition from the viewpoint of classical physics.

\vspace*{3mm}

\noindent \textbf{Definition.} The ``NEC" implies that  for all null vectors $n^\mu$:
 \begin{equation}
T^{\mu\nu} n_\mu n_\nu \geq 0.
\end{equation}
This condition is implicit in the WEC. That is, if we assume that the WEC holds, then the NEC follows by continuity as $u^\mu$ approaches a null vector.

\vspace*{3mm}

\noindent \textbf{Definition.} The ``dominant energy condition" (DEC) implies that,  in addition to the WEC holding true, for every future-pointing causal vector field (either timelike or null) $v^\mu$, 
the vector field  $-T^\nu_\mu v^\mu$ must be a future-pointing causal vector. That is, mass-energy can never be observed to be flowing faster than light.

The first singularity theorem was Penrose's theorem \cite{Penrose1965}, which implies that singularities must occur when a black hole is formed by gravitational collapse. In addition to some technical assumptions, the proof of this theorem relies upon an energy condition and on the assumption of the existence of a trapped surface (see the following definition). Such a surface arises when the gravitational field of the collapsing body becomes so strong that outgoing light rays are pulled back toward the body. The essence of the theorem is that as long as either of these energy conditions is obeyed, once gravitational collapse proceeds to the point of formation of a trapped surface, then the formation of a singularity is inevitable. Let us be more precise on the definition of the trapped surfaces.

\vspace*{3mm}

\noindent \textbf{Definition.} A ``trapped surface" is defined to be a compact, 2-dimensional smooth spacelike submanifold  such that the families of outgoing as well as ingoing future-pointing null normal geodesics are contracting \cite{Hawking1974}. A geodesic family is defined by specifying a transversal vector field on a submanifold which uniquely determines a family of geodesic curves through each point of the submanifold in a direction given by the vector field at that point. For a compact 2-surface, we can use the inward pointing and outward pointing null normals as those vector fields, defining the ingoing and outgoing null geodesic families. 
The ingoing family of normal geodesics is usually contracting in the sense that its cross-section area decreases, but conversely we intuitively expect the outgoing family to be expanding. 
A trapped surface requires even the outgoing family to be contracting and thus occurs only under special circumstances.
The `expansion' of a null geodesic, with the tangent vector field $u^\mu$, is defined as $\theta:=\nabla_\mu u^\mu$.  

A key result which is used to link the energy conditions to the properties of space-time is the Raychaudhuri equation for the expansion $\theta$ along a bundle of timelike or null rays. It takes the form
 \begin{equation}
\frac{d\theta}{d\tau}\  =\   -{\cal R}^{\mu\nu} u_\mu u_\nu  + \ \text{(other terms)},
\end{equation}
where the `other terms' are non-positive for rotation-free geodesics. If the stress tensor satisfies the SEC, then Einstein's equations,
${\cal R}^{\mu\nu} = 8\pi G (T^{\mu\nu}- \frac{1}{2} g^{\mu\nu} T)$,
imply that $\frac{d\theta}{d\tau}\leq 0$. This is the condition that the bundle of rays is being \emph{focused} by the gravitational field, and is consistent with our intuition that gravity is attractive.

In this thesis, we will 
mainly focus on  some specific  ``spherically symmetric" solutions of Einstein's field equations, which describe the space-times of gravitational collapse and dark energy late time cosmologies.

\section{Motivation and work plan}

The main purpose of this dissertation is to investigate  space-time singularities. 
In particular we will  study singularities that may arise at the final state of the gravitational collapse of a particular class of  star and at the late time evolution of the universe.

After a star has exhausted its nuclear fuel, it can no longer remain in equilibrium and must ultimately undergo a gravitational collapse.
It is believed that, if the mass of the collapsing core is heavier than about 5 solar masses the collapse will end as a black hole.
However, there is another alternative possibility allowed by the theory which is a naked singularity forming. 
In chapter~\ref{collapse-class} we will discuss  the formation of black holes and naked singularities, by means of our 
model (cf. Ref.~\cite{Tavakoli2013a}) for gravitational collapse. In particular, we will consider a spherically symmetric gravitational collapse, whose matter content includes a homogeneous  tachyon field.

When general relativity breaks down at small scales (such as at collapse endstate), the suggestion is to replace it at around classical singularities with a more fundamental theory.
It is generally believed that quantization will remove the gravitational singularities.
Criteria for the avoidance of a space-time singularity, in the presence of a quantized model for gravity,
has been studied within several frameworks; string theory \cite{Horowitz1990}, covariant approaches (effective theories \cite{Donoghue1994},
the renormalization group \cite{Kawai1990}, path integrals \cite{Hawking1978}), canonical approaches  (such as quantum
geometrodynamics \cite{Wheeler1984, Kiefer2009,Kiefer:2011}).

Another candidate within the canonical approach to quantum gravity, which provides a fruitful ground to investigate the removal of the space-time singularities, is loop quantum gravity (LQG)  \cite{Ashtekar2004,Rovelli2004,Thiemann2007}. The construction of a full non-perturbative quantum theory of gravity is among the main open problems in theoretical physics. 
In addition,  loop quantum cosmology (LQC) is a framework based on LQG, to study cosmological scenarios whereby  a `symmetry reduction' is applied to the minisuperspace after the quantization procedure  \cite{Bojowald2000,Bojowald2003}. Furthermore,  results  suggest that the classical singularities of gravitational collapse may be avoided in LQG \cite{Modesto2004,Modesto2004b,Modesto2008,Bojowald2005b,Goswami2006}. 
 Our aim in chapter~\ref{QuantumG} will be,  to investigate how  LQG  induced effects can alter the outcome of gravitational collapse, 
namely, some of the classical singularities  discussed previously in chapter~\ref{collapse-class} (cf. see Refs. \cite{Tavakoli2013b,Tavakoli2013d,Tach-holonomy:2013}).

The other possible scenario in which singularities may appear is at the late time evolution of the universe. 
Several astrophysical observations confirm that the universe is undergoing a state of accelerating expansion.
Such observations indicate that the matter content of the universe, leading to the accelerating expansion, must contain an exotic energy which is characterized with negative pressure, and constitutes 
$68\%$ of the total matter content of the universe.
When the universe is expanding, the density of dark matter decreases more quickly than the density of dark energy, and eventually the matter content of the universe becomes dark energy dominant at 
late times\footnote{In an expanding universe, fluids  (whose energy densities are given by $\rho\propto a^{3(1+w)}$) with larger $w$, dilute more quickly than those with smaller $w$. 
For non-relativistic matter (e.g., dust)  $w_{\rm m}=0$, and for ultra-relativistic matter (e.g. radiation)  $w_{\rm m}=1/3$, which give, respectively, the densities $\rho\propto a^{-3}$ and $\rho\propto a^{-4}$.  
On the other hand, the accelerated expansion of the universe, which is characterized by the equation of state of dark energy, implies that $w_{\rm DE}<-1/3$.  Therefore, dark matter and radiation redshift  much faster  than  dark energy.}. 
Therefore, it is expected that the dark energy scenarios play an important role on the implications for the fate of the universe. 
There have been many recent investigations into the theoretical possibility that expanding universes can come to a violent end at a finite cosmic time,  experiencing a singular fate. 
In the context of dark energy cosmology (see for example Ref. \cite{Bamba2012}), the study of  space-time singularities at late times has made a considerable progress in the last years \cite{Bouhmadi-Lopez:2005,Elizalde:2004,Abdalla:2005,Dabrowski:2006,Kamenshchik:2007}. This constitutes our main goal in chapter~\ref{cosmology}:
In this chapter we will study  the status of late time singularities within models of dark energy cosmology such as the Dvali-Gabadadze-Porrati (DGP) brane-world model with a Gauss-Bonnet (GB) term on the bulk and  phantom matter on the brane \cite{Mariam2010a}, the holographic Ricci dark energy (HRDE) scenario \cite{Tavakoli2013e,Tavakoli2013c}, and the generalized running vacuum energy (GRVE) model \cite{Tavakoli2013c}.
We will further investigate whether or not modifications provided by these models of dark energy can alter such singularities, with smoother ones.
\chapter{Gravitational collapse and space-time singularities}
\label{collapse-class}

\lhead{Chapter 2. \emph{Gravitational collapse and space-time singularities}} 

Under a variety of circumstances \cite{Hawking-Penrose1996,Hawking1974},
singularities will inevitably emerge (geodesic incompleteness in space-time) in classical general relativity,
where matter densities and space-time curvatures diverge. Albeit the singularity
theorems \cite{Hawking1974} 
state that  space-time singularities exist
in a generic gravitational collapse, they provide no information on
the nature of singularities: In particular, the problem of whether these regions
are hidden by a space-time event horizon or can actually be observed,
remains unsolved (see also Ref.~\cite{Joshi2007}). 

The cosmic censorship conjecture (CCC), as hypothesized
by Penrose \cite{Penrose1969}, conveys that the singularities appearing
at the collapse final outcome must be hidden within an event horizon
and thus no distant observer could detect them; a black hole forms.
Although the  CCC  plays a crucial role in the physics of black holes,
there is yet no proof of it, due to the lack of adequate tools to
treat the global characteristics of the field equations (see Ref.~\cite{Hawking1974}). 
Nevertheless,
in the past thirty years many solutions to the field equations have been
discovered, which exhibit the occurrence of naked singularities, where
the matter content has included perfect and imperfect fluids \cite{Harada1998,Harada-Maeda2001,Goswami2002,Giambo2004,Villas2000,Giambo2003,Szekeres1993,Barve2000,Coley1984,Lake1982}, 
scalar fields \cite{Giambo2005,Bhattacharya2011,Bhattacharya}, scalar tensor \cite{Ziaie2010}
self-similar models \cite{Ori1987,Ori1990,Foglizzo1993} 
and null
strange quarks \cite{Mena2004,Gosh2003,Harko2000}.
Basically, it is the geometry of trapped
surfaces  that decides the visibility or otherwise of the space-time
singularity  \cite{Hawking1974, Joshi2007}. In case the collapse terminates into a naked singularity,
the trapped surfaces do not emerge early enough, allowing (otherwise
hidden) regions to be visible to the distant observers.

The gravitational collapse of scalar fields is of relevance,
\cite{Gundlach2007,Schunck2003}, 
owing to the fact that they are able to mimic other
kinds of behaviours, depending on the choice of the potentials. 
Scalar field models have been extensively examined for studying CCC \cite{Shapiro1991}
in spherically symmetric models \cite{Giambo2008,Giambo2009}, 
non-spherically symmetric
models \cite{Ganguly2011} 
and also for static cases \cite{ Janis1968,Wyman1981,Xanthop oulos1989,Bergmann1957,Buchdahl1959}. 
Their role in understanding the machinery governing the causal structure
of space-time was obvious since the 90's, when  solutions
exhibiting black holes versus naked singularities were found numerically by Choptuik
\cite{Choptuik1993} and analytically by Christodoulou \cite{Christodoulou1994}.
There are in the literature a few papers discussing gravitational
collapse in the presence of a scalar field joined by a fluid for the
matter content  \cite{Pereira:2008yq,Jankiewicz:2006fh}: In summary,
a black hole forms in these collapsing situations. 

Recently, we have considered the gravitational collapse of a tachyon scalar field regarding whether
a black hole or naked singularity forms, that is to say, in the CCC
context, together with a fluid \cite{Tavakoli2013a}. Tachyon fields arise in the framework
of string theory \cite{Mazumdar2001} 
and have been of recent use in cosmology.
\cite{Song Piao2002,Fairbairn2002}. 
The action for the tachyon field has a non-standard
kinetic term \cite{ASen2005}, 
enabling for several effects whose dynamical
consequences are different from those of a standard scalar field \cite{RLazkoz2004}.
Namely, other (anti-)friction features that can alter the outcome
of a collapsing scenario. 
This constitutes a worthy motivation to investigate: 
On the one hand, the fluid will play the role of conventional
matter from which a collapse can proceed into, whereas, on the other
hand, the tachyon would convey, albeit by means of a simple framework,
some intrinsic features from a string theory setting.

In this chapter, we
 study the gravitational collapse in the presence  of 
a  tachyon field 
together with a barotropic fluid.   We determine the types
of final state that can occur (i.e., whether a black hole or a naked
singularity emerges, in the context of the CCC), which matter component
will be dominant. 
More precisely, in section \ref{collapse-gen}
we give a brief review on the gravitational collapse of a spherically symmetric space-time. 
In section \ref{collapse-tachyon} we investigate, by means of a dynamical system analysis,
the gravitational collapse by employing a tachyon and a barotropic fluid
as the matter content.

\section{Spherically symmetric collapse}
\label{collapse-gen}

An isotropic FLRW metric, in comoving coordinates,
will be considered as the interior space-time for the gravitational
collapse. However, in order to study the whole space-time, we must
match this interior region to a suitable exterior geometry. We explain in more detail these issues in the following.

\subsection{Interior space-time geometry}
\label{collapse-gen11}

The interior space-time of a collapsing cloud
can be parametrized by a flat FLRW line element as \cite{Joshi2007}
\begin{align}
g^{\rm (int)}_{\mu\nu}dx^{\mu}dx^{\nu}=-dt^{2}+a^{2}(t)dr^{2}+R^{2}(t,r)d\Omega^{2},
\label{metric}
\end{align}
 where $t$ is the proper time for a falling observer whose geodesic
trajectories are labeled by the comoving radial coordinate $r$, $R(t,r)=ra(t)$
is the area radius of the collapsing volume and $d\Omega^{2}=d\theta^2+\sin^2\theta d\varphi^2$, being
the standard line element on the unit two-sphere. The Einstein's field
equations for this model reads \cite{Joshi2007}  
\begin{equation}
8\pi G \rho=\frac{F_{,r}}{R^{2}R_{,r}}\ ,\ \ \  \  \  \  \  \ \ \ \ 8\pi Gp=-\frac{\dot{F}}{R^{2}\dot{R}}\ ,\label{einstein-1}
\end{equation}
where $\rho$ and $p$ are the energy density and pressure of the collapsing matter, respectively;  $F(t,r)$ is the mass function, satisfying the relation
\begin{equation}
\frac{F(t,r)}{R}=\dot{R}^{2},\label{einstein2}
\end{equation}
which is the total gravitational mass within the shell labelled by $r$, 
with ``$,r\equiv\partial_{r}"$ and ``$\cdot\equiv\partial_{t}$".
Since we are interested in a continuous collapsing scenario, we take
$\dot{R}=r\dot{a}<0$, implying that the area radius of the collapsing shell,
for a constant value of $r$, decreases monotonically. 

Splitting the
metric (\ref{metric}) for a two-sphere and a two dimensional hypersurface,
normal to the two-sphere, we have
\begin{equation}
g^{\rm (int)}_{\mu\nu}dx^{\mu}dx^{\nu}=h_{ab}dx^{a}dx^{b}+R^2d\Omega^{2}, \label{metric1}
\end{equation}
 with $h_{ab}={\rm diag}\left[-1,a^2(t)\right]$, whereby the Misner-Sharp energy \cite{Misner1964} is defined to be
\begin{align}
E(t,r) := \frac{R(t,r)}{2}\left[1-h^{ab}\partial_{a}R\partial_{b}R\right]=\frac{R\dot{R}^{2}}{2}\ .\label{Misner-Sharp}
\end{align}
To discuss the final state of gravitational collapse it is important
to study the conditions under which the trapped surfaces form. From
the above definition, it is the ratio $2E(t,r)/R(t,r)$ that controls
the formation of trapped surfaces  (note that the mass function
defined above is nothing but twice the Misner-Sharp mass). Introducing
the null coordinates
\begin{align}
d\xi^{+} := -\frac{1}{\sqrt{2}}\left[dt-a(t)dr\right],\  \  \  \  \  \  \   \   \  d\xi^{-} := -\frac{1}{\sqrt{2}}\left[dt+a(t)dr\right],\label{doublenull}
\end{align}
 the metric (\ref{metric1}) can be recast into the double-null form
\begin{align}
g^{\rm (int)}_{\mu\nu}dx^{\mu}dx^{\nu}=-2d\xi^{+}d\xi^{-}+R^2(t,r)d\Omega^{2}.\label{metricdnull}
\end{align}
The radial null geodesics are given by $g^{\rm (int)}_{\mu\nu}dx^{\mu}dx^{\nu}=0$. We then may deduce
that there exists two kinds of null geodesics which correspond to
$\xi^{+}=constant$ and $\xi^{-}=constant$, the expansions of which
are given by \cite{Hayward1996}
\begin{align}
\Theta_{\pm}\  =\  \frac{2}{R}\partial_{\pm}R,\label{expansion}
\end{align}
 where
\begin{align}
\partial_{+} := \frac{\partial}{\partial\xi^{+}}=-\sqrt{2}\left[\partial_{t}-\frac{\partial_{r}}{a(t)}\right],\  \  \  \   \  \  \  \   \  \  \  \partial_{-} := \frac{\partial}{\partial\xi^{-}}=-\sqrt{2}\left[\partial_{t}+\frac{\partial_{r}}{a(t)}\right].\label{p+p_}
\end{align}
The expansion parameter measures whether the bundle of null rays
normal to the sphere is diverging $(\Theta_{\pm}>0)$ or converging
$(\Theta_{\pm}<0)$. The space-time is referred to as trapped, untrapped
and marginally trapped if
\begin{align}
\Theta_{+}\Theta_{-}>0,\ \ \ \ \ \ \Theta_{+}\Theta_{-}<0, \  \  \  \  \  \  \  \Theta_{+}\Theta_{-}=0, 
\label{sp}
\end{align}
respectively,
where we have noted that $h^{ab}\partial_{a}R\partial_{b}R=-(R^{2}\Theta_{+}\Theta_{-})/2$ \cite{Hayward1996}.
The third case characterizes the outermost boundary of the trapped
region, the apparent horizon equation being $\dot{R}^{2}=1$. 

From
the point of view of Eq. (\ref{Misner-Sharp}), we see that at
the boundary of the trapped region, $2E(t,r)=R(t,r)$. Thus,  in the region in which the mass function is less than the area radius i.e., $F(t, r) < R(t, r)$, no trapped surface forms, whereas
in the regions where the mass function satisfies the relation $F(t, r) > R(t, r)$, the trapping of light does occur. 
In other words, in the former there
exists a congruence of future directed null geodesics emerging from
a past singularity and reaching to distant observers, causing the
singularity to be exposed (a naked singularity forms). In the latter,
no such families exist and trapped surfaces will form early enough
before the singularity formation; regions will be hidden behind the
event horizon, which lead to the formation of a black hole \cite{Joshi2007}.

For the interior metric (\ref{metric}), introducing a new parameter  $\Theta(t,r)\equiv\Theta_+\Theta_-$, where  $\Theta_+$ and $\Theta_-$ are given by Eq. (\ref{expansion}),
we can write $\Theta$ as 
\begin{equation}
\Theta\  =\  8\left(\frac{\dot{R}^{2}}{R^{2}}-\frac{1}{R^{2}}\right).\label{theta1-q}
\end{equation}
This is to be thought of as a function of the phase space, for
every fixed shell $r$. 
For the boundary shell, $r=r_{\rm b}$, 
we define 
\begin{align}
\Theta_{\rm b}(t):=\Theta(t,r_{\rm b})=8\left(\frac{\dot{a}^2}{a^2}-\frac{1}{a^{2}r_{\rm b}^{2}}\right).
\label{THETAboundary}
\end{align}
Therefore, the trapped region is determined by the relation $\Theta(t)>0$, with $\Theta(t)=0$ being the equation for apparent horizon.

In order to further illustrate this  gravitational collapse
process, let us employ a very particular class of solutions. We assume
that the behavior of the matter density, for $a\ll1$, is given by
the following ansatz \cite{Goswami2004}:
\begin{equation}
\rho(t)\approx\frac{c}{a^{n}}\  ,\label{energyapprox}
\end{equation}
 where $c$ and $n$ are positive constants. Integration of the first equation in (\ref{einstein-1}) gives
the following relation for the mass function as
\begin{equation}
F(R)=\frac{8\pi G}{3}\rho(t)R^{3}.\label{massfunction}
\end{equation}
Then, by substituting the energy density ansatz (\ref{energyapprox}) in Eq.~(\ref{massfunction}) we have
\begin{equation}
\frac{F}{R}=\frac{8\pi G}{3}cr^{2}a^{2-n}.\label{ratio}
\end{equation}
It is therefore possible to identify the following outcomes:
\begin{itemize}
\item For $0<n<2$, the ratio $F/R<1$ stands less than unity 
throughout the collapse and if no trapped surfaces exist 
initially, due to the regularity condition ($F(t_{0},r)/R(t_{0},r)<1$
for a suitable value of $c$ and $r$), then no trapped surfaces would
form until the epoch $a(t_{s})=0$. More precisely, there exists a
family of radial null trajectories emerging from a naked singularity.
\item For $n\geq2$, the ratio $F/R$ goes to infinity and trapped surfaces
will form as the collapse evolves, which means that singularity will
be covered and no radial geodesics can emerge from it; thus, a black
hole forms.
\end{itemize}
From Eq. (\ref{energyapprox}), we may easily find a relation
between the energy density and pressure as
\begin{equation}
p=\frac{n-3}{3}\rho,\label{pressapprox}
\end{equation}
 which shows that for $n<2$, where a naked singularity occurs, the
pressure gets negative values \cite{Joshi2007,Goswami2004}. From Eqs. (\ref{einstein2}) and 
(\ref{energyapprox}),  bearing in mind the continuous collapse
process $(\dot{a}<0)$, we can solve for the scale factor as
\begin{equation}
a(t)=\left(a_{0}^{\frac{n}{2}}+\frac{n}{2}\sqrt{\frac{c}{3}}(t_{0}-t)\right)^{\frac{2}{n}},\label{scalefactor}
\end{equation}
 where $t_{0}$ is the time at which the energy density begins increasing
as $a^{-n}$ and $a_{0}$ corresponds to $t_{0}$. Assuming that the
collapse process initiates at $t_{0}=0$, the time at which the scale
factor vanishes corresponds to $t_{s}=\frac{2}{n}\sqrt{\frac{3}{c}}a_{0}^{n/2}$,
implying that the collapse ends in a finite proper time.

\subsection{Energy conditions}

In order to have a physically reasonable matter content for the collapsing cloud, the matter are taken to satisfy  the WEC and DEC\footnote{We note that satisfying the WEC  implies that the NEC  is held as well (see chapter \ref{Intro}) \cite{Hawking1974,Joshi2007}.}: For any time-like
vector $u^{\mu}$, the energy-momentum tensor $T_{\mu\nu}$ satisfies $T_{\mu\nu}u^{\mu}u^{\nu}\geq0$, 
and $T^{\mu\nu}u_{\nu}$ must be non-space-like. 

Since for the symmetric
metric (\ref{metric}) we have $T_{t}^{t}=-\rho(t)$ and $T_{r}^{r}=T_{\theta}^{\theta}=T_{\varphi}^{\varphi}=p(t)$,
then, for the energy densities and the pressures of matter the energy conditions amount to the following relations:
\begin{align}
 \rho\geq0,\ \ \ \ \ \  \ \rho+p\geq0,\  \  \  \  \ \ \ |p|\leq\rho.  \label{EC1}
\end{align}
The first two expressions in Eq. (\ref{EC1}) denote the WEC, and the last
expression denotes the DEC.

Using the energy density (\ref{energyapprox}) and pressure (\ref{pressapprox}), it is straightforward to check 
the WEC as follows
\begin{equation}
\rho+p=\frac{nc}{3a^{n}}>0.\label{WEC}
\end{equation}
Therefore, the WEC  is satisfied throughout the collapse process in this specific case of matter density.%

\subsection{Exterior geometry}
\label{collapse-exterior}

In order to complete our discussion of space-time structure, concerning the formation of horizon, the interior space-time of the dynamical collapse should be matched to a suitable exterior geometry. 
Let us consider the interior metric $g_{\mu\nu}^{\rm (int)}$ in Eq. (\ref{metric}). The exterior metric $g_{\mu\nu}^{\rm (ext)}$ is written in advanced Eddington-Finkelstein-like coordinates $({\mathrm{v}}, r_{\mathrm{v}})$ as \cite{Joshi2007}:
\begin{equation}
g_{\mu\nu}^{\rm (ext)}dx^\mu dx^\nu=-{\cal F}({\mathrm{v}},r_{\mathrm{v}})d{\mathrm{v}}^2-2d{\mathrm{v}}dr_{\mathrm{v}}+r_{\mathrm{v}}^2d\Omega^2.
\label{metric2}
\end{equation}
In this coordinates system, the trapping horizon is given simply by the relation ${\cal F}({\mathrm{v}},r_{\mathrm{v}})=0$.
It should be noticed that, in general the formation of the singularity at $a=0$ is independent of matching the interior to the exterior space-time.

Nevertheless, from a realistic astrophysical perspective, the exterior of any collapsing star would not be  completely empty, it would always be surrounded by a radiation zone, as well as possible matter emissions from the star. Moreover, in the case which will be of use in chapter \ref{QuantumG}, where quantum induced effects in the semiclassical interior will be considered, the pressure may not  vanish at the boundary, so that, the interior can not be matched to a Schwarzschild exterior at the boundary surface.

We match the interior to an  exterior region, described by a general class of non-stationary metrics which is the generalized Vaidya space-time  \cite{Vaidya1943,Vaidya1953,Vaidya1957},  
at the boundary hypersurface $\Sigma$ given by $r=r_{\text{b}}$. More precisely, this exterior geometry plays two important roles in a collapsing process: Firstly, it allows the collapsing cloud to be radiated away as collapse evolves, and secondly, it enables us to study the formation and evolution of horizon during the collapse \cite{Joshi2007}. The generalized Vaidya geometry is given by the line element  \cite{Vaidya1957}:
\begin{equation}
g_{\mu\nu}^{\rm (ext)}dx^\mu dx^\nu=-\left(1-\frac{2M({\mathrm{v}},r_{\mathrm{v}})}{r_{\mathrm{v}}}\right)d{\mathrm{v}}^{2} - 2d{\mathrm{v}}dr_{\mathrm{v}}
+r_{\mathrm{v}}^2d\Omega^2,
\label{Vaidya}
\end{equation}
where $M(r_{\mathrm{v}},{\mathrm{v}})$ is the usual Vaidya mass, with ${\mathrm{v}}$ and $r_{\mathrm{v}}$ being the retarded null coordinate and the Vaidya radius, respectively. Comparing Eq. (\ref{Vaidya}) with the metric (\ref{metric2}) we get ${\cal F}({\mathrm{v}},r_{\mathrm{v}})=1-2M({\mathrm{v}},r_{\mathrm{v}})/r_{\mathrm{v}}$, is the so-called boundary function.

For the required matching, the Israel-Darmois \cite{Darmois1927,Israel1966} matching conditions are used, where the first and second fundamental forms are matched \cite{Joshi2007}. That is, the metric coefficients and the extrinsic curvature are matched at the boundary of the cloud; matching the area radius at the boundary results in 
\begin{equation}
R(t, r_{\text{b}})=r_{\text{b}}a(t)\ =\ r_{\mathrm{v}}({\mathrm{v}}).
\label{V1-a}
\end{equation}
Then, on the hypersurface $\Sigma$, the interior and exterior metrics are given by
\begin{equation}
g_{\mu\nu}^{\Sigma (\rm int)}dx^\mu dx^\nu=-dt^2+a^2(t)r_{\mathrm{b}}^2d\Omega^2,
\end{equation}
and
\begin{equation}
g_{\mu\nu}^{\Sigma (\rm ext)}dx^\mu dx^\nu=-\left(1-\frac{2M({\mathrm{v}},r_{\mathrm{v}})}{r_{\mathrm{v}}} +2 \frac{dr_\mathrm{v}}{d\mathrm{v}}\right) d{\mathrm{v}}^{2} +r_{\mathrm{v}}^2d\Omega^2.
\end{equation}
Matching the first and second fundamental forms (i.e., matching the extrinsic curvatures)  for the interior and exterior  metrics on this hypersurface, gives the following equations  \cite{Joshi2007}:
\begin{align}
F(t, r_{\text{b}})\ =\ 2M({\mathrm{v}},r_{\mathrm{v}}),
\label{V2}
\end{align}
\begin{align}
\left(\frac{dv}{dt}\right)_{\Sigma}\ =\ \frac{1+r_{\text{b}}\dot{a}}{1-\frac{F}{R}}\ ,
\label{V3-a}
\end{align}
\begin{align}
M({\mathrm{v}},r_{\mathrm{v}})_{,r_{\mathrm{v}}}\ =\ \frac{F}{2R}+r_{\text{b}}^2a\ddot{a}.
\label{V4-a}
\end{align}
Since the matter content in the exterior is not specified in our model, thus the Vaidya mass is not determined herein. However, any generalized Vaidya mass $M({\mathrm{v}}, r_{\mathrm{v}})$ satisfying Eq. (\ref{V4-a}) gives a unique exterior space-time with required equations of motion
which is provided by  Eqs. (\ref{V2}), (\ref{V3-a}), and (\ref{V1-a}).
Furthermore, from this matching analysis, we can get the information regarding the behavior of trapped surfaces close to the matter shells. For the particular case in Eq. (\ref{V2}) where the condition $2M({\mathrm{v}}, r_{\mathrm{v}})=r_{\mathrm{v}}$ is satisfied, trapped surfaces will form in the exterior region close to the boundary surface \cite{Joshi2007}.

In the case of solutions presented in \ref{collapse-gen11},
 we can see that at the singular epoch $t=t_s$,
\begin{equation}
\lim_{r_{\mathrm{v}}\rightarrow0}\frac{2M(r_{\mathrm{v}}, \mathrm{v})}{r_{\mathrm{v}}}\rightarrow0.
\end{equation}
Thus, the exterior metric (\ref{Vaidya})
smoothly transform to
\begin{equation}
g^{\rm (ext)}_{\mu\nu}dx^\mu dx^\nu=-d{\mathrm{v}}^{2} - 2d{\mathrm{v}}dr_{\mathrm{v}} +r_{\mathrm{v}}^2d\Omega^2.
\label{Vaidyacc}
\end{equation}
The above metric describes a Minkowski space-time in retarded null coordinate. Hence we see that the exterior generalized Vaidya metric, together with the singular point at $(t_s, 0)$, can be smoothly extended to the Minkowski space-time as the collapse completes.

\section{Gravitational collapse with tachyon field and barotropic fluid}
\label{collapse-tachyon}

The model we consider in this section to study the gravitational collapse,  has an interior flat
FLRW geometry (\ref{metric}). 
We will consider henceforth  a  \emph{homogeneous}
tachyon field\footnote{In subsection \ref{pert}, we clarify our assumption of homogeneity, by including linear inhomogeneous tachyon perturbations, in order to determine the stability and validity of the solutions presented for a homogeneous case in this section.}, $\Phi(t)$, together with a barotropic fluid as the matter involved in
the physical process of gravitational collapse. 
The main ingredient is the \emph{non-standard}
kinetic term for the tachyon scalar field \cite{Tavakoli2013a}.

\subsection{Tachyon matter}

To investigate the  tachyon field's dynamics, we
make use of the line element (\ref{metric}), employing A. Sen's
effective action, whose Hamiltonian  can  be written as\footnote{In this section, we will work in the unit $\kappa=8\pi G=1$.}
\cite{ASen2002a,ASen2002b}
\begin{equation}
C_{\Phi}\ =\  a^3\sqrt{V^2(\Phi)+a^{-6}\pi_\Phi^2}\ . \label{Lagrangianfield}
\end{equation}
where $\pi_\Phi$ 
is the conjugate momentum for the tachyon field\footnote{The conjugate momentum for the tachyon field can be obtained as 
\begin{equation}
\pi_{\Phi} = \frac{\partial L_\Phi}{\partial \dot{\Phi}} = \frac{a^3V(\Phi)\dot{\Phi}}{\sqrt{1-\dot{\Phi}^2}}\ ,
\end{equation}
where $L_\Phi=-a^3V(\Phi)\sqrt{1-\dot{\Phi}^2}$  is the Lagrangian of the tachyon field.}, and $V(\Phi)$ is the tachyon potential.
The classical energy density and pressure of a scalar field $\Phi$ can be written as \cite{Hossain2005}
\begin{equation}
\rho_{\Phi} =   a^{-3}C_{\Phi},\  \  \  \  \  \  p_{\Phi} = - a^{-3}\left(\frac{a}{3}
\frac{\partial {C_{\Phi}}}{\partial{a}}\right).
\label{Henergy-sf-a-Tach}
\end{equation}
Therefore,  the corresponding energy density and pressure of tachyon matter  are obtained as
\begin{equation}
\rho_{\Phi}=\frac{V(\Phi)}{\sqrt{1-\dot{\Phi}^{2}}}\ , \label{energytach}
\end{equation}
\begin{equation}
p_{\Phi}=-V(\Phi)\sqrt{1-\dot{\Phi}^{2}}\ , \label{momentum2} 
\end{equation}
i.e., satisfying
\begin{equation}
p_{\Phi} = (\dot{\Phi}^2 - 1) \rho_{\Phi}\ . \label{nonbara}
\end{equation}
Using the equation for the
energy conservation, $\dot{\rho}_{\Phi}+3H(\rho_{\Phi}+p_{\Phi})=0$, we get the equation
of motion for the tachyon field $\Phi$ as
\begin{equation}
\ddot{\Phi}=-(1-\dot{\Phi}^2)\left[3H\dot{\Phi}+\frac{V_{,\Phi}}{V}\right],
\label{field}
\end{equation}
where ``$,\Phi$" denotes the derivative  with respect to the tachyon
field. 

A definitive form for the tachyon potential has not
yet been  established \cite{Feinstein:2002,Guo:2003,Ambramo:2003,Copeland:2005,Quiros:2010}, but
there are few proposals satisfying the (qualitative) requirements
indicated by open string  theory. Being more concrete, $V(\Phi)$
should have a maximum at $\Phi \sim 0$ and it should go to zero as
$\Phi\rightarrow\infty$. 
It has been argued \cite{ASen2002a,ASen2002b,ASen:1998a,Ghate1999,Garousi:2000,Bergshoeff:2000,Feinstein:2002} that $V(\Phi)$
would have an exponential behavior, $V \sim e^{-\Phi}$, or even a
$\cosh^{-1} \Phi $ form. An alternative is  to use an inverse
power law, i.e.,
\begin{equation}
V \  \sim\   \Phi^{-{\mathcal{P}}}, \label{INV-SQ-POT}
\end{equation}
with  $\mathcal{P}$   being a positive
constant (allowing for scaling solutions), from which the
exponential potential can be suitably built
\cite{Feinstein:2002,Guo:2003,Ambramo:2003,Copeland:2005,Quiros:2010}. Simple exact analytical
solutions were found in Ref.~\cite{Feinstein:2002}; nevertheless, it can be
argued  that with this potential there is a divergence at $\Phi \sim
0$, although,  it can be mentioned  that it has the asymptotic
behavior for  a required tachyon potential. A thorough discussion on
other potentials and contexts can be found in Refs.~\cite{Copeland:2005} and
\cite{Gorini:2004}. We will henceforth focus
only on an  inverse power law potential for the tachyon, 
where  $\mathcal{P} =2$, adopting a purely phenomenological
perspective, likewise recent papers
\cite{Feinstein:2002,Guo:2003,Ambramo:2003,Copeland:2005,Quiros:2010,RLazkoz2004}. 

\subsection{Linear perturbations in tachyon matter collapse}
\label{pert}

In this section, we check the dynamics  of tachyon fluctuations around the background
solution for collapsing scenario.

A general inhomogeneous tachyon field $\varphi(t,\textbf{r})$, satisfies the field
equation \cite{ASen2002a,ASen2002b,Frolov,Ghate1999}:
\begin{align}
\nabla_{\mu}\nabla^{\mu}\varphi-
\dfrac{\nabla_{\mu}\nabla_{\nu}\varphi}{1+\nabla_{\alpha}
\varphi\nabla^{\alpha}\varphi}\nabla^{\mu}\varphi\nabla^{\mu}\varphi-\dfrac{V_{,\varphi}}{V}=0.
\label{general-Field}
\end{align}
For a homogeneous tachyon
field $\varphi(t,\textbf{r})=\Phi(t)$, this equation  reduces to
Eq.~(\ref{field}). To study the  stability of our  marginal bound
case collapsing system 
with respect to
small spatially inhomogeneous fluctuations, we take  the background
scalar field as homogeneous,  i.e., $\varphi(t, \textbf{r})=\Phi(t)$ where  the  energy density and pressure depend
only on time, being given by Eqs.~(\ref{energytach}) and
(\ref{momentum2}), respectively; then,  small inhomogeneous perturbations of
the tachyon field $\varphi(t, \textbf{r})$ around the time-dependent
background solution, $\Phi(t)$, are expressed by
\begin{equation}
\varphi(t, \textbf{r})=\Phi(t)+\delta\varphi(t, \textbf{r}).
\label{PertTach}
\end{equation}
Using the perturbed inhomogeneous tachyon field (\ref{PertTach}),
and substituting in the
general equation of motion (\ref{general-Field}), ignoring
 the  coupling of tachyon
fluctuations with the metric perturbations, we
 can extract
\begin{equation}
\frac{\ddot{\varphi}_k}{1-\dot{\Phi}^2}+\frac{2\dot{\Phi}\ddot{\Phi}}{(1-\dot{\Phi}^2)^2}\dot{\varphi}_k+\left[k^2+(\log V)_{,\Phi\Phi}\right]\varphi_k=0~,
\label{PertFieldEq}
\end{equation}
where we have used,  for the
small fluctuations $\delta\varphi$, a  Fourier decomposition,
\begin{equation}
\delta\varphi(t, \textbf{r})= \int d^3k~\varphi_k(t)e^{i\textbf{k}.\textbf{r}},
\label{Fourier}
\end{equation}
to linearize the field equation (without Hubble friction term).

Let us now include small
scalar metric perturbations, which,  around our  background (\ref{metric}) can be written in terms of a
Newtonian gravitational potential, $\chi=\chi(t,\textbf{r})$, as
\begin{align}
ds^2_{\rm pert}  & \  =\ 
\left[g_{\mu\nu}^{\rm (int)}+\delta g_{\mu\nu}^{\rm (int)}\right]dx^\mu dx^\nu \notag \\
&\  =\  -(1+2\chi)dt^2+(1-2\chi)a(t)^2[dr^2+r^2d\Omega^2].
\label{metricPert}
\end{align}
From Eq.~(\ref{metricPert}) the linearized Einstein's equations
and the field equation with respect to small fluctuations
$\delta\varphi$ and $\chi$ lead then to two equations for the time-dependent Fourier
amplitudes $\varphi_k(t)$ and $\chi_k(t)$ as \cite{Frolov, Garriga1999, Armendariz-Picon1999}
\begin{equation}
\frac{\partial}{\partial t}\left(\frac{\varphi_k}{\dot{\Phi}}\right)=\left(1-\frac{k^2}{2a^2}\frac{(1-\dot{\Phi}^2)^{3/2}}{V\dot{\Phi}^2}\right)\chi_k~,
\label{Pert-E1}
\end{equation}
\begin{equation}
\frac{2}{a}\frac{\partial}{\partial t}(a\chi_k)=\frac{V\dot{\Phi}^2}{(1-\dot{\Phi}^2)^{1/2}}\frac{\varphi_k}{\dot{\Phi}}~,
\label{Pert-E2}
\end{equation}
where $\chi_k$ is also given through the Fourier decomposition $\chi(t,\textbf{r})=\int d^3k~\chi_k(t)e^{i\textbf{k}.\textbf{r}}$.

By introducing Mukhanov's variable $v_k$ \cite{Garriga1999, Armendariz-Picon1999}:
\begin{equation}
\frac{v_k}{z}=\frac{5\rho+3p}{3(\rho+p)}\chi_k+\frac{2\rho}{3(\rho+p)}\frac{\dot{\chi}_k}{H}~,
\label{v-z}
\end{equation}
where
\begin{equation}
z=\frac{\sqrt{3}a\dot{\Phi}}{(1-\dot{\Phi}^2)^{1/2}}~,
\end{equation}
we  can rewrite Eqs.~(\ref{Pert-E1}) and (\ref{Pert-E2})
 in terms of a second order equation for $v_k$ as
\begin{equation}
v''_k+\left[(1-\dot{\Phi}^2)k^2-\frac{z''}{z}\right]v_k=0,
\label{v-Eq}
\end{equation}
where here a prime denotes a derivative with respect to the conformal time $d\eta=dt/a(t)$.

In the following,  we will
study the solutions near the background homogeneous solutions
where the
tachyon field is pressureless (cf. see subsection \ref{Classic1} for the homogeneous tachyon dominated solution). More precisely,  we will consider
linear inhomogeneous perturbations around those background
solutions, with the aim to investigate how inhomogeneity can affect
the stability and therefore the validity of the background solution
(and thus the approximation) in the collapse process. Moreover, we
will consider for $\varphi$ the  tachyon potential as
\begin{equation}
V(\varphi)\  =\  V_0\varphi^{-2}~,
\label{pot-inhom}
\end{equation}
where $V_0$ is a constant.

\subsubsection*{Pressureless tachyon matter collapse}

Let us  now consider a tachyon matter field fluctuation
(\ref{PertTach}), ignoring
the space-time perturbation. It can be seen from (\ref{Pert-E1})
that, in the absence of metric
perturbations (i.e. when $\chi=0$), the tachyon field is
fully homogeneous, that is $\delta\varphi=\delta\varphi(t)$.
For this case, the collapsing cloud enters very quickly into the
regime where the tachyon is rolling very fast and $\dot{\Phi}^2$
approaches unity. Considering $\Phi(t)=t+\vartheta(t)+\Phi_0$, where
$\delta\varphi=|\vartheta|\ll (t+\Phi_0)$ with $\dot{\vartheta}<0$
and substituting in equation of field (\ref{field}), we may study
the behavior of trajectories near the fixed point solutions. For the
first order of $\vartheta$, the field equation reduces to
\cite{Frolov}
\begin{align}
\frac{\ddot{\vartheta}}{\dot{\vartheta}}\ =\ 6\frac{\dot{a}}{a}+2\frac{V'}{V}|_{\Phi=t+\Phi_0}~.
\label{apr-theta}
\end{align}
For the inverse square potential (\ref{pot-inhom}), Eq.~(\ref{apr-theta}) can be
integrated to
\begin{align}
\dot{\vartheta}\ =\ -\frac{1}{2}\left(\frac{a}{a_0}\right)^6\frac{V^2}{V_0^2}|_{\Phi=t+\Phi_0},\ \ \   \       \   \   \   \   \   \  \   \ \ \dot{\Phi}^2\ =\ 1+2\dot{\vartheta}~.
\end{align}
For this case, the background tachyon field can be obtained as
\begin{align}
\Phi(t)\ =\ t+\frac{1}{4}\left(\frac{a}{a_0}\right)^6\frac{\Phi_1}{(t+\Phi_0)^4}+\Phi_0~,
\label{phi-pert}
\end{align}
where $\Phi_1$ is a constant of integration. Then, from the Eq.~(\ref{phi-pert}) one can get the
energy density and pressure of the collapsing system as
\begin{align}
\rho_{\Phi}(t)\ =\  V_0\left(\frac{a}{a_0}\right)^{-3}, \        \   \   \   \   \   \   \   \  p_{\Phi}(t)\ =\ \left(\frac{a}{a_0}\right)^{3}\frac{V^2}{V_0}|_{\Phi=t+\Phi_0},
\label{rho-pert}
\end{align}
From Eq.~(\ref{PertFieldEq}) for the inverse square potential,
neglecting the space-time perturbations, it follows that the tachyon
fluctuations are not growing with time, but become a constant as
$\varphi_k=const$.

\subsubsection*{Tachyon fluctuations in the presence of space-time perturbations}

Now, let us consider
tachyon fluctuations  coupled to the potential $\chi$. For a
pressureless tachyon matter, the coefficient of $k^2$ in Eq.~(\ref{v-Eq}) vanishes quickly as the collapse evolves, that is, the
growth of linear tachyon fluctuations is scale free. Then, the
solution of (\ref{v-Eq}) is $v_k=z$, and the left hand side of the
Eq.~(\ref{v-z}) is constant. This means that $\chi_{k}$ must be
constant as $\chi_k=C_k$, and more importantly,  that in our case
(where classically the effective pressure $p_\Phi\rightarrow0$) it is a
\emph{positive} constant given by $C_k=3/5$; the gravitational
potential $\chi(t,\textbf{r})$ has just spatial dependence and is
given by
\begin{align}
\chi(\textbf{r})=\int d^3k~C_ke^{i\textbf{k}.\textbf{r}}.
\label{pot-gr-pert}
\end{align}
In this limit, the second term in the  right hand side of Eq.~(\ref{Pert-E1}) is given by
$(1-\dot{\Phi}^2)^{3/2}/(a^2\dot{\Phi}^2)\approx a^7$,  vanishing  as
$a\rightarrow0$. From this, the time derivative of the tachyon field
can be obtained as $\dot{\varphi}_k=\pm C_k$, where `$+$' and `$-$'
signs correspond  to the background solutions for time derivative
of homogeneous field $\dot{\Phi}=+1$ and $\dot{\Phi}=-1$ (cf. see subsection \ref{Classic1} for the homogeneous tachyon dominated solution),
respectively. On the other hand, Eq.~(\ref{Pert-E2}) for
constant $\chi_k=C_k$ reduces to
\begin{equation}
\frac{2}{a}\frac{\partial}{\partial t}(a\chi_k)=2\frac{\dot{a}}{a}\chi_k=\frac{V\dot{\Phi}^2}{(1-\dot{\Phi}^2)^{1/2}}\frac{\varphi_k}{\dot{\Phi}}<0 \  .
\label{Pert-E2-1}
\end{equation}
Since $\dot{a}<0$  for the continuous  collapse herein, Eq.~(\ref{Pert-E2-1})  implies
that $\varphi<0$ for $\dot{\Phi}>0$ and $\varphi>0$ for
$\dot{\Phi}<0$. Thus, time evolution of  $\varphi_k$ are obtained
as
\begin{align}
\varphi_k(t)\ &=\ C_kt-\varphi_0 <0,  \   \   \  \   \  \   \  \  \  \   (\dot{\Phi}>0), \notag \\  \varphi_k(t)\ &=\ -C_kt+\varphi_0>0,   \   \   \  \  \  \  \   (\dot{\Phi}<0),
\label{c-k}
\end{align}
where $\varphi_0>0$ is a constant of integration. Let us discuss in
the following the collapsing scenario for the case
$\dot{\Phi}\rightarrow+1$ of the background solutions (cf.  subsection \ref{Classic1}) for which, the
amplitude of Fourier transformation $\varphi_k<0$, i.e. the
first relation in Eq.~(\ref{c-k}). In this case, the full tachyon field
in the presence of the fluctuation $\delta\varphi(t,\textbf{r})$ in a collapsing space-time
is given by
\begin{equation}
\varphi(t,\textbf{r})\ =\ t+\frac{1}{4}\left(\frac{a}{a_0}\right)^6\frac{\Phi_1}{(t+\Phi_0)^4}+\delta\varphi+\Phi_0,
\label{pert-tach}
\end{equation}
where $\delta\varphi$ is obtained by substituting
$\varphi_k=C_kt-\varphi_0$ from Eq.~(\ref{c-k});  from Eq.~(\ref{Fourier}) for the
fluctuations  it follows that
\begin{align}
\delta\varphi(t,\textbf{r})\ =\ \int d^3k\ \left(\frac{3}{5}t-\varphi_0\right) e^{i\textbf{k}.\textbf{r}},
\label{pert-tach2}
\end{align}
in which, the constant $\varphi_0>0$ determines the initial amplitude of the fluctuation, that is
\begin{align}
\delta\varphi(t=0, \textbf{r})\ =\ -\varphi_0\int d^3k\  e^{i\textbf{k}.\textbf{r}}.
\label{init-pert}
\end{align}

To study the dynamics of the  inhomogeneous perturbations, we need to study the behavior of the
full spatially gravitational potential $\chi(\textbf{r})$. In this
case small fluctuations in tachyon field and the metric are related
by $\varphi_{k}=C_{k}t-\varphi_{0}$; since $\varphi_{k}$ is always
negative, then as time evolves, it increases  negatively and approaches to zero.  Hence, the
linear perturbation (\ref{Fourier}),  given by
\begin{equation}
\delta\varphi(t,\textbf{r})=\int d^{3}k\:\varphi_{k}(t)e^{i\textbf{k}.\textbf{r}}<\int d^{3}k\:\left|\varphi_{k}(t)\right|,
\end{equation}
decreases when $|\varphi_{k}|$ is decreasing towards the singularity. More interestingly,
the larger is the constant $\chi_{k}=C_k$ (related to the metric
perturbation),  the faster
the tachyon fluctuations vanish. This may indicate that a higher
perturbations in the metric strongly damp the tachyon field
fluctuations during the collapse. To be more precise, by assuming a
finite and small fluctuation,  for
$C_k=3/5$ in Eq.~(\ref{pot-gr-pert}), the gravitational potential $\chi(\textbf{r})$ is
constant in time and thus, can be roughly estimated as
\begin{align}
\chi(\textbf{r})&\approx\frac{3}{5}\int_{k_0}^{k_0+\Delta k}d^3k\ e^{i\textbf{k}.\textbf{r}}=\frac{3}{5}\Upsilon(k_x,k_y,k_z;x,y,z),
\end{align}
where $\Upsilon(k_x,k_y,k_z;x,y,z)$ is defined as
\begin{align}
\Upsilon := \left(\frac{2\sin(\Delta k_x x)}{\Delta k_x}\right)\left(\frac{2\sin(\Delta k_y y)}{\Delta k_y}\right)\left(\frac{2\sin(\Delta k_z z)}{\Delta k_z}\right).
\label{Upsilon}
\end{align}
Using this approximation in equation of fluctuation (\ref{pert-tach2}) it follows that
\begin{align}
\delta\varphi(t,\textbf{r})\ =\ \left(\frac{3}{5}t-\varphi_0\right)\Upsilon(k_x,k_y,k_z;x,y,z),
\end{align}
where the initial amplitude of the fluctuation (\ref{init-pert}) can be approximated as
\begin{align}
\delta\varphi(t=0, \textbf{r})\ =\ -\varphi_0\Upsilon(k_x,k_y,k_z;x,y,z).
\end{align}
This analysis reveals that the tachyon fluctuation $|\delta\varphi(t,\textbf{r})|$ has a maximum value at $t=0$, and then, as the collapse evolves, it starts to  decrease due to both the tachyon amplitude fluctuations and the metric perturbations;
this clearly implies that the tachyon field inhomogeneity decays and after a finite time $t_p=\frac{\varphi_0}{C_k}$ it vanishes\footnote{It is worthy to compare our model herein to a FLRW  expanding universe with tachyon matter field \cite{Garriga1999, Armendariz-Picon1999}. Therein, unlike our collapsing model (where $\delta\varphi=(t-t_p)\chi$),  the tachyon fluctuation is given by $\delta\varphi=\chi(\textbf{r})t$  in which the gravitational potential $\chi$ increases during the expansion of the universe, and thus,  tachyon field inhomogeneity will be increasing. Furthermore, the
instability of the tachyon field has a linear dependence on time.}.
Since the inhomogeneity is not increasing in the system, the linear approximation remains valid during the dynamical evolution of collapse. In some sense, the above discussion clarifies what can be the  contribution  of an inhomogeneous tachyon matter.

In order to find the speed for decaying of inhomogeneity, let us
assume that the initial amplitude is of the order $\varphi_0\sim
10^{-5}$ (cf. Ref.~\cite{Frolov}). Then, the time necessary for fluctuations
$\varphi\sim 3/5t-10^{-5}$ to vanish can be estimated as $t\sim
10^{-5}$.

The energy density of the tachyon matter in the presence of inhomogeneity in a perturbed collapsing space-time reads
\begin{equation}
\rho_\varphi \ =\ \frac{V_0\varphi(t,\textbf{r})^{-2}}{\sqrt{1-(1-2\chi)\dot{\varphi}^2+a^{-2}(\nabla\varphi)^2}}.
\label{energy-pert}
\end{equation}
The gradient term in Eq.~(\ref{energy-pert}) can be approximated as
\begin{align}
(\nabla\chi)^2\ \approx\ (C_kt-\varphi_0)^2\left(\nabla\Upsilon\right)^2,
\end{align}
which rapidly vanishes as the collapse evolves, and the term $(1-2\chi)\dot{\varphi}^2$ tends to
$\dot{\Phi}^2$. Consequently near the center where $a\rightarrow0$,
the energy density will diverge and in fact
reproduce the results obtained by the background homogeneous
solutions.

\subsection{Tachyon field coupled with barotropic fluid: Phase space analysis}

\label{Classic1}

Let us consider now a tachyon field $\Phi$ with the potential $V(\Phi)$, and a barotropic fluid with the energy density $\rho_{b}$ for the collapsing matter content.
The total energy density of the collapsing system is therefore $\rho=\rho_{\Phi}+\rho_{b}$,
with 
\begin{equation}
\rho=3H^{2}=\frac{V(\Phi)}{\sqrt{1-\dot{\Phi}^{2}}}+\rho_{b}.  \label{energy}
\end{equation}
The barotropic matter with the density  $\rho_{b}$
whose pressure is given by $p_{b}$, in terms of the barotropic parameter,
$\gamma_b$, satisfies the relation $p_{b}=(\gamma_b-1)\rho_{b}$, where
the barotropic parameter is taken to be positive, $\gamma_b>0$. Then, the Raychadhuri
equation for the collapsing system can be written as 
\begin{equation}
-2\dot{H}=\frac{V(\Phi)\dot{\Phi}^{2}}{\sqrt{1-\dot{\Phi}^{2}}}+\gamma_b\rho_{b}.\label{Raychad1baro}
\end{equation}
The energy conservation of the barotropic matter is 
\begin{equation}
\dot{\rho}_{b}+3\gamma_b H\rho_{b}=0.\label{consenrbaro}
\end{equation}
Integrating (\ref{consenrbaro}), the energy density for barotropic
matter can be presented as $\rho_{b}=\rho_{0b}a^{-3\gamma_b}$,
where $\rho_{0b}$ is a constant of integration. The total
pressure is given by 
\begin{equation}
p=p_{\Phi}+(\gamma_b-1)\rho_{0b}a^{-3\gamma_b}.\label{press-a}
\end{equation}

Concerning the energy conditions, for a tachyon  field it is straightforward
to show that the WEC is satisfied: 
\begin{align}
\rho_{\Phi}=\frac{V(\Phi)}{\sqrt{1-\dot{\Phi}^{2}}}\geq0,\ \ \ \ \text{and}\ \ \ \ \rho_{\Phi}+p_{\Phi}=\frac{V\dot{\Phi}^{2}}{\sqrt{1-\dot{\Phi}^{2}}}\geq0.\label{EC1a}
\end{align}
In addition, since $\dot{\Phi}^{2}<1$, then $V(\Phi)(1-\dot{\Phi}^{2})^{-1/2}\geq V(\Phi)\sqrt{1-\dot{\Phi}^{2}}$ from
which it follows that $|p_{\Phi}|\leq\rho_{\Phi}$, and hence, the DEC
is held for the tachyon matter.

For the barotropic fluid considered herein, we assume
a regular initial condition for the collapsing system, such as $\rho_{0b}>0$, indicating that $\rho_{b}=\rho_{0b}a^{-3\gamma_b}>0$.
In addition, since $p_{b}=(\gamma_b-1)\rho_{b}$, then $p_{b}+\rho_{b}=\gamma_b\rho_{b}\geq0$
provided that $\gamma_b>0$. We will consider only the
positive barotropic parameter $\gamma_b$, hence, the WEC  is satisfied.
Furthermore, considering the DEC  ($\rho_{b}\geq|p_{b}|$) it follows
that the barotropic parameter $\gamma_b$ must satisfy the range $\gamma_b\leq2$.

\begin{figure*}[t]
\begin{minipage}[c]{1\textwidth}%
\begin{center}
\includegraphics[width=3in]{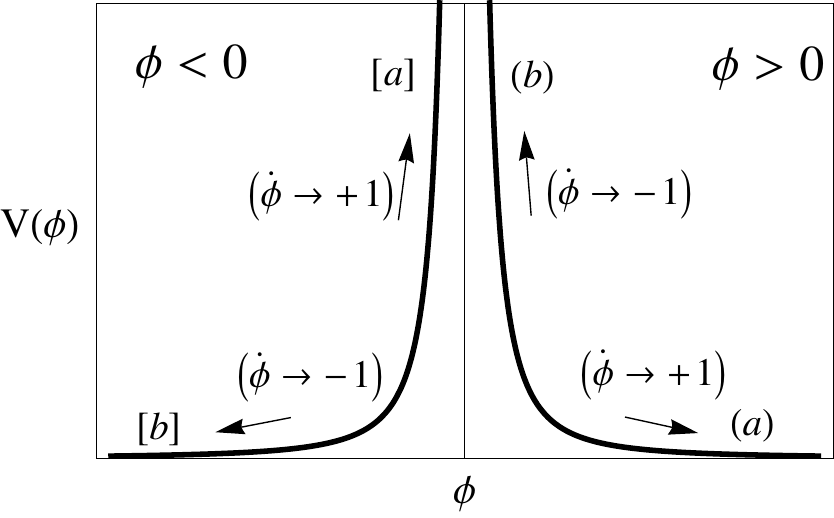} 
\par\end{center}%
\end{minipage}\caption{The potential $V=V_{0}\Phi^{-2}$, denoting the $\Phi<0$
and $\Phi>0$ branches as well as the asymptotic stages.}

\label{pot-v2} 
\end{figure*}

We analyze  the gravitational collapse of the
above setup by employing a dynamical system perspective \cite{Khalil1996}. 
We use an inverse square potential for the tachyon field given by 
\begin{equation}
V(\Phi)\  =\  V_{0}\Phi^{-2},\label{potential}
\end{equation}
where $V_{0}$ is a constant. Note that we can consider the potential
(\ref{potential}) as two (mirror) branches (see figure \ref{pot-v2}),
upon the symmetry $\Phi\rightarrow-\Phi$, treating the $\Phi>0$
and $\Phi<0$ separately.

In order to employ the dynamical system approach, it is convenient to introduce
a new time variable $\tau$ (instead of the proper time $t$ present
in the comoving coordinate  $\{t,r,\theta,\varphi\}$), defined
as \cite{Tavakoli2013a}
\begin{equation}
\tau \  :=\  -\log\left(\frac{a}{a_{0}}\right)^{3},\label{n-timeclass}
\end{equation}
where $0<\tau<\infty$ (the limit $\tau\rightarrow0$ corresponds
to an initial condition of the collapsing system ($a\rightarrow a_{0}$)
and the limit $\tau\rightarrow\infty$ corresponds to $a\rightarrow0$).
For any time dependent function $f=f(t)$, we can therefore write 
\begin{equation}
\frac{df}{d\tau}\  =\  -\frac{\dot{f}}{3H}\ . \label{dyn1}
\end{equation}
We further introduce a new set of dynamical variables $x$, $y$ and
$s$ as follows \cite{RLazkoz2004} 
\begin{equation}
x:= \dot{\Phi},\qquad y:= \frac{V}{3H^{2}},\qquad s:=\frac{\rho_{b}}{3H^{2}}~.\label{DyClXY}
\end{equation}

From the new variables (\ref{dyn1}), (\ref{DyClXY}) and the system
(\ref{Raychad1baro})-(\ref{consenrbaro}) with Eq.~(\ref{potential}),
an autonomous system of equations is retrieved: 
\begin{align}
\frac{dx}{d\tau} & =:  f_{1}=\left(x-\frac{\lambda}{\sqrt{3}}\sqrt{y}\right)(1-x^{2}),\label{DyClXbaro}\\
\frac{dy}{d\tau} & =: f_{2}=-y\left[x\left(x-\frac{\lambda}{\sqrt{3}}\sqrt{y}\right)+s(\gamma_b-x^{2})\right],\label{DyClYbaro}\\
\frac{ds}{d\tau} & =:  f_{3}=s(1-s)(\gamma_b-x^{2}).\label{DyClsbaro}
\end{align}
Note that $\lambda$ is given by 
\begin{equation}
\lambda \ :=\  -\frac{V_{,\Phi}}{V^{3/2}}~,\label{lambdanew}
\end{equation}
which for the potential (\ref{potential}) brings $\lambda=\pm2/\sqrt{V_{0}}$
as a constant; for the $\Phi>0$ branch, $\lambda>0$, and for $\Phi<0$
branch, $\lambda<0$. Equation (\ref{energy}) in terms of the new
variables reads, 
\begin{equation}
\frac{y}{\sqrt{1-x^{2}}}+s=1.\label{DyClFriedbaro}
\end{equation}
The dynamical variables have the range $y>0,\:  0\leq s\leq1$, and $-1<x<1$.
This brings an effective two-dimensional phase space%
\footnote{In the absence of a barotropic fluid, the effective phase space is
one-dimensional.%
}.

Setting $(f_{1},f_{2},f_{3})|_{(x_{c},y_{c},s_{c})}=0$, we can obtain
the critical points $(x_{c},y_{c},s_{c})$ for the autonomous system.
The stability can be subsequently discussed by using the eigenvalues
of the Jacobi matrix $\mathcal{A}$, defined at each fixed point $(x_{c},y_{c},s_{c})$,
as 
\begin{equation}
\mathcal{A}=\left(\begin{array}{ccc}
\frac{\partial f_{1}}{\partial x} & \frac{\partial f_{1}}{\partial y} & \frac{\partial f_{1}}{\partial s}\\
\frac{\partial f_{2}}{\partial x} & \frac{\partial f_{2}}{\partial y} & \frac{\partial f_{2}}{\partial s}\\
\frac{\partial f_{3}}{\partial x} & \frac{\partial f_{3}}{\partial y} & \frac{\partial f_{3}}{\partial s}
\end{array}\right)_{|(x_{c},y_{c},s_{c})}.\label{matrix2D}
\end{equation}
Solutions, in terms of the dynamical variables, in the neighborhood
of a critical point $q_{i}^{c}$ can be extracted by making
use of 
\begin{align}
q_{i}(t)\ =\ q_{i}^{c}+\delta q_{i}(t), \label{pertQ1-A}
\end{align}
with the perturbation $\delta q_{i}$ given by 
\begin{align}
\delta q_{i}\ =\ \sum_{j}^{k}(q_{0})_{i}^{j}\exp(\zeta_{j}\tau),\label{pertQ1-cl}
\end{align}
where $q_{i}\equiv\{x,y,s\}$, $\zeta_{j}$ are the eigenvalues of
the Jacobi matrix, and the $(q_{0})_{i}^{j}$ are constants of integration.
We have summarized the fixed points for the autonomous system and
their stability properties in table \ref{T1}.

\begin{table}[h!]

\begin{center}

\caption{Summary of critical points and their properties for
the $\Phi<0$ (denoted with square brackets, i.e., as, for example,
$\left[a\right]$) and $\Phi>0$ branches (denoted with regular brackets,
i.e., as, for example, $\left(a\right)$).}

\begin{tabular}{cccccc}
\hline  \hline 
Point  & $x$  &  $y$  & $s$  & \qquad{}Existence\qquad{}  & Stability \tabularnewline
\hline 
$(a),\ [a]$  & $1$  & $0$  & $0$  & All $\lambda$; $\gamma_b<1$  & Stable node\tabularnewline
 &  &  &  & All $\lambda$; $\gamma_b>1$  & Saddle point\tabularnewline
$(b),\ [b]$  & $-1$  & $0$  & $0$  &  All $\lambda$; $\gamma_b<1$  & Stable node\tabularnewline
 &  &  &  & All $\lambda$; $\gamma_b>1$  & Saddle point\tabularnewline
$(c),\ [c]$  & $\frac{\lambda}{\sqrt{3}}\sqrt{y_{0}}$  & $y_{0}$  & $0$  & All $\lambda$; $\gamma_b>\gamma_{1}$  & Unstable node\tabularnewline
 &  &  &  & All $\lambda$; $\gamma_b\leq\gamma_{1}$  & Saddle point\tabularnewline
$(d),\ [d]$  & $0$  & $0$  & $1$  &  All $\lambda$; $\gamma_b\neq0$  & Saddle point \tabularnewline
 &  &  &  &  All $\lambda$; $\gamma_b=0$  & Unstable node\tabularnewline
$(e)$  & $-\sqrt{\gamma_b}$  & $\frac{3\gamma_b}{\lambda^{2}}$  & $s_{0}$  &  For $\lambda>0$; $\gamma_b<\gamma_{1}<1$  & Unstable node\tabularnewline
$[e]$  & $\sqrt{\gamma_b}$  & $\frac{3\gamma_b}{\lambda^{2}}$  & $s_{0}$  & For $\lambda<0$;  $\gamma_b<\gamma_{1}<1$  & Unstable node\tabularnewline
$\left(f\right),\ [f]$  & $1$  & $0$  & $1$  &  All $\lambda$; $\gamma_b>1$  & Stable node\tabularnewline
 &  &  &  &  All $\lambda$; $\gamma_b<1$  & Saddle point\tabularnewline
$\left(g\right),\ [g]$  & $-1$  & $0$  & $1$  &  All $\lambda$; $\gamma_b>1$  & Stable node\tabularnewline
 &  &  &  &  All $\lambda$; $\gamma_b<1$  & Saddle point\tabularnewline
\hline \hline  
\end{tabular}

\label{T1} 
\end{center}
\end{table}

\par

Table~\ref{T1} is complemented with figure \ref{3D-stream}. The
autonomous system equations (\ref{DyClXbaro})-(\ref{DyClsbaro})
unfolds a three dimensional phase space constrained by Eq.~(\ref{DyClFriedbaro}).
Therefore the main fixed points (and the trajectories nearby) are
located on surfaces of this three dimensional phase space. Furthermore,
from table \ref{T1}, we can focus our attention on three situations
characterized by having $s=0$, $s=1$ and $s=s_{0}$. Briefly, in
more detail: 
\begin{itemize}
\item Point $(a)$: The eigenvalues are $\zeta_{1}=-2$, $\zeta_{2}=-1$
and $\zeta_{3}=(\gamma_b-1)$. For $\gamma_b<1$, all the characteristic
values are real and negative, then the trajectories in the neighborhood
of this point are attracted towards it. Hence, $(a)$ is a stable
node (attractor). For $\gamma_b>1$, all characteristic values
are real, but one is positive and two are negative, the trajectories
approach this point on a surface and diverge along a curve: this is
a \emph{saddle} point. 
\begin{figure*}[h]
\begin{centering}
\includegraphics[width=2.5in]{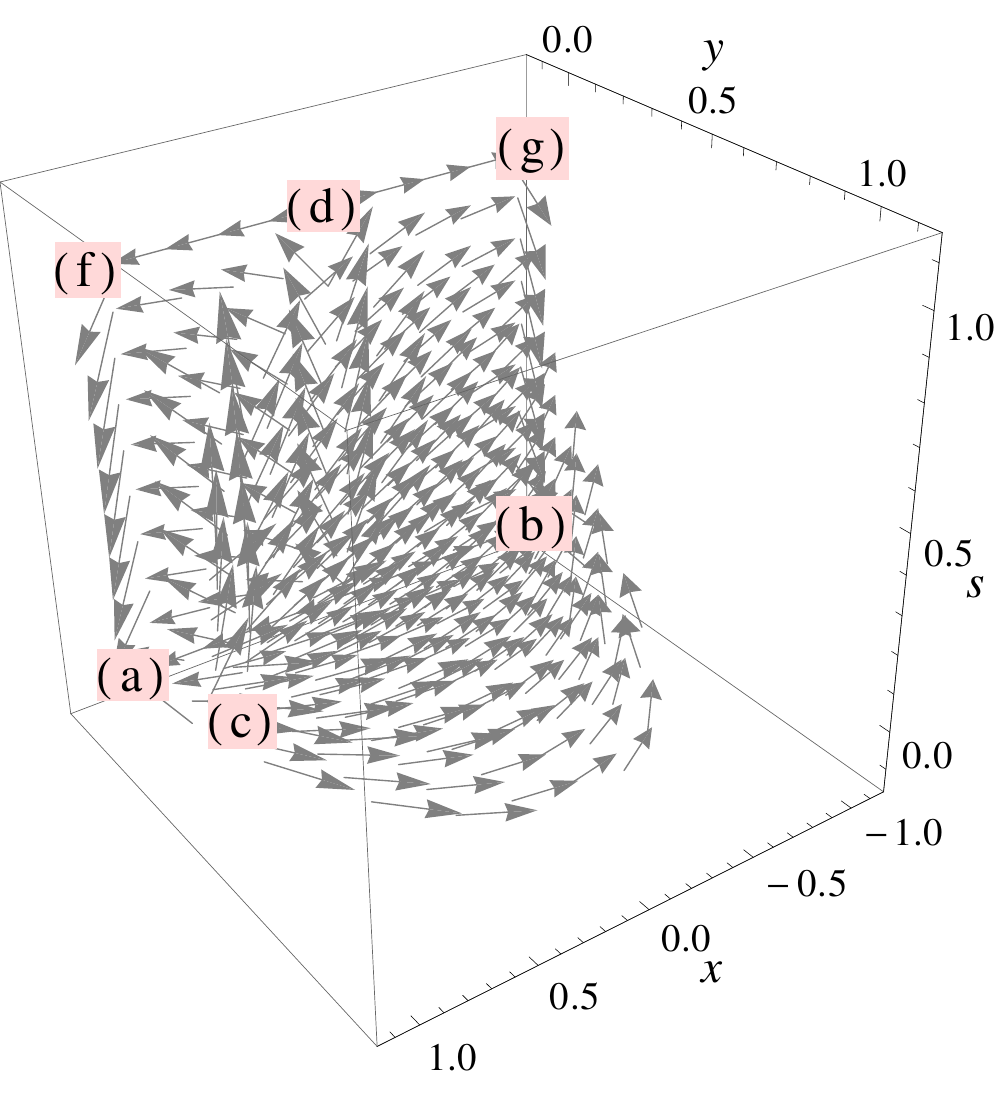}\quad{}\includegraphics[width=2.5in]{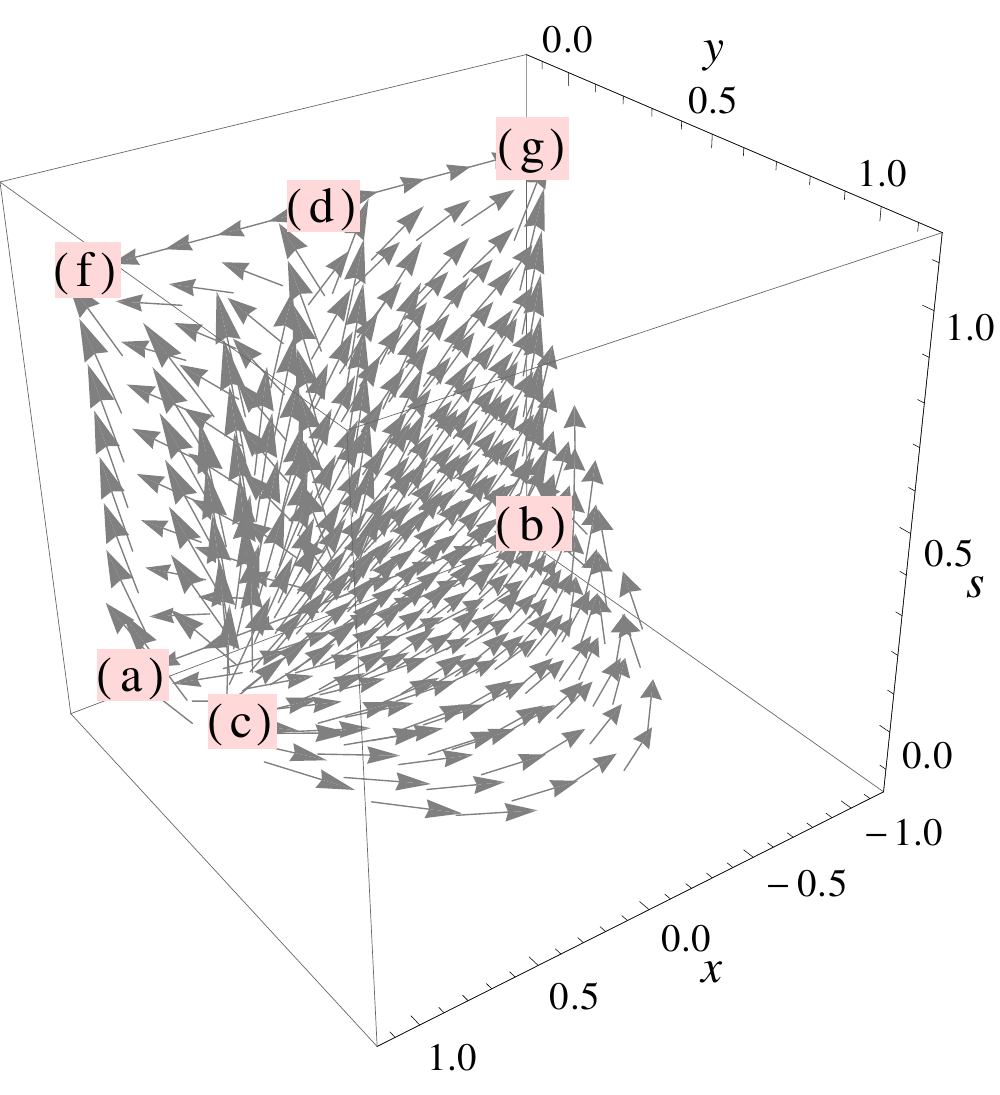} 
\par\end{centering}

\caption{Trajectories in phase space and critical points: (i)
Left plot represents the phase space region $\left(x,y,s\right)$
constrained by Eq.~(\ref{DyClFriedbaro}). Therein we also depicted all
the fixed points (see table \ref{T1}) except point $\left(e\right)$,
for $\gamma_b<1$. (ii) In the right plot we considered
the conditions as $\gamma_b>1$. We can illustrate that
going from $\gamma_b<1$ to $\gamma_b>1$ (from left to the right plot)
reverses the direction of the trajectories, i.e., in the left plot
the vector field is directed towards points $\left(a\right)$ or $\left(b\right)$.
In the right plot the vector field is directed towards $\left(f\right)$
or $\left(g\right)$.}

\label{3D-stream} 
\end{figure*}

\item Point $(b)$: The eigenvalues are $\zeta_{1}=-2$, $\zeta_{2}=-1$
and $\zeta_{3}=(\gamma_b-1)$. This point has the same eigenvalues of
point $(a)$ and similar asymptotic behavior, being also a stable
node for $\gamma_b<1$ (see the left plot in figure \ref{2D-stream})
and a saddle point for $\gamma_b>1$. 
\item Point $(c)$: This fixed point has eigenvalues $\zeta_{1}=0$,
$\zeta_{2}=(\lambda^{2}y_{0}+6y_{0}^{2})/6>0$ and $\zeta_{3}=(\gamma_b-\lambda^{2}y_{0}/3)$,
which all are real and $y_{0}=-\lambda^{2}/6+\sqrt{1+(\lambda^{2}/6)^{2}}$.
For $\gamma_b>\gamma_{1}\equiv\lambda^{2}y_{0}/3$, two components are
positive. However, from a numerical investigation we can assert that
this corresponds to an unstable saddle (see left plot in figure \ref{2D-stream}).
On the other hand, for $\gamma_b<\gamma_{1}$, one component is negative
and other is positive, and hence, a saddle point configuration would
emerge. 
\item Point $(d)$: The eigenvalues are $\zeta_{1}=1$, $\zeta_{2}=-\gamma_b$
and $\zeta_{3}=-\gamma_b$. As $\gamma_b>0$, trajectories approach this
point on a surface (the in-set) and diverge along a curve (the out-set).
This is a saddle\emph{ }point (see left plot in figure \ref{3D-stream}). 
\item Point $(e)$: This point is located at $(-\sqrt{\gamma_b},\frac{3\gamma_b}{\lambda^{2}},s_{0})$,
where $s_{0} := (1-\frac{3\gamma_b}{\lambda^{2}\sqrt{1-\gamma_b}})$. From
the constraint $0\leq s_{0}\leq1$ we get 
\begin{equation}
0\leq1-\frac{3\gamma_b}{\lambda^{2}\sqrt{1-\gamma_b}}\leq1,\label{const-track}
\end{equation}
 i.e., 
\begin{equation}
0\leq\gamma_b\leq\gamma_{1}<1\ .\label{const-track4}
\end{equation}
 When $\gamma_b\rightarrow0$, points $\left(e\right)$ and $\left(d\right)$
become coincident (see table \ref{T1}). When we consider $s_{0}=0$,
we obtain $\gamma_b=\gamma_{1}$, $x=\frac{\lambda}{\sqrt{3}}\sqrt{y_{0}}$
and $y=y_{0}$. Consequently, points $\left(e\right)$ and $\left(c\right)$
become coincident. Therefore, point $\left(e\right)$ can be found
along a curve that joins points $\left(c\right)$ (on the surface
with $s=0$) and $\left(d\right)$ (on the surface with $s=1$). The
eigenvalues are $\zeta_{1}=0$, with $\zeta_{2}$ and $\zeta_{3}$
given by, 
\begin{align}
\begin{aligned}\zeta_{2,3}\  =\  \frac{1}{4}\left(2-\gamma_b\pm\sqrt{\left(1-\gamma_b\right)\left(4-16s_{0}\gamma_b\right)+\gamma_b^{2}}\right)\end{aligned}
.\label{eigencle}
\end{align}
 The eigenvalues (\ref{eigencle}) are non-negative for $\gamma_b<\gamma_{1}$,
and for $\gamma_b=\gamma_{1}$, $\zeta_{2}>0$ and $\zeta_{3}=0$.\medskip{}

\begin{figure*}[h]
\begin{minipage}[c]{1\textwidth}%
\begin{center}
\includegraphics[width=1.8in]{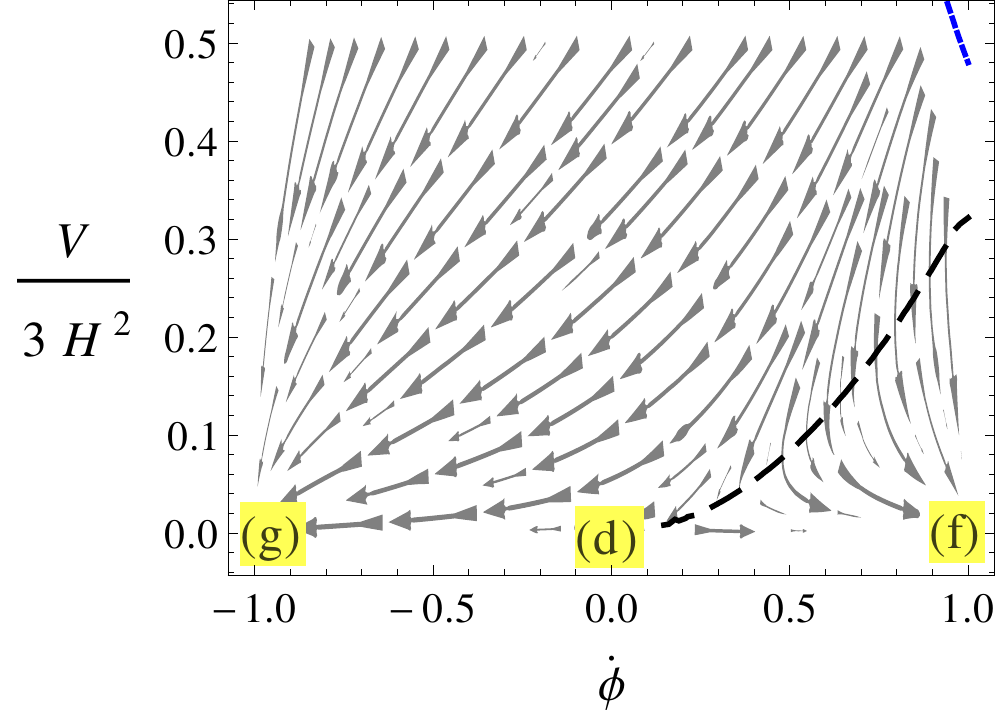}\quad{}\includegraphics[width=1.8in]{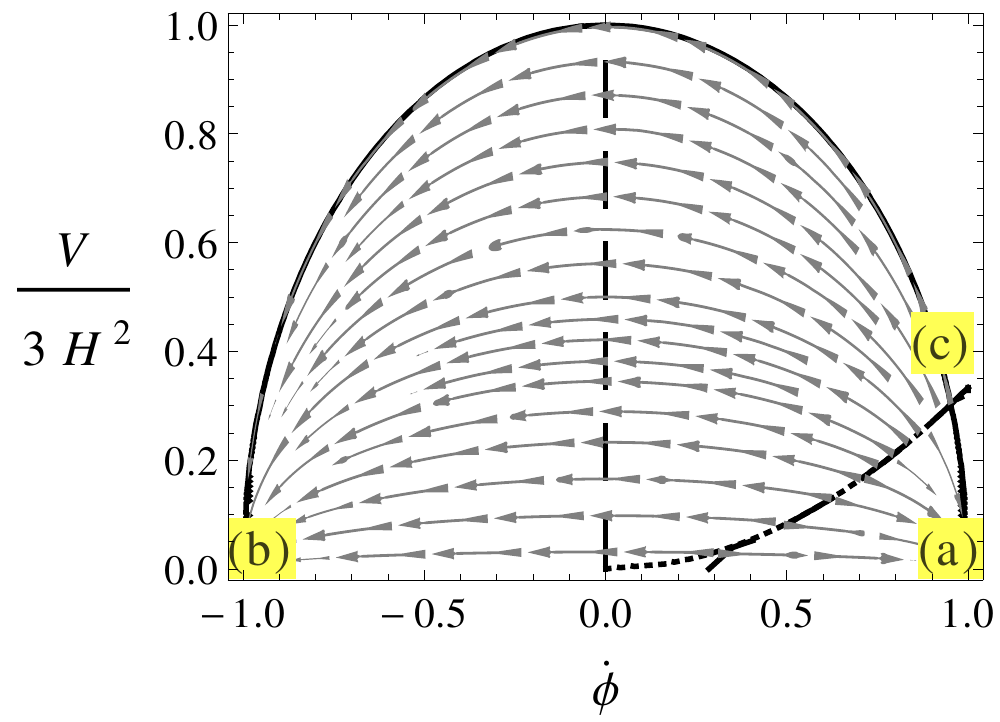}
\includegraphics[width=1.8in]{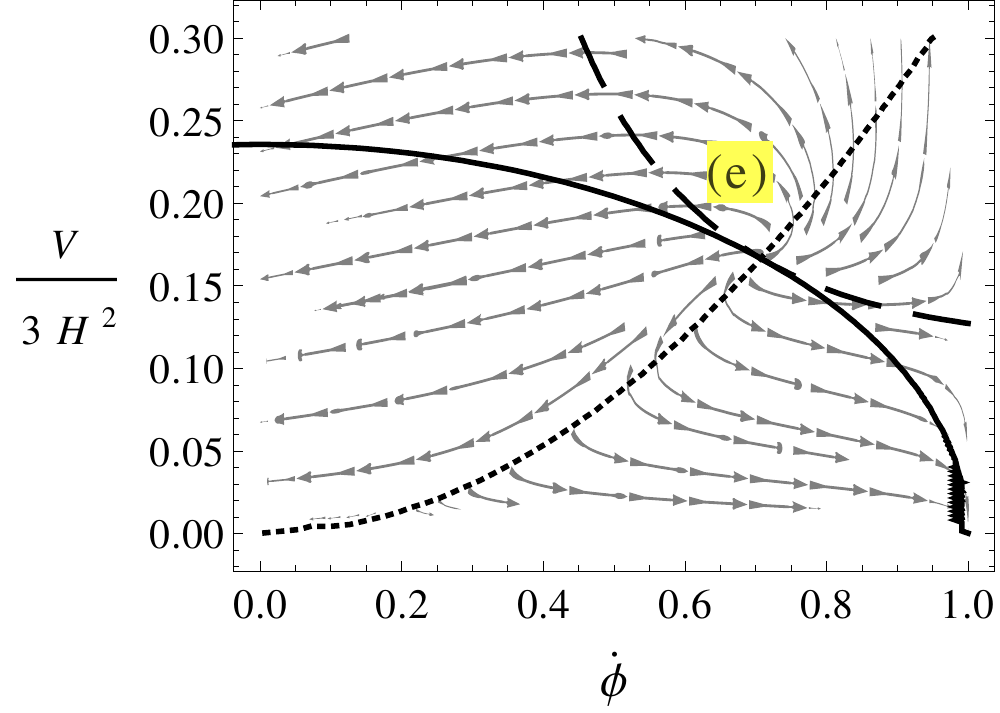} 
\par\end{center}%
\end{minipage}\caption{(i) The top left plot represents a section, of the
three dimensional phase space presented in figure \ref{3D-stream},
labeled with $s=0$. The dashed and dotted lines are the zeros of
Eqs.~(\ref{DyClXbaro})-(\ref{DyClsbaro}). The solid line represents
constraint (\ref{DyClFriedbaro}). Therefore, the fixed points are
found on their intersections ($\gamma_b<1$). In this example we can locate the position of fixed
points $\left(a\right)$, $\left(b\right)$ and $\left(c\right)$
according to table \ref{T1}. (ii) The middle plot represents a section
showing the fixed point $\left(e\right)$, labeled with $s=0.76$. The
dashed (and also dotted) lines are the zeros of Eqs.~(\ref{DyClXbaro})-(\ref{DyClsbaro})
and the solid line represents constraint (\ref{DyClFriedbaro}), as
before ($\gamma_b<1$).
(iii) Finally, in the top right plot we represent a section, labeled
with $s=1$. We can identify fixed points $\left(g\right)$, $\left(d\right)$
and $\left(f\right)$. We considered the conditions as
$\gamma_b>1$.}

\label{2D-stream} 
\end{figure*}

\medskip{}
In the middle plot  of figure \ref{2D-stream} we provide a
section of the phase space showing point $\left(e\right)$, where
it can be seen that all trajectories are divergent from it. This behavior
is characteristic of an unstable node and expected whenever point
$\left(e\right)$ is found along the curve that connects $\left(c\right)$
and $\left(d\right)$.

\item Point $(f)$: The eigenvalues are $\zeta_{1}=-2$, $\zeta_{2}=-\gamma_b$
and $\zeta_{3}=(1-\gamma_b)$. For $\gamma_b>1$, all the characteristic
values are real and negative, then the trajectories in the neighborhood
of this point are attracted towards it. Hence, $(f)$ is a stable
node (attractor). Finally, for $\gamma_b<1$, all characteristic values
are real, but one is positive and two are negative. The trajectories
approach this point on a surface and diverge along a curve; this is
a saddle point. 
\item Point $(g)$: The eigenvalues are $\zeta_{1}=-2$, $\zeta_{2}=-\gamma_b$
and $\zeta_{3}=(1-\gamma_b)$. This point has the same eigenvalues of
point $(f)$ and similar asymptotic behavior, being also a stable
node for $\gamma_b>1$ (see the right plot in figure \ref{2D-stream})
and a saddle point for $\gamma_b<1$. 
\item Point $(h)$: This point is found along the line segment
with $x=1$, $y=0$ and $s\in\left]0,1\right[$. The eigenvalues are
$\zeta_{1}=-2$, $\zeta_{2}=-1$ and $\zeta_{3}=2(s_{1}-1)$, where
$s_{1}\in\left]0,1\right[$. In this case, all the characteristic
values are real and negative, then the trajectories in the neighborhood
of this point  are attracted towards it. Hence, $(h)$ is a stable
node (attractor).
\item Point $(i)$: This point is found along the line segment
with $x=-1$, $y=0$ and $s\in\left]0,1\right[$. The eigenvalues
are $\zeta_{1}=-2$, $\zeta_{2}=-1$ and $\zeta_{3}=2(s_{1}-1)$.
This point has the same eigenvalues of point $(h)$ and similar asymptotic
behavior, being also a stable node. 
\end{itemize}
Before proceeding, let us add two comments. On the one hand, note
the transition that occurs, when going from $\gamma_b<1$ to $\gamma_b>1$,
an intermediate state as discussed (by means of a numerical study)
in subsection \ref{Tracking-solutions}. This bifurcation behaviour
is made explicit through the numerical methods employed. These allowed
us to confirm the results on the dynamical system analysis and to
further assess in regions like the transition from tachyon dominance
to fluid dominance, verifying all possible scenarios with those two
tools. On the other hand, figure \ref{3D-stream} deserves some attention
when we consider the trajectories approaching the stable nodes $(a)-(b)$
(left plot) and $(f)-(g)$ (right plot). Therein, the trajectories
approach $x\rightarrow\pm1$ along a segment line containing the stable
nodes. This situation implies that the energy density (\ref{energy})
diverges after reaching a point where $x\rightarrow\pm1$, $y\rightarrow0$
and $s\rightarrow s_{1}$. This observation will be important in the
next sections concerning the possible outcomes of the gravitational
collapse.


\subsection{Analytical and numerical solutions}
\label{Classic1-0}

Let us herewith discuss this subsection possible outcomes regarding the
collapsing system, employing elements from both our analytical as
well as numerical studies.

\subsection*{A.~Tachyon dominated solutions}
\label{Tachyon dominated}

From the trajectories in the vicinity of $(a)$ and $(b)$, attractor
solutions can be described.

\begin{figure*}[h]
\begin{minipage}[c]{1\textwidth}%
\centering \includegraphics[width=1.8in]{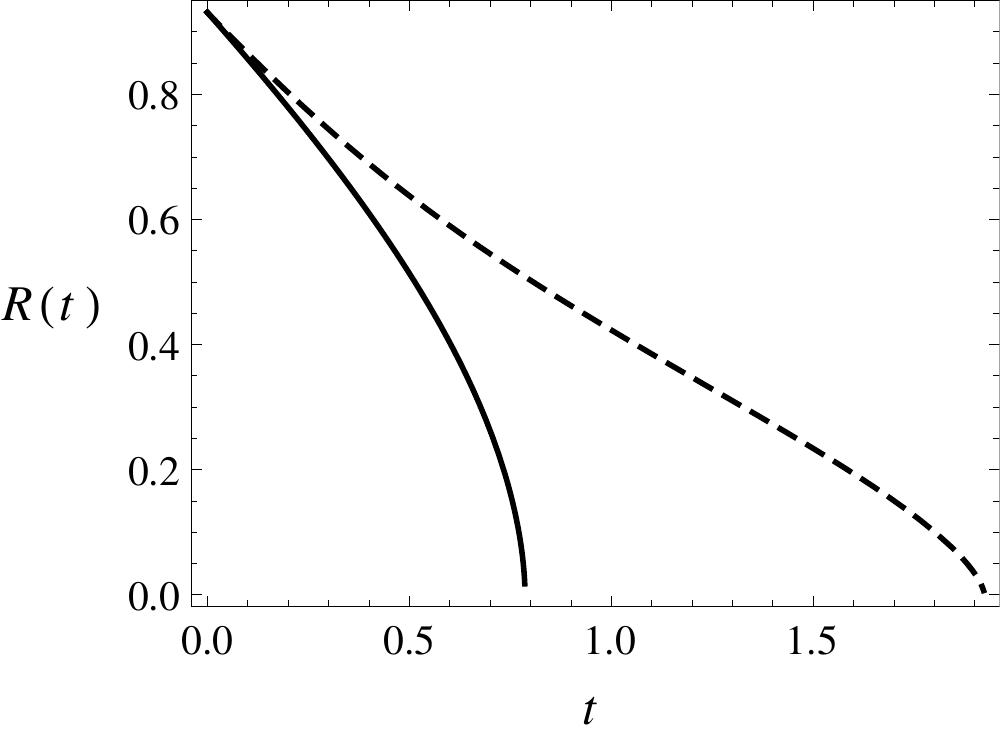}\quad{}\includegraphics[width=1.8in]{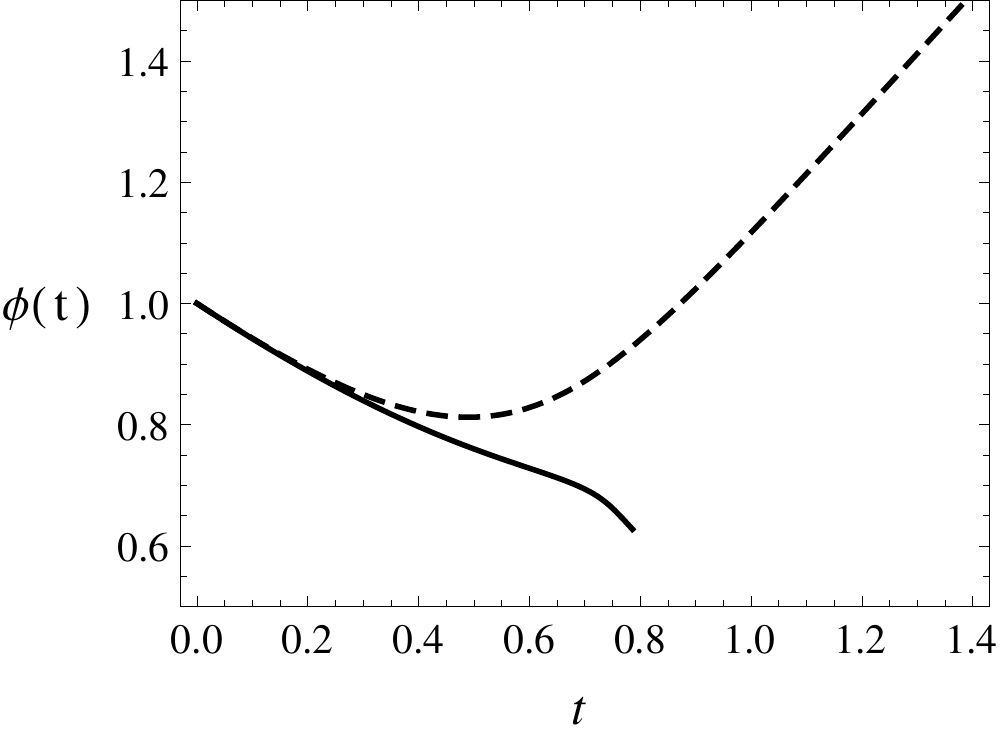}$\quad$\includegraphics[width=1.8in]{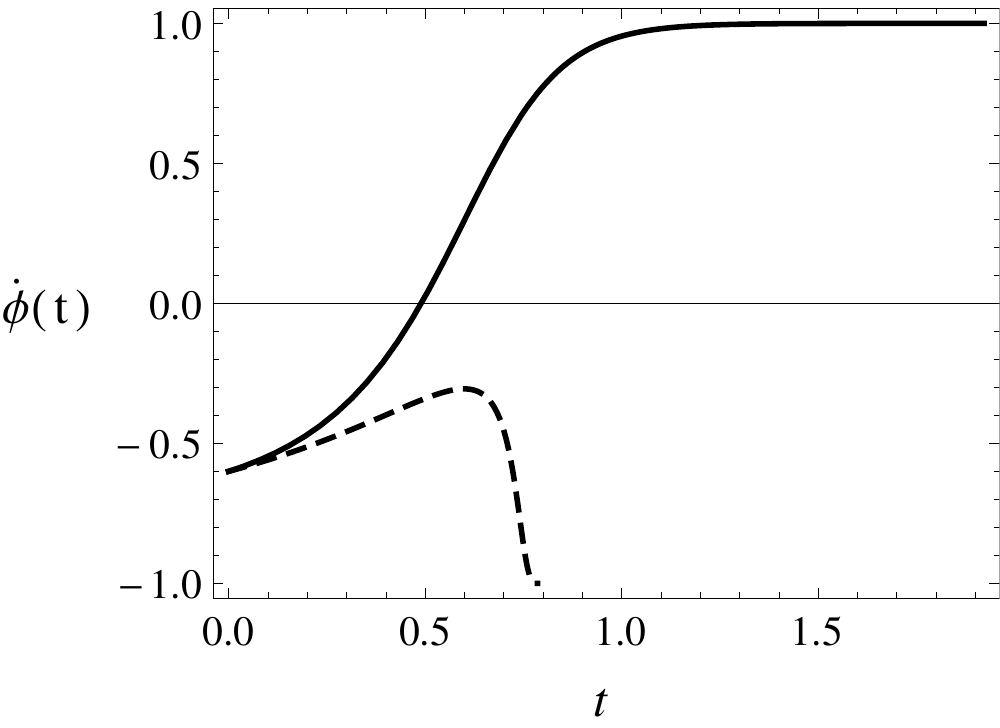} %
\end{minipage}\caption{Behavior of the area radius, tachyon field and its
time derivative over time. We considered the initial conditions as $\dot{\Phi}\left(0\right)<0$
and $\Phi_{0}>0$. We also have $\gamma_b<1$ (solid lines) and $\gamma_b>1$
(dashed lines).}

\label{F3b-class} 
\end{figure*}

The asymptotic behavior of $s$ near the point $(a)$ can be approximated
as $s\approx s_{c}+\exp(-\tau)=\exp(-\tau)$; hence, as $\tau\rightarrow\infty$
(i.e., $a\rightarrow0$), $s$ vanishes. Moreover, the time derivative
of the tachyon field is given by $\dot{\Phi}\simeq1$, that is, the
tachyon field $\Phi(t)$ has a linear time dependence and can be approximated
as (see figure \ref{F3b-class}) 
\begin{equation}
\Phi(t)\ \simeq\ t+\Phi_{0}.
\label{tachdom-1}
\end{equation}
 \medskip{}

It should be noted that, for the $\Phi>0$ branch, $\Phi_{0}$ is
positive on the initial configuration of the collapsing system (where
$t=0$) and hence, the tachyon field increases with time, proceeding
downhill the potential. Within a finite amount of time, the tachyon
field reaches its maximum, 
but finite value $\Phi(t_{s})=\Phi_{s}$ at $t_{s}<\Phi_{0}$, with
the minimum (but non-zero value) $V\propto\Phi_{s}^{-2}$. As the
tachyon potential decreases, the dynamical variable $y=\frac{V}{3H^{2}}$
vanishes.\medskip{}

\begin{figure*}[h]
\begin{minipage}[c]{1\textwidth}%
\centering \includegraphics[width=1.8in]{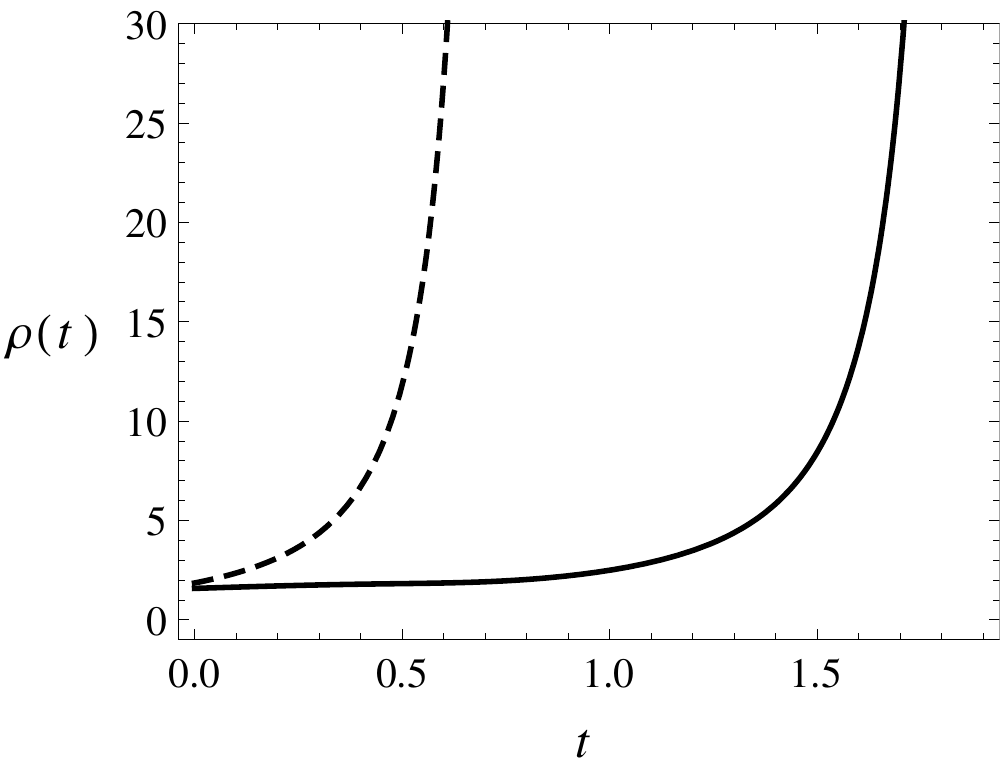}\quad{}\includegraphics[width=1.8in]{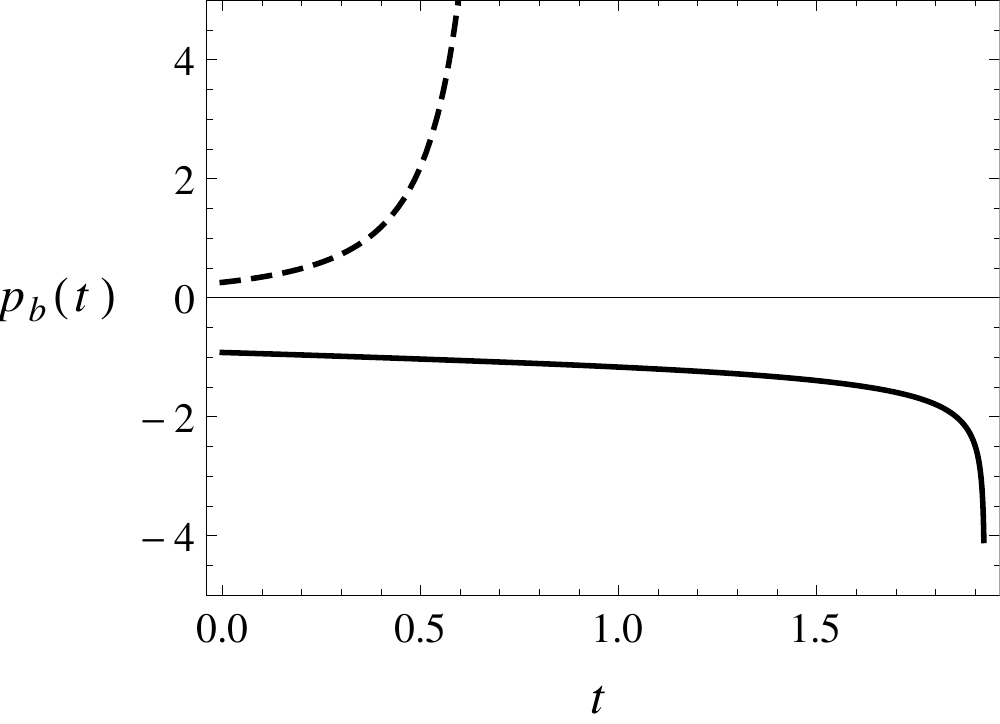}\quad{}\includegraphics[width=1.8in]{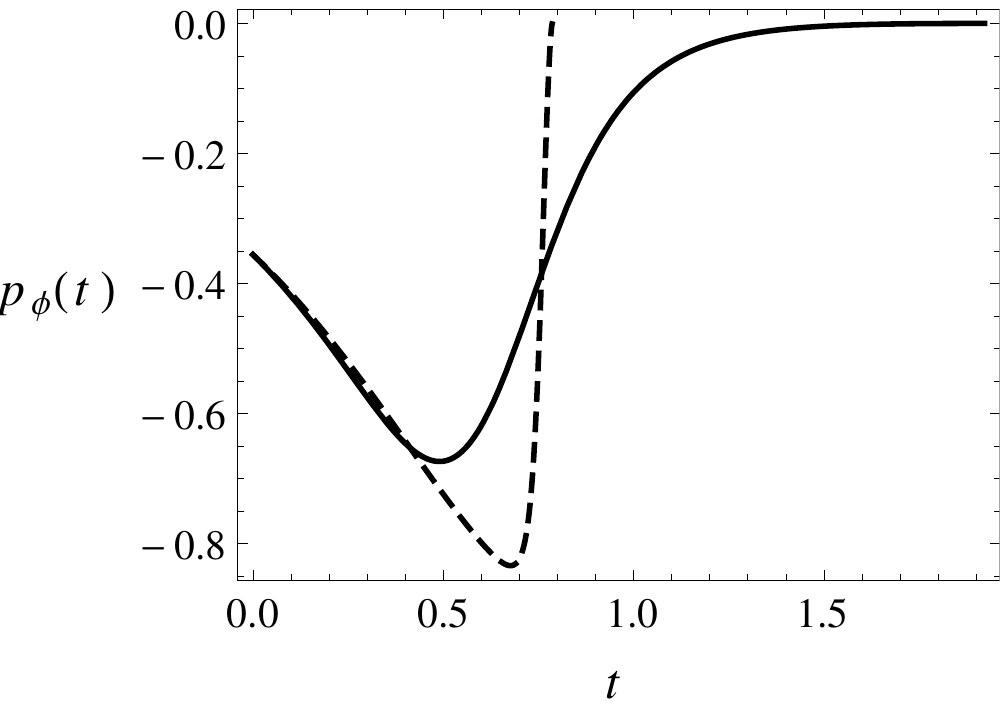} %
\end{minipage}\caption{The energy density, barotropic pressure and tachyon
field pressure. We considered 
$\Phi_{0}>0$ with $\gamma_b<1$ (solid lines) and $\gamma_b>1$
(dashed lines). The total effective pressure ($p_{b}+p_{\Phi}$) is
divergent and negative (for $\gamma_b<1$) in the final stage of the
collapse.}

\label{F3-class} 
\end{figure*}

Thus, as $\dot{\Phi}\rightarrow1$, the energy density of the system
diverges. Furthermore, the tachyon pressure $p_{\Phi}=-V(\Phi)(1-\dot{\Phi}^{2})^{\frac{1}{2}}$
vanishes asymptotically%
\footnote{For the $\Phi<0$ branch, $\Phi_{0}$ is negative at the initial condition.
Thus, the absolute value of the tachyon field starts to decrease from
the initial configuration as $\Phi(t)=t-|\Phi_{0}|$ until the singular
point at time $t_{s}<|\Phi_{0}|$, where tachyon field reaches its
minimum but non-zero value $\Phi_{s}$. This leads it uphill the potential
until the singular epoch, where the potential becomes maximum but
finite.%
} (see figure \ref{F3-class} for plots of the energy density, barotropic
pressure and tachyon field pressure).\medskip{}

\begin{figure*}[h]
\begin{minipage}[c]{1\textwidth}%
\centering \includegraphics[width=2.7in]{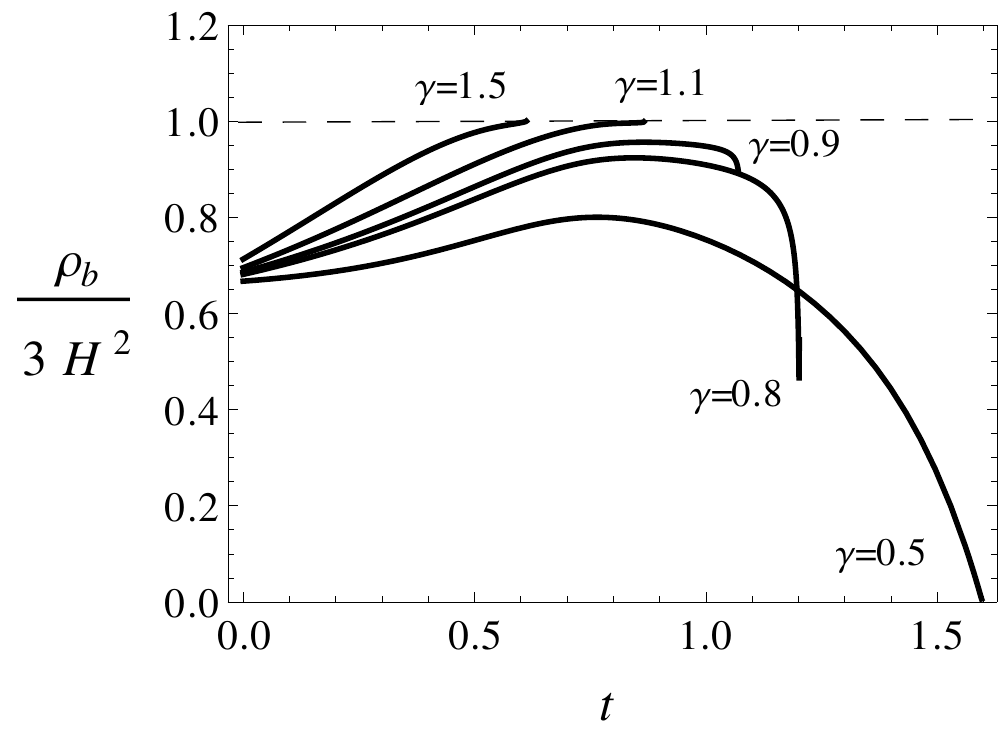}\quad{}\includegraphics[width=2.75in]{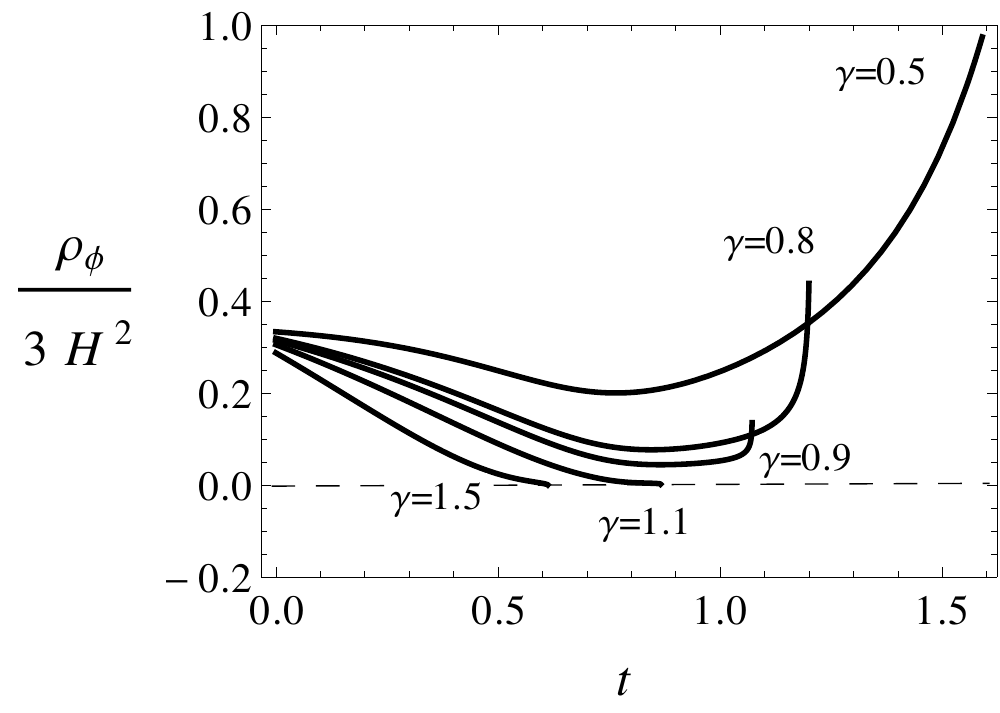} %
\end{minipage}\caption{Behavior of the ratios $\frac{\rho_{b}}{3H^{2}}$ (left)
and $\frac{\rho_{\Phi}}{3H^{2}}$ (right). We considered the initial
conditions as: 
$\dot{\Phi}\left(0\right)<0$ and $\Phi_{0}>0$. We illustrate
the transition between tachyon and fluid dominated solutions as a
function of the barotropic parameter. In these collapsing cases the
potential does not diverge and the tachyon field does not vanish.}

\label{dominance} 
\end{figure*}

The total energy density of the collapsing system is given approximately
by the energy density of tachyon field (see figure \ref{dominance}
for numerical solutions), which corresponds to a dust-like matter
near the singularity as $\rho_{\Phi}\propto1/a^{3}$ . From this relation
for the tachyon energy, we may write the energy density of the system
near point $(a)$, 
\begin{equation}
\rho\ \approx\ \rho_{0}a^{-3}.\label{Tenerg-a}
\end{equation}
 This induces a black hole at the collapse final state.

As far as $\left(b\right)$ is concerned, the time dependence of the
tachyon field can be obtained as 
\begin{equation}
\Phi(t)\simeq-t+\Phi_{0}.
\end{equation}
 For the $\Phi>0$ branch, the tachyon field decreases from its initial
value at $t=0$ (where $\Phi_{0}>0$), moving uphill the potential.
Then, the potential will reach a maximum (but finite) value when the
tachyon field reaches its minimum and nonzero value (i.e. $\Phi\rightarrow\Phi_{s}$
as $t\rightarrow t_{s}\leq\Phi_{0}$). In the limit case $t_{s}=\Phi_{0}$,
the tachyon field vanishes and the potential diverges. I.e., $y$
and $s$ vanish, the Hubble rate increases faster than the potential
and the barotropic fluid energy density diverges, that is, the total
energy density of the system asymptotically diverges%
\footnote{On the other hand, for the $\Phi<0$ branch, the tachyon field increases
from its initial condition as time evolves, proceeding to ever less
negative values as $\Phi\rightarrow0^{-}$. In this case, it proceeds
downhill the tachyon potential till the system reaches $a=0$ at $t=t_{s}$,
where the Hubble rate and hence the total energy density of the system
diverge. This implies that the time at which the collapse system reaches
the singularity is always $t_{s}\leq\Phi_{0}$.
}. Likewise, when $\dot{\Phi}\rightarrow1$ and $a\rightarrow0$, the
tachyon matter behaves as dust matter. The fate of the collapse for
this fixed point is as well a black hole formation.

The asymptotic solutions provided by the fixed points $(a)$ and $(b)$
correspond to a dust-like solution with a vanishing pressure for the
tachyon field, whose energy density reads $\rho_{\Phi}\propto1/a^{3}$.
This is consistent with the  WEC and  DEC  for the tachyon matter being
satisfied, as was mentioned before. Concerning the status of the energy
conditions for the barotropic fluid, as we indicated before, regularity
of the initial data for the collapsing matter respects the  WEC. On
the other hand, stability of the solution in this case ensures that
$\gamma_b<1$, which satisfies the sufficient condition for the  DEC.

\subsection*{B.~Fluid dominated solutions}
\label{Fluid dominated}

From the trajectories in the vicinity of $(f)$ and $(g)$, solutions
can be described with some having an attractor behaviour.

The asymptotic behavior of $s$ near the point $(f)$ can be approximated
as $s\approx1+\exp(-\tau)$; hence, as $\tau\rightarrow\infty$ (i.e.,
$a\rightarrow0$), $s\rightarrow1$. Moreover, the time derivative
of the tachyon field is given by $\dot{\Phi}\simeq1$; that is, the
tachyon field $\Phi(t)$ has a linear time dependence and can be approximated
as (see figure \ref{F3b-class}), 
\begin{equation}
\Phi(t)\ \simeq\ t+\Phi_{0}.
\end{equation}

The total energy density of the collapsing system is given approximately
by the energy density of the barotropic fluid (see figure \ref{dominance},
where $\rho_{b}/\left(3H^{2}\right)\rightarrow1$ while $\rho_{\Phi}/\left(3H^{2}\right)\rightarrow0$
for $\gamma_b>1$ ), which goes, near the singularity, as $\rho_{b}\propto1/a^{3\gamma_b}$.
From this relation for the fluid energy, we may write the energy density
of the system near point $(f)$, 
\begin{equation}
\rho\ \approx\ \rho_{0b}a^{-3\gamma_b}.\label{Tenerg-a-1}
\end{equation}
 For $\gamma_b<2/3$, the ratio $F/R=\frac{1}{3}r^{2}\rho_{0b}a^{2-3\gamma_b}$,
converges as the singularity is reached leading to the avoidance of
trapped surfaces, but since the corresponding fixed point $(f)$ turns
to be a saddle, then the resulting naked singularity is not stable.
For $2/3<\gamma_b<1$ the ratio $F/R$ diverges and the trapped surfaces
do form. But still the point $(f)$ is saddle and the resulting black
hole is not stable. The case $\gamma_b>1$ corresponds to a stable solution
for which the ratio $F/R$ goes to infinity as the collapse advances.
Then, the trapped surface formation in the collapse takes place before
the singularity formation and thus the final outcome is a black hole.

As far as $\left(g\right)$ is concerned, the time dependence of the
tachyon field can be obtained as 
\begin{equation}
\Phi(t)\simeq-t+\Phi_{0}.
\end{equation}
 For the $\Phi>0$ branch, the tachyon field decreases from its initial
value at $t=0$ (where $\Phi_{0}>0$), moving uphill the potential,
the potential will reach a maximum (but finite) value when the tachyon
field reaches its minimum and nonzero value (i.e. $\Phi\rightarrow\Phi_{s}$
as $t\rightarrow t_{s}\leq\Phi_{0}$). In the limit case $t_{s}=\Phi_{0}$,
the tachyon field vanishes and the potential diverges. I.e., $y$
and $s$ vanish, the Hubble rate increases faster than the potential
and the barotropic fluid energy density diverges, that is, the total
energy density of the system, is given by Eq.~(\ref{Tenerg-a-1}),
and asymptotically diverges. Similar to point $(f)$ the mass function
of the system for the fixed point solution $(g)$ is given by $F/R=\frac{1}{3}r^{2}\rho_{0b}a^{2-3\gamma_b}$,
which diverges for $\gamma_b>1$. Therefore, the resulting singularity
in this case will be covered by a black hole horizon.

As far as the energy conditions are concerned for the fixed point
solutions $(f)$ and $(g)$, we find that the tachyon field satisfies
the  WEC. Also the  DEC  remains valid case as well. On the other hand,
for the barotropic fluid, the  WEC  is satisfied initially and will
hold until the endstate of the collapse. The stable solution in this
case corresponds to the range $1<\gamma_b<2$ which satisfies  DEC  as well.

\subsection*{C.~Tracking solution}
\label{Tracking-solutions}

Now we discuss a different type of solution,
where the fluid and tachyon appear with a tracking behaviour. Let
us introduce this situation as follows. 

Interesting and physically reasonable tracking solutions can be found,
where $\dot{\Phi}\rightarrow\pm1$, when we consider $\gamma_b\rightarrow1$,
i.e., a situation whereby the emergence of points $\left(h\right)$
and $\left(i\right)$ will be of relevance as attractors. The transition
from tachyon dominated to fluid dominated scenarios, like those described
before, is not straightforward. In this situation, the
tachyon field and barotropic fluid compete to establish the dominance
in the late stage of the collapse. In figure \ref{dominance} we have
an illustration of this kind of solutions. The mentioned dominance
seems to depend strongly on the initial ratio $\rho_{\Phi}/\rho_{b}$
at an earlier stage of the collapse and also on the value of $\gamma_b$.
Such a dependence on the initial conditions can lead to a set of solutions
between those provided by fixed points $\left(a\right)$, $\left(b\right)$
(tachyon dominated solutions) and by points $\left(f\right)$, $\left(g\right)$
(fluid dominated solutions). From a dynamical system point of view,
this corresponds to have trajectories asymptotically approaching $x\rightarrow\pm1$
in sections where $s$ is between 0 and 1 (see figure \ref{3D-stream}).
At the end of the collapse we observe that $\frac{\rho_{\Phi}}{3H^{2}}\sim\frac{\rho_{b}}{3H^{2}}<1$,
as illustrated in figure \ref{dominance}. Therefore, in this scenario,
the trajectories would convey a collapsing case in which the energy
density of the tachyon field and of the barotropic fluid are given
by $\rho_{\Phi}\propto\rho_{b}\propto a^{-3\gamma_b}$. This
shows a tracking behavior for the collapsing system \cite{RLazkoz2004,Liddle:1999}.
Moreover, the total energy density of the collapse, in terms of $a$,
reads 
\begin{equation}
\rho\ \propto\; a^{-3\gamma_b}.\label{energ-e}
\end{equation}
Equation (\ref{energ-e}) shows that the energy densities of the tachyon
field, the barotropic matter and hence, the total energy density (for the collapsing
system), diverge as $a\rightarrow0$. The ratio of the total mass
function over the area radius is given by 
\begin{equation}
\frac{F}{R}\ \propto\ \frac{1}{3}r^{2}a^{2-3\gamma_b}.\label{massd}
\end{equation}
 Equation (\ref{massd}) subsequently implies that, for an adequate
choice of values $(\gamma_b,\Phi_{0})$, trapped surfaces can form as
the collapse evolves and a few scenarios can be extracted. More precisely,
for the range $\gamma_b>\frac{2}{3}$, for both $\Phi>0$ and $\Phi<0$
branches, the final fate of the collapse is a black hole. For the
case in which $\gamma_b<\frac{2}{3}$, the ratio $F/R$ remains finite
as the collapse proceeds and an apparent horizon is delayed or fails
to form; the final state is a naked singularity (a solution for the
choice of `$-$' sign in Eq.~(\ref{eigencle})). The tracking
solution indicates $\gamma_b=\frac{2}{3}$ as the threshold (illustrated
in figure \ref{2D-stream}), which distinguishes a black hole and  a
naked singularity forming. 

\begin{figure*}[h]
\begin{minipage}[c]{1\textwidth}%
\centering \includegraphics[width=2.5in]{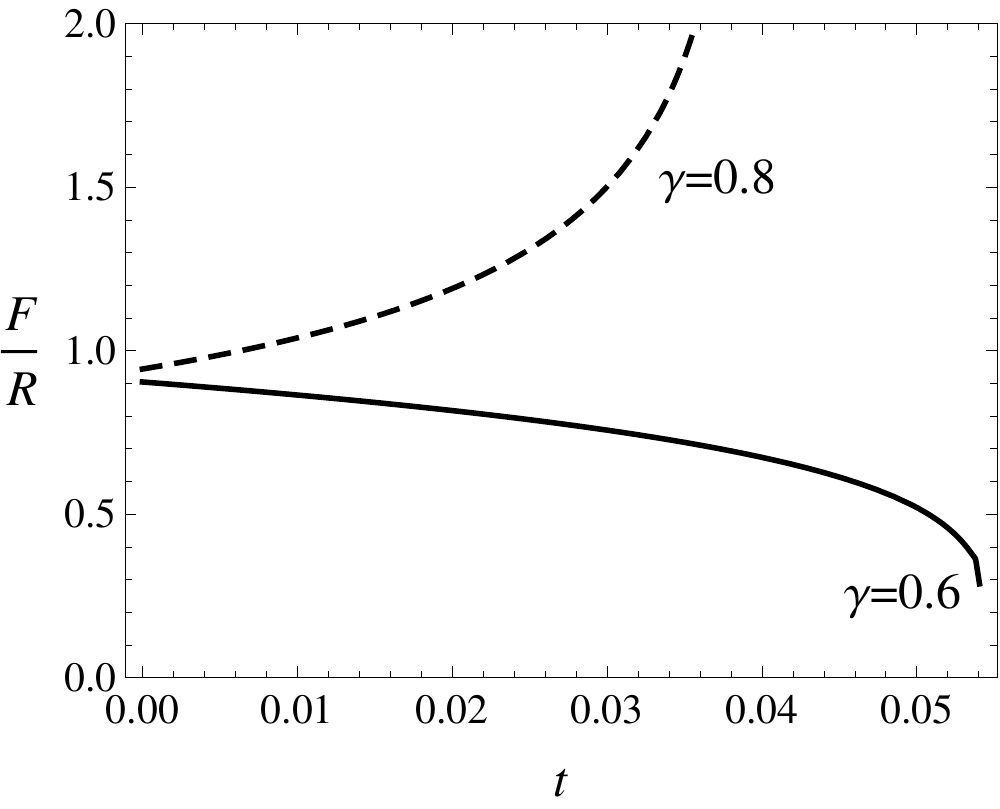}\quad{}\includegraphics[width=2.5in]{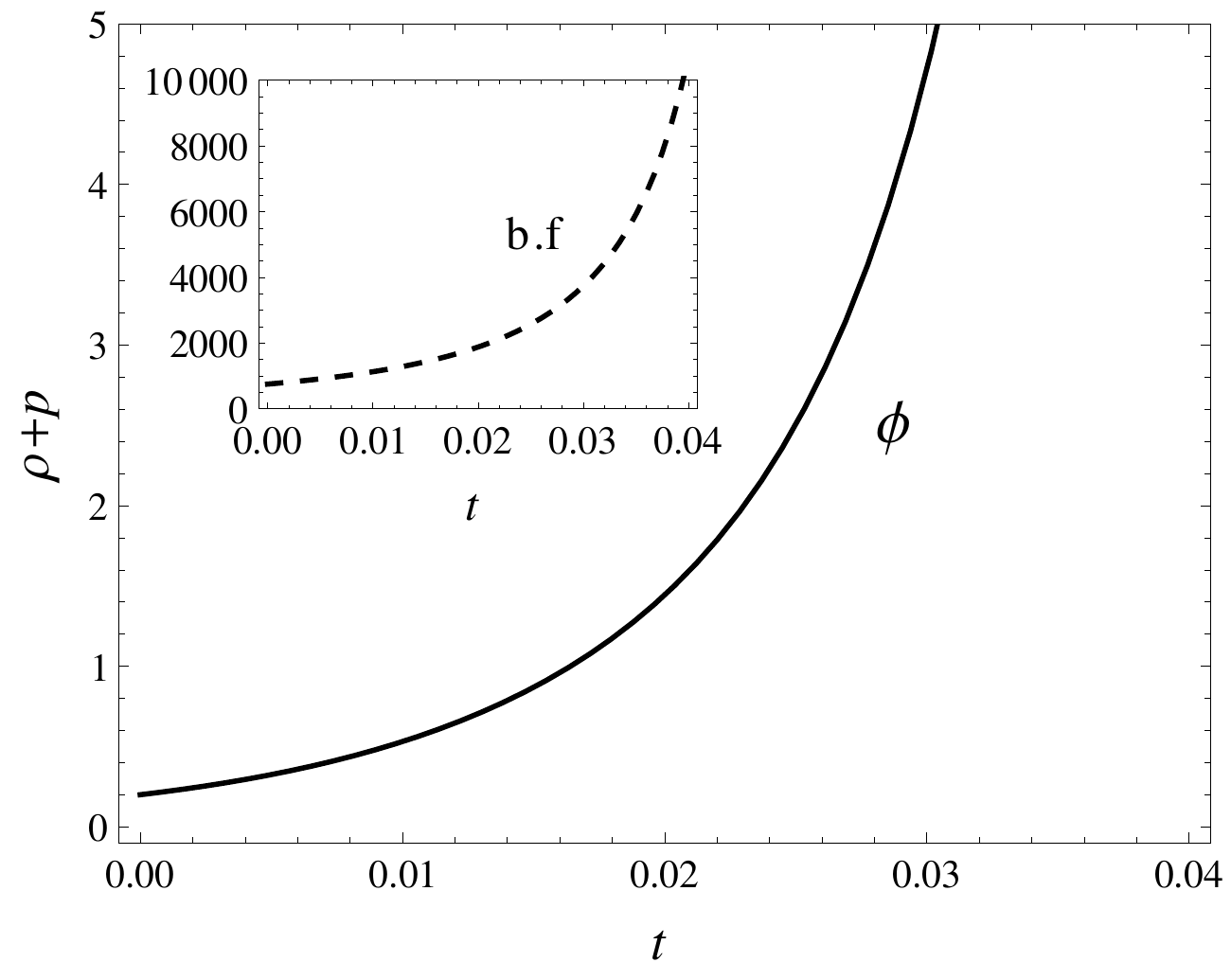}%
\end{minipage}\caption{In the left plot it is shown the ratio of the total
mass function over the area radius.  It is shown that for $\gamma_b>2/3$ we have black hole
formation and for $\gamma_b<2/3$ we have naked singularity formation.
The right plot shows the  WEC  over time for the barotropic
fluid (b.f) and for the tachyon field.}

{\footnotesize \label{naked singularity} }
\end{figure*}

Therefore, under suitable conditions, we can determine whether it
is possible to have the formation of a naked singularity. In fact,
if we assume a very unbalanced initial ratio $\rho_{\Phi}/\rho_{b}$
with $\rho_{0\Phi}\ll\rho_{0b}$ and a barotropic fluid having $\gamma_b<\frac{2}{3}$,
then we can have a situation where the ratio (\ref{massd}) is converging.
The set of initial conditions described by $\rho_{0\Phi}\ll\rho_{0b}$
are equivalent to consider the barotropic fluid as initially dominant.
If this specific unbalanced distribution of matter is allowed to evolve
into a regime where the tachyon dominates then the system will evolve
until $\rho_{\Phi}$ becomes comparable to $\rho_{b}$. The singularity
is reached in finite time and it can happen before the tachyon can
effectively dominate. In figure \ref{naked singularity} we have a
graphical representation of the ratio $F/R$. It can be seen that
the ratio $F/R$ remains finite for $\gamma_b<2/3$, while the energy
density is diverging, as the collapse proceeds and apparent horizon
is delayed or fails to form till the singularity formation. As the
right plot shows, the validity of  WEC  is guaranteed throughout the
collapse scenario for both barotropic fluid and tachyon field. Also
the DEC  is valid for the solutions that exhibit naked singularity,
i.e., those for which $\gamma_b<2/3$. \medskip{}

\subsection{Exterior geometry}
\label{Classic3}

For a perfect fluid gravitational
collapse set up, with equation of state $p_b=\left(\gamma_b-1\right)\rho_b$,
the pressure does not necessarily vanish at the boundary. E.g., matching
the internal geometry filled with matter (and radiation) to a boundary
layer (which is crossed with the radiation), which could in turn be matched to an exterior geometry, is not completely empty (e.g., is filled by radiation). More concretely, we match  the interior with 
a generalized exterior Vaidya space-time (as we discussed in section \ref{collapse-exterior}) across the boundary given by $r=r_{\mathrm{b}}$.

Similarly to the section \ref{collapse-exterior}, we proceed by considering the interior
metric which describes the collapsing cloud given by Eq.~(\ref{metric}),
and the exterior one in advanced null coordinates $(\mathrm{v},r_\mathrm{v})$ given by Eq.~(\ref{metric2}).
In order to find
a suitable exterior metric function ${\cal F}(\mathrm{v},r_\mathrm{v})$ we resort to a  Hamiltonian
perspective of the model. The total Hamiltonian constraint is given
by 
\cite{Tsamparlis, ASen2006} 
\begin{align}
C(a,\pi_{a};\pi_{\Phi}) & = C_{a} + C_{\Phi} + C_{b}\notag\\
 & =\frac{\pi_{a}^{2}}{12a}\ +\ a^{3}\sqrt{V^{2}+{a}^{-6}\pi_{\Phi}^{2}}\ +\ \rho_{0b}a^{-3\left(\gamma_b-1\right)},\label{Hamiltonian-cl}
\end{align}
where $\pi_{a}$ and $\pi_{\Phi}$ are the conjugate momentums for
the scale factor $a$ and for the tachyon field $\Phi$, respectively.
Furthermore, the Hamilton equations for parameters $a$, $\pi_{a}$
and $\pi_{\Phi}$ can be obtained by using  Eq.~(\ref{Hamiltonian-cl})
as follows: 
\begin{align}
\dot{a} & =\frac{\partial{C}}{\partial\pi_{a}}=\frac{\pi_{a}}{6a},\ \ \ \ \ \ \dot{\Phi}=\frac{\partial{C}}{\partial\pi_{\Phi}}=\frac{a^{-3}\pi_{\Phi}}{\sqrt{V^{2}+a^{-6}\pi_{\Phi}^{2}}}\ ,\label{eqs-1}\\
\dot{\pi}_{a} & =-\frac{\partial{C}}{\partial a}=\frac{\pi_{a}^{2}}{12a^{2}}+3\left(\gamma_b-1\right)\rho_{0b}a^{-(1+3\left(\gamma_b-1\right))}-\frac{3a^{5}V^{2}}{\sqrt{a^{6}V^{2}+\pi_{\Phi}^{2}}}\ ,\label{eqs-2}\\
\dot{\pi}_{\Phi} & =-\frac{\partial{C}}{\partial\Phi}=-\frac{a^{3}VV_{,\Phi}}{\sqrt{a^{6}V^{2}+\pi_{\Phi}^{2}}}\ .\label{eqs-3}
\end{align}
On the other hand, since the Hamiltonian constraint, $C(a,\pi_{a};\pi_{\Phi})=0$,
must be held across the boundary $\Sigma$, Eq.~(\ref{Hamiltonian-cl})
reduces to 
\begin{equation}
\pi_{\Phi}^{2}=\frac{\pi_{a}^{4}}{144a^{2}}+\rho_{0b}^{2}a^{-6\left(\gamma_b-1\right)}+\rho_{0b}\frac{\pi_{a}^{2}a^{-3\left(\gamma_b-1\right)}}{6a}-a^{6}V^{2},\label{pphi}
\end{equation}
whereby, substituting for $\pi_{a}$ and $\pi_{\Phi}$ from Eqs.~(\ref{eqs-1})-(\ref{eqs-3}), we get 
\begin{equation}
9a^{2}\dot{a}^{4}+\rho_{0b}^{2}a^{-6\left(\gamma_b-1\right)}+6\rho_{0b}\dot{a}^{2}a^{2-3\gamma_b}=\frac{a^{6}V^{2}}{1-\dot{\Phi}^{2}}\ .\label{fex}
\end{equation}
Furthermore, it is required for the junction condition $r_{\rm b}a(t)=R(t)$
and the equation of motion for scale factor $r_{\rm b}\dot{a}=\dot{R}$
to be satisfied at the boundary of two regions. By substituting these
conditions in the Hamiltonian constraint equation, $C(r_{\rm v}, {\rm v})=0$,
given by Eq.~(\ref{fex}), we get 
\begin{align}
9(1-{\cal F})^2\left(\frac{R}{r_{\rm b}}\right)^{-4}+6\rho_{0b}(1-{\cal F})\left(\frac{R}{r_{\rm b}}\right)^{-3\gamma_b}=\frac{V^{2}}{1-\dot{\Phi}^{2}}-\rho_{0b}^{2}\left(\frac{R}{r_{\rm b}}\right)^{-6\gamma_b}~.\label{AREA}
\end{align}
Notice that, we have substituted the four-velocity of the boundary
being seen from the exterior by $\dot{R}=-\sqrt{1-{\cal F}}$.
Then, by solving Eq.~(\ref{AREA}) for ${\cal F}$, the boundary function
is obtained simply as 
\begin{equation}
{\cal F}(R)=1-\frac{2}{3}\frac{V}{\sqrt{1-\dot{\Phi}^{2}}}\left(\frac{R}{r_{\rm b}}\right)^{2}+\frac{1}{3}\rho_{0b}\left(\frac{r_{\rm b}}{R}\right)^{1+3\left(\gamma_b-1\right)}.\label{exteriormf}
\end{equation}

Let us now study the behavior of the boundary function for the stable
fixed point solutions we obtained in the previous subsections. We
note that the above expression is valid when both the tachyon field
and barotropic fluid are present, thus in the regimes where the tachyon
field is dominant the contribution of the fluid to Hamiltonian constraint
is set aside. For the dust-like solutions, described by points $\left(a\right)$
and $\left(b\right)$, the energy density of tachyon field at the
boundary $r_{\rm b}$, included in second term of Eq.~(\ref{exteriormf}),
is given by $\rho_{\Phi}\approx\rho_{0}r_{\rm b}^{3}/R^{3}$. So, the
boundary function ${\cal F}$ in Eq.~(\ref{exteriormf}) reduces to:
\begin{equation}
{\cal F}(R)=1-\frac{M}{R}~,\qquad\text{where}\qquad M\equiv\frac{4\pi}{3}r_{\rm b}^{3}\tilde{\rho}_{0}~,\label{exteriormf-dust}
\end{equation}
and $\tilde{\rho}_{0}\equiv\rho_{0}/2\pi r_{\rm b}^{2}=\text{const}$.
Equation (\ref{exteriormf-dust}) for ${\cal F}(R)$ constrains the exterior
space-time to have a Schwarzschild geometry  in advanced null coordinates,
providing an interpretation of the collapsing system as a dust ball
with the radius $r_{\rm b}$ and the density $\tilde{\rho}_{0}$.

In order to get a possible class of dynamical exterior solutions,
we proceed by considering the following metric at the boundary \cite{Tsamparlis}:
\begin{equation}
ds^{2}=-\left(1-\frac{2M(R,v)}{R}\right)dv^{2}+2dvdR+r^{2}d\Omega^{2},\label{bm}
\end{equation}
where 
\begin{equation}
M(R,v)=m(v)-\frac{g(v)}{2(2\gamma_b-3)R^{2\gamma_b-3}}\ ,\label{mass}
\end{equation}
is the total mass within the collapsing cloud. Matching the boundary function for the exterior
metric at $r_b$  gives 
\begin{align}
 & 1-\frac{2m(v)}{R}+\frac{g(v)}{(2\gamma_b-3)R^{2\gamma_b-2}}=1-\frac{\tilde{M}}{R^{3\gamma_b-2}}\ ,\notag\\
 & \frac{2\dot{m}(v)}{R}=\frac{\dot{g}(v)}{(2\gamma_b-3)R^{2\gamma_b-2}}\ ,\label{matching}
\end{align}
in which, the second part stands for matching for the extrinsic curvature.
Differentiation of the first expression in Eq.~(\ref{matching}) with
respect to time and using the second one, we get 
\begin{equation}
\frac{2m(v)}{R^{2}}-\frac{2(\gamma_b-1)g(v)}{(2\gamma_b-3)R^{2\gamma_b-1}}=\frac{\tilde{M}(3\gamma_b-2)}{3R^{3\gamma_b-1}}\ .\label{1}
\end{equation}
Multiplying the first expression in Eq.~(\ref{matching}) by
$R^{-1}$ and after adding the result with the above equation, we
get 
\begin{equation}
g(v)=-\frac{\tilde{M}(\gamma_b-1)}{R(v)^{\gamma_b}}\ .\label{g}
\end{equation}
Now, by substituting for $g(v)$ into Eq.~(\ref{matching}),
we obtain 
\begin{equation}
m(v)=\frac{\tilde{M}\gamma_b}{6(3-2\gamma_b)R(v)^{3\gamma_b-3}}\ .\label{m}
\end{equation}
Then, as seen from Eq.~(\ref{massd}) for $\gamma_b<2/3$, where
the trapping of light has failed to occur, the exterior geometry is
dynamical in contrast to the tachyon dominated regime in which a space-like
singularity forms with a static exterior space-time. Finally, for
fluid dominated solutions, depending on the value of $\gamma_b$, both
naked singularities and black holes may form, the mass being different
to those of tracking solutions and the exterior geometry being static
or dynamical, respectively.

\section{Summary}

 In this chapter we started with a brief review of gravitational collapse of an isotropic and homogeneous interior space-time. The interior region included 
 a flat FLRW   space-time of gravitational collapse constitutes
 a tachyon field with a non-standard kinetic term together with a barotropic fluid, as matter content. 
For simplicity, the
tachyon potential was assumed to be of  inverse square form  i.e.,
$V(\Phi)\sim\Phi^{-2}$. Our purpose, by making use of the non-standard
kinematical features of the tachyon, which are rather different from
a standard scalar field, was to establish the different types of asymptotic
behavior that our matter content induces. Employing a dynamical system
analysis, complemented by a thorough numerical study, we found classical
solutions corresponding to a naked singularity or a black hole formation.
In particular, there was a subset where the fluid and tachyon participate in an
interesting tracking behaviour, depending sensitively on the initial conditions of  the energy densities of  tachyon field and barotropic fluid.

We think it is fair to indicate that we employed a $\Phi^{-2}$ potential
for the tachyon, whereas for $\Phi\rightarrow0$, the tachyon should
not induce a divergent behavior as far as string theory advises \cite{Feinstein:2002,Guo:2003,Ambramo:2003,Copeland:2005,Quiros:2010}.
In fact, an exponential-like potential for the tachyon could bring
a richer set of possible outcomes \cite{Garousi:2000,Bergshoeff:2000,Feinstein:2002,ASen2002a,ASen2002b,ASen:1998a,Ghate1999},
including a better behaved and possibly a more realistic evolution
when dealing with $\Phi\rightarrow0$.


Finally, let us add that it will be of interest to investigate (i)~other scenarios for the geometry of the interior space-time region,
(ii)~specific couplings between the tachyon and the fluid, within
e.g., a chamaleonic scenario for gravitational collapse and black
hole production (broadening the scope in Refs.~\cite{Folomeev,Folomeev:2012a,Folomeev:2012b}),
(iii)~adding either axionic, dilatonic or other terms (e.g., curvature
invariants) that could be considered from a string setup, but at the
price of making the framework severely less workable and (iv) whether
other induced quantum effects can alter the outcomes presented in this
model. To this purpose  we point to our  chapter \ref{QuantumG}. 
\chapter{Semiclassical  collapse with tachyon field and barotropic fluid}
\label{QuantumG}

\lhead{Chapter 3. \emph{Semiclassical collapse with tachyon field and barotropic fluid}} %

A fundamental quantum theory for gravity must provide a general formulation which shows how evolution can continue through  classical singularities. 
LQG is a candidate within the canonical approach to quantum gravity  that provides a fruitful ground to investigate the removal of the space-time singularities \cite{Bojowald2007,Bojowald2010}.
The picture offered  is that the smooth geometry underlying general relativity is, on very small length scales, to be replaced by a discrete structure resembling a lattice. When this discreteness becomes relevant because the universe itself reaches such small scales, classical continuum equations break down \cite{Bojowald2003}.

In addition to the ingredients and settings mentioned on the issue of gravitational collapse in the previous chapter, it is therefore of pertinence to investigate whether elements induced from LQG  may alter the asymptotic states of gravitational collapse \cite{Bojowald2010,Modesto2004,Modesto2008}.
In LQG, by means of a suitable operator (actually, the inverse volume and holonomy), in some concrete configurations, the resulting effective equations provide significant differences with respect to the classical setting.

Inverse triad modifications  \cite{Bojowald2008,Bojowald2002}, in the presence of a homogeneous massless scalar field,  are expected to replace a classical singularity with a bounce at the semiclassical limit (cf. \cite{Bojowald2005b}), or
to lead to an outward flux of energy, when a general potential is considered for the scalar field (cf. \cite{Goswami2006}).  
However, it still remains an open issue whether  or not  this specific correction
 resolves those singularities in the presence of different matter sources. 
Eg., with regards to the tachyon equation of motion, 
we widen the  settings being explored\footnote{See chapter \ref{collapse-class} for the classical analysis of tachyon field collapse} \cite{Tavakoli2013b}
in order to fully investigate whether a gravitational collapse bears a  non singular nature or not. 

There is another type of  modification,
motivated by the effective constraint approach used in the LQG program, which comes from the  ``holonomy" operator  \cite{Ashtekar:2011};
this, namely the \emph{holonomy correction},  provides an upper limit  for the energy density of matter, which leads to a non singular bounce for the matter density of the system \cite{Ashtekar:2011,A-P-S:2006a,A-P-S:2006b,Taveras2006}. Therefore, it is also of interest to study the nature of the classical singularities and the evolution of the apparent horizon during the collapsing scenario within this context (see Ref.~\cite{Tach-holonomy:2013,Tavakoli2013d}).

In section \ref{LQG-tach},  three concrete elements are present: Loop
effects (inverse triad type), tachyonic dynamics \cite{Tavakoli2013a} and (barotropic) fluid
pressure. We address whether those loop modifications can modify the
classical outcome (cf. chapter \ref{collapse-tachyon}) of tachyon gravitational collapse, by means of presenting a set of thorough numerical and analytical solutions.
In section~\ref{Holonomy-Tach}, we will investigate the semiclassical collapse (of  tachyon field and barotropic fluid) within  an effective scenario provided by a `holonomy correction'. 
By employing a dynamical system analysis, we will study the asymptotic  behaviour of the matter fields during the collapse;
 we will further provide a comparison with the solutions obtained in chapter \ref{collapse-class} for the corresponding general relativistic collapse.
Finally in section~\ref{Summary-3} we will provide our conclusion and summary of this chapter.

\section{Semiclassical collapse: Inverse triad correction}
\label{LQG-tach} 

We will study  in this section\footnote{This section is mainly written based on our paper with the Ref. \cite{Tavakoli2013b}. The unit $\kappa=8\pi G=1$ is used in this section.}, the semiclassical behavior of  gravitational
collapse setting including a tachyon field together with a fluid  (which is rather different from the case of a
standard scalar field\footnote{See appendix \ref{LQG-scalar} for a brief review of the (standard) scalar field collapse in the semiclassical regimes.}) as matter content.

\subsection{Modified tachyonic model}
\label{triad-tachyon}

In section \ref{collapse-tachyon} we conveyed the classical behavior of a tachyon collapse in classical setting. 
However, the question arises here is what will be the outcome
 in the presence of quantum effects? We will investigate herein
this section how non-perturbative  loop  modifications   may provide an answer; we consider solely
inverse triad corrections here. 

Let us then proceed and start with  the Hamiltonian constraint (\ref{Lagrangianfield}) for the
tachyon field  \cite{ASen2002a,ASen2002b}:
\begin{equation}
C_\Phi := C_{\rm matt}(\Phi,
\pi_{\Phi})={|p|}^{3/2}\sqrt{V^2+{|p|}^{-3}\pi_{\Phi}^2} \  ,
\label{Hamiltonian}
\end{equation}
where the conjugate momentum for the tachyon field $\Phi(t)$ reads
\begin{equation}
\pi_{\Phi}=\frac{\partial
L}{\partial\dot{\Phi}}=\frac{a^3V(\Phi)\dot{\Phi}}{\sqrt{1-\dot{\Phi}^2}}~.
\label{momentumfield}
\end{equation}


We now consider the inverse triad modifications 
based on LQG  for the interior.  
In particular, there exists a critical scale $a_* =\sqrt{|p_*|} = \sqrt{8\pi G j \gamma/3} \ell_{\mathrm{Pl}}$ 
below which the eigenvalues of the inverse scale factor become proportional 
to the positive powers of scale factor \cite{Bojowald2002a}; $\gamma=0.13$ is the Barbero-Immirzi parameter, and $\ell_{\mathrm{Pl}}$ is the Planck length.
Thus, for scales $\sqrt{\gamma}  \ell_{\mathrm{Pl}} =: a_i   \lesssim a \lesssim a_*$, the dynamics of the interior  can be described by modifications
to the  Friedmann dynamics   \cite{Bojowald2005a, Singh2005a}.
More precisely, the term ${|p|}^{-1}$ associated to the momentum
operator $\pi_{\Phi}$ is replaced by $d_{j,l}(p)=D_l(q)|p|^{-3/2}$, the eigenvalues of inverse triad operator.  
Notice that, the value of the half-integer free parameter $j$  is arbitrary and shall be constrained 
by phenomenological considerations.
For the choice of parameter $l=3/4$, the relation for $D_l(q)$ leads to the following equation \cite{Bojowald2002a, Date2004}: 
\begin{align} 
D(q) & \ =\   \left({8/ 77}\right)^6  q^{3/2}   \left\{7 \left[(q+1)^{11/4} -|q-1|^{11/4}\right]  \right. \notag \\
& \  \  \  \  \  \  \    \left.  -11q\left[(q+1)^{7/4}-{\rm sgn}\,(q-1) |q-1|^{7/4}\right] \right\}^4,
\label{defD}
\end{align}
where $q := a^2/a_*^2$, so that, $d_{j}(a)=D(q)a^{-3}$.
For  $a \gg a_*$ this function behaves classically with $d_{j} \approx a^{-3}$ (where $D(q)\rightarrow 1$),
and for the semiclassical regime, $a_i<a\ll a_{*}$, Eq. (\ref{defD}) reduces to
\begin{equation}
D(q)\ \approx \ \left(\frac{12}{7}q\right)^4. \label{redD2}
\end{equation}
The scale at which transition in the behavior of the geometrical 
density takes place is determined by the parameter $j$.
Furthermore, $a_i=\sqrt{\gamma} \ell_{\mathrm{Pl}}$ is the scale above which a classical continuous space-time can be defined and below which the space-time is discrete.

Now, substituting ${|p|}^{-1}$ in Eq.~(\ref{Hamiltonian}) by $d_{j,l}(p)$, given in Eq. (\ref{defD}),
the effective Hamiltonian for the tachyon field is re-written as \cite{ASen2006}
\begin{equation}
C_{\Phi}^{\rm sc}={|p|}^{3/2}\sqrt{V^2+|d_{j,l}({|p|})|^3\pi_{\Phi}^2}\  .
\label{loopH}
\end{equation}
From Eq.~(\ref{Henergy-sf-a-Tach}), the classical energy density and pressure of a scalar matter field can be written as 
\begin{equation}
\rho_{\Phi} =   |p|^{-3/2}C_{\Phi},\  \  \  \  \  \  p_\Phi = - |p|^{-3/2}\left(\frac{2}{3}|p|
\frac{\partial {C_{\Phi}}}{\partial{p}}\right).
\label{Henergy-sf}
\end{equation}
Using Eq. (\ref{Henergy-sf}), bearing in mind that $\dot{\Phi}=\partial C_{\Phi}^{\rm sc}/\partial \pi_{\Phi}$,
the equation for the effective energy density and pressure with  this
(loop)  modification leads to
\begin{equation}
\rho_{\Phi}^{\rm sc}=\frac{Vq^{3/2}|D(q)|^{3/2}}{\sqrt{q^{3} |D(q)|^3-\dot{\Phi}^2}}~,
\label{looprho2}
\end{equation}
and
\begin{equation}
p_{\Phi}^{\rm sc}=-\frac{Vq^{3/2}|D(q)|^{3/2}}{\sqrt{q^{3}
|D(q)|^3-\dot{\Phi}^2}}\left[1+\frac{\dot{\Phi}^2 p_{j}
|D(q)|_{, p}}{q^2|D(q)|^4}\right], \label{loopP2}
\end{equation}
where ``$,p$" stands for a differentiation with respect to the argument $p$.  Consequently, we can also retrieve a
modified equation of state by means of
\begin{equation}
w_{\Phi}^{\rm sc}=-\left[1+\frac{\dot{\Phi}^2
p_{j}|D(q)|_{,p}}{q^2|D(q)|^4}\right]. \label{stateE}
\end{equation}

If we further take, within the regime $a_i<a\ll a_{*}$, that  $q\ll
1$, or $D(q)\ll 1$, the modified equation of motion for the  tachyon
field is then given by 
\begin{equation}
\ddot{\Phi}-12H\dot{\Phi}\left[\frac{7}{2}-A^{-1}q^{-15}\dot{\Phi}^2\right]+
\left[Aq^{15}-\dot{\Phi}^2\right]\frac{V_{,\Phi}}{V}=0,
\label{loopField}
\end{equation}
where $A\equiv (12/7)^{12}$. Then, the effective energy density (\ref{looprho2}) becomes
\begin{equation}
\rho_{\Phi}^{\rm sc}=\frac{V(\Phi)}{\sqrt{1-A^{-1}q^{-15}\dot{\Phi}^2}}~,
\label{hamiltonianL}
\end{equation}
and for the effective pressure (\ref{loopP2}),  we get
\begin{equation}
-p_{\Phi}^{\rm sc}=\frac{V(\Phi)}{\sqrt{1-A^{-1}q^{-15}\dot{\Phi}^2}}\left[1+\frac{4\dot{\Phi}^2}{Aq^{15}}\right].
\label{RayChad}
\end{equation}
Subsequently, an effective equation of state can be suggested for
this limit, where
\begin{equation}
w_{\Phi}^{\rm sc}\ =\
-\left[1+\frac{4\dot{\Phi}^2}{Aq^{15}}\right]. \label{stateE2}
\end{equation}

Using the classical energy density
$\rho_{\Phi}={|p|}^{-3/2}C_{\Phi}$ in Eq.
(\ref{Henergy-sf}) and substituting in Eq. (\ref{einstein2}),
the classical mass function can be given in terms of the classical
Hamiltonian of the system as
\begin{equation}
F(t,r)\ =\ \frac{1}{3}r^3C_{\Phi}(t)~.
\label{massfuncH}
\end{equation}
From $C_{\Phi}$ in Eq. (\ref{Hamiltonian}) and
substituting in Eq. (\ref{massfuncH}) we get
\begin{equation}
F=\frac{1}{3}r^3{|p|}^{3/2}\sqrt{V^2+{|p|}^{-3}\pi_{\Phi}^2}~.
\label{massfuncH2}
\end{equation}
Replacing the classical geometrical density term
$|p|^{-1}$ by $d_{j,l}({\mathsf{p}})$ in Eq.
(\ref{massfuncH2}) the modified mass function in the semiclassical
regime for a tachyon field collapse is now determined as follows
\begin{equation}
F_{\rm sc}=\frac{1}{3}r^3{|p|}^{3/2}\sqrt{V^2+|d_{j,l}(|p|)|^3\pi_{\Phi}^2}~.
\label{massfuncH3}
\end{equation}
Using equation $\dot{\Phi}=\partial C_{\Phi}^{\rm sc}/\partial
\pi_{\Phi}$, we  may easily get
\begin{equation}
F_{\rm sc}=\frac{1}{3}r^3a^3\frac{Vq^{3/2}|D(q)|^{3/2}}{\sqrt{q^{3} |D(q)|^3-\dot{\Phi}^2}}~,
\label{massfuncH4}
\end{equation}
where,  using  Eq. (\ref{looprho2}),  it is rewritten as
$F_{\rm sc}=1/3\rho_{\Phi}^{\rm sc}R^3$, with $R(t,r)=a(t)r$. Moreover,
within the regime $a\ll a_*$, the modified mass function
(\ref{massfuncH4}) takes the form
\begin{equation}
F_{\rm sc}=\frac{1}{3}r^3\frac{a^{3}V(\Phi)}{\sqrt{1-A^{-1}q^{-15}\dot{\Phi}^2}}~.
\label{massfunc}
\end{equation}
The behavior of this effective mass function determines whether
trapped surfaces will form during  the collapse procedure, within
the semiclassical regime.

In order to solve  the modified Klein-Gordon equation (\ref{loopField}), we need to introduce the potential $V(\Phi)$ for the tachyon field. 
Let us follow the  Hamilton-Jacobi formulation to discuss the solutions for tachyon collapse in semiclassical regime. 
If the tachyon field is a monotonically varying
function of the proper time, then Eq.~(\ref{hamiltonianL}) can be written in a Hamilton-Jacobi form: 
\begin{equation}
V^2(\Phi)=9H^4\left[1-\frac{1}{16\tilde{\beta}}a^{30}H_{\Phi}^2\right],
\label{HJ}
\end{equation}
where $H_{\Phi}$ is defined as  \cite{ASen2006}, 
\begin{equation}
H_{\Phi}\  :=\   \frac{2H_{,\Phi}}{3H^2}\  .
\label{HJ2}
\end{equation}
Let us  therefore assume a Hubble parameter of the form 
\begin{equation}
H(\Phi)\  =\  H_1\Phi^n \ , 
\label{HubbleS1-tach}
\end{equation}
(where $H_1<0$, and $n<0$ are constants), describing a collapsing model  (see e.g., \cite{Lidsey:2004}).
Then, integrating $H_{,\Phi}$, the scale factor
is obtained as a function of the tachyon field:
\begin{equation}
a^{30}(\Phi)=B\Phi^{2n+2},  \label{Hubble2}
\end{equation}
where $n\neq-1$ and $B$ 
is a positive
constant. 
Then, Eq.~(\ref{HJ}) implies that the potential, as a function of the
tachyon field, has the form
\begin{equation}
V(\Phi)\ =\  \bar{V}_0\Phi^{2n}, \label{Hubble3}
\end{equation}
with the constant $\bar{V}_0$.

For $n=-1$,  the Hubble parameter 
takes instead
the form
$H=H_1\Phi^{-1}$, 
for which the scale factor
can be obtained as
\begin{equation}
a^{30}(\Phi)\ =\  180\tilde{\beta} H_1^2\ln\Phi, 
\label{Hubble2}
\end{equation}
where $1<\Phi<\Phi_0$ and $\tilde{\beta}$ is a constant. 
The potential of the system can be
established from Eq.~(\ref{HJ}),  as
\begin{equation}
V(\Phi)=\frac{3H_1^2}{\Phi^2}\left(1-5\ln\Phi\right)^{\frac{1}{2}}.
\label{HJPotential}
\end{equation}
In this case, the initial value of the potential is given by
$\bar{V}_0=3H_1^2(1-5\ln\Phi)^{1/2}/\bar{\Phi}_0^2$, as
$a\rightarrow a_0$. On the other hand, the potential of the system
at the semiclassical limit (i.e., $a\ll a_*$) behaves as an inverse square function of
the tachyon field: 
\begin{equation}
V(\Phi)\  \approx \  3H_1^2~\Phi^{-2}. 
\label{HJ-Pot}
\end{equation}
This result implies that, at the semiclassical limit, the choice of  inverse square potential is a good approximation for the tachyon potential.
Meanwhile, for tachyon potential of the general form  $V(\Phi)\simeq \Phi^n$,  the dynamical system is much more complicated. 
If the potential is of inverse square form, it  allows  a three-dimensional autonomous system to be extracted, whereas for more general cases of the potential, the number of dimensions would become higher if the system is to remain autonomous.
Therefore, in order to make the phase space analysis  tractable, we assume an inverse square potential for the tachyon field.
In addition, exact solutions can be found for a classical purely tachyonic matter content  (cf. Ref.~\cite{Padmanabhan:2002a}), for cases which combine tachyonic and barotropic fluids 
(cf. Refs.~\cite{RLazkoz2004,Tavakoli2013a}),  and for the cases with the loop quantum correction terms (cf. Ref.~\cite{Huang}), with an inverse square potential are known. On the next subsection, we study semiclassical collapse in the presence of tachyon field and a barotropic fluid by employing a dynamical system analysis.

\subsection{Tachyon matter coupled with barotropic fluid: Phase space analysis}

We follow the section \ref{collapse-gen} 
and take for space-time geometry an homogeneous
interior, matched to a suitable (inhomogeneous) exterior
 geometry to provide the whole space-time structure. More precisely,   the interior space-time is the
marginally bound case  and is parameterized by
the line element 
(\ref{metric}).
We will consider as matter content for interior a spherically symmetric homogeneous tachyon field together
with a barotropic fluid. We use an inverse
square potential $V = V_0~\Phi^{-2}$  for the tachyon field, given by Eq.~(\ref{HJ-Pot}).  
We designate by  $\rho_{b}$  the energy density of the
classical barotropic matter, whose pressure $p_{b}$, in terms
of barotropic parameter $\gamma_b$, satisfies the relation
$p_{b}=(\gamma_b-1)\rho_{b} \equiv w_b \rho_{b}$, where $\gamma_b>0$
and $\rho_{b} = \rho_{0b}a^{-3(1+w_{b})}$. The total energy density  is therefore, without loop elements
$\rho=\rho_{\phi}+\rho_{b}$, given by Eq.~(\ref{energy}).

Let us now  add a specific type of non-perturbative modifications to the dynamics,
motivated by LQG. In the semiclassical framework of loop  corrections of the inverse
triad type, we have an effective energy density for
the barotropic fluid given as $\rho_{b}^{\rm sc}=D(q)^{w_{b}}\rho_{b}$ 
\cite{Singh2005b},  or
\begin{equation}
\rho_{b}^{\rm sc}\ =\ \rho_{0b}D^{(\gamma_b-1)}a^{-3\gamma_b}. \label{baroloop}
\end{equation}

 Now, the total energy density will be
 given by the loop modified energy density of the tachyon
field plus that for the  fluid, i.e.,
$\rho_{\rm sc}=\rho_{\Phi}^{\rm sc}+\rho_{b}^{\rm sc}$. As a
consequence, the corresponding constraint equation follows that
\begin{equation}
3H^{2}\ =\
\rho_{\Phi}^{\rm sc}+\rho_{b}^{\rm sc}.\label{baroFried}
\end{equation}
The Raychaudhury equation becomes
\begin{equation}
2\dot{H}+3H^{2}\ =\ p_{\Phi}^{\rm sc}-
w_{b}\left(1-\frac{1}{3}\frac{d\ln D}{d\ln
a}\right)\rho_{b}^{\rm sc}.\label{baroRayChad}
\end{equation}
Moreover, from Eq.~(\ref{baroRayChad}), considering that
$2\dot{H}+3H^{2}=-p_{b}^{\rm sc}$, an effective  equation of
state  in this semiclassical regime will be
\begin{equation}
w_{b}^{\rm sc}\ =\ w_{b}\left(1-\frac{1}{3}\frac{d\ln D}{d\ln a}\right).\label{baroRayChad3}
\end{equation}
 From Eq.~(\ref{baroRayChad3}), it is seen that, if we rewrite
$w_{b}^{\rm sc}$ similar to the classical expression, as
$w_{b}^{\rm sc}=(\tilde{\gamma}_b-1)$,  we get
\begin{equation}
\tilde{\gamma}_b \ \simeq\ \frac{8-5\gamma_b}{3}\  .
\end{equation}
In the semiclassical region, where $D(q)\ll1$, from Eq.~(\ref{baroRayChad}) we further have
\begin{equation}
2\dot{H}\ =\
\frac{4V(\Phi)\dot{\Phi}^{2}/(Aq^{15})}{\sqrt{1-A^{-1}q^{-15}\dot{\Phi}^{2}}}-
\tilde{\gamma}_b\rho_{b}^{\rm sc}.\label{baroRayChad2}
\end{equation}

In what follows, we will study our collapsing model, using a
dynamical system description \cite{JCar1981, Khalil1996, Guckenheimer1983} for the Eqs.~(\ref{loopField})-(\ref{baroRayChad2}). We use a new time variable
$N$ (instead of the proper time $t$ in the comoving coordinate
system $\{t,r,\theta,\varphi\}$). In more concrete terms, we choose
\begin{equation}
N\ := \ -\log q^{3/2},\label{n-time}
\end{equation}
with $q$  being defined in the interval $0<N<\infty$;
the limit $N\rightarrow0$ corresponds to the initial condition of
the collapsing system ($a\rightarrow a_{*}$) and the limit
$N\rightarrow\infty$ corresponds to $a\rightarrow0$. For any time
dependent function $f$,
\begin{equation}
\frac{df}{dN}\ =\ -\frac{\dot{f}}{3H}~.\label{dyn1-a}
\end{equation}
We further  use a set of new dynamical variables:
\begin{align}
X\  & := \ \frac{\dot{\Phi}}{A^{\frac{1}{2}}q^{\frac{15}{2}}}\ , \  \  \  \ \  \  \ \  \  \  Y\ := \ \frac{V}{3H^{2}}\ ,\ \  \  \  \  \ \  \  \ \ Z\ := \ A^{\frac{1}{2}}q^{\frac{15}{2}}\ ,\notag\\
S\  & := \ \frac{\rho_{b}^{\rm sc}}{3H^{2}}\  ,\ \  \  \  \  \ \  \  \  \  \ \  \ \xi\ := \ -\frac{V_{,\Phi}}{V^{\frac{3}{2}}}\  ,\  \  \ \  \  \  \  \  \   \ \Gamma\ := \ \frac{VV_{,\Phi\Phi}}{(V_{,\Phi})^{2}}\ , \label{Var-Q}
\end{align}
 in which $S$ can also be written
as
\begin{equation}
S\  =\ D^{w_{b}}\left(\frac{\rho_{b}}{3H^{2}}\right),\label{baroDyn}
\end{equation}
such that in the limit $D(q)\ll1$, it reduces to
\begin{equation}
S \simeq
\left(\frac{12}{7}\right)^{4w_{b}}q^{4w_{b}}\left(\frac{\rho_{b}}{3H^{2}}\right).\label{baroDyn2}
\end{equation}
An autonomous system of equations, in terms of the  dynamical
variables (\ref{Var-Q}), for Eqs. (\ref{loopField}) and
(\ref{baroRayChad2}), is then retrieved:
\begin{align}
\frac{dX}{dN}  & \  =\  X(4X^{2}-9)+\frac{1}{\sqrt{3}}\xi Z\sqrt{Y}(1-X^{2}),\label{dynXbaro}\\
\frac{dY}{dN} & \  =\  Y\left[4X^{2}-\frac{\xi}{\sqrt{3}} XZ\sqrt{Y}-(\tilde{\gamma}_b+4X^{2})S\right],\label{dynYbaro}\\
\frac{dZ}{dN} & \ =\  -5Z,\label{dynZbaro}
\end{align}
\begin{align}
\frac{dS}{dN} & \  =\  S(1-S)(\tilde{\gamma}_b+4X^{2}),\label{baroDynSys1}\\
\frac{d\xi}{dN} & \ =\  -\frac{1}{\sqrt{3}}\xi^{2}XZ\sqrt{Y}\left(\Gamma-\frac{3}{2}\right).\label{dynLambdabaro}
\end{align}
For  $V(\Phi)$ as in Eq.~(\ref{potential}), it brings
$\xi=\pm2/\sqrt{V_0}$ and $\Gamma=3/2$. i.e., as constants. Then,
the dynamical system reduces to four differential equations with
variables $(X,Y,Z,S)$, namely Eqs.~(\ref{dynXbaro})-(\ref{baroDynSys1}).
Equation (\ref{baroFried}), in terms of the new variables, can be
written as
\begin{equation}
\frac{Y}{\sqrt{1-X^{2}}}+S=1,\label{Constraintbaro}
\end{equation}
 in which $Y\geq0$ and $-1\leq X\leq1$ and $0\leq S\leq1$.

A discussion of the autonomous system of Eqs.~
(\ref{dynXbaro})-(\ref{baroDynSys1}) requires to identify the
critical points $(X_{c},Y_{c},Z_{c},S_{c})$; the properties of each critical point (and
associated stability features) are determined by the eigenvalues
of the corresponding $(4\times4)$-Jacobi matrix
${\cal B}$. Setting therefore
$(f_{1},f_{2},f_{3},f_{4})|_{(X_{c},Y_{c},Z_{c},S_{c})}=0$, we can
obtain them, where we have defined $f_1\equiv dX/dN$, $f_2 \equiv  dY/dN$, $f_3\equiv dZ/dN $, $f_4\equiv dS/dN $. The
eigenvalues, defined at each fixed point
$(X_{c},Y_{c},Z_{c},S_{c})$, are then brought from
\begin{equation}
{\cal B}=\left(\begin{array}{cccc}
\frac{\partial f_{1}}{\partial X} & \frac{\partial f_{1}}{\partial Y} & \frac{\partial f_{1}}{\partial Z} & \frac{\partial f_{1}}{\partial S}\\
\frac{\partial f_{2}}{\partial X} & \frac{\partial f_{2}}{\partial Y} & \frac{\partial f_{2}}{\partial Z} & \frac{\partial f_{2}}{\partial S}\\
\frac{\partial f_{3}}{\partial X} & \frac{\partial f_{3}}{\partial Y} & \frac{\partial f_{3}}{\partial Z} & \frac{\partial f_{3}}{\partial S}\\
\frac{\partial f_{4}}{\partial X} & \frac{\partial f_{4}}{\partial Y} & \frac{\partial f_{4}}{\partial Z} & \frac{\partial f_{4}}{\partial S}
\end{array}\right)_{|(X_{c},Y_{c},Z_{c},S_{c})}.\label{matrix2D}
\end{equation}
Physical solutions in the neighborhood
of a critical point, $Q_{i}^{c}$, can be extracted by
making use of
\begin{align}
Q_{i}(t)\ =\ Q_{i}^{c}+\delta Q_{i}(t),
\label{stab-1}
\end{align}
 with the perturbation $\delta Q_{i}$ given by
\begin{align}
\delta Q_{i}\ =\ \sum_{j}^{k}(Q_{0})_{i}^{j}\exp(\zeta_{j}N),\label{pertQ1} 
\end{align}
 where $Q_{i}\equiv\{X,Y,Z,S\}$, and $\zeta_{j}$ are the eigenvalues
of the Jacobi matrix; the $(Q_{0})_{i}^{j}$ are constants of integration.
We have summarized the fixed points for the autonomous system and
their stability properties in the following table.
\begin{table} [h!]
\begin{center}
\caption{Critical points and their properties.} \label{fixed-points}
\begin{tabular} { ccccccccc }
\hline \hline
~Point~ & ~$X$~ &~$Y$~ & ~$Z$~ & $S$~ & ~$\Omega_\Phi$~ & ~$\gamma_{\Phi}$~ & ~~~~~~~~~Existence~~~~~~~~~ & ~~Stability~~ \\
\hline

$(A)$ & $ 0$  & $1$ & $0$ & $0$ &$1$ & $0$ & {\small All $\xi$; ~$\tilde{\gamma}_b\leq0$ (i.e. $\gamma_b\geq\frac{8}{5}$)} &  {\small Stable}\\

$$ & $ $  & $$ & $$ & $$ &$$ & $$ &  {\small All $\xi$;  ~$\tilde{\gamma}_b>0$ (i.e. $\gamma_b<\frac{8}{5}$)} &  {\small Saddle point}\\

$(B^{+})$ & $\frac{3}{2}$ & $0$ & $0$ & $1$  & $1$ & $-9$ &  {\small All $\xi,~ \tilde{\gamma}_b$} &  {\small Saddle point}\\

$(B^{-})$ & $-\frac{3}{2}$ & $0$ & $0$ & $1$ & $1$ & $-9$ &  {\small All $\xi,~\tilde{\gamma}_b$} &  {\small Saddle point}\\

$(C)$ & $ 0$  & $0$ & $0$ & $1$ &$0$ & $0$ &   {\small All $\xi$; ~ $\tilde{\gamma}_b=0$} (i.e. $\gamma_b=\frac{8}{5}$) &  {\small Stable} \\

$$ & $ $  & $$ & $$ & $$ &$$ & $$ &   {\small All $\xi$; ~$\tilde{\gamma}_b\neq0$ (i.e. $\gamma_b\neq\frac{8}{5}$)} &  {\small Saddle point}\\

\hline \hline
\end{tabular}
\end{center}
\end{table}
In more detail:

For point $(A)$ as $(0,1,0,0)$, we have to determine the eigenvalues (and eigenvectors) of the
matrix (\ref{matrix2D}).
Using equations (\ref{dynXbaro})-(\ref{dynLambdabaro}) and (\ref{matrix2D}),
matrix ${\cal B}$ becomes
\begin{equation}
{\cal B}=\left(\begin{array}{cccc}
-9 & 0 & \frac{\xi}{\sqrt{3}} & 0\\
0 & 0 & 0 & -\tilde{\gamma}_b\\
0 & 0 & -5 & 0\\
0 & 0 & 0 & \tilde{\gamma}_b \\
\end{array}\right).
\label{matrixA2}
\end{equation}
Eigenvalues of the matrix (\ref{matrixA2}) are $\sigma_1=-9$, $\sigma_2=0$, $\sigma_3=-5$, and $\sigma_4=\tilde{\gamma}_b$. Therefore, all eigenvalues are real but one  is  zero, and the rest being negative when $\gamma>8/5$; this implies  that this is a nonlinear autonomous system
with a non-hyperbolic point.   The  asymptotic
properties cannot be simply determined by  linearization  and we
need to resort to the \emph{center manifold theorem} \cite{JCar1981}. For convenience, we transform the critical point
$A(X_c=0, Y_c=1, Z_c=0, S_c=0)$ to $\tilde{A}(X_c=0, \tilde{Y}_c=Y_c-1=0,
Z_c=0, S_c=0)$. The autonomous system (\ref{dynXbaro})-(\ref{dynLambdabaro}) is
rewritten as
\begin{align}
\frac{dX}{dN} &\  =\   X\left(4X^2-9\right)+\frac{1}{\sqrt{3}}\xi Z\sqrt{\tilde{Y}+1}\left(1-X^2\right),
\label{dynX1} \\
\frac{d\tilde{Y}}{dN} &
\  =\   -\frac{1}{\sqrt{3}}\xi XZ\left(\tilde{Y}+1\right)^{3/2} + \left(\tilde{Y}+1\right)\left[4X^{2} -\left(\tilde{\gamma}_b+4X^{2}\right)S\right], \label{dynY1} \\
\frac{dZ}{dN} &\   =\  -5Z, \label{dynZ1} \\
\frac{dS}{dN} & \  =\   S(1-S)\left(\tilde{\gamma}_b+4X^2\right). \label{dynS1}
\end{align}
Let  ${\cal M}$ be a matrix whose columns are the
eigenvectors of ${\cal B}$; whence, for the matrix (\ref{matrixA2}) we obtain  ${\cal M}$ and its inverse matrix,  ${\cal M}^{-1}$, as
\begin{equation}
{\cal M}=\left(\begin{array}{cccc}
1 & 0 & \frac{\xi}{4\sqrt{3}} & 0\\
0 & 1 & 0 & -1\\
0 & 0 & 1 & 0\\
0 & 0 & 0 & 1\\
\end{array}\right), \  \  \  \  \   \   \   {\rm and}\  \  \  \   \  \  \
{\cal T} := {\cal M}^{-1}=\left(\begin{array}{cccc}
1 & 0 & -\frac{\xi}{4\sqrt{3}} & 0\\
0 & 1 & 0 & 1\\
0 & 0 & 1 & 0\\
0 & 0 & 0 & 1\\
\end{array}\right).
\end{equation}
Using the similarity transformation $ {\cal T}{\cal B}{\cal T}^{-1}$, the matrix ${\cal
B}$ can be rewritten as a block diagonal form:
\begin{equation}
\widetilde{{\cal B}} := {\cal T}{\cal B}{\cal T}^{-1}=\left(\begin{array}{cccc}
-9 & 0 & 0 & 0\\
0 & 0 & 0 & 0\\
0 & 0 & -5 & 0\\
0 & 0 & 0 & \tilde{\gamma}_b\\
\end{array}\right)=\left(\begin{array}{cc}
\widetilde{{\cal B}}_1 & 0 \\
0 & \widetilde{{\cal B}}_2 \\
\end{array}\right),
\label{matrixDA}
\end{equation}
where $\widetilde{{\cal B}}_1$ is the matrix whose all eigenvalues have zero real parts, and all eigenvalues of $\widetilde{{\cal B}}_2$ have negative real parts.   Let us change the variables as bellow:
\begin{equation}
\left(\begin{array}{c}
X'\\
Y'\\
Z'\\
S'\\
\end{array}\right)\ := \ {\cal T}\left(\begin{array}{c}
X\\
\tilde{Y}\\
Z\\
S\\
\end{array}\right)=\left(\begin{array}{c}
X+\frac{\xi}{4\sqrt{3}}Z\\
\tilde{Y}+S\\
Z\\
S\\
\end{array}\right).
\label{NewVar}
\end{equation}
Then, the dynamical system (\ref{dynXbaro})-(\ref{dynLambdabaro}) in terms of the new
variables $(X', Y', Z', S')$ becomes,
\begin{align}
\frac{dX'}{dN} &\ =\  \frac{dX}{dN}+\frac{\xi}{4\sqrt{3}}\frac{dZ}{dN} = f'_1(X', Y', Z', S'), \label{dynX2} \\
\frac{dY'}{dN} &\ =\  \frac{d\tilde{Y}}{dN}+\frac{dS}{dN} = f'_2(X', Y', Z', S'), \label{dynY2} \\
\frac{dZ'}{dN} &\ =\  \frac{dZ}{dN}= -5Z' , \label{dynZ2} \\
\frac{dS'}{dN} &\  =\   S'\left(1-S'\right)\left[\tilde{\gamma}_b+4\left(X'-\frac{\xi}{4\sqrt{3}}Z'\right)^2\right], \label{dynS2}
\end{align}
where $f'_1(X', Y', Z', S')$ and $f'_2(X', Y', Z', S')$ are given by
\begin{align}
f'_1 & :=  \left(X'-\frac{\xi}{4\sqrt{3}}Z'\right)\left[4\left(X'-\frac{\xi}{4\sqrt{3}}Z'\right)^2-9\right] - \frac{5\xi}{4\sqrt{3}}Z'  \notag \\
&\ \ \ \ \ \  + \frac{\xi}{\sqrt{3}} Z'\left(Y'-S'+1\right)^{\frac{1}{2}} \left[1-\left(X'-\frac{\xi}{4\sqrt{3}}Z'\right)^2\right],
\label{dynXprime} \\
f'_2 & := \frac{1}{\sqrt{3}}\xi Z'(Y'-S'+1)^{3/2}\left(X'-\frac{\xi}{4\sqrt{3}}Z'\right) + 4(Y'-S'+1)\left(X'-\frac{\xi}{4\sqrt{3}}Z'\right)^2 \notag \\
&\ \ \  \  \  \  \ \ + S'(1-S')\left[\tilde{\gamma}_b+4\left(X'-\frac{\xi}{4\sqrt{3}}Z'\right)^2\right].
\label{dynYprime}
\end{align}
In particular, Eqs.~(\ref{dynX2})-(\ref{dynS2}) have the point of
equilibrium $X'_c=Y'_c=Z'_c=S'_c=0$. Let us now suppose that we take
initial data with $Z'=0$ and $S'=0$. Then, $Z'(t)$ and $S'(t)$ are zero for all times and we
examine the stability
of the equilibrium $X'_c=0$ and
$Y'_c=0$:
A description of the center manifold
(cf.  Ref. \cite{Khalil1996} for more details), is usually retrieved with the
assistance of functions  $h(Y')$, $g(Y')$ and $f(Y')$, where $X' \equiv
h(Y')$, $Z' \equiv g(Y')$ and $S' \equiv  f(Y')$. Let us  assume the solution for $h$, $g$ and $f$
to be approximated arbitrary closely as a Taylor series in $Y'$.
Using this method, we can start with the simplest approximation
$h(Y')\approx 0$ and $f(Y')\approx 0$. Then the reduced system, obtained by substituting
$h(Y')={\cal O}(|Y|^2)$, $g(Y')\approx 0$ and $f(Y')\approx 0$ in Eqs.
(\ref{dynX2})-(\ref{dynZ2}), is $dY'/dN={\cal O}(|Y'|^4)$. Clearly,
this cannot assist in reaching any conclusion about the stability.
Therefore, we calculate the coefficient $h_2$ in the series $h(Y')=
h_2Y'^2+{\cal O}(|Y'|^3)$ and study the stability of the origin. The
reduced system then reads
\begin{equation}
\frac{dY'}{dN}\ =\ 4h_2^2Y'^4+{\cal O}(|Y'|^5)~. \label{ReddynY2}
\end{equation}
To get the value of $h_2$, we should solve the equation
\cite{Khalil1996}
\begin{equation}
{\cal N}\left(h(Y')\right)\ :=\  (\partial h/\partial Y')
f'_2-f'_1=0. \label{ReddynY3}
\end{equation}
But this leads to  all coefficients  being zero valued in the series
$h(Y')$. Considering instead  $Z'\equiv g(Y')\neq 0$, this  will not
reach any result, because all coefficients $g_i$ in the series
$g(Y')$ are zero as well. Hence we just focus on an analysis of the
stability of this fixed point from a qualitative perspective  for
the reduced system:
\begin{equation}
\frac{dY'}{dN}\  =\  4X'^2(1+Y').
\label{ReddynY}
\end{equation}
Clearly (for initial data near the origin) $X'(t)$ converges
exponentially  to zero, say, approximately as $\exp(-9N)$. Since
$Y'$ is monotonous with $N$, then the reduced system $dY'/dN \approx
4Y'\exp(-18N)$  will converge as $N$ gets larger, but in general the
limit is not zero. Thus, the critical point $(A)$ is asymptotically
\emph{stable}.

On the other hand, for the case of barotropic parameter $\gamma=8/5$, the set of eigenvalues of point $(A)$ includes two negative real parts and two zero real parts. In order to analyse the stability of the system in this case, we can consider an additional condition for non-vanishing initial data $S'\neq0$ where $Z'\approx0$; so that, the corresponding reduced system for $S'(t)$ can be written from (\ref{dynS2}):
\begin{align}
\frac{dS'}{dN}\ =\  4 S'\left(1-S'\right)X'^2.
\label{dynS2-2}
\end{align}
Since $S'$ is monotonous with $N$, and $0\leq S'\leq1$, then the reduced system $dS'/dN \approx 4S'\exp(-18N)$  will converge as $N$ gets larger; therefore, the fixed point $(A)$ for the case  $\gamma=8/5$ is asymptotically \emph{stable}.
On the other hand, for $\tilde{\gamma}_b>0$ (i.e. $\gamma_b<\frac{8}{5}$), one eigenvalue is zero, another is positive and two others are negative; hence, a \emph{saddle} point setting will be recovered.
In the limit case in which $\tilde{\gamma}_b=0$, a \emph{stable} point behavior can likewise be shown to emerge. 

For the fixed point $(B^{+})$ the eigenvalues  are
$\sigma_{1}=18$, $\sigma_{2}=9$, $\sigma_{3}=-5$, and $\sigma_{4}=\tilde{\gamma}_b+9$.
For all values of $\tilde{\gamma}_b$, two characteristic values are
positive and others are negative. Thus, this fixed point is a \emph{saddle}
point.

For the fixed point $(B^{-})$, the characteristic
values are $\sigma_{1}=18$, $\sigma_{2}=9$, $\sigma_{3}=-5$ and
$\sigma_{4}=\tilde{\gamma}_b+9$, which are the same eigenvalues as
the fixed point $(B^{+})$, and thus, similarly to $B^{+}$, this
is a \emph{saddle} point.
  

 Finally, for the fixed point $(C)$ the eigenvalues are $\sigma_1=-9$, $\sigma_2=-\tilde{\gamma}_b$, $\sigma_3=-5$,
  and $\sigma_4=\tilde{\gamma}_b$. For all $\tilde{\gamma}_b\neq0$, $\sigma_2$ and $\sigma_4$ have always opposite signs: This
corresponds to a \emph{saddle} point. For this case
 (i.e. $\gamma_b\neq\frac{8}{5}$), the power of
the exponential term $\delta S$ has a different sign with respect to
the others and then, by assuming $\tilde{\gamma}_b>0$, the term
$\delta S$ increases as $N$ increases (i.e., $a$ decreases).
For the  case $\tilde{\gamma}_b=0$  (i.e. for the corresponding
 barotropic parameter $\gamma_b=8/5$),
$\sigma_{2}$ and $\sigma_{4}$ are zero whereas two others are
negative. Using the centre manifold theorem (cf. point $(A)$)\footnote{It should be noticed that, in the case of parameter  $\gamma=8/5$, the eigenvalues of two fixed points $(A)$ and $(C)$ become equal. Therefore, by employing a similar analysis for the fixed point $(C)$ we can show that, this fixed point is asymptotically \emph{stable} for  the case of barotropic parameter $\gamma=8/5$}, it can be shown that it
corresponds to a \emph{stable} fixed point: Using Eq.~(\ref{pertQ1}), solutions in terms of the dynamical variables $X$,
$Y$, $Z$ and $S$ are given, respectively, by $\delta
X\approx\exp(-9N)=(a/a_{*})^{27}$, $\delta
Y\approx\exp(-\tilde{\gamma}_bN)=(a/a_{*})^{3\tilde{\gamma}_b}$, $\delta
Z\approx\exp(-5N)=(a/a_{*})^{15}$ and $\delta
S\approx\exp(\tilde{\gamma}_bN)=(a/a_{*})^{-3\tilde{\gamma}_b}$.

In summary, our phase space analysis showed that, there are two solutions which are of
particular interest concerning  gravitational collapse  whose asymptotic behaviors are stable at the late stages of the collapse (where scale factor is small):
They are the case $\gamma_b>\frac{8}{5}$ (for the fixed point $(A)$),
and the case in which $\gamma_b=\frac{8}{5}$ (for the fixed point $(C)$). In
the following, we will discuss these two solutions.

\subsubsection{Tachyon dominated solutions}

Let us now study the behavior of the system near the asymptotic solution (for point $(A)$) when $\gamma\geq 8/5$.
From Eq.~(\ref{pertQ1}), we can
find the perturbation around the fixed point, by using
$X(t)=X_{c}+\delta X$, $Y(t)=Y_{c}+\delta Y$, $Z(t)=Z_{c}+\delta Z$,
$S(t)=S_{c}+\delta S$, from which we can  write
\begin{align}
X(t)\ &\approx\ \left(\frac{a}{a_{*}}\right)^{27},\ \ \ \ \ \  \ \ \ \ \  \ \ \ \ \    \ \ \ \ \  \ \ \ \ \    Y(t)\approx1, \notag \\
Z(t)\ & \approx\ A^{1/2}\left(\frac{a}{a_{*}}\right)^{15},\ \ \  \ \ \ \ \  \ \ \ \ \    \ \ \ \ \   S(t)\approx\left(\frac{a}{a_{*}}\right)^{-3\tilde{\gamma}_b}.
\label{pertQB}
\end{align}
 In this neighborhood, the effective energy density of tachyon field
is given by
\begin{align}
\rho_{\Phi}^{\rm sc}\ \approx\ V(\Phi)\left(1+\frac{1}{2}\frac{\dot{\Phi}^{2}}{Aq^{15}}\right).\label{sem-tach}
\end{align}
In addition, the energy density of barotropic matter is modified by the loop
parameter $D(q)^{(\gamma_b-1)}$ (cf. Eq.~(\ref{baroloop})) and can be
approximated as
\begin{align}
\rho_{b}^{\rm sc}\  \approx\   \left(\frac{a}{a_{*}}\right)^{(5\gamma_b-8)}.
\label{baro-sem}
\end{align}

For this solution,  $\dot{\Phi}$
decreases very fast as $\dot{\Phi}\propto (a/a_*)^{42}$ as the scale factor decreases; in this approximation, for very small values of  $a$, the second term in right
hand side of Eq.~(\ref{sem-tach}) evolves as
$\dot{\Phi}^{2}/Aq^{15}\propto (a/a_*)^{54}$ and decreases faster than
$\dot{\Phi}$ and becomes negligible during the final stages of the collapse. We can then analyze as follows.
%

\begin{figure*}
\includegraphics[width=0.32\textwidth]{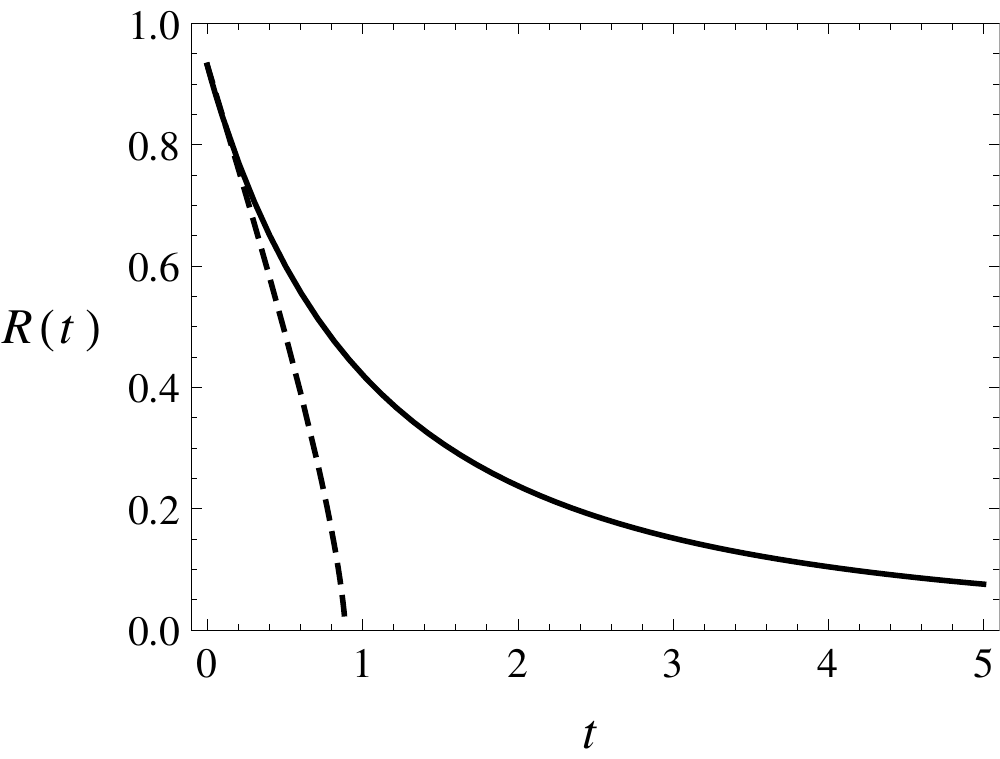}\quad{}\includegraphics[width=0.31\textwidth]{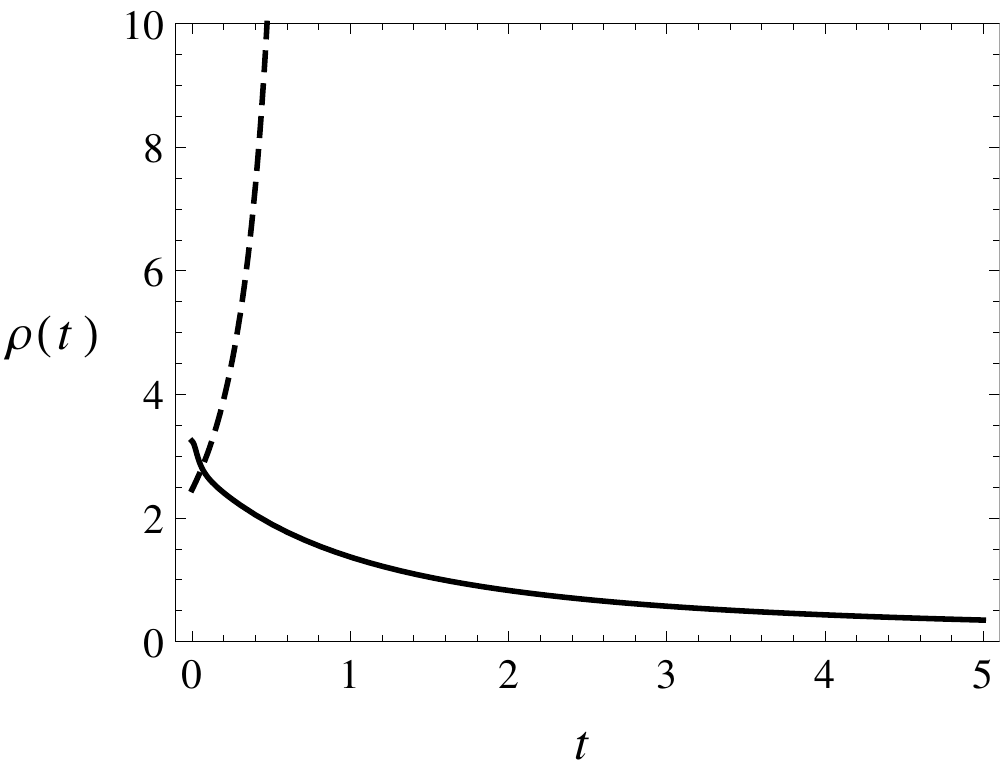}\quad{}\includegraphics[width=0.3\textwidth]{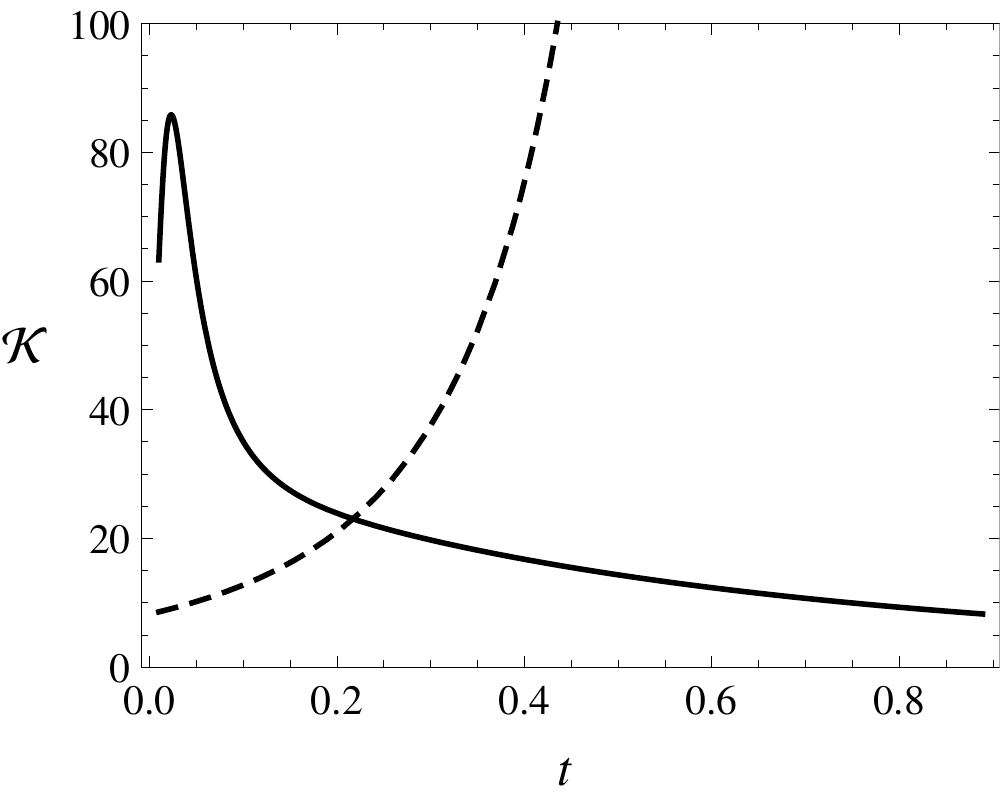}
\caption{ Behaviours of the area radius, energy density and Kretschmann
scalars (with loop corrections - solid line) against classical ones
(dashed line). We considered $t_{0}=0$, $a(0)=a_{*}$, $V_{0}=1/3$,
$\Phi_{0}=-0.6$ and $\gamma_b=\frac{8}{5}$~.}
\label{F2}
\end{figure*}

This solution presents semiclassical effects that modify the energy density of the
barotropic matter as well as the tachyon field: The energy density
of the tachyon field is determined by Eq.~(\ref{sem-tach}), and
the energy density of barotropic matter is given by Eq.~(\ref{baro-sem}).
This solution  shows that the loop correction term
$D(q)$ scales down the effect of the barotropic fluid and avoid its
divergence towards the center of star (i.e. for $\gamma_b>\frac{8}{5}$,
then $\rho_{b}^{\rm sc}$ becomes negligible when $a$ is close to the Planck scale). Then,
the collapsing matter content at this point is tachyon dominated,
and the total energy density of the collapse is determined by the
effective energy density of tachyonic matter, given by
Eq.~(\ref{sem-tach}).

On the other hand,  $Y\rightarrow1$ at this point, with
$V\approx3H^{2}$ as well, implying that, in this regime (differently
from its classical counterpart; cf. see chapter \ref{collapse-tachyon}) the potential of the tachyon field
has the main role in determining the effective energy density of the
system $\rho_{\rm sc}\approx V(\Phi)$, where
$\dot{\Phi}^{2}/Aq^{15}\ll1$.
Substituting the potential by $V=V_0\Phi^{-2}$ in $V\approx3H^{2}$ we get
$H(\Phi) \simeq -\sqrt{V_0/3}|\Phi|^{-1}$ for the both $\Phi>0$ and
$\Phi<0$ branches. Integrating for $H(\Phi)$, we can obtain
\begin{align}
a^{42}(\Phi)\ =\
42\sqrt{\frac{V_0}{3}}\ln\left|\frac{\Phi_{f}}{\Phi}\right|,\label{scal-Sem}
\end{align}
 where $\Phi_{f}$ is a constant of integration. For an initial condition
such as $\Phi(0)=\Phi_{0}$, and
$a(0)=a_{*}\approx(\sqrt{V_0/3}\ln|\Phi_{f}/\Phi_{0}|)^{\frac{1}{42}}$,
then the tachyon field approaches the finite value as
$\Phi\rightarrow\Phi_{f}$, when the scale factor is small at about the Planck scale. Thus, the potential of
tachyon field decreases from its initial value, and approaches a
finite value.

Then, using Eqs.~(\ref{sem-tach}) and (\ref{baro-sem}), the total energy of the system in this regime reads
\begin{equation}
\rho_{\rm sc}\ \approx\ \frac{V_0}{\Phi^{2}} + \left(\frac{a}{a_{*}}\right)^{(5\gamma_b-8)}.
\label{ener-sem}
\end{equation}
When the scale factor is small, where $\Phi\rightarrow\Phi_{f}$, the effective energy density of the
fluid, Eq.~(\ref{baro-sem}), is very small, and thus the second term in Eq.~(\ref{ener-sem}) is negligible.
Then, the total energy of the system in this regime is given by the
effective energy density of tachyon field.
It should be noticed that, since $\dot{\Phi}\approx (a/a_*)^{42}>0$, then for a $\Phi>0$ branch, the
tachyon field increases from the initial value $\Phi_0>0$ reaches its maximum at $\Phi_f$.
From Eq.~(\ref{ener-sem}), it is seen that, when the tachyon field changes
in the interval $\Phi_{0}<\Phi<\Phi_{f}$, the total energy density
of the system decreases from its initial value $\rho^{\rm sc}_0$ and reaches its
minimum and finite value at $\rho_{\rm sc} \rightarrow
 V_0/\Phi_{f}^{2}$ for a very small $a$. Thus, (differently from the
classical counterpart in chapter \ref{collapse-tachyon}) the total energy density does not blow up,
becoming finite.

The total mass function
in this regime, can be approximated as
\begin{equation}
\frac{F_{\rm sc}}{R}\ \approx\ \frac{V_0}{\Phi^{2}}\ r^{2}a^{2} + \left(\frac{a}{a_{*}}\right)^{(5\gamma_b-6)}~.
\label{mass-sem}
\end{equation}
Since $\gamma_b\geq\frac{8}{5}$, therefore, for very small values of $S$ the second term in Eq.~(\ref{mass-sem}) is negligible, the mass function behaves as $F_{\rm sc}/R\propto a^{2}$
and decreases towards the center. Moreover, in this region, the total pressure of the
system is approximately given by
\begin{align}
p_{\rm sc}\approx-V(\Phi)\left(1+\frac{9}{2}\frac{\dot{\Phi}^{2}}{Aq^{15}}\right)+\left(\frac{8-5\gamma_b}{3}\right)\rho_{b}^{\rm sc}~,
\label{press-sem}
\end{align}
 which is negative for the semiclassical collapse. The effective pressure
(\ref{press-sem}) evolves asymptotically such that
$p_{\rm s}\approx-V_0/\Phi^{2}$ at scales $a_i$. Thus, it
remains finite towards the late-time stages of the collapse, inducing an outward flux of
energy in the semiclassical regime.

A thorough numerical study allows the following to be additionally mentioned about the case $\gamma_b=\frac{8}{5}$. In figure \ref{F2} the semiclassical area radius (solid line) shows
some deviation from what could be expected classically (see chapter \ref{collapse-tachyon}) in the early
stages of the collapse; the energy density slowly converges to zero
as the area radius gets smaller.

\begin{figure}
\begin{center}
\includegraphics[width=0.5\textwidth]{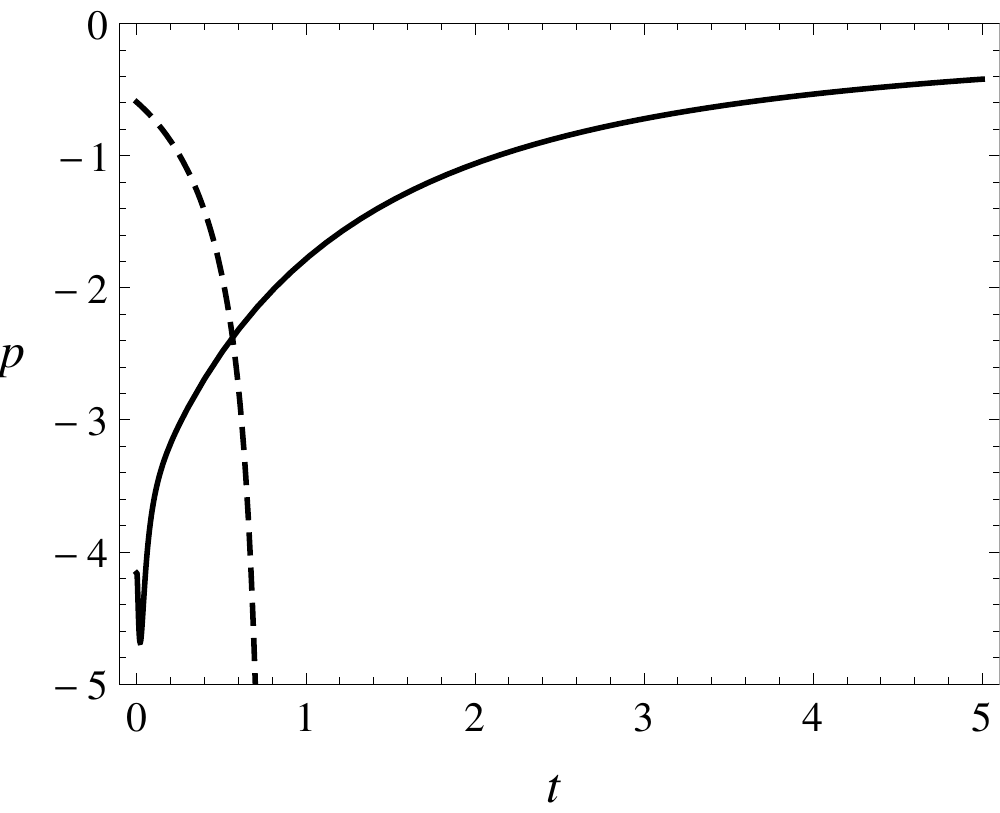}
\end{center}
\caption{ Behaviors of the effective pressure (with loop corrections - solid line) against classical one
(dashed line). We considered $t_{0}=0$, $a(0)=a_{*}$, $V_{0}=1/3$,
$\Phi_{0}=-0.6$ and $\gamma_b=\frac{8}{5}$~.}
\label{F3}
\end{figure}

In order to further investigate  curvature  singularities, we can use scalar polynomials
constructed out of the metric and the Riemann tensors.
An appropriate example is provided by the Kretschmann scalar
$\mathcal{K}=R_{\mu\nu\sigma\eta}R^{\mu\nu\sigma\eta}$ \cite{Joshi2007}, which for the line element
(\ref{metric}), is given by
$\mathcal{K}=12[(\ddot{a}/a)^{2}+(\dot{a}/a)^{4}]$.
The right plot in figure \ref{F2} shows the semiclassical behavior of
Kretschmann scalar (solid line) as a function of proper time.
Therein we observe that in the semiclassical regime, this
quantity remains finite as the physical area radius decreases,
consequently signaling the avoidance of a curvature singularity.
This, together with the
regularity of the energy density, seems to suggest  that the corresponding
space-time of the setting in this subsection is regular as long as this specific semiclassical  scenario is valid.

Figure \ref{F3} shows the semiclassical behavior of the effective pressure indicating that the pressure remains negative during the semiclassical regime.

\begin{figure}
\begin{center}
\includegraphics[width=0.5\textwidth]{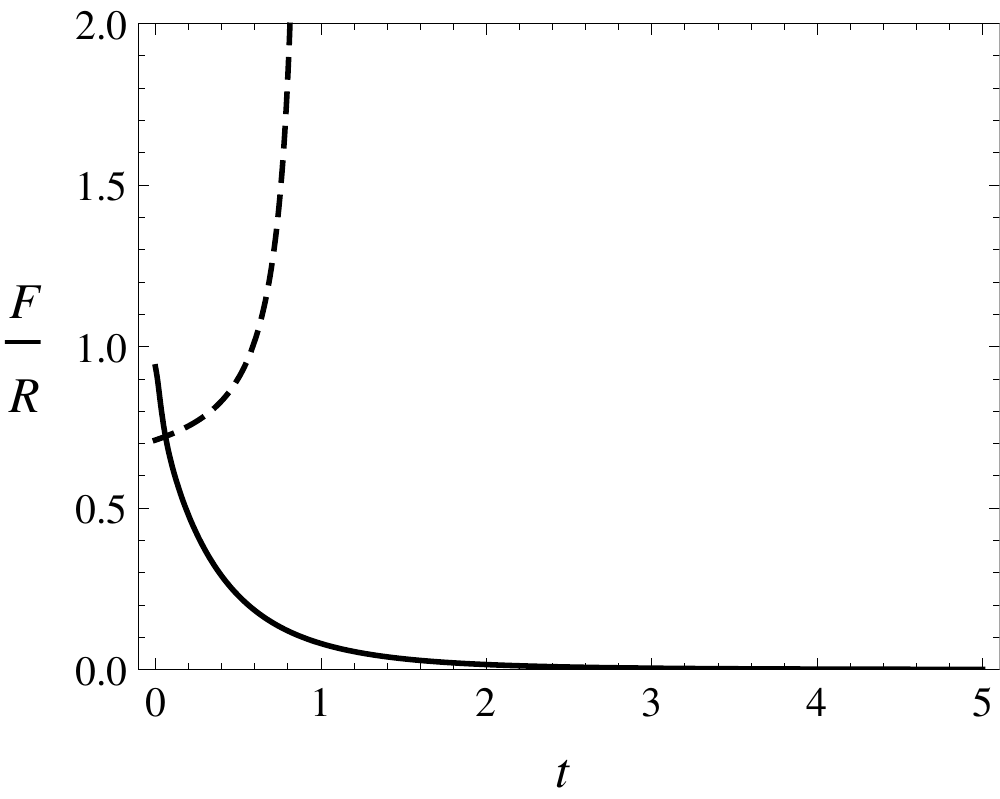}
\end{center}
\caption{ Behaviors of the mass function (with loop corrections - solid line) against classical one
(dashed line). We considered $t_{0}=0$, $a(0)=a_{*}$, $V_{0}=1/3$,
$\Phi_{0}=-0.6$ and $\gamma_b=\frac{8}{5}$~.}
\label{F3b}
\end{figure}

We further depicted the semiclassical (full
line) behavior of the mass function in figure \ref{F3b}, showing that from the early
stages of the collapse this quantity stays
smaller than the area radius, and converges to zero when it approaches the final stage of
the collapse. Therefore, there is no trapped surfaces forming.
Moreover,
inverse triad corrections appear to induce
an outward flux of energy
at the final state of the collapse (cf. figure \ref{F4}) which we will discuss in section \ref{exteriorR}.

From Eq.~(\ref{stateE2}) it is seen that,
$w_\Phi^{\rm sc}<-1$, so the effective equation of state behaves as a phantom matter for which the energy conditions are violated (see similar behavior for a standard scalar field in Ref.~\cite{Bojowald2005b}). However, satisfying the energy conditions is not expected in quantum gravity, but, these conditions must be held in classical collapse (see chapter \ref{collapse-gen}). Furthermore, in order to have a well-defined initial condition for our collapsing model, the initial data for the (classical featured) barotropic parameter $\gamma_b$ must respect the energy conditions. So that,   for the barotropic matter 
in    this case with the parameter $\gamma_b \geq \frac{8}{5}$, the energy conditions are satisfied, which demands $\gamma_b$ to hold 
the range $\frac{8}{5}\leq \gamma_b<2$.

\subsubsection{Barotropic dominated solutions}

For the case with $\tilde{\gamma}_b=0$ (or $\gamma_b=\frac{8}{5}$),
nearby the fixed point solution $(C)$,  the time derivative of
tachyon field vanishes towards the center (i.e. $\dot{\Phi}\rightarrow0$
for very small scale factor $a$), and hence, asymptotically the tachyon field
and its potential remain constant. Furthermore, the energy density
of tachyon field in this regime is essentially dominated  by the
tachyon potential, i.e., $\rho_{\Phi}^{\rm sc} \simeq V(\Phi)$. On the
other hand,  $S \rightarrow 1$, implying  that the total energy
density of the collapse is dominated by the energy density of
barotropic matter as $\rho_{b}^{\rm sc}=3H^2$. From Eq.~(\ref{baroloop}), we get
$\rho_{\rm sc}\approx\rho_{b}^{\rm sc}=\rho_{0b}$, and
thus $3H^2=\rho_{0b}$, which gives an expression as $H=-(\rho_{0b}/3)^{1/2}<0$.

The ratio of the effective mass function to the area radius in this case can be obtained as
$F_{\rm sc}/R=r^2\rho_{0b}/3$, which, for any shell $r$,
remains finite. For an adequate  choice of $\rho_{0b}$  in
the the semiclassical regime, the effective mass function in this
regime remains smaller than the area radius, and no trapped surface will form as the
collapse proceeds.

The effective equation of state for the tachyon field in this case reads $w_\Phi^{\rm sc}=-1$, which satisfies the energy condition. Moreover,  barotropic parameter satisfies the range $0<\gamma_b=\frac{8}{5}<2$, for which the energy conditions are satisfied as well.

\subsection{Outward flux of energy in tachyon dominated collapse}
\label{exteriorR}

Since the effective pressure in the interior, Eq. (\ref{press-sem}), is negative at the boundary, and furthermore, there are no trapped surfaces forming as the collapse evolves, such region cannot
be matched to a Schwarzschild exterior space-time. Therefore,  we consider a generalized Vaidya geometry (\ref{Vaidya}) for the exterior region to be matched to the semiclassical interior at the boundary hypersurface $\Sigma$ given by $r=r_{\rm b}$ (see section  \ref{collapse-exterior}).

Let us designate the mass function at scales $a\gg a_{*}$ (i.e., in
the classical regime), as  $F$, whereas for $a<a_{*}$ (in the
semiclassical regime) we use $F_{\rm sc}$. The  mass loss is provided
by the following expression,
\begin{equation}
\frac{\Delta
F}{F} \  :=\  \frac{F-F_{\rm sc}}{F}=\left(1-\frac{\rho_{\rm sc}}{\rho}\right).\label{massloss1}
\end{equation}

\begin{figure}
\begin{center}
\includegraphics[width=0.5\textwidth]{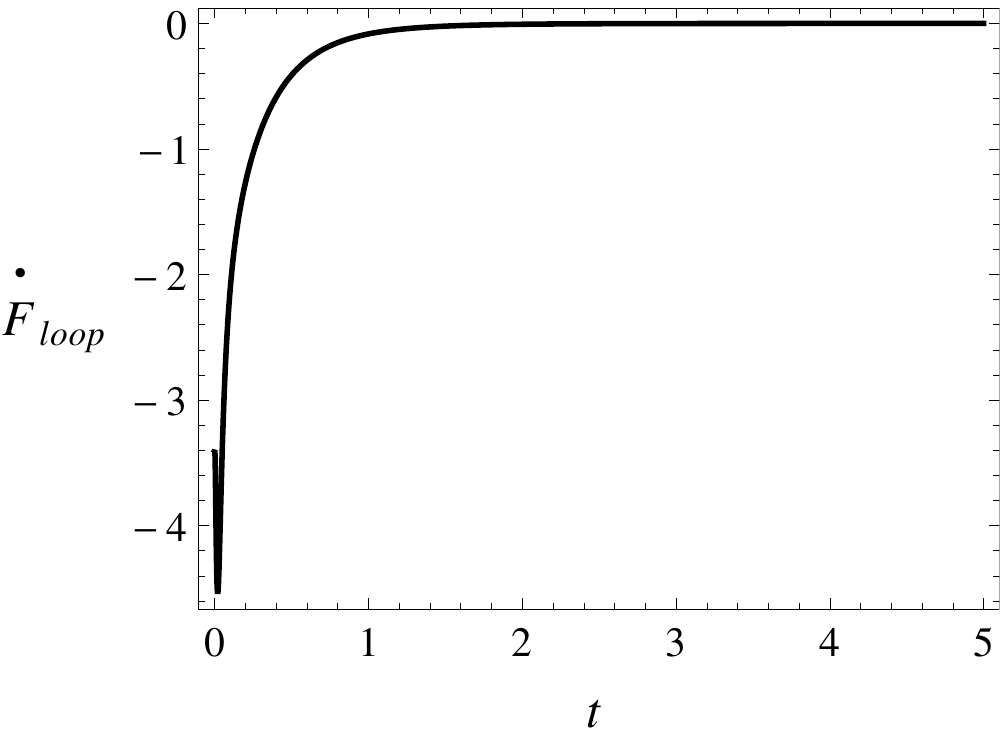}
\end{center}
\caption{ Behaviors of the derivative of the mass function
over time (with loop corrections). We consider $t_{0}=0$, $a(0)=a_{*}$, $V_{0}=1/3$,
$\Phi_{0}=-0.6$ and $\gamma_b=\frac{8}{5}$~.}
\label{F4}
\end{figure}

In order to  understand this, let us consider the geometry
outside a spherically symmetric matter, as given by the
Vaidya metric, Eq. (\ref{Vaidya}), with $\textrm{v}=t-r_{\textrm{v}}$ and $M(\textrm{v})$ being the
retarded null coordinate and the Vaidya mass, respectively. From Eq. (\ref{V2}) we can
further take the relation $F_{\rm sc}=2M(\textrm{v})$ between mass function
 and the Vaidya mass. Let us also assume the energy
density of the flux to be measured locally by an observer with
a four-velocity vector $\xi^{\mu}$. Then,  the energy flux as well as
the energy density of radiation measured in this local frame, is given by
$\sigma\equiv T_{\mu\nu}\xi^{\mu}\xi^{\nu}$,  which, for only radially moving observers with the
radial velocity  $\vartheta \equiv \xi^{r_{\textrm{v}}}=\frac{dr_{\rm v}}{dt}$, becomes
\begin{equation}
\sigma\  := \  -\frac{1}{(\gamma+\vartheta)^{2}}\left(\frac{1}{4\pi
r_{\rm v}^{2}}\frac{dM(\mathrm{v})}{d\mathrm{v}}\right),
\label{sigma}
\end{equation}
where $\gamma=(1+\vartheta^{2}-2M(\mathrm{v})/r_{\mathrm{v}})^{-1}$. The total luminosity for an observer with radial velocity $\vartheta$ and
for the radius $r_{\mathrm{v}}$, is given by
$L(\mathrm{v})=4\pi r_{\mathrm{v}}^{2}\sigma$ \cite{Lindquist1965}. 
Therefore, by substituting $\sigma$ from Eq.~(\ref{sigma}) into the equation of luminosity,
we can establish
\begin{equation}
L(\mathrm{v})\  =\  \frac{1}{(\gamma+\vartheta)^{2}}\frac{dM(\mathrm{v})}{d\mathrm{v}}\ .\label{luminosity2}
\end{equation}
Then, from Eq. (\ref{luminosity2}), the
luminosity in terms of the mass function $F_{\rm sc}$ is written as
\begin{equation}
L(\mathrm{v})\  =\  \frac{\dot{F}_{\rm sc}}{2(\gamma+\vartheta)}~.\label{luminosity3}
\end{equation}
For an observer being at rest ($\vartheta=0$) at infinity
($r_{\mathrm{v}}\rightarrow\infty$), the total luminosity of the energy flux
can be obtained by taking the limit of Eq. (\ref{luminosity3}):
\begin{equation}
L_{\infty}(\mathrm{v})\  =\  -\frac{\dot{F}_{\rm sc}}{2}\ .\label{luminosity4}
\end{equation}
As long as $dM/dv\leq0$, the total luminosity of the energy flux is
positive; since the effective energy density near the center decreases very slowly, the effective mass function given by Eq. (\ref{mass-sem}) can be approximated as $F_{\rm sc}\approx a^{2}$
in the loop modified regime, and is decreasing as the collapse evolves (see Fig. \ref{F4}).
Then, its time derivative is always negative, pointing to the
positiveness of the luminosity; this indicates that there exist an
energy flux radiated away from  interior space-time reaching the
distance observer.

\section{Improved dynamics and gravitational  collapse: Holonomy correction}
\label{Holonomy-Tach}

In this section  we present a semiclassical description for the gravitational collapse of   tachyon field and barotropic fluid, 
by employing a holonomy correction induced from LQG \cite{A-P-S:2006a,Willis:2004,Taveras:2008,Taveras2006}.

\subsection{Effective interior geometry}

The gravitational sector of the reduced phase space of LQG, for the interior line element (\ref{metric}), is coordinatized by the pair $(c, p)$,  where $c := \gamma\dot{a}$ and $p:=a^2$ 
which are, respectively, conjugate connection\footnote{Notice that, for a continuous collapse herein, we have $c<0$.} and triad satisfying the non-vanishing  Poisson bracket  
$\{c, p\}= 8\pi G\gamma/3$ \cite{Bojowald2003}. 
The Hamiltonian constraint for this system reads
\begin{equation}
C\  =\  C_{\mathrm{grav}}+C_{\mathrm{matt}}=0\  ,
\label{Ham-tot}
\end{equation}
where $C_{\mathrm{matt}}=\rho V$ is the Hamiltonian of the matter (with $V=|p|^{3/2}$ being the volume of the fiducial  cell), and $C_{\mathrm{grav}}$ is the gravitational part of the Hamiltonian constraint, which is given by \cite{Bojowald2003}
\begin{equation}
C_{\mathrm{grav}} =  -\frac{6}{8\pi G\gamma^2}c^2\mathrm{sgn}(p)\sqrt{|p|}\  .
\label{Hamil-gr}
\end{equation}
As described in the introduction, a certain pertinent scenario  to investigate semiclassically, the promising effects of LQG (as a gravitational collapse is concerned) is to employ the so-called \emph{holonomy correction}.
The algebra generated by the holonomy of phase space variables $c$ is just the algebra of the almost periodic function of $c$, i.e., 
$e^{i\bar{\mu} c/2}$  (where $\bar{\mu}$ is inferred as kinematical length of the square loop since its order of magnitude is similar to that of length), which  together with $p$, constitutes the fundamental canonical  variables  in quantum theory \cite{Bojowald2003}.
This consists in replacing $c$ in Eq.~(\ref{Hamil-gr}) with the phase space function, by means of  
\begin{align}
\frac{1}{2i\bar{\mu}}\left(e^{i\bar{\mu} c}-e^{-i\bar{\mu} c}\right)=\frac{\sin(\bar{\mu} c)}{\bar{\mu}}~. 
\label{HolonomyCorrection1}
\end{align}
It is expected that the classical theory is recovered for
small $\bar{\mu}$;  we therefore obtain the effective semiclassical Hamiltonian \cite{Taveras2006,A-P-S:2006a}
\begin{align}
C_{\rm eff}\  =\  -\frac{3}{4\pi G\gamma^2\bar{\mu}^2} \sqrt{|p|} \sin^{2}(\bar{\mu} c) + C_{\mathrm{matt}} \ .\label{EFFham-1-tach}
\end{align}
The dynamics of the fundamental variables 
is then obtained by solving the system of Hamilton equations; i.e., 
\begin{align}
\dot{p}  \ =\  \{p, C_{\rm eff}\} & \ =\  - \frac{8\pi G\gamma}{3}\frac{\partial C_{\rm eff}}{\partial c}   \notag \\
&  \  =\   \frac{2a}{\gamma\bar{\mu}} \sin(\bar{\mu} c) \cos(\bar{\mu} c) .  \label{HAMeqs-1-tach}
\end{align}
Furthermore, the vanishing Hamiltonian constraint (\ref{EFFham-1-tach}) implies that
\begin{equation}
\sin^2(\bar\mu c)\ = \ \frac{8\pi G \gamma^2 \bar{\mu}^2}{3a}  C_{\rm matt}\  .
\label{Hamilton-Eq2-tach}
\end{equation}
Thus, using  Eqs. (\ref{HAMeqs-1-tach}) and (\ref{Hamilton-Eq2-tach}),  we subsequently obtain 
the modified Friedmann equation, $H=\dot{a}/a=\dot{p}/2p$~:
\begin{equation}
H^{2}\  =\  \frac{8\pi G}{3}\rho\left(1-\frac{\rho}{\rho_{\text{crit}}}\right),
\label{Friedmann-eff-1a-tach}
\end{equation}

where $\rho_{\rm crit}=3/(8\pi G\gamma^{2}\lambda^{2})\approx0.41\rho_{\rm Pl}$, and 
$\rho$ is the total (classical) energy density of the collapse matter content.
Eq. (\ref{Friedmann-eff-1a-tach}) implies that the classical energy density
$\rho$ is limited to the interval $\rho_{0}<\rho<\rho_{\rm crit}$
having an upper bound at $\rho_{\rm crit}$, where  $\rho_{0}\ll\rho_{\rm crit}$
is the energy density of the star at the initial configuration, $t=0$.
Hence, the effective energy density reads 
\begin{equation}
\rho_{\rm eff}\  :=\  \rho\left(1-\frac{\rho}{\rho_{\rm crit}}\right).
\label{newFriedmann}
\end{equation}
We see that the effective scenario, provided by holonomy corrections, leads to a $-\rho^{2}$ modification of the energy density, 
 which becomes important when the energy density becomes
comparable to $\rho_{\rm crit}$. In the limit $\rho\rightarrow\rho_{\rm crit}$,
the Hubble rate vanishes; the classical singularity is thus replaced
by a bounce. 

The time derivative of the Hubble rate can be written in the effective dynamics as bellow,
\begin{equation}
\dot{H}\  =\  -\frac{8\pi G}{2}(\rho+p)\left(1 - 2\frac{\rho}{\rho_{\rm crit}}\right).
\label{LFriedmann2}
\end{equation}
The classical Raychaudhuri equation, $\frac{\ddot{a}}{a}=-\frac{4\pi G}{3}(\rho+3p)$,
in effective theory can be modified by using Eqs.~(\ref{Friedmann-eff-1a-tach}) and (\ref{LFriedmann2}) as follows:
\begin{align}
\frac{\ddot{a}}{a}=-\frac{4\pi G}{3}\left\{\rho\left(1-\frac{\rho}{\rho_{\mathrm{crit}}}\right) + 3\left[p\left(1-2\frac{\rho}{\rho_{\mathrm{crit}}}\right) -\frac{\rho^2}{\rho_{\rm crit}}\right] \right\}.\label{Raych-eff-tach}
\end{align}
Then, by rewriting the modified Raychaudhuri equation as 
\begin{equation}
\frac{\ddot{a}}{a}\  =\  -\frac{4\pi G}{3}(\rho_{\mathrm{eff}}+3p_{\mathrm{eff}}),
\label{Raychaudhuri-improv}
\end{equation}
the corresponding pressure of the system is given by  
\begin{equation}
p_{\text{eff}}\  =\  p\left(1-2\frac{\rho}{\rho_{\rm crit}}\right)-\frac{\rho^{2}}{\rho_{\rm crit}}\  .
\label{Peff}
\end{equation}
The effective energy conservation for $\rho_{\rm eff}$ is given by the relation $\dot{\rho}_{\text{eff}}=-3H(\rho_{\text{eff}}+p_{\text{eff}})$,
so that,  we can define the effective equation of state as
\begin{equation}
w_{\rm eff}\  := \ \frac{p_{\rm eff}}{\rho_{\rm eff}}\  =\  \frac{p}{\rho}\left(\frac{\rho_{\rm crit}-2\rho}{\rho_{\rm crit}-\rho}\right) - \frac{\rho}{\rho_{\rm crit}-\rho} \ . 
\label{Eq-State-eff}
\end{equation}

In classical general relativity, the equation for the mass function  is given by Eq.~(\ref{einstein2}), which can be written as  $F(R)=(8\pi G/3)\rho R^3$.
Since in the effective scenario herein, the energy density $\rho$  is modified as $\rho_{\rm eff}$, given by Eq.~(\ref{newFriedmann}),  
hence, the mass function  $F(R)$ is modified in the semiclassical regime as \cite{Tavakoli2013d}
\begin{equation}
F_{\rm eff}\  =\  \frac{8\pi G}{3}\rho_{\rm eff} R^3 = F\left(1-\frac{\rho}{\rho_{\rm crit}}\right).
\label{massF-eff}
\end{equation}
This means that the phase space trajectories are considered classical, whereas the matter content is assumed to be effective due to the semiclassical effects.
The $\rho/\rho_{\rm crit}$ term in Eq.~(\ref{massF-eff}) can be written as
\begin{equation}
\frac{\rho}{\rho_{\rm crit}}\  = \  \frac{a_{\rm crit}^3}{a^3} \frac{F}{F_{\rm crit}}\  ,
\label{massF-eff-2}
\end{equation}
where $a_{\rm crit}=a(t_b)$ and $F_{\rm crit}:=(8\pi G/3)\rho_{\rm crit}r^3a_{\rm crit}$ are, respectively, the values of scale factor and the mass function at the bounce, at the time $t=t_b$.
It is seen from Eq.~(\ref{massF-eff-2})  that, the mass function $F$ changes in the interval $F_0\leq F \leq F_{\rm crit}$ along with the collapse dynamical evolution, so that, it remains finite during the semiclassical regime;
$F_0=(8\pi G/3)\rho_0r^3a_0^3$ is the initial data for the mass function at $t=0$. Furthermore,  the effective mass function (\ref{massF-eff}) vanishes at the bounce.


Let us  follow  section \ref{Classic1} and consider the total energy density, $\rho$, of the collapse  to be 
\begin{equation}
\rho\  = \  \rho_\Phi +  \rho_b \  ,
\label{total-tach-bar}
\end{equation}
which constitutes the classical energy densities of the tachyon field  and the barotropic fluid.
In a strictly classical setting (cf. section \ref{Classic1}), the energy densities of tachyon field and barotropic fluid  
are given by
\begin{equation}
\rho_{\Phi}=\frac{V(\Phi)}{\sqrt{1-\dot{\Phi}^{2}}}\  , \   \   \   \   \   \   \  \  \  \   \   \   \rho_b \ = \  \rho_{b0} \left(\frac{a}{a_0}\right)^{-3\gamma_b}\ .
\label{energyTach-hol}
\end{equation}
Furthermore, the equation of state $w_\Phi$ for tachyon field is given by
\begin{equation}
w_\Phi\  := \  \frac{p_\Phi}{\rho_\Phi} = -\left(1-\dot{\Phi}^2\right).
\label{Eq-State-Tach}
\end{equation}
In addition, one can define a barotropic index for the tachyon fluid:
 $\gamma_\Phi:=(\rho_\Phi+p_\Phi)/\rho_\Phi=\dot{\Phi}^2$.

\subsection{Dynamics of tachyon matter and barotropic fluid: Phase space analysis}

The use of  dynamical system techniques to analyse a  classical  tachyon field in gravitational collapse has been considered in
section \ref{Classic1}.  In what follows, a dynamical system analysis of the tachyon field gravitational collapse within the improved dynamics approach
of LQG will be studied.

We consider the time variable defined by  Eq.~(\ref{n-timeclass}).
To analyze the dynamical behaviour of the collapse, we further introduce the following variables (we set $\kappa:=8\pi G$):
\begin{align}
& x := \dot{\Phi}\ \  \ \  \   \  \  \   \  \  \  \  \   \  \   y :=  \frac{\kappa V}{3H^{2}}\ ,\  \  \  \  \  \  \  \   \  \   \  \   \    \   \   z  :=  \frac{\rho}{\rho_{\rm crit}}\ , \notag \\ 
&  s:= \frac{\kappa\rho_{b}}{3H^2}\ , \  \  \  \   \  \   \   \   \   \lambda := -\frac{V_{,\Phi}}{\sqrt{\kappa}V^{\frac{3}{2}}}\ ,\  \  \  \  \   \  \    \  \   \Gamma := \frac{VV_{,\Phi\Phi}}{(V_{,\Phi})^{2}}\ . 
\label{DynV2-a}
\end{align}
The Friedmann constraint (\ref{Friedmann-eff-1a-tach}), in terms of the new variables (\ref{DynV2-a}), can be rewritten as
\begin{equation}
1 =  \left(\frac{y}{\sqrt{1-x^{2}}}+s\right)(1-z),
\label{DynFrid}
\end{equation}
in which, the dynamical variables $x$, $y$ and $z$  must satisfy the  constraints  $-1\leq x\leq 1$, $y\geq 0$ and $0\leq z\leq 1$.
Furthermore,  the time derivative of the Hubble rate, Eq.~(\ref{LFriedmann2}), in terms of  variables (\ref{DynV2-a}), becomes
\begin{equation}
\frac{\dot{H}}{3H^2}\ = \  - \frac{1}{2}(1-2z) \left[\frac{x^2y}{\sqrt{1-x^2}}+\gamma_b s\right].
\label{H-der}
\end{equation}

Using the Eq. (\ref{DynV2-a}) and the constraint (\ref{DynFrid}),  the classical equation of state of tachyon field, Eq. (\ref{Eq-State-Tach}), and the effective equation of state (\ref{Eq-State-eff}), in terms of  dynamical variables 
read
\begin{align}
w_\Phi \ & = \  -\left(1-x^{2}\right)\ ,  \\
w_{\rm eff}\  & =\  -\left(1-x^2\right)\left(\frac{1-2z}{1-z}\right) + s\left(\gamma_b-x^2\right)(1-2z) - \frac{z}{1-z}\  .  \label{DynState}
\end{align}
Moreover,  the fractional densities of the two fluids are respectively defined as:
\begin{equation}
\Omega_{\Phi} := \frac{\kappa\rho_\Phi}{3H^2} = \frac{y}{\sqrt{1-x^2}}\ , \  \  \   \  \  \   \  \   \  \  \   \   \    \  \Omega_b := \frac{\kappa\rho_b}{3H^2}= s\ .
\end{equation}

An autonomous system of equations, in terms of the dynamical variables of Eq. (\ref{DynV2-a}), together with  Eqs. (\ref{DynFrid}) and (\ref{H-der}), is then retrieved:
\begin{align}
\frac{dx}{d\tau} &  \  =\     (1-x^{2})\left(x - \frac{\lambda}{\sqrt{3}} \sqrt{y} \right),  \label{Dynx} \\
\frac{dy}{d\tau} & \  =\    \frac{\lambda}{\sqrt{3}} x y^{\frac{3}{2}} - y(1-2z)  \left[\frac{x^2}{1-z} + s\left(\gamma_b-x^2\right)\right].  \label{Dyny} \\   
\frac{dz}{d\tau} &  \  =\   z\left[ x^2 +  s(1-z)\left(\gamma_b-x^2\right) \right] \  ,  \label{Dynz} \\
\frac{ds}{d\tau} & \   =\   s \left[\gamma_b - \left(1-2z\right)\left(\frac{x^2}{1-z} +  s\left(\gamma_b-x^2\right) \right)\right]\ . \label{Dyns}
\end{align}
We will assume that the tachyon potential has an inverse square form (\ref{potential}),
so that, we get  $\lambda=\pm 2/\sqrt{V_0}$ and $\Gamma=3/2$, i.e., as constants.

Let $g_1:= dx/d\tau$, $g_2 := dy/d\tau$, $g_3 := dz/d\tau$ and $g_4 := ds/d\tau$. Then, the critical points $q_{c}=(x_c, y_c, z_c, s_c)$ are obtained by setting the condition
$(g_1, g_2, g_3, g_4)|_{q_{c}}=0$. Next we will study the stability of our dynamical system at each critical point by  using the standard linearization and stability analysis in Eqs. (\ref{stab-1}) and (\ref{pertQ1}).
In fact, to determine the stability of critical points, we need to perform linear perturbations around $q_{c}$ by using the form $q(t)=q_{c}+\delta q(t)$
which results in the equations of motion $\delta q'={\cal M} \delta q$, where ${\cal M}$ is the Jacobi matrix of each point whose components are ${\cal M}_{ij}=(\partial g_i/\partial q_j)|_{q_{c}}$:
\begin{equation}
{\cal M}=\left(\begin{array}{cccc}
\frac{\partial g_{1}}{\partial x} & \frac{\partial g_{1}}{\partial y} & \frac{\partial g_{1}}{\partial z} & \frac{\partial g_{1}}{\partial s} \\
\frac{\partial g_{2}}{\partial x} & \frac{\partial g_{2}}{\partial y} & \frac{\partial g_{2}}{\partial z} & \frac{\partial g_{2}}{\partial s} \\
\frac{\partial g_{3}}{\partial x} & \frac{\partial g_{3}}{\partial y} & \frac{\partial g_{3}}{\partial z} & \frac{\partial g_{3}}{\partial s}\\
\frac{\partial g_{4}}{\partial x} & \frac{\partial g_{4}}{\partial y} & \frac{\partial g_{4}}{\partial z} & \frac{\partial g_{4}}{\partial s}
\end{array}\right)_{| q_{c}}. \label{matrixA-a}
\end{equation}
We have summarized the fixed points for the autonomous system (\ref{Dynx})-(\ref{Dyns}) and their stability properties in Table \ref{fixed-points}.


\begin{center}

\begin{table}[h!]

\caption{Summary of critical points and their properties.}

\begin{tabular}{cccccccc}
\hline \hline 
Point  & $x$  & $y$  & $z$  & $s$ &  $\Omega_\Phi$ & Existence  &  Stability  \tabularnewline
\hline 

$(\tilde a)$   &  $1$  & 0  & $0$ & $0$ &  $1$ &  {\small All $\lambda$ and $\gamma_b$}   & {\small  Saddle point}  \tabularnewline
$(\tilde b)$   &  $-1$  & 0 &$0$   & $0$ & $1$ &  {\small All $\lambda$ and $\gamma_b$}  &   {\small Saddle point}   \tabularnewline
$(\tilde c)$  & $\frac{\lambda}{\sqrt{3}}\sqrt{y_0}$ & $y_0$ & $0$ & 0 & $1$ &  {\small All $\lambda$ and $\gamma_b$}   & {\small Unstable point for $\gamma_b\ge\gamma_1$} \tabularnewline
   &    &   &  &  &   &   &  Saddle point for $\gamma_b < \gamma_1$   \tabularnewline
$(\tilde d)$ & $0$ & $0$ & $0$ & $1$ & 0 &  {\small All $\lambda$ and $\gamma_b$}  &  {\small Saddle  point for $\gamma_b\neq0$}   \tabularnewline
   &    &   &  &  &   &   & {\small Unstable point for $\gamma_b=0$} \tabularnewline
$(\tilde e)$  & $-\sqrt{\gamma_b}$  & $\frac{3\gamma_b}{\lambda^{2}}$ & $0$ & $s_0$ & $1-s_0$ &  {\small All $\lambda$ and $\gamma_b<\gamma_1<1$}  & {\small Unstable point}  \tabularnewline
$[\tilde e]$  & $\sqrt{\gamma_b}$  & $\frac{3\gamma_b}{\lambda^{2}}$  & $0$  & $s_0$ & $1-s_0$ &  {\small All $\lambda$ and $\gamma_b<\gamma_1<1$}  &  {\small Unstable point} \tabularnewline
$(\tilde f)$   &  $1$  & 0 &$0$   & $1$ &  $0$ &  {\small All $\lambda$ and $\gamma_b$}   & {\small Saddle point}  \tabularnewline
$(\tilde g)$   &  $-1$  & 0 &$0$   & $1$ &  $0$ &  {\small All $\lambda$ and $\gamma_b$}   & {\small Saddle point} \tabularnewline
\hline \hline 
\end{tabular}

\label{fixed-points} 
\end{table}

\par\end{center}


\noindent Point  $(\tilde a)$:  The eigenvalues of this fixed point are $\zeta_1=-2$, $\zeta_2=-1$, $\zeta_3=+1$ and $\zeta_4=\gamma_b-1$.
All characteristic values of this point are real, but at least one is positive and two are negative, thus, the trajectories approach this point on a surface and diverge along a curve; this is a \emph{saddle} point.

\noindent Point  $(\tilde b)$:  For this fixed point, the characteristic values are $\zeta_1=-2$, $\zeta_2=-1$, $\zeta_3=+1$ and $\zeta_4=\gamma_b-1$, which are the same eigenvalues as the fixed point $(\tilde a)$, and thus, similar to  $(\tilde a)$, this is a  \emph{saddle} point.

\noindent Point  $(\tilde c)$:   This fixed point has eigenvalues $\zeta_1=0$,  $\zeta_2=y_0^2+\lambda^2y_0/6>0$, $\zeta_3=\gamma_1$ and $\zeta_4=(\gamma_b-\gamma_1)$, where, 
$y_0 := -\lambda^2/6+\sqrt{(\lambda^2/6)^2+1}$ and $\gamma_1:=\lambda^2y_0/3$. For $\gamma_b>\gamma_1$ this point  is an unstable, and for $\gamma_b<\gamma_1$ this is a saddle point.

\noindent Point  $(\tilde d)$:  The eigenvalues read  $\zeta_1=+1$, $\zeta_2=-\gamma_b$, $\zeta_3=+\gamma_b$ and $\zeta_4=-\gamma_b$. 
For $\gamma_b\neq0$, this point possesses eigenvalues with opposite signs; therefore, this point is saddle. For the case $\gamma_b=0$, this point has one real and positive eigenvalues, and others are zero, so $(\tilde d)$ is an unstable point.

\noindent Point  $(\tilde e)$:  This point is located at $(-\sqrt{\gamma_b},  3\gamma_b/\lambda^2, s_0)$, where $s_0:=\left(1-\frac{3\gamma_b}{\lambda^2\sqrt{1-\gamma_b}}\right)$.  The eigenvalues for this fixed point are $\zeta_1=0$, $\zeta_3=\gamma$ and
\begin{align}
\zeta_{2,4}\  =\  \frac{1}{4}\left(2-\gamma_b\pm\sqrt{\left(1-\gamma_b\right)\left(4-16s_{0}\gamma_b\right)+\gamma_b^{2}}\right).
\label{eigencle-eff}
\end{align}
For $\gamma<\gamma_1$, all eigenvalues are non-negative, and for $\gamma=\gamma_1$, we have $\zeta_2>0$ and $\zeta_4=0$. Therefore, this point is an unstable fixed point.
Notice that, since $0<s_0<1$, the barotropic parameter $\gamma_b$ must hold the range $0\leq\gamma_b\leq\gamma_{1}<1$, given by Eq.~(\ref{const-track4}) for this fixed point. 

\noindent Point  $[\tilde e]$: The eigenvalues of this point are the same as the point $(\tilde e)$, so that this is an unstable point.

\noindent Point  $(\tilde f)$: The eigenvalues for this fixed point are $\zeta_1=-2$, $\zeta_2=-\gamma_b$, $\zeta_3=\gamma_b$ and $\zeta_4=1-\gamma_b$. At least one characteristic value is negative and one is positive, so this point is a saddle point.

\noindent Point  $(\tilde g)$:  For this point, the eigenvalues are similar to those of point $(\tilde f)$, i.e., $\zeta_1=-2$, $\zeta_2=-\gamma_b$, $\zeta_3=\gamma_b$ and $\zeta_4=1-\gamma_b$. Therefore, this is a saddle point.


\subsection{The fate of the classical singularities}

\begin{figure}
\centering
\includegraphics[height=2.8in]{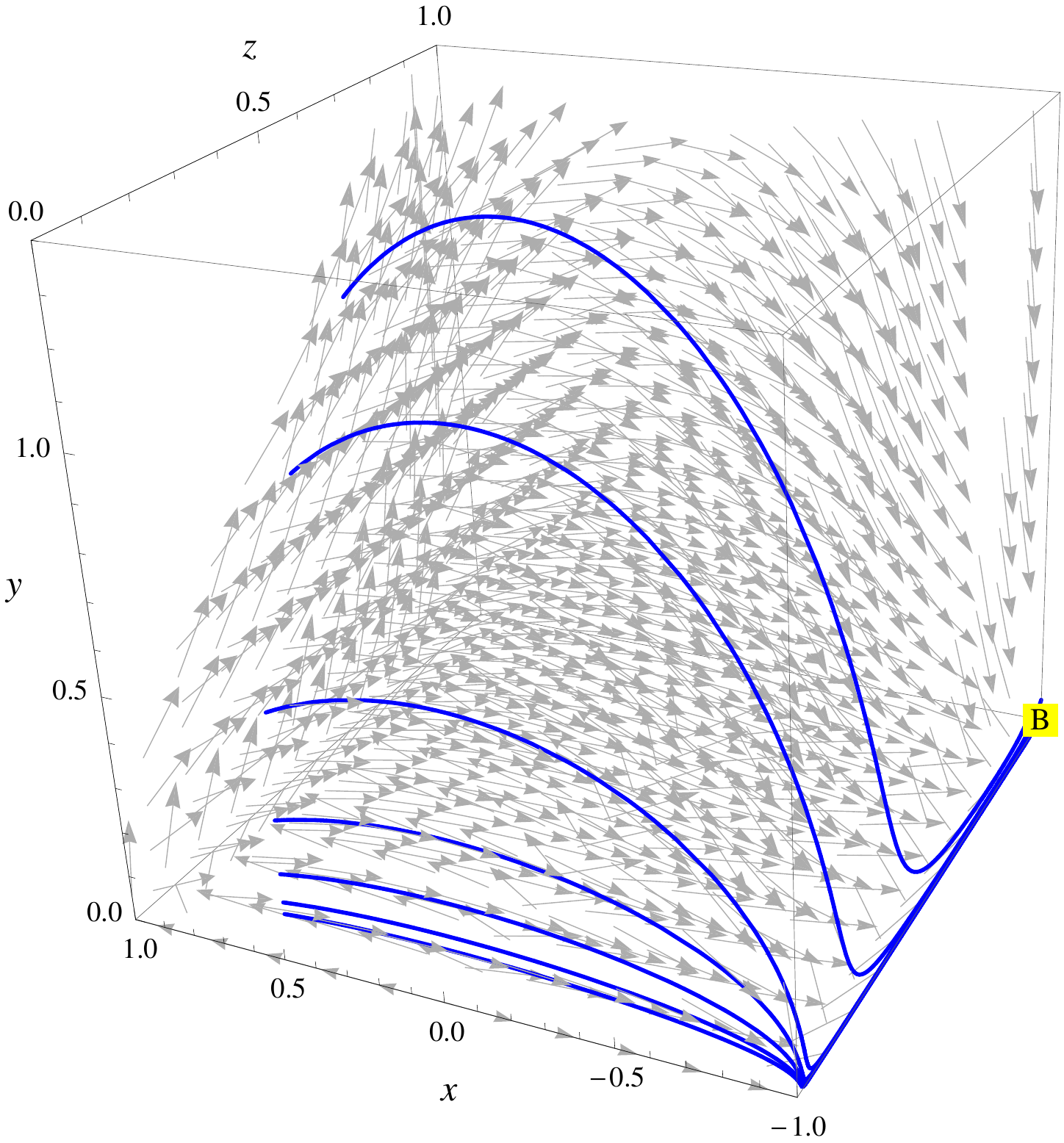} 
\caption{This plot represents a set of trajectories evolving
in the three dimensional phase space $\left(x,\: y,\: z\right)\equiv\left(\dot{\Phi},\:\kappa V/3H^{2},\:\rho/\rho_{\textrm{crit}}\right)$;
also the complete vector field generated by the dynamical system is
shown. All the possible trajectories are tangent to this vector field.
The initial conditions for solving the dynamical system Eqs. (\ref{Dynx})-(\ref{Dyns})
are chosen such that the trajectories start from locations near the
$x-y$ plane. In this plane, $z\approx0$, consequently $\rho\ll\rho_{\textrm{crit}}$.
The different curves are obtained varying the initial value of the
tachyon field $\Phi_{0}$. Point B correspond to the location in the
phase space where the semiclassical bounce is defined.}

\label{F-stream3D} 
\end{figure}

In the standard general relativistic collapse of a tachyon field with
barotropic fluid (see table~\ref{T1} in chapter \ref{collapse-tachyon}), the fixed points $(x_{c},y_{c},s_{c})=(1,0,0)$
and $(x_{c},y_{c},s_{c})=(-1,0,0)$ are stable fixed points (attractors)
and correspond to a tachyon dominated solution (see section~\ref{Tachyon dominated}.A); therein, the collapse
matter content behaves, asymptotically, as a homogeneous dust-like
matter which leads to a black hole formation at late times.
Nevertheless, in the semiclassical regime herein, in the presence
of the loop (quantum) holonomy correction term $z\neq0$, these fixed
points become a saddle, so that the stable points (i.e., the singular
black hole solution) of the classical collapse disappears here.

The points $(x_{c},y_{c},s_{c})=(1,0,1)$ and $(x_{c},y_{c},s_{c})=(-1,0,1)$,
in the classical regime (in the absence of the $z$ term), correspond
to the stable fixed points (attractors), namely the fluid dominated
solutions, and lead to the black hole formation as the collapse end
state (see section~\ref{Fluid dominated}.B). Nevertheless, holonomy effects, in the presence
of $z$ term induce respectively, the corresponding saddle points
$(x_{c},y_{c},s_{c},z_{c})=(1,0,1,0)$ and $(x_{c},y_{c},s_{c},z_{c})=(-1,0,1,0)$
for the collapsing system, instead. This means that the classical
singular black holes are absent in the semiclassical regime herein.

In figure \ref{F-stream3D} we show a selection of numerical solutions
of the dynamical system Eqs. (\ref{Dynx})--(\ref{Dyns}), in terms
of the variables $\left(x,\: y,\: z,\: s\right)$. This figure represents
trajectories which start from the lower $x-y$ plane and evolve in
the phase space. These trajectories will initially converge to a point
where $\dot{\Phi}\rightarrow-1$, along the $x-y$ plane; however,
in the vicinity of this point, they diverge along the $y-z$ plane
and move away from it. This point is identified to be the saddle
fixed fixed points $(\tilde{b})$ or $(\tilde{g})$. 

However, it is pertinent to point the following. Figure \ref{F-stream3D}
envolves parametric functions $x\left(\tau\right)$, $y\left(\tau\right)$
and $z\left(\tau\right)$. The numerical solution shows that the variable
$N$ is only defined on a finite interval $\left[0,\: \tau_{{\rm bounce}}\right]$;
this can be seen from Eq. (\ref{n-timeclass}) in which the scale factor
is bounded from below, i.e., $a_{{\rm min}}<a<a_{0}$. 
 In fact, and contrasting with the classical solution (cf. chapter \ref{collapse-tachyon}),
where $x\left(\tau\rightarrow\infty\right)\rightarrow\pm1$ and $y\left(\tau\rightarrow\infty\right)\rightarrow0$
are asymptotic limits, in the herein semiclassical scenario, the variable
$\tau$ is bounded at the bounce. This boundary is shown in figure~\ref{F-stream3D}
where the curves end at a region where $z\rightarrow1$ (identified
as point B in the plot), which consequently, cannot be classified
as a fixed point of the dynamical system. The numerical study supports
the analytical discussion that the solutions in section~\ref{collapse-tachyon} for
points $(\tilde{a})$, $(\tilde{b})$, $(\tilde{f})$ and $(\tilde{g})$, is now avoided on
the semiclassical trajectories. In addition, for all the trajectories
on the phase space shown in figure \ref{F-stream3D}, point $B$ corresponds
to a bouncing scenario which we will analyse it on the next section.


\subsection{Semiclassical collapse end state}

In this section we present additional results related to the numerical
studies of model.

\subsection*{A.~Tracking solutions: Tachyon versus barotropic fluid}

\begin{figure}
\centering
\includegraphics[height=2.3in]{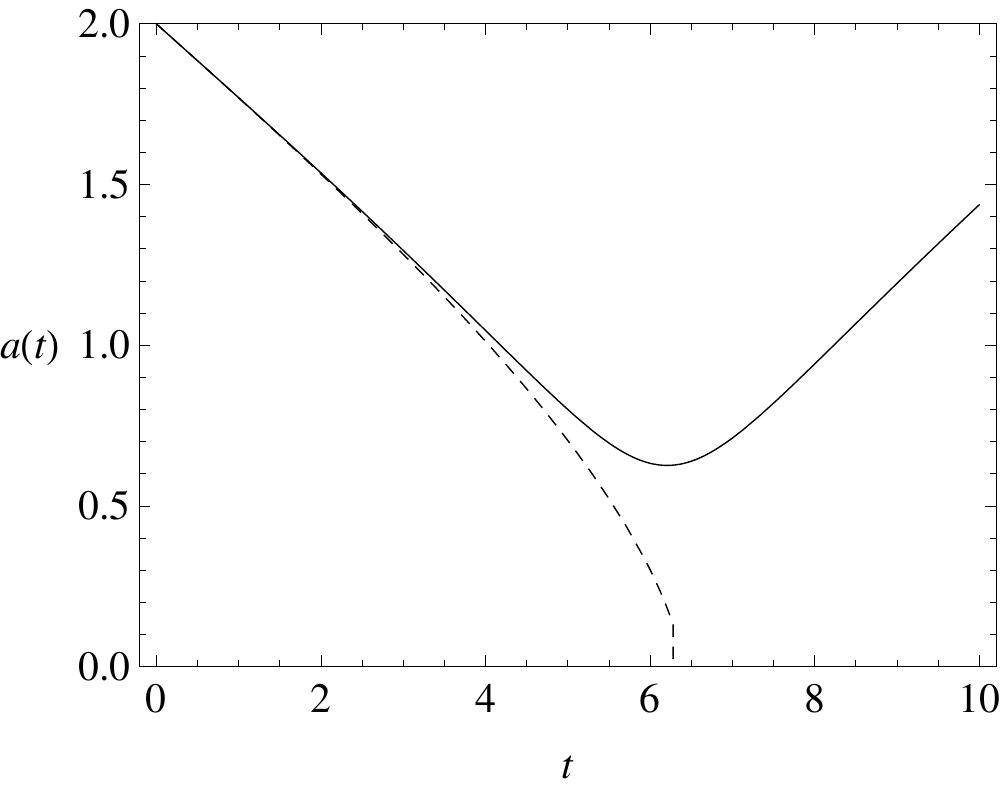}
\caption{Behaviours of the scale factor $a(t)$ in the semiclassical
(solid line), and classical (dashed curve) regime. We considered the
initial conditions as: $\rho_{0}=0.04$, $V_{0}=0.001$, $a(0)=2$,
$\dot{\Phi}\left(0\right)=0.5$ and $\Phi_{0}=0.6$. We also have
$\gamma_b=0.1$.}
\label{F-scalef3} 
\end{figure}

\begin{figure}
\includegraphics[height=2in]{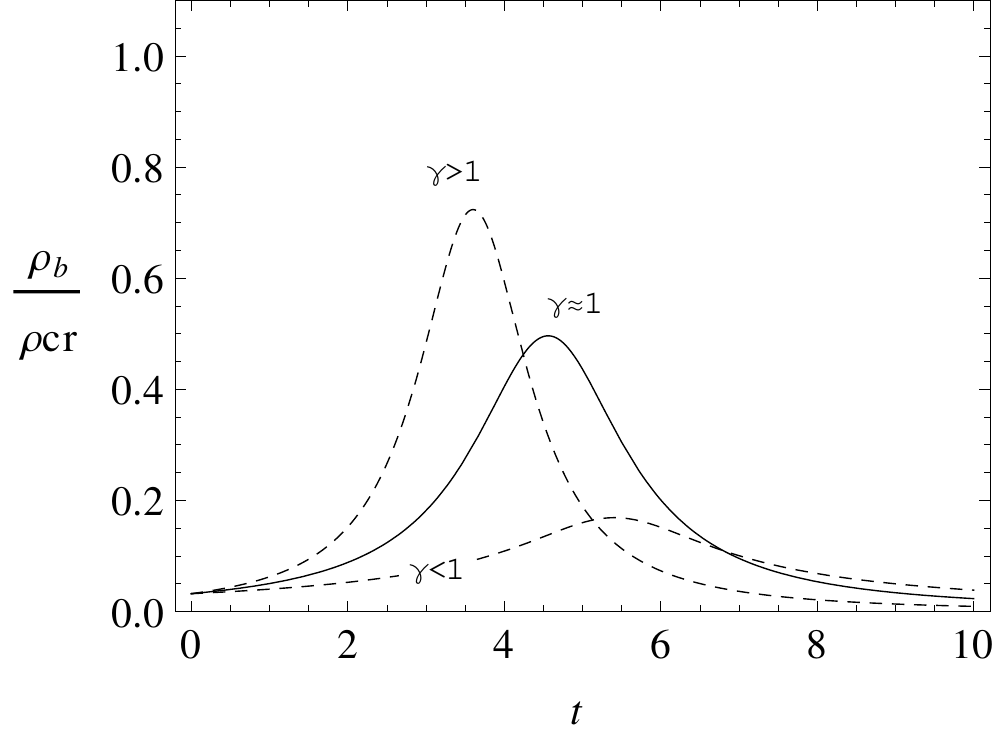}\quad{}\includegraphics[height=2in]{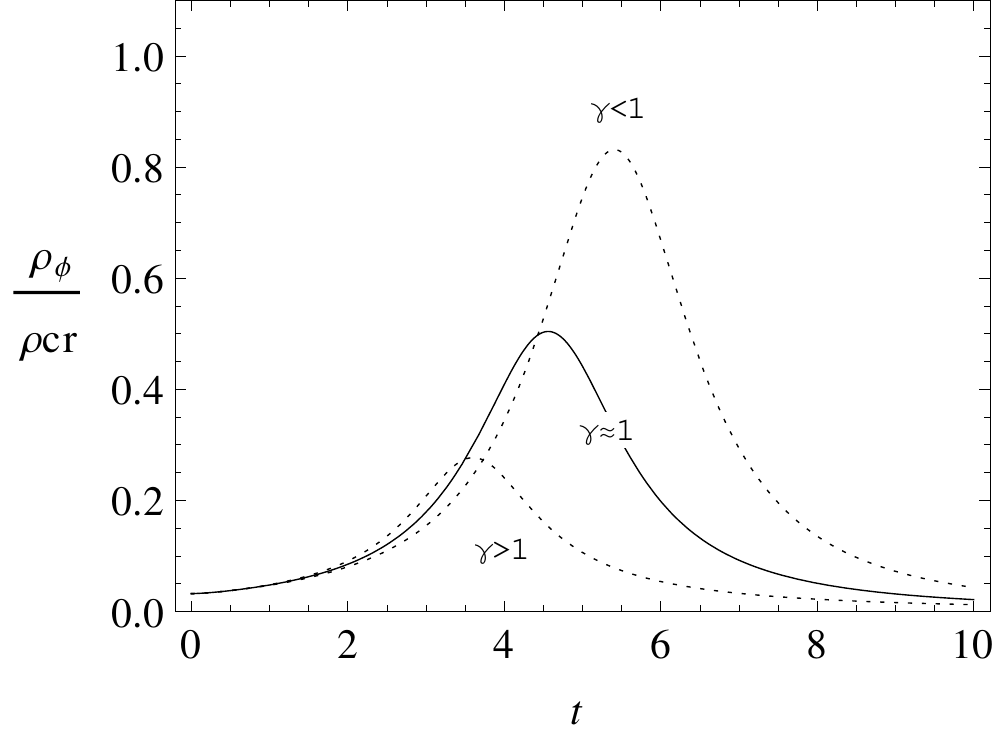} 
\caption{Behaviours of the energy densities $\rho_{b}(t)$ (left
plot) and $\rho_{\Phi}(t)$ (right plot) for different values of
the barotropic parameter $\gamma_b$. We have considered the same initial
energy densities for the barotropic fluid and the tachyon field. At
the bounce we have three different scenarios concerning the matter
content dominance. When $\gamma_b\approx1$ the system exhibits a tracking
solution. If $\gamma_b>1$ the fluid is dominant, whereas for $\gamma_b<1$
it is the tachyon field that is dominant. 
}
\label{F-Endens} 
\end{figure}

Figure \ref{F-scalef3} shows the behaviour of the scale factor. Therein
we observe that in the limit $\rho\rightarrow\rho_{\text{crit}}$, when
the Hubble rate vanishes, the classical singularity is replaced by
a bounce (cf. figure \ref{F-scalef3}). In figure \ref{F-Endens} we
represent the energy densities, $\rho_{b}(t)$ (left plot) and $\rho_{\Phi}(t)$
(right plot), for different values of the barotropic parameter $\gamma_b$
at the bounce. We see that three scenarios have to be considered.
When the energy densities of the tachyon and of the fluid scale exactly
at the same power of the scale factor, namely
\begin{equation}
\rho_{\Phi}\approx\rho_{b}\approx\rho_{0}\, a^{-3\gamma_b},\label{trackb-1}
\end{equation}
then the semiclassical solutions display a tracking behaviour.
Numerical analysis shows that this happens when the barotropic parameter
is approximately $\gamma_b\sim1$, that is, the collapse matter content
acts like dust. From Eq. (\ref{trackb-1}), we have $a_{\textrm{crit}}=\left[\rho_{\textrm{crit}}/\left(2\rho_{0}\right)\right]^{-1/3}$
at the bounce for the tracking solution. In the case where $\gamma_b>1$
the solution at the bounce is fluid dominated, whereas for $\gamma_b<1$,
the tachyon field is the dominant component of the energy density
content of the system.

From figure \ref{F-Endens} we also observe that, starting
from very low values of the energy density (classical regime), a system
that is fluid dominated reaches the bounce faster than a system that
is tachyon dominated. This seems to point to the fact that a fluid
dominant solution will drive the energy density until its critical
value more efficiently that when the tachyon field is dominant. In
order to explain this result, let us consider what happens to the
total pressure $p_{\Phi}+p_b$ for each solutions discussed
in this section. When the tachyon field is dominant, for $\gamma_b<1$,
the total pressure
\begin{equation}
p=-V(\Phi)\sqrt{1-\dot{\Phi}^{2}}+\left(\gamma_b-1\right)\rho_{b},\label{total press.}
\end{equation}
is negative until the collapsing body reaches the bounce. In fact,
near the bounce $\dot{\Phi}\rightarrow1$ and Eq. (\ref{total press.})
becomes $p\thickapprox\left(\gamma_b-1\right)\rho_{b}\:\left(<0\right)$.
For the tracking solution $\left(\gamma_b\sim1\right)$, we have $p\sim0$
and the matter content behaves as dust. Finally, for the fluid dominated
solutions, the total pressure is approximately $p\thickapprox\left(\gamma_b-1\right)\rho_{b}$,
which is positive because in this case $\gamma_b>1$. Consequently,
in this last scenario, the positive pressure drives the fluid dominant
content of the energy density rapidly towards its critical value $\rho\rightarrow\rho_{\text{crit}}$
at the semiclassical bounce. 

\begin{figure}
\begin{center}
\includegraphics[height=2.5in]{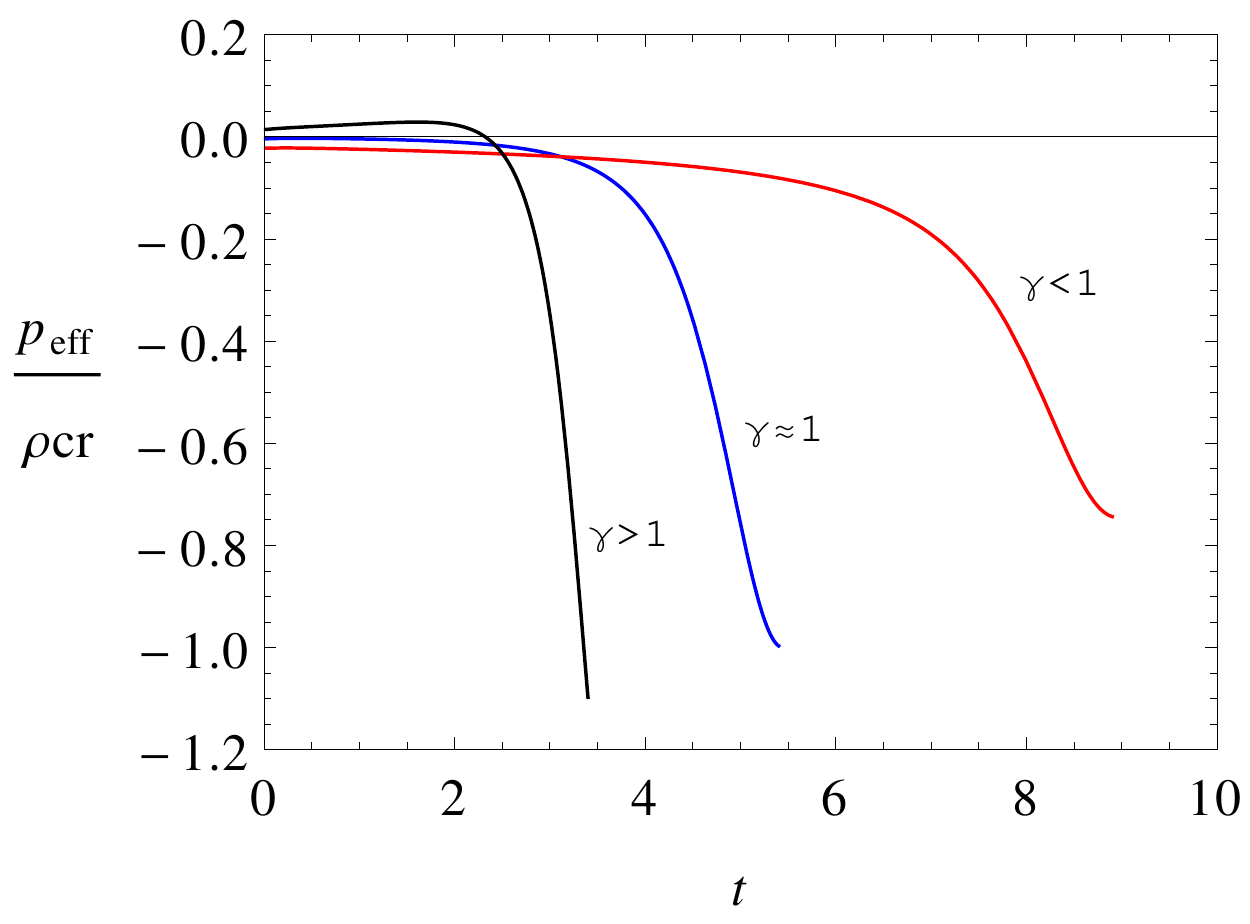}
\caption{ 
%
In this plot we represent
the evolution of effective pressure, for different values of $\gamma_b$ until
the bounce is reached.}
\label{F-Endens-b} 
\end{center}
\end{figure}

In addition, when we consider Eq. (\ref{Peff}) for the effective
pressure, in particular its value at the bounce (where $\rho\rightarrow\rho_{\textrm{crit}}$),
\begin{equation}
\begin{cases}
p_{\textrm{eff}}^{\Phi}\approx-\left(\gamma_b-1\right)\rho_{b}-\rho_{\textrm{crit}} & \quad\gamma_b<1\ ,\\
p_{\textrm{eff}}^{tr}\approx-\rho_{\textrm{crit}} & \quad\gamma_b\approx1 \ ,\\
p_{\textrm{eff}}^{\gamma_b}\approx-\left(\gamma_b-1\right)\rho_{b}-\rho_{\textrm{crit}} & \quad\gamma_b>1\ ,
\end{cases}
\end{equation}
we can establish that $p_{\textrm{eff}}^{\gamma_b}<p_{\textrm{eff}}^{tr}<p_{\textrm{eff}}^{\Phi}<0$
(see  figure \ref{F-Endens-b}). In this plot we have that
for the fluid dominated solution, the effective pressure start at
a positive value (assisting the collapsing system energy density rapidly
towards its critical value $\rho\rightarrow\rho_{\text{crit}}$).
However, near the bounce, the effective pressure rapidly switches
to negative values. In contrast, for the tachyon dominated solution,
the effective pressure starts from negative values from the beginning;
this is related to the fact that the initial energy densities of both
the tachyon and barotropic fluid are approximate. Moreover, the change
near the bounce is less pronounced in this last case. Therefore the
evolution of the collapse is slower and the bounce is delayed when
compared to the fluid dominated scenario. It is straightforward to
verify that the tracking solution provides an intermediate context
between the fluid and tachyon dominated solutions.

\subsection*{B.~Horizon formation}

\begin{figure}
\includegraphics[height=1.4in]{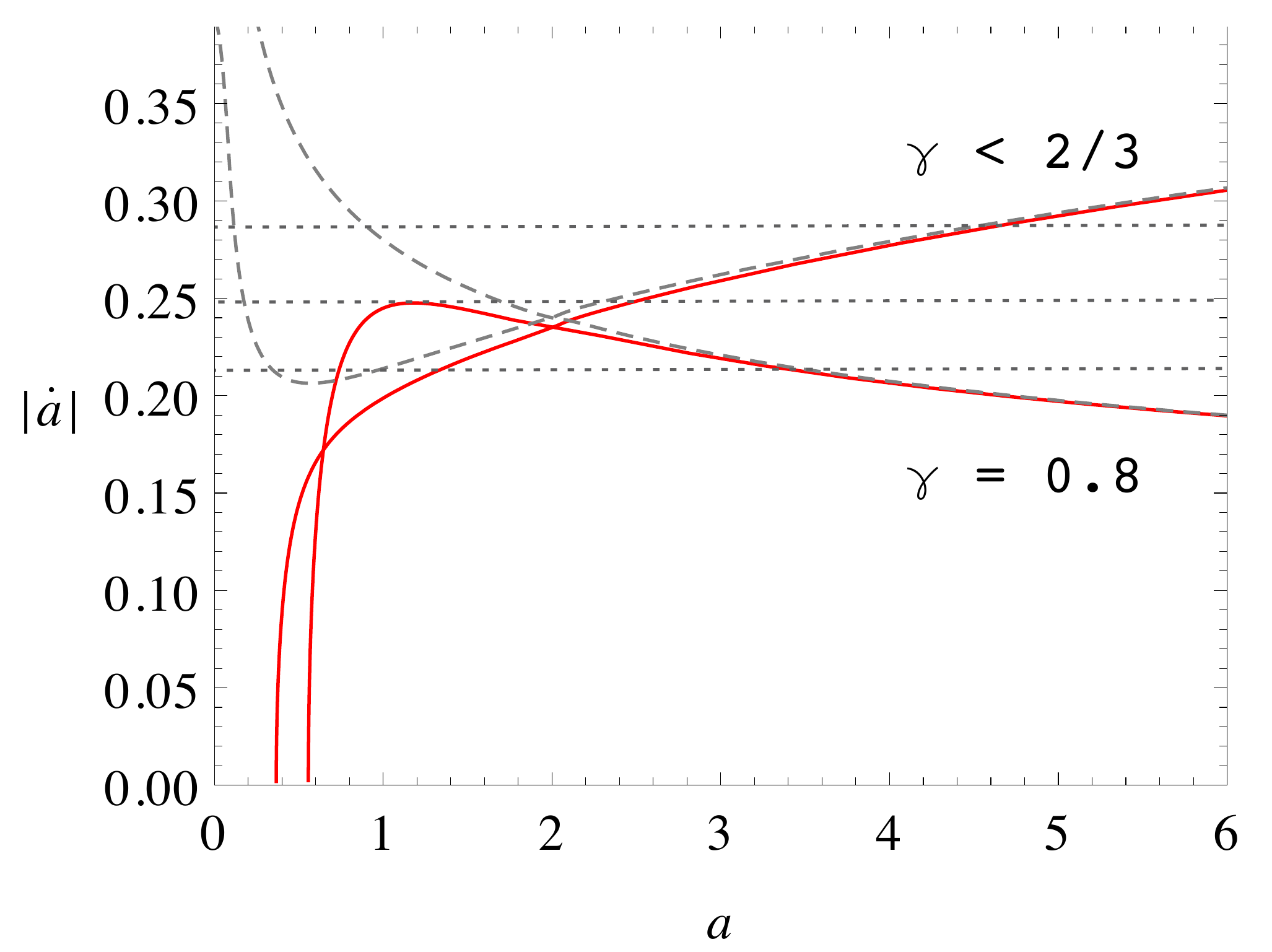}\enskip{}\includegraphics[height=1.4in]{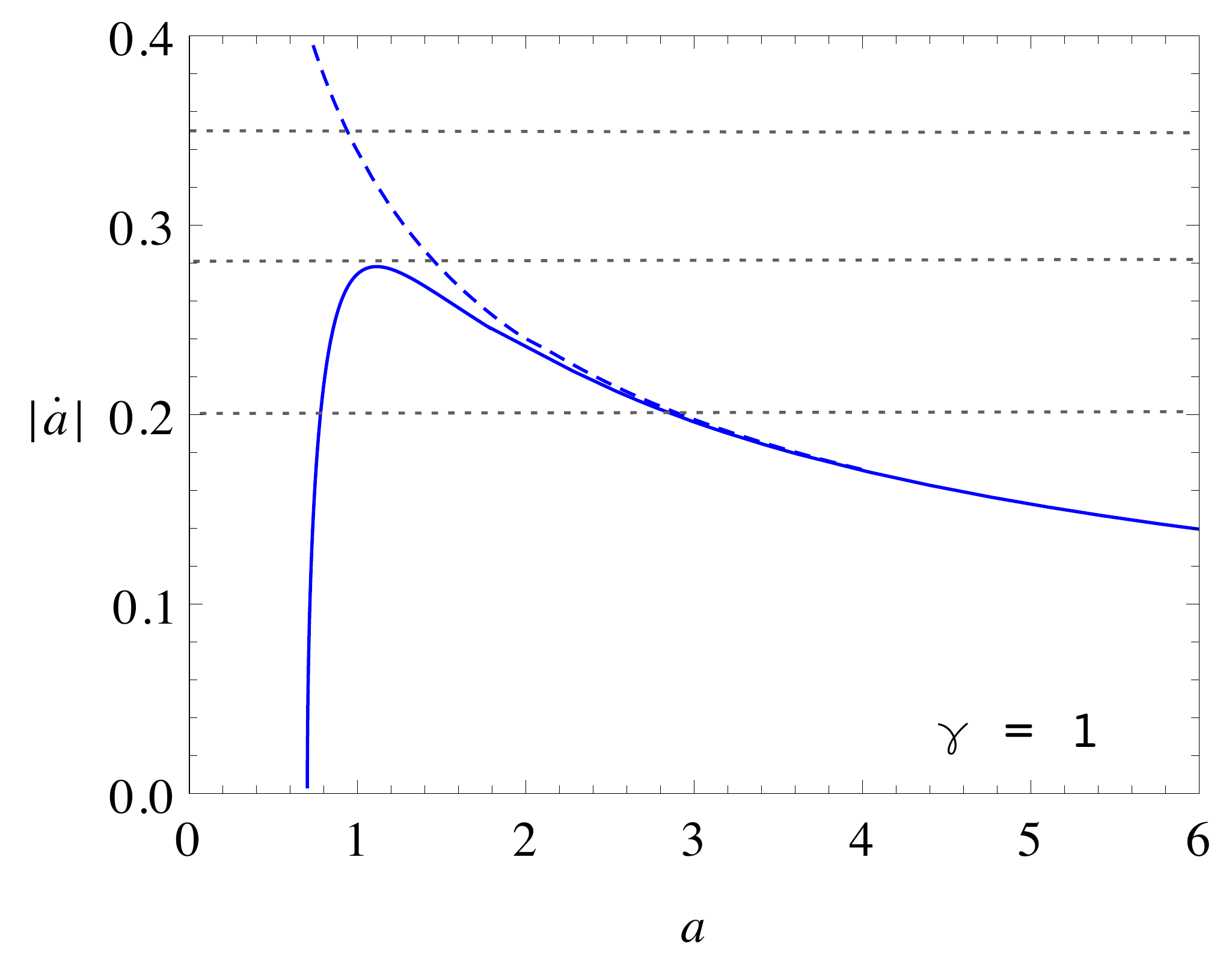}\enskip{}\includegraphics[height=1.4in]{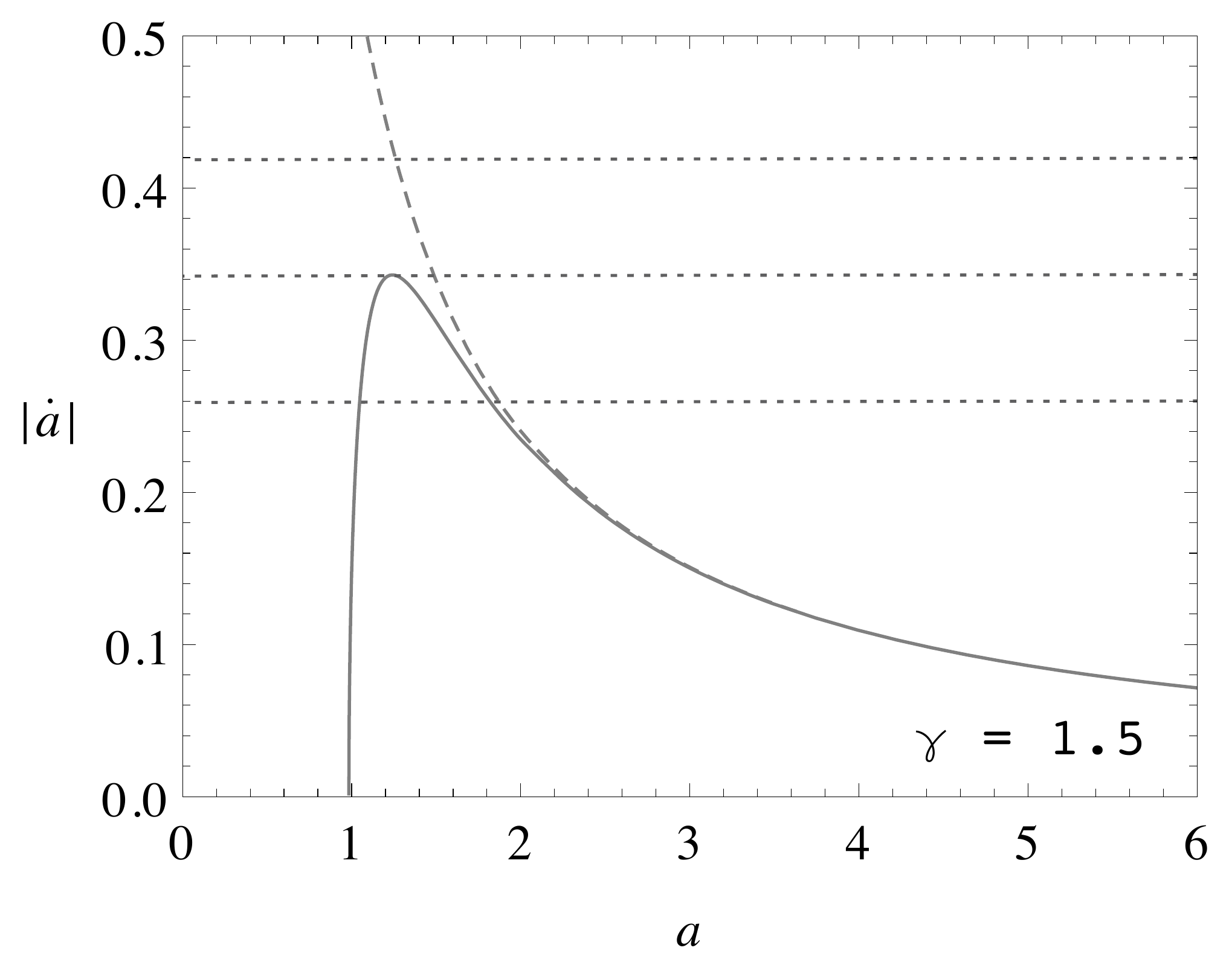}
\caption{The speed of collapse, $|\dot{a}|$, with respect to
the scale factor $a$, in the semiclassical (solid line), and classical
(dashed curve) regime. We considered the initial conditions as: $\rho_{0}=0.04$,
$V_{0}=0.001$, $a(0)=2$, $\dot{\Phi}\left(0\right)=0.5$ and $\Phi_{0}=0.6$.
Left plot is for a tachyon dominated solution with $\gamma_b=0.8$ and
$\gamma_b<2/3$. Center plot is for a tracking solution with $\gamma_b=1$.
Finally, right plot is for a fluid dominated solution with $\gamma_b=1.5$
The dotted lines are for the different values of $r_{b}<r_{\star}$
(upper line), $r_{b}=r_{\star}$ (inner line) and $r_{b}>r_{\star}$
(lower line).}

\label{F-speed-1} 
\end{figure}

From the equation $\dot{R}^{2}(t,r_{b})=1$ (where $r_{b}$ is the
radius of the boundary shell) we can determine the speed of the collapse,
$|\dot{a}|_{\mathrm{AH}}$, at which horizons form, i.e., $|\dot{a}|_{\mathrm{AH}}=\frac{1}{r_{b}}$.
When the speed of collapse, $|\dot{a}|$, reaches the value $1/r_{b}$,
then an apparent horizon forms. Thus, if the maximum speed $|\dot{a}|_{\text{max}}$
is lower than the critical speed $|\dot{a}|_{\mathrm{AH}}$, no horizon
can form. More precisely, in order to discuss the dynamics of the
trapped region in the perspective of the effective dynamics scenario,
we consider $|\dot{a}|$ from Eq.~(\ref{Friedmann-eff-1a-tach}) to be
equal to $|\dot{a}|_{\mathrm{AH}}=1/r_{b}$. Solving this new equation
for $\rho$ and $a$ we get scale factors and energy densities at
which the horizon forms. Figure \ref{F-speed-1} represents the speed
of the collapse, $|\dot{a}|$, as a function of the scale factor,
reaching the maximum value $|\dot{a}|_{\textrm{max}}$.

The tachyon field equation (\ref{field}) implies that $\phi\equiv\phi\left(a\right)$.
Therefore, from Eqs.~(\ref{total-tach-bar})--(\ref{energyTach-hol}) we can also
establish that the total energy density can be expressed as a function
$\rho\equiv\rho\left(a\right)$. Then, we can rewrite $|\dot{a}|=\frac{1}{r_{b}}$
by setting $X:=\rho/\rho_{{\rm crit}}$ and $a^{2}:=f(X)$ as 
\begin{equation}
f\left(X\right)X\left(1-X\right)-A=0\ ,\label{theta2-q2-X-tach}
\end{equation}
where $A:=3/\left(8\pi G\rho_{{\rm crit}}r_{b}^{2}\right)$ is a constant.
The study of roots of the Eq. (\ref{theta2-q2-X-tach}) enables us to get
the values of energy density at which an apparent horizons form. Considering
more closely Eq.~(\ref{theta2-q2-X-tach}), we need to estimate the behaviour
of the function $f(X)$. In figures \ref{F-scalef3} and \ref{F-Endens},
we have that $f(X)$ is minimum when $X$ is maximum. It is also expected
that, since $f(X)$ is a monotonically decreasing function near the
bounce, Eq.~(\ref{theta2-q2-X-tach}) is essentially described as a second
order polynomial. Therefore, depending on the initial conditions,
in particular on the choice of the $r_{b}$, three cases can be evaluated,
which correspond to \emph{no} apparent horizon formation ($A/f\left(X\right)>1/4$),
one and two horizons formation ($A/f\left(X\right)\leq1/4$).

Let us introduce a radius $r_{\star}$, defined by 
\begin{equation}
r_{\star}\ :=\ \frac{1}{|\dot{a}|_{\text{max}}}\ .
\end{equation}
We see that $r_{\star}$ determines a \emph{threshold radius} for
the horizon formation; if $r_{b}<r_{\star}$, then no horizon can
form at any stage of the collapse. The case $r_{b}=r_{\star}$ corresponds
to the formation of a dynamical horizon at the boundary of the two
spacetime regions \cite{Ashtekar-Horizon:2002,Hayward1994}. Finally, for the case $r_{b}>r_{\star}$
two horizons will form, one inside and the other outside of the collapsing
matter.

The behaviour of the three possible scenarios (tracking solution,
tachyon and fluid dominated solutions) are also represented in figure
\ref{F-speed-1}. Therein, we note that only one horizon forms for
some particular tachyon dominated solutions. Therefore, for these
solutions the bounce will be covered by a horizon. In order to further
clarify this aspect, we note that when more than one horizon forms,
the speed of the collapse $\dot{a}$ must have a local maximum. In
that case, the acceleration must be $\ddot{a}=0$, and from Eqs.~(\ref{Friedmann-eff-1a-tach})--(\ref{LFriedmann2}) we can determine that
this local maximum can be found by imposing
\begin{equation}
\ddot{a}\equiv\dot{H}+H^{2}=-\dfrac{\rho_{\textrm{eff}}}{6}-\dfrac{p_{\textrm{eff}}}{2}=0\:.\label{accel-1}
\end{equation}
This last condition, being equivalent to $\rho_{\textrm{eff}}=-3\, p_{\textrm{eff}}$,
must be closely monitored for the three different solutions discussed
herein this section. For the fluid dominated solution, and since the
effective pressure starts from positive values and evolve to negatives
one near the bounce, it is straightforward to verify that the function
$-3\, p_{\textrm{eff}}$ must intersect $\rho_{\textrm{eff}}$ at
some point before reaching the bounce. For the tracking solution we
can use the same argument but with an initial effective pressure starting
near zero and reaching $3\,\rho_{\textrm{crit}}$ at the bounce. Finally,
the case of the tachyon dominated solution depends on the value of
the barotropic parameter $\gamma_b$. When the initial values for the
effective pressure and energy densities are $p_{\textrm{eff}}^{\Phi}\approx-\left(\gamma_b-1\right)\rho_{0b}$
and $\rho_{\textrm{eff}}^{\Phi}\approx\rho_{0b}>\rho_{\Phi0}$,
respectively; then, if $\gamma_b>2/3$, the argument given for the tracking
and fluid dominated solution is also valid for this case. However,
if $\gamma_b<2/3$, there will be always one horizon forming. Besides
taking $\gamma_b>2/3$, if we consider an unbalanced initial energy
density, with the tachyon being slightly dominant, i.e., $\rho_{\Phi0}\geq\rho_{0b}$,
a local maximum for $\dot{a}$ will also be present.

\subsection*{C.~Exterior geometry}

\begin{figure}
\includegraphics[height=2in]{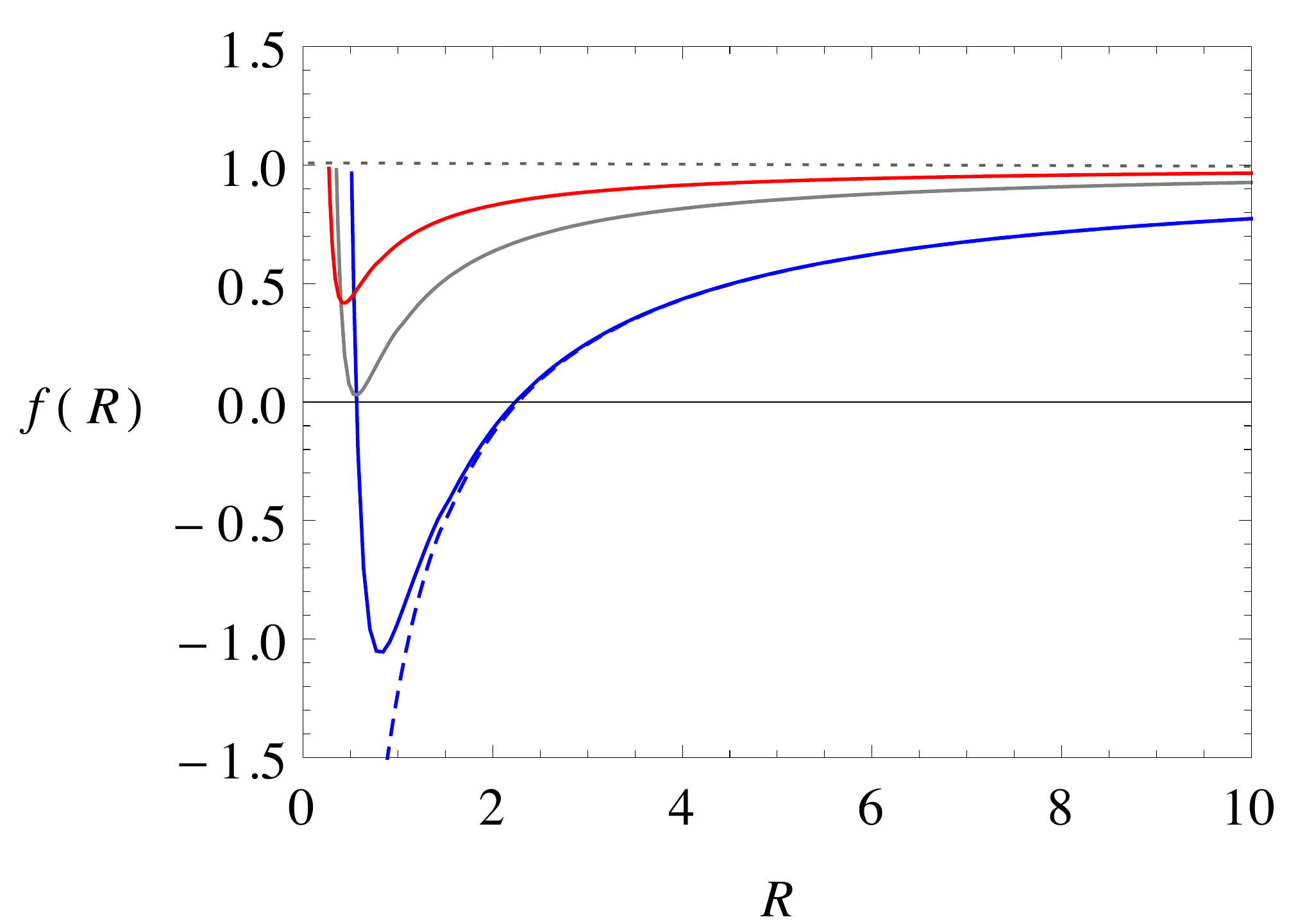}\includegraphics[height=2in]{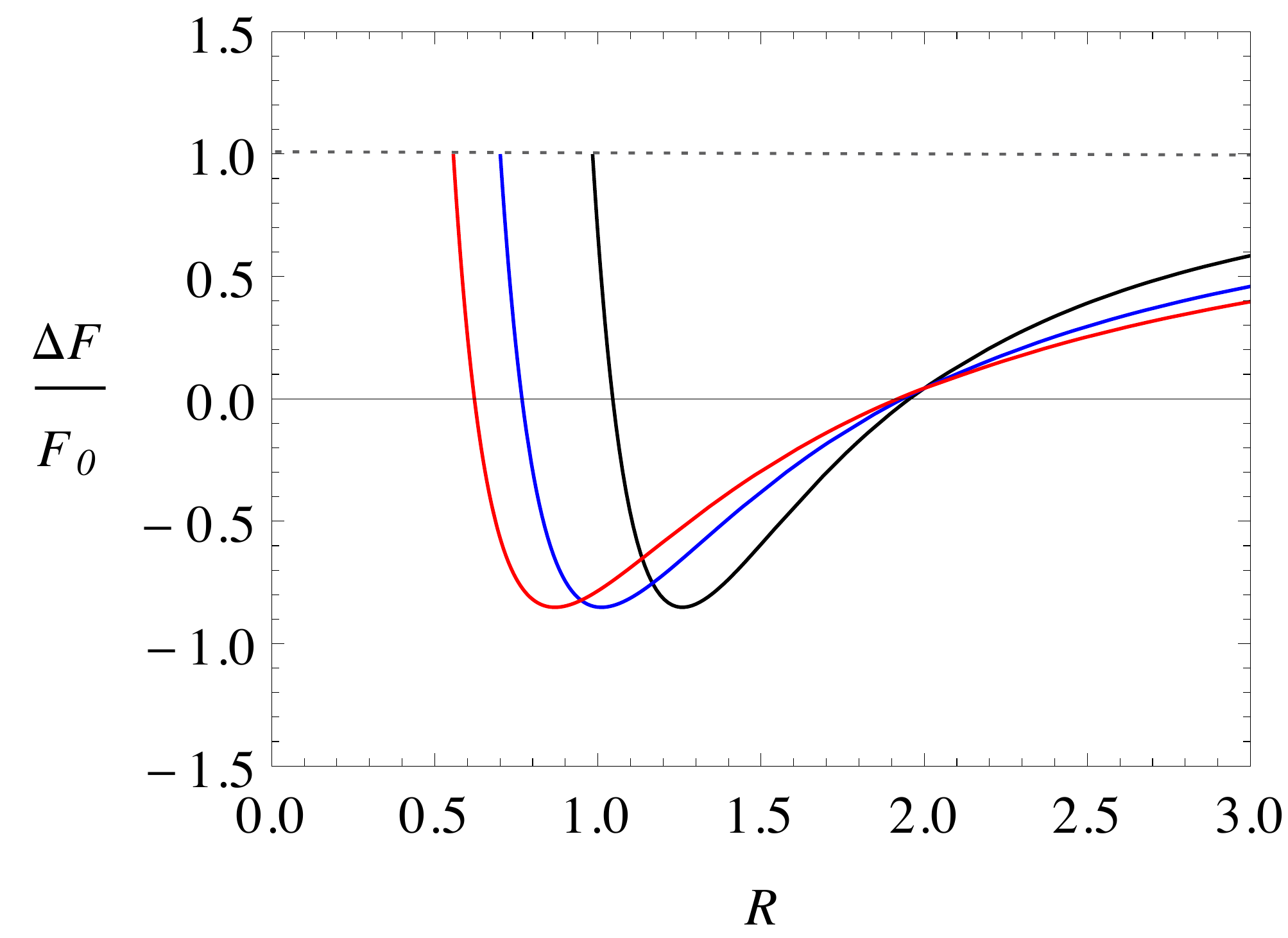}
\caption{The left plot represents the tracking solution ($\gamma_b\approx1$)
boundary function, $f(R)$, with respect to the radius $R$, in the
semiclassical (solid line), and classical (dashed curve) regime. We
considered the initial conditions as: $\rho_{0}=0.04$, $V_{0}=0.001$,
$a(0)=2$, $\dot{\Phi}\left(0\right)=0.5$ and $\Phi_{0}=0.6$. The
right plot shows the behaviour of the mass loss $\Delta F/F$, as
a function of area radius $R$. In this last plot we present the behaviour
of the tachyon dominated solution (reaching $\Delta F/F\rightarrow1$
at the smaller $R$), the tracking solution (reaching $\Delta F/F\rightarrow1$
at the intermediate $R$) and the fluid dominated solution (reaching
$\Delta F/F\rightarrow1$ at the bigger $R$).}

\label{Bound-f(R)} 
\end{figure}

Finally, the discussion of the final outcomes related to the herein
semiclassical solution follows the one made in appendix \ref{effectiveLQC}. Therein,
it is described that the fate of the collapsing star whose shell radius
is less than the threshold radius $r_{\star}$ points to the existence
of an energy flux radiated away from the interior spacetime and reaching
the distant observer. Herein, for a collapsing system whose initial
boundary radius $r_{b}$ is less than $r_{\star}$, we analyze the
resulting mass loss due to the semiclassical modified interior geometry.
In particular, this analysis is only carried for the tracking solution
or fluid dominated scenario, since the tachyon dominated solution
develops an horizon, for $\gamma_b<2/3$ and $\rho_{0b}\geq\rho_{\Phi0}$,
before reaching the bounce. Let us designate the initial mass function
at scales $\rho\ll\rho_{{\rm crit}}$, i.e, in the classical regime,
as $F_{0}=(8\pi G/3)\rho_{0}R_{0}^{3}$, with $\rho_{0}=\rho_{\Phi0}+\rho_{0b}$.
For $\rho\lesssim\rho_{{\rm crit}}$ (in the semiclassical regime)
we have, instead, an effective mass function $F_{\text{eff}}$ given
by Eq. (\ref{massF-eff}). Then, the (quantum geometrical) mass loss,
$\Delta F/F_{0}$ (where $\Delta F=F_{0}-F_{\text{eff}}$), for any
shell, is provided by the following expression: 
\begin{align}
\frac{\triangle F}{F(a_{0})}\   =\ 1-\frac{F_{\text{eff}}}{F_{0}} \  =\ 1-\sqrt{\frac{\rho}{\rho_{0}}}\left(1-\frac{\rho}{\rho_{{\rm crit}}}\right)\ .
\label{mass-loss-eff}
\end{align}
As $\rho$ increases, the mass loss decreases positively until it vanishes
at a point. Then, $\Delta F/F$ continues decreasing (negatively)
until it reaches to a minimum at $\rho=\rho_{{\rm crit}}/3$. Henceforth,
in the energy interval $\rho_{{\rm crit}}/3<\rho<\rho_{{\rm crit}}$,
the mass loss increases until the bouncing point at $\rho\rightarrow\rho_{\text{crit}}$,
where $\Delta F/F\rightarrow1$; this means that the quantum gravity
corrections, applied to the interior region, give rise to an outward
flux of energy near the bounce in the semiclassical regime. The previously
described behaviour for the mass loss will be qualitatively identical
with respect to the solution considered. Therefore the tachyon (when
$\gamma_b>2/3$ or the initial energy densities are $\rho_{\Phi0}>\rho_{0b}$),
fluid dominated or tracking solutions will exhibit the same profile
for the mass loss. The only difference between these three cases,
shown in the right plot of figure \ref{Bound-f(R)}, is the value
of the radius where the mass loss reaches the maximum $\Delta F/F\rightarrow1$.
In the last section we discussed the fact that the bounce occurring
in the tachyon dominated solution is delayed compared with the other
solutions. Consequently, the bounce (where $\Delta F/F\rightarrow1$)
take place for a smaller value of the radius $R$.

In the other case, where $r_{b}\geq r_{\star}$, in which one or two
horizon form, the exterior geometry can be obtained by matching the
interior to a generalized Vaidya exterior geometry at the boundary
$r_{b}$ of the cloud. 
Following section \ref{collapse-exterior}
we can write the exterior metric in advanced null coordinates $(v,R)$, given by Eq.~(\ref{metric2})
where the exterior function is given by $f(v,R)=1-2Gm(v,R)/R$. By
applying the matching conditions at the boundary $r_{b}$ we have
that 
\begin{equation}
m(v,R)\ =\ M-\frac{3}{4\pi\rho_{{\rm crit}}}\frac{M^{2}}{R^{3}}
\end{equation}
 where we have defined $M$ as  $M:=(4\pi/3)\rho R^{3}$ being the mass within
the volume $R^{3}$. For the tracking solution (\ref{trackb-1}) we
have 
\begin{equation}
M\ =\ \frac{4\pi}{3}\rho_{0}R_{0}^{3\gamma_b}R^{-3(\gamma_b-1)}.\label{mass-2}
\end{equation}
In the limit case $\gamma_b\sim1$, Eq.~(\ref{mass-2}) reduces to
$M=M_{0}=\left(4\pi/3\right)\rho_{0}R_{0}^{3}$. Figure \ref{Bound-f(R)}
shows the numerical behaviour of the boundary function $f(R)$ in
the classical (dashed curve) and semiclassical regime (solid curves)
for the cases of the initial masses $r_{b}<r_{\star}$ and $r_{b}\geq r_{\star}$.
The later shows the behaviour of an exotic non singular black hole
geometry which is different than its classical counterpart.

\section{Summary}
\label{Summary-3}

In  this  chapter  we   studied semiclassical models of gravitational collapse whose interior space-times include a tachyon scalar field and a barotropic fluid as matter content. 
By employing  quantum corrections,  induced by the inverse triad and holonomy operators of LQG, 
within a semiclassical description,  we analysed the final state of the collapse. Our studies can be summarized as follows.

The purpose of  section  \ref{LQG-tach}  
was to investigate a tachyon field, by means of investigating how (a modification from an inverse triad type) loop quantum effect can alter the outcome of gravitational collapse. To be more concrete, a particular class of spherically symmetric space-time was considered with a tachyon field $\Phi$ and a barotropic fluid constituting the matter content.
So, it was  of interest to investigate, by employing a phase space analysis, whether with an inverse square potential of the tachyon field, the space-time at the final stages of the collapse is regular or not. 
Within an inverse triad correction, we  obtained, for a semiclassical description, several classes of analytical as well as numerical solutions;  these were also subject to a study
involving a numerical appraisal, which added a clearer description of the dynamics.

We subsequently found a class of solutions which showed that the matter effective energy density remained finite as collapse evolves.
More precisely, using a dynamical system analysis, it was shown that the energy density of the tachyon was governed by its potential. Then 
the tachyonic energy density becomes regular during the collapse. On the other hand, the classical fluid with the energy density $\rho_\mathrm{b}\approx (a/a_*)^{-3\gamma_b}$ (which is singular at $a=0$ for $\gamma_b>0$) was modified in the loop quantum regime by a $D^{\gamma_b-1}$ factor, as $\rho_\mathrm{b}^{\rm sc}\approx (a/a_*)^{5\gamma_b-8}$. The phase space analysis allowed that, stable solutions  exist only for the range of  
$\gamma_b\geq 8/5$. Indeed, the semiclassical effects prevent a fluid with the barotropic parameter $\gamma_b<8/5$ from contributing to the gravitational collapse. Therefore, the barotropic energy density remains always finite and never blows up, as long as the semiclassical regime is valid.
Furthermore, the ratio of the effective mass function over the area radius stayed less than $1$; thus, no trapped surfaces form. In addition, a thorough numerical analysis showed that the Kretschmann scalar remains finite, suggesting the regularity of the geometry. This, together with the regularity of the energy density, indicated that the space-time in this semiclassical (inverse-triad-corrected) collapse does not lead to any naked singularity formation as long as the semiclassical approximation holds. Moreover, those corrections  induce an outward flux of energy  at the final state of the
collapse.

In section \ref{Holonomy-Tach} 
we employed another correction, namely the ``holonomy"  correction, imported from LQG to the dynamics of the gravitational collapse with a tachyon field and barotropic fluid as matter source.
Differently from the inverse triad corrected regime  (cf.  section \ref{LQG-tach}), the corresponding effective Hamiltonian constraint leaded to a quadratic density modification $H^2\propto \rho(1-\rho/\rho_{\rm crit})$. 
It is expected that the quadratic density modification can dominate over the inverse volume correction \cite{TPSingh:2006}.
This modification provides an upper limit $\rho_{\rm crit}$ for energy density $\rho$ of the collapse matter, 
indicating that the gravitational collapse includes a non singular bounce at the critical density $\rho=\rho_{\rm crit}$ (see also Refs. \cite{Tavakoli2013d,Xiong2007}). 
Our aim was to enlarge the discussion on (classical) tachyon field gravitational collapse, extending the scope analysed in
chapter \ref{collapse-class}, by investigating how  the  quantum gravity correction term $-\rho^2/\rho_{\rm crit}$, can alter the fate of the collapse.
Using a dynamical system analysis, we subsequently found a class of solutions corresponding to those introduced  in  section \ref{collapse-tachyon}.
Our analysis showed that, the corresponding stable fixed point solutions in the classical general relativistic collapse, are only \emph{saddle} points in our semiclassical regime;
hence, the  classical  black hole and naked singularities  provided by   section \ref{collapse-tachyon}  are no longer present within this loop semiclassical collapse. 
In addition, thorough numerical studies showed that there exists a threshold scale which distinguishes an outward energy 
flux from a non singular black hole forming at the collapse  final stages.

It is worthy to mention that, in the case of  a standard scalar field  (cf. appendix \ref{effectiveLQC}),  
a holonomy correction  to the semiclassical interior regime,  predicted a threshold scale below which no horizon forms during the collapse. 
This further leads to a modification to the mass function  resulted in an effective exterior Vaidya geometry.
Below this scale, by studying the effective mass loss of collapsing shells, it was shown that 
the quantum gravity induced effects give rise to an outward flux of energy reaching the distant observer; however, above
this threshold scale,  an exotic non singular black hole is predicted on the exterior whose geometry  is rather different than the Schwarzschild line element.
\chapter{Late time  singularities in dark energy cosmologies}
\label{cosmology}

\lhead{Chapter 4. \emph{Late time  singularities in dark energy cosmologies}} %



Several astrophysical observations (cf.  for example  SN-Ia \cite{type-Ia:1998,type-Ia:1999,type-Ia:2008},
LSS  \cite{LSS:2004,LSS:2008}, and  the CMB  \cite{CMB:2003,CMB:2007,CMB:2009} as observed by the  WMAP nine-year data (or  WMAP-9) \cite{WMAP2013} and Planck mission \cite{Planck2013}) 
confirm that the universe is undergoing a state of accelerating expansion.
The simplest setup to describe this acceleration is by
means of a cosmological constant, with an equation of state
$p_\Lambda=-\rho_\Lambda$, where $p_\Lambda$ is the pressure and
$\rho_\Lambda$ is the energy density of such a cosmological
constant.

There are, however, other candidates that astronomical
observations allow.
Indeed, such experiments indicate that  the matter content  of the universe, leading to the accelerating expansion, must contain an exotic energy which is characterized with negative pressure,  
and constitutes $68.3\%$ of the total matter content of the universe \cite{Planck2013}. 
The nature of this dark energy is still among the long-standing problems in theoretical physics.

There are several promising candidates  to alleviate  the dark energy problem which are inspired in fundamental physics. 
Among those models of dark energy, an interesting attempt for probing the nature of dark energy within the framework of quantum gravity  is the 
so-called ``holographic dark energy" proposal \cite{Hsu:2004ri,Holographich,Li:2004rb,holograph-1}. 
The holographic principle \cite{Bousso:2002} regards black holes as the maximally entropic objects
of a given region and implies the Bekenstein entropy bound of $S_\lambda = (L\lambda)^3 \leq S_{\rm BH} = M_{\rm Pl}^2L^2/8$, 
where $\lambda$ and $L$ are the UV (ultra-violet) cutoff and IR (infra-red) cutoff of the given system \cite{Bekenstein:1973,Bekenstein:1981,Joos:1992,Gonzalez:1983}. 
On the other hand,  it was suggested in Ref.~\cite{holograph-1} that 
the total energy of a system should not exceed the mass of the same-size black hole, $E_\lambda = \rho_\lambda L^3 \leq E_{\rm BH}=LM_{\rm Pl}^2/8\pi$ with the energy density $\rho_\lambda=\lambda^4$. 
This implies the maximum Bekenstein entropy bound of $S_\lambda \leq S^{3/4}_{\rm BH}$ \cite{Hooft:1993}. Based on the energy bound, it has been proposed in Ref.~\cite{Li:2004rb} that the holographic dark energy density may be inversely proportional to the square of  $L$ (namely, $\rho_{\rm DE}\propto L^{-2}$),  that characterises the size of the universe and representing the IR cutoff of it.
It turns out that there are many different ways of characterising the size of the universe and one of them is related to the inverse of the Ricci curvature, ${\cal R}$, of the universe. 
When the size of the universe is charaterised in such a way, namely $L^{-2}\propto {\cal R}$, we end up with the HRDE model \cite{Holographich}. 
This model is a promising candidate to explain the present accelerating universe.

Another very interesting approach to explain the late time speed up of the universe is to invoke an evolving vacuum energy which was named the running vacuum energy \cite{Sola2,Sola3,Shapiro-Sola}. 
This approach is related to the cosmological constant $\Lambda$ (problem), where  $\Lambda$ would be no longer constant along the expansion of the universe but changing as predicted from the renormalisation group equation for the vacuum energy \cite{Sola2}. This model has been recently generalized in Ref.~\cite{GRVE},  being almost undistinguishable from the standard $\Lambda$-CDM model  and renamed the GRVE.
Phenomenologically, this model has a formal analogy to the HRDE, where a combination of $\dot{H}$ and $H^2$ terms  is present on its energy density (cf. Eq.~(\ref{energy-GRVE})) with $\dot{H}$ the cosmic time derivative of the Hubble rate. Despite this analogy, the GRVE contains an additional constant term in order to allow for a transition from a decelerated to an accelerated expansion \cite{GRVE}. Furthermore, in the HRDE model, the energy density of every matter component is conserved while in the GRVE model, only the total energy density of matter is conserved.

In an expanding universe, the density of dark matter  decreases more quickly than the density of dark energy, and eventually the matter content of the universe becomes dark energy dominant at late times. 
Therefore, it is expected that the dark energy scenarios play an important role on implications for the fate of the universe.
In the context of dark energy cosmology, the study of  gravitational
theories that entail space-time singularities at late times has made
a considerable progress in the last years. Being more specific, within a FLRW
framework, the following classification of the late time singularities related to dark energy has been established  \cite{Nojiri:2005sx,Nojiri:2008a,Bamba:2008}: 
\begin{itemize}
\item Big rip singularity:  A singularity
at a finite cosmic time where the scale factor, the Hubble rate and
its cosmic time derivative diverge \cite{Caldwell:2003vq,O1,O2}.

\item Sudden singularity: A singularity
at a finite scale factor and in a finite cosmic time where the
Hubble rate is finite but its cosmic time derivative diverges
\cite{Barrow:2004xh,Nojiri:2004ip}.

\item Big freeze singularity:  A singularity at a
finite scale factor and in a finite cosmic time where the Hubble
rate and its cosmic time derivative diverge. \cite{BigFreeze1,BigFreeze2,BigFreeze3,BigFreeze4,BigFreeze5,BigFreeze6}.

\item Type IV singularity: A singularity at a finite
scale factor and in a finite cosmic time where the Hubble rate and
its cosmic derivative are finite but higher derivative of the Hubble
rate diverges. These singularities can appear in the  framework
of modified theories of gravity \cite{Nojiri:2005sx,Nojiri:2008a,Bamba:2008} (for a review on modified theories of gravity, see Ref.~\cite{Capozziello:2011}).
\end{itemize}
Subsequently, it has become of interest  to determine under which
conditions any of the above singularities can be removed. 
There exists a significant range of  approaches
to deal with dark energy related singularity: Modified theories of gravity \cite{Abdalla:2005,Bamba2012,Capozziello:2009}, WDW quantum cosmology  \cite{Kiefer:1,Kiefer:2,Kiefer:2011,Kiefer:2010,Kiefer-Moniz:2009}, 
and LQC  \cite{Future-sing-LQC,DE-LQC:2012,DE-LQC:2008,DE-LQC:2012b,DE-LQC:2008b,DE-LQC:2013}.

Moreover, there are at least two ways of implying late time acceleration in a FLRW universe:  Modified theories of gravity like brane-world models, and dark energy models like GRVE and HRDE models. In addition, both setups offer an arena to analyse dark energy related singularities. 

Higher-dimensional brane-world models \cite{Randall:1999,Brane-wolrd} which contain both bulk and brane curvature terms in the action admit late time cosmological singularities of rather unusual form and nature \cite{Brown:2005,BouhmadiLopez:2008nf,preparation}. 
So that, there has been great interest in brane-world cosmological models  in attempt to understand the dynamics of the universe at late times  \cite{Deffayet:2002a}.  
Brane-world models have a different qualitative behaviour than their general relativistic counterpart. 
A simple and effective model of brane  gravity is the DGP brane-world model \cite{DGP,Deffayet:2000uy} that has low energy (or IR) modifications through an induced gravity term. 
The GB higher order curvature term in the bulk action can also be used to construct brane models  \cite{GB-Brane}.  
These models modify gravity at high energy (or UV), unlike the DGP model.

It is known that the big rip singularity occurs at high energy \emph{and} in the future, so, it is of interest to 
investigate  whether  a combination of IR and UV effects can remove the big rip singularity or change its fate.  
This constitutes our main goal in  section~\ref{DGP}: We will employ a  DGP brane-world model, where the five dimensional bulk is characterized by a GB higher curvature term, and 
a `phantom matter fluid' (that emulates the dark energy dynamics) is  present on the brane.
We then investigate the late time cosmology of this DGP-GB brane-world model and analyse if the big rip can be removed or somehow appaised  \cite{Mariam2010a}.

In section \ref{GRVE} we study the GRVE scenario as an interesting theoretical model for dark energy cosmology.  
We first review  very briefly the GRVE model within the context of a FLRW universe.  
The Friedmann equation of the GRVE model looks pretty much similar to that of a homogeneous and isotropic universe filled with a  HRDE component. 
Despite the analogy between these two models, it turns out that one of them,  GRVE, is singularity-free in the future while the other, HRDE, is not \cite{Tavakoli2013c}. Indeed, a universe filled with a  HRDE component can hit, for example, a big rip singularity. This constitutes our motivation to study, in section  \ref{holographic},  the HRDE model and the issues of singularities therein. Finally we will provide a comparison between these two models, i.e., GRVE and HRDE.

\section{Smoothing the  big rip singularity in a brane-world  model of cosmology}
\label{DGP}

At high enough energies, 
the classical singularities predicted by general relativity  are expected to be removed by quantum gravity effects. 
However, even below the fundamental energy scale that marks the transition to a quantum regime, significant corrections to general relativity will arise, 
that could have a major impact on the process corresponding to gravitational collapse, black holes physics, and the early and late time (for some dark energy models) cosmology.
So that, quantum gravity effects could leave a trace  in some observations or even experiments. 
In this way, quantum gravity can begin to be tested by astrophysical and cosmological observations.

Two of the main contenders to a consistent quantum gravity theory are LQG
(see chapter \ref{QuantumG} where we explored some of its astrophysical predictions) 
and  M theory (for reviews see, e.g., \cite{M-theory}). 
In this section  we consider only  models that are inspired in string/M theories, namely,  brane-world gravity, and  explore some of its cosmological predictions.

The central idea in brane-world cosmology is that, 
the observable universe could be a $1+3$-Surface (the ``brane") embedded in a $1+3+d$-dimensional space-time (the ``bulk"),
with the standard model of particles and fields trapped on the brane while gravity is free to access the bulk.
At least one of the $d$ extra spatial dimensions could be very large relative to the Planck scale, which lowers the fundamental gravity scale, possibly even down to the electroweak ($\sim$TeV) level \cite{ADD:1998}.

Brane-world models offer a phenomenological way to test some of the novel predictions and corrections to general relativity,
including two simple models for cosmology:  The first is mainly based on warped 5D geometries and  based on the Randall–Sundrum models \cite{Randall:1999}. 
The second is the DGP model in which the extra volume can be infinite and induce mainly IR corrections to general relativity  \cite{DGP}.
We focus herein this section on the DGP model
modified by the GB effects (characterizing the five-dimensional bulk), due to the fact that this can provide a framework to investigate the intertwining of late time dynamics and high energy effects. 
Furthermore, we consider  the brane's matter content to be a `phantom matter fluid',
since this matter,  in a standard FLRW setting, induces a big rip singularity, and what we will analyse is the possibility of smoothing this singularity on the setup of brane-world models.

\subsection{The  DGP brane-world model with a GB term in the bulk}

A simple and effective brane-world model for late time cosmology is the DGP model \cite{DGP} that has IR modifications through an induced gravity term which corresponds to a scalar curvature on the brane action. 
It models our 4-dimensional world as a FLRW brane embedded in a 5-dimensional Minkowski bulk. 
Furthermore, this model explains the origin of the late time acceleration on the brane as gravity leaks into the bulk at late times. Nevertheless, this model has severe theoretical problems like the ghost problem as well as some issues with observations  \cite{Koyama:2005}. But it is still the simplest modified model of gravity to explain the late time acceleration of the universe, and therefore, can be used as a guiding light for constructing modified theories of gravity for late time cosmology.

On the DGP model, the 4-dimensional brane action  is proportional to $M_{\rm Pl}^2$ (with $M_{\rm Pl}$ being the 4-dimensional Planck mass) whereas in the bulk it  is proportional
to the corresponding quantity $\mathbf{M}_{\rm Pl}$ in 5-dimensions. 
More concretely,  the action for the DGP model is given by \cite{DGP}
\begin{equation}
S_{\rm  DGP} = \int d^5x\sqrt{|\det g^{(5)}|}~ \mathbf{M}_{\rm Pl}^3{\cal R}_{(5)} + \int d^4x\sqrt{|\det g|}\left( M_{\rm Pl}^2 {\cal R}_{(4)} - \tau + {\cal L}_{\rm matt} \right),
\label{DGP-action}
\end{equation}
where ${\cal R}_{(5)}$ is the bulk Ricci scalar, $g^{(5)}$ and $g$ denote the  metric in the bulk and the induced metric on the brane, 
${\cal R}_{(4)}$ is the Ricci scalar of the brane geometry, ${\cal L}_{\rm matt}$ is the matter Lagrangian on the brane, and $\tau$ is the brane tension which is zero in the DGP model. This model admits two solutions with the line element of the form \cite{Deffayet:2000uy}
\begin{equation}
ds^{2} = dy^2 + a^2(y) ds^2_{(4)}~,
\end{equation}
where $ds^2_{(4)}=-dt^2+a_0^2e^{2Ht}[dx_1^2+dx_2^2+dx_3^2]$  is the metric on a de Sitter space for $k=0$ FLRW model, and 
\begin{equation}
a(y)\ = \ H^{-1} \pm |y|~,
\end{equation}
where it was assumed a $\mathbb{Z}_2$ symmetry across the brane. The parameter $H$ is the Hubble constant of the brane which is located at $y=0$. Throughout this section, the upper sign corresponds to the  self-accelerating branch (or simply `$+$' branch) \cite{Deffayet:2000uy}. The `$-$' or `normal' branch can be visualized as the interior of the space bounded by a hyperboloid (the brane), whereas the $+$ branch corresponds to the exterior.

In order to obtain the Friedmann equation at the brane we need to use the five-dimensional bulk Einstein equations and the junction condition at the brane \cite{Davis:2003,Deruelle:2003,Maeda:2000}. 
For simplicity, we will assume a $\mathbb{Z}_2$ symmetry across the brane. 
For  a spatially flat brane without tension, in a Minkowski bulk, the Friedmann equation is then  given by \cite{Deffayet:2000uy}
\begin{equation}
\frac{H}{r_c} \left(Hr_c\mp1\right)\ =\  \frac{\kappa^2_4}{3}\rho\  ,
\label{DGP-Fried}
\end{equation}
where $\kappa^2_4=M_{\rm Pl}^{-2}$, and $\rho$ stands for the total energy density of the brane. Furthermore, we introduce the crossover scale $r_c$ as 
\begin{equation}
r_c\  :=\  \frac{M_{\rm Pl}^2}{2\mathbf{M}_{\rm Pl}^3}~,
\end{equation}
which marks the transition from 4-dimensional to 5-dimensional cosmology. More precisely, the model is characterized by the crossover length 
$r_c$ such that gravity is a 4-dimensional theory at scales $H^{-1}\ll r_c$, 
but it leaks out into the bulk at scales $H^{-1} \gg r_c$. 
Moreover it has been shown that the standard Friedmann cosmology can be embedded in DGP brane at least at relatively high energy.

The DGP model provides  IR modifications to gravity at large scales which is convenient  to study late time cosmologies. 
However, in a FLRW universe, a big rip singularity occurs at high energies and at late time,
so that, we need to consider a  brane model which includes as well   UV correction to general relativity for such a brane-world model to be able to smooth the big rip singularity. 
It turns out that a GB  curvature terms in the bulk action can be used to construct brane models that modify gravity at high energy, unlike the DGP model.
Therefore, we consider a brane-world model in which a GB term is present in the 5-dimensional Minkowski bulk containing a FLRW brane with an induced gravity term \cite{Kofinas:2003,Brown:2005}. 
This combination of IR and UV modifications leads to an intriguing cosmological scenario \cite{Brown:2005}. Therefore, it is assumed that  the gravitational action contains a GB term in the bulk, on top of the Einstein-Hilbert term, and an induced gravity term (see the second term in Eq.~(\ref{DGP-action})) is considered on the brane action. Therefore, the total gravitational action can be written as  \cite{Kofinas:2003}
\begin{equation}
S_{\rm  grav}= S_{\rm  DGP} + \frac{\alpha}{2\kappa_5^2}\int d^5x \sqrt{|\det g^{(5)}|} \left({\cal R}_{(5)}^2 - 4 {\cal R}_{(5)ab} {\cal R}_{(5)}^{ab}  + {\cal R}_{(5)abcd}{\cal R}_{(5)}^{abcd}\right),
\label{DGP-GB-action}
\end{equation}
where $\alpha>0$ is the GB coupling constant and $\kappa^2_5=\mathbf{M}_{\rm Pl}^{-3}$. Notice that, the DGP model is the special case $\alpha=0$, and in this case the crossover scale defines an effective 4-dimensional gravitational constant via $\kappa_4^2=\kappa_5^2/2r_c$.

The generalized Friedmann equation for the model (\ref{DGP-GB-action})
can be written as \cite{Kofinas:2003,Brown:2005}
\begin{equation}
H^2=\frac{\kappa_5^2}{6r_c}\rho\pm
\frac{1}{r_c}\left(1+\frac83\alpha H^2\right)H,
\label{modifiedfriedmann}
\end{equation}
where,  for  $\alpha=0$ this equation reduces to Eq.~(\ref{DGP-Fried}).

\subsection{The DGP-GB cosmological scenario with phantom matter}

In this section, we will consider the DGP-GB cosmological scenario for a universe whose (brane's) matter content  has a phantom fluid constituting the dark energy component.
Before proceeding into a more technical discussion, it could be of interest to further add the following about having a phantom matter component on the budget of the universe.

On the one hand, dark energy component with
$w<-1$, i.e. a phantom energy component is still a reasonable possibility as pointed out by 
the  recent WMAP-9 data \cite{WMAP2013} as well as by the Planck mission \cite{Planck2013}. For example, the WMAP-9 data (in
combination with other data) for a standard FLRW universe with
spatially flat sections, filled with CDM and a dark
energy component with a constant equation of state parameter, $w$,
predicts that $- 0.12<1+w<0.16$~ when combined with BAO data which gives
the most stringent limit, while WMAP-9 data alone predicts
that $-0.71<1+w<0.68$ for this model. For more details, see Ref.~\cite{Cos-parameters} and the recent Planck results for the cosmological parameters \cite{Planck2013}.
On the other hand, to investigate future singularities,  a perfect
fluid is satisfactory\footnote{Notice that the matter distribution is satisfactorily described by perfect fluids due to the large scale distribution of galaxies in our universe. However, realistic treatment of the problem requires the consideration of matter distribution other than the perfect fluid. It is well known that when neutrino decoupling occurs, the matter behaves as a viscous fluid in an early stage of the universe \cite{Misner1967}.}  and therefore we have not given an explicit
action for the phantom matter in terms of a minimally coupled scalar
field (with the opposite kinetical term) or through more general
scalar field actions like a k-essence action. Let us also add that
it is well known that the DGP brane has  an unstable
branch solution whose  instability is due to a ghost problem at the perturbative level \cite{Gregory:2008bf}. It may
therefore be questionable why to initiate a study within a DGP
setting or even insert phantom matter. Our point is that, in spite of these open
lines, the features characterizing the DGP as well as GB elements
can provide a framework to investigate the intertwining of late time
dynamics and high energy effects.  An interesting ground to test it
is with a phantom fluid, since this matter, in a standard relativistic FLRW
setting, induces a big rip singularity. Eventually, the unresolved
issues for the DGP brane will be eliminated and results such as the
one we bring here will increase in interest.

Therefore, for a late time evolving brane the energy density
is well described by
\begin{equation}
\rho=\rho_{\rm B}+\rho_{\rm CDM}+\rho_{\rm DE}~,
\end{equation}
where $\rho_{\rm B}$, ~$\rho_{\rm CDM}$ and $\rho_{\rm DE}$ corresponds to the
energy density of baryons, CDM and dark energy,
respectively. As $\rho_{\rm B}$ and $\rho_{\rm CDM}$ are both proportional to
$a^{-3}$, we will define their sum as $\rho_{\rm m}$; i.e.
$\rho_{\rm m}=\rho_{\rm B}+\rho_{\rm CDM}$. On the other hand, we will consider dark
energy to correspond to phantom energy. Finally, the total energy
density on the brane can be written as\footnote{In this chapter, quantities with the
subscript $0$ denote their values as observed today.}
\begin{equation}
\rho=\frac{\rho_{m0}}{a^3}+\frac{\rho_{d0}}{a^{3(w+1)}} \  ,
\label{mattercontent}
\end{equation}
where $w$,~$\rho_{m0}$,~$\rho_{d0}$ are constants and $1+w<0$.

It should be noted that from Eq.~(\ref{modifiedfriedmann}),  the
known  self-accelerating  DGP solution
($+$ sign in Eq.~(\ref{modifiedfriedmann}) with $\alpha = 0$) can be obtained, while the
\emph{normal} branch is retrieved for the $-$ sign with $\alpha =
0$ (cf. Ref. \cite{BouhmadiLopez:2008nf} for more details and notation).

Let us then address  Eq.~(\ref{modifiedfriedmann})
analytically,  selecting the $+$ sign. This equation with the matter content (\ref{mattercontent}) can be
expressed as
\begin{eqnarray}\label{eq5}
E^2(z) = \Omega_m(1+z)^3+\Omega_{d}(1+z)^{3(1+w)} + 2\sqrt{\Omega_{r_c}}\left [1+\Omega_\alpha E^2(z)\right ]E(z),
\end{eqnarray}
where $E(z) := H/H_0$, is the dimensionless Hubble parameter,  $z$ being the redshift and
\begin{eqnarray}\label{eq6}
 &\,&\Omega_m  := \frac{\kappa_4^2 \rho_{m0}}{3H_0^2}\  ,\ \ \  \  \   \   \   \   \   \   \ \  \  \   \   \   \   \   \ 
\Omega_{d} := \frac{\kappa_4^2\rho_{d0} }{3H_0^2}\  ,\,\,\,\,\nonumber \\
&\,&\Omega_{r_c} := \frac{1}{4r_c^2H_0^2}\  ,\ \  \  \  \  \   \  \  \  \    \ \  \  \   \   \   \   \   \ 
\Omega_{\alpha} := \frac{8}{3}\alpha H_0^2\  .
\end{eqnarray}

Evaluating the Friedmann equation (\ref{eq5}) at $z=0$ gives a
constraint on the cosmological parameter of the model
\begin{equation}
1=\Omega_{m}+\Omega_{d}+2\sqrt{\Omega_{r_{c}}}(1+\Omega_{\alpha}).\label{constraint}
\end{equation}
For $\Omega_{\alpha}=0$, we recover the constraint in the DGP model
without UV corrections. Coming back to our model, if we assume the
dimensionless crossover factor $\Omega_{r_{c}}$ to be  the same as
in the self-acceleration DGP model, then the similarities with a
spatially open universe are made more significant from the GB
effect, since $\Omega_{\alpha}>0$.

In order to obtain the evolution of the Hubble rate as a function of
the total energy density of the brane, we introduce the following
dimensionless variables:
\begin{eqnarray}\label{eq8}
\bar H& :=  &\frac83 \frac{\alpha}{r_c}H =
2\Omega_\alpha\sqrt{\Omega_{r_c}}E(z), \label{eq8a}\\
\bar\rho&   :=   &\frac{32}{27} \frac{\kappa_5^2\alpha^2}{r_c^3} \rho=4\Omega_{r_c}\Omega_\alpha^2\left[\Omega_{d}(1+z)^{3(1+w)}+\Omega_m(1+z)^3\right], \label{dimensionlessrho}\\
b&  :=   &\frac83\frac{\alpha}{r_c^2}=4\Omega_\alpha\Omega_{r_c}.
\label{defb} 
\end{eqnarray}
In terms of these variables, the modified Friedmann equation then
reads \cite{Mariam2010a} (see also Ref.~\cite{Abramowitz:1980}):
\begin{equation}
{\bar H}^3-{\bar H}^2+b\bar H+\bar\rho=0~. \label{Friedmannnb}
\end{equation}
The number of real roots is
determined by the sign of the discriminant function $N$ defined
as \cite{Abramowitz:1980},
\begin{equation}\label{eq10}
N=Q^3+R^2~,
\end{equation}
where $Q$ and $R$ are,
\begin{equation}\label{eq11}
Q=\frac{1}{3}\left(b-\frac{1}{3}\right)~,\ \ \ \ \ \   \   \ \  \  \   \   \   \   \   \   \ \ R=-\frac{1}{6}b-\frac{1}{2}\bar{\rho}+\frac{1}{27} ~ .
\end{equation}
For the analysis of the number of physical solutions of the modified Friedmann
equation (\ref{Friedmannnb}), it is helpful to rewrite $N$ as
\begin{equation}\label{eq12}
N=\frac{1}{4}(\bar{\rho}-\bar{\rho}_{1})(\bar{\rho}-\bar{\rho}_{2}),
\end{equation}
where
\begin{align}
\bar{\rho}_{1}\  &:=\   -\frac{1}{3}\left\{b-\frac{2}{9}[1+\sqrt{(1-3b)^3}]\right\},\label{rhoone} \\ 
\bar{\rho}_{2} \  & :=\   -\frac{1}{3}\left\{b-\frac{2}{9}[1-\sqrt{(1-3b)^3}]\right\}.\label{rhotwo}
\end{align}
Hence,  if $N$ is positive then there is a unique real solution. If
$N$ is negative, there are three real solutions, and finally, if $N$
vanishes, all roots are real and at least two are equal \cite{Abramowitz:1980}.

An approximated bound for the value of $b$ can be established noticing
 that  $b$  is proportional to
$\Omega_{r_{c}}$ through Eq.~(\ref{defb}). Hence, from the
equivalent quantity for the $\Omega_{r_{c}}$ in the DGP scenario
for the  self-accelerating branch \cite{Maartens:2006,Lazkoz:2007} 
and the constraint on the curvature of the 
universe\footnote{On this respect, notice that at $z=0$, the term $2\sqrt{\Omega_{r_c}}(1+\Omega_{\alpha})$ mimics a curvature term on the modified Friedmann equation (\ref{Friedmannnb}).}
(e.g., see Refs.~\cite{CMB:2003,CMB:2007,CMB:2009}),  
its value should be small. These physical solutions can be included on the set of mathematical solutions with  $0<b<\frac{1}{4}$. 
Therefore, for the remaining of this
section,  we shall study  in detail this setup. For completeness, the other cases are summarized in table \ref{DGP-table}.

\begin{table}[h!]

\begin{center}

\caption{Solutions for the algebraic equation (\ref{eq10}) with different ranges for $b$;    see also Eqs.~(\ref{highenergy})-(\ref{lowenergytwo}).}

\begin{tabular}{ccccc}
\hline  \hline 
     {\small$b$} &  {\small $\bar{\rho_{1}}$ and $\bar{\rho_{2}}$}  &  {\small Solutions for $\bar{H}$} &  {\small $\alpha$ and $\vartheta$}  &  {\small Description}  \tabularnewline    \hline 
     &  &   
   & {\small $\cosh(\alpha)\equiv-\frac{R}{\sqrt{-Q^3}}$},    &  {\small $\bar{H}_{1}<0$;}  \tabularnewline
    {\small $\frac14\leq b<\frac13$}   &  {\small $\bar{\rho_{1}}\leq0$, $\bar{\rho_{2}}<0$}  &   {\small $\bar{H}_{1}=-\frac{1}{3}[2\sqrt{1-3b}\cosh(\frac{\alpha}{3})-1]$ } &     {\small $\sinh(\alpha)\equiv\sqrt{\frac{N}{-Q^3}}$}~,     &  {\small contracting}  \tabularnewline 
     &  &    &     {\small $\alpha_0<\alpha$, where }   &  {\small brane} \tabularnewline 
      &  &    &    {\small $\cosh(\frac{\alpha_{0}}{3})=\frac{1}{2\sqrt{1-3b}}$}   &  \tabularnewline  
  \hline 
     & &  & &   {\small $\bar{H}_{1}<0$;}  \tabularnewline
     {\small $b=\frac13$} &    {\small $\bar{\rho_{1}}=\bar{\rho_{2}}=- \frac{1}{27}$}  &   {\small$\bar{H}_{1}=-\frac{1}{3}[(1+27\bar{\rho})^{-\frac{1}{3}}-1]$ } &        & {\small contracting}  \tabularnewline 
     &  &    &       & {\small brane}  \tabularnewline 
 \hline 
     & {\small $\bar{\rho_{1}}$ and   {\small $\bar{\rho_{2}}$ }} &  &  {\small $\cosh(\vartheta)\equiv\sqrt{\frac{N}{Q^3}}$}~,  &  {\small $\bar{H}_{1}<0$;} \tabularnewline
     {\small $b>\frac13$}  & {\small are complex}  &    {\small $\bar{H}_{1}=-\frac{1}{3}[2\sqrt{3b-1}\sinh(\frac{\vartheta}{3})-1]$} &   $\sinh(\vartheta)\equiv-\frac{R}{\sqrt{Q^3}}$~,     &  {\small contracting}  \tabularnewline 
     & {\small conjugates} &    &    {\small $\vartheta_0<\vartheta$, where }    & brane  \tabularnewline 
      &  &    &   {\small  $\sinh(\frac{\vartheta_{0}}{3})=\frac{1}{2\sqrt{1-3b}}$ }   &  \tabularnewline 
\hline \hline  
\end{tabular}

\label{DGP-table} 
\end{center}
\end{table}

For $0<b<\frac{1}{4}$ the values of $\bar{\rho}_{1}$ and $\bar{\rho}_{2}$ in Eqs.~(\ref{rhoone}) and (\ref{rhotwo}) are real. More precisely, in this case
$\bar{\rho}_{1}>0$ and $\bar{\rho}_{2}<0$. The number of solutions of the cubic
Friedmann equation (\ref{Friedmannnb}) will depend on the values of the energy
density with respect to $\bar{\rho}_{1}$. As the (standard) energy density redshifts
backward in time (i.e. it  grows), we can distinguish three regimes: (i) High
energy regime: $\bar{\rho}_{1}<\bar{\rho}$, (ii) limiting regime:
$\bar{\rho}=\bar{\rho}_{1}$, (iii) low energy regime: $\bar{\rho}<\bar{\rho}_{1}$:
\begin{itemize}
\item During the high energy regime, the energy density of the brane is
bounded from below by $\bar{\rho}_{1}$. There is a unique solution for the
cubic Friedmann equation (\ref{Friedmannnb}), because
 $N$ is a positive function for this case. So, the solution reads,
\begin{equation}
\bar{H}_{1}=-\frac{1}{3}\left[2\sqrt{1-3b}\cosh\left(\frac{\eta}{3}\right)-1\right],\label{highenergy}
\end{equation}
where $\eta$ is defined by
\begin{equation}\label{eq16}
\cosh(\eta)\equiv \frac{-R}{\sqrt{-Q^3}}~, \  \  \  \  \  \  \  \   \    \  \  \  \  \   \   \  \sinh(\eta)\equiv
\sqrt{\frac{N}{-Q^3}}~,
\end{equation}
and $\eta>0$. When $\eta\rightarrow0$, the energy density of the
brane approaches $\bar{\rho}_{1}$. This solution has a negative Hubble rate and
therefore it is unphysical for late time cosmology.
\item During the limiting regime, $\bar{\rho}=\bar{\rho}_{1}$, the
function $N$ vanishes, and there are two real solutions
\begin{align}
\bar{H}_{1}\ & =\  -\frac{1}{3}\left(2\sqrt{1-3b}-1\right),\label{limitingone} \\ 
\bar{H}_{2}\  &=\  \frac{1}{3}(\sqrt{1-3b}+1).\label{limitingtwo}
\end{align}
The solution $\bar{H}_{1}$ is negative and so it is also  not relevant
physically.
\item For the low energy regime,
$\bar{\rho}<\bar{\rho}_{1}$. In this case the function $N$ is
negative, and there are three different solutions. One of these
solutions is negative and corresponds to a contracting brane,
while the other two positive solutions correspond to  expanding
branes. Let us be more concrete:

The solution that describes the
contracting brane is similar to the corresponding solution of the
high energy regime:
\begin{equation}
\bar{H}_{1}=-\frac{1}{3}\left[2\sqrt{1-3b}\cos\left(\frac{\theta}{3}\right)-1\right],\label{lowenergy}
\end{equation}
where
\begin{equation}
\cos(\theta)\equiv\frac{-R}{\sqrt{-Q^3}}~,\  \   \   \   \  \    \  \  \  \  \   \    \  \   \   \   \sin(\theta)\equiv\sqrt{1+\frac{R^2}{Q^3}}~, \label{theta}
\end{equation}
and $0<\theta<\theta_{0}$. The parameter $\theta=0$ is defined as in Eq.~(\ref{theta}) in which the value of $\bar{\rho}$ reaches $\bar{\rho}_{1}$, and
the parameter $\theta_{0}$ corresponds to $\bar{\rho}=0$. For this  solution $\bar{H}_1$ is
negative and hence not suitable for the late time cosmology. In addition, the solution approaches the same Hubble rate at
$\theta=0$ as the limiting solution~(\ref{limitingone}).

On the other hand, the  two expanding branches are
described by
\begin{align}
\bar{H}_{2}\  &=\  \frac{1}{3}\left[2\sqrt{1-3b}\cos\left(\frac{\pi+\theta}{3}\right)+1\right],  \label{lowenergyone} \\
\bar{H}_{3}\  &=\  \frac{1}{3}\left[2\sqrt{1-3b}\cos\left(\frac{\pi-\theta}{3}\right)+1\right].  \label{lowenergytwo}
\end{align}
For $\theta\rightarrow0$ the energy density $\bar{\rho}$ approaches
$\bar{\rho}_{1}$ and  the low energy regime connects at the limiting
regime with the solution (\ref{limitingtwo}) where both solutions coincide;
it can be further shown that $\bar{H}_{2}\leq \bar{H}_{3}$.
The more interesting solution for us is the brane expanding solution described by
Eq.~(\ref{lowenergyone}): This solution constitutes a
generalization of the  self-accelerating DGP solution with GB effects 
\cite{Brown:2005,Kofinas:2003}, but within  our herein\footnote{It should be clear that we have 
a DGP-GB setting with phantom matter and therefore our
brane does not have, technically speaking, a self-accelerating phase,
asymptotically approaching a de Sitter stage in the future.} model.
\end{itemize}

\subsection{Asymptotic behaviour and nature of the future singularity}

\begin{figure*}
\includegraphics[width=1.0\textwidth]{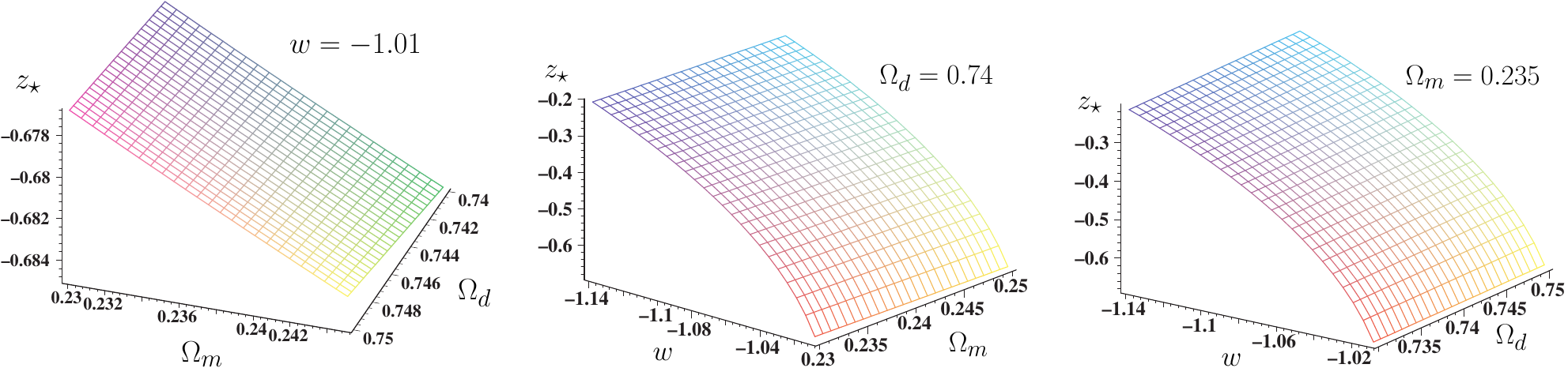}
\caption{Plot of the redshift  $z_{\star}$ at which the total energy density of the brane reaches its minimum value.}
\label{zstar}
\end{figure*}

The usual self-accelerating DGP-GB solution is known to have a sudden
singularity in the \emph{past}, when the  brane is filled by standard matter \cite{Brown:2005,Kofinas:2003}. 
Herein, we consider instead  the brane filled with CDM
plus phantom energy, performing an analysis aiming at the asymptotic future of the universe.
Starting from the modified Friedmann equation (\ref{Friedmannnb}), it  can be
shown that the first derivative of the Hubble rate is
\begin{equation}
\dot{H}=\frac{\kappa_{4}^2\dot{\rho}}{3[2H-\frac{1}{r_{c}}(1+8\alpha
H^2)]}\ ,\label{hderivative}
\end{equation}
where a `dot' stands for a  derivative with respect to the cosmic time $t$; $\dot{\rho}$ is given by the  energy density conservation equation (because there is no flow of energy from the brane to the bulk)
\begin{equation}\label{eq23}
\dot{\rho}+3H[\rho_{m}+(1+w)\rho_{d}]=0,
\end{equation}
for the total energy density of the brane. Therefore, $\rho$ decreases initially until the redshift  reaches the value $z_{\star}$,
\begin{equation}
z_{\star}=-1+\left[-(1+w)\frac{\Omega_{d}}{\Omega_{m}}\right]^{-\frac{1}{3w}},\label{redshiftmin}
\end{equation}
where $\dot\rho=0$. Afterwards, the phantom matter starts dominating the expansion of the brane; indeed the brane starts super-accelerating
($\dot{H}>0$) and the total energy density of the brane starts growing as the brane expands. In figure \ref{zstar}, we plot the redshift $z_{\star}$  versus reasonable observational  values of the parameters 
$\Omega_d$, $\Omega_m$ and $w$. As can be noticed in this figure, the total energy density of the brane would start increasing only in the future. Furthermore,  the larger  is $|w|$,  the sooner the phantom matter would dominate the expansion of the brane. The parameter  $\Omega_m$ has the opposite effect on $z_\star$ while $\Omega_d$ has a much milder effect on  $z_{\star}$.

Substituting  Eq.~(\ref{eq23}) into Eq.~(\ref{hderivative}), the first time derivative of the Hubble rate reads,
\begin{equation}
\dot{H}=-\frac{\kappa_{4}^2
H[(1+w)\rho_{d}+\rho_{m}]}{2H-\frac{1}{r_{c}}(1+8\alpha H^2)}\  .
\label{eq24}
\end{equation}
This equation shows that  when the Hubble rate approaches the constant
value\footnote{The Hubble rate given in Eq.~(\ref{eq25}) can be mapped into the dimensionless Hubble parameter  given in
Eq.~(\ref{limitingtwo}) for the limiting regime, reached at
the constant dimensionless energy density $\bar{\rho}=\bar{\rho}_{1}$.},
\begin{equation}
H\ =\  \frac{r_{c}}{8\alpha}\left(\sqrt{1-3b}+1\right),
\label{eq25}
\end{equation}
the first time derivative of the Hubble parameter, $\dot{H}$, diverges, while
 the energy density of the brane remains finite. Thus, instead of a scenario where  the energy density on the brane
blueshifts and eventually diverges, with a big rip singularity
eventually emerging, we find that $(i)$  a finite value of the (dimensionless) energy density
$\bar{\rho}=\bar{\rho}_{1}$ (the same applies for the pressure, i.e., a constant pressure), and $(ii)$ a finite value for the (dimensionless) Hubble
parameter $\bar{H}=\bar{H}_{2}$. Those values  are reached in the limiting regime described by Eq.~(\ref{limitingtwo}). And (iii) the first derivative of the Hubble parameter with respect to the cosmic time diverges at that point.
Therefore, the energy density is bounded, i.e. the limit
$z\rightarrow -1$ or $a\rightarrow \infty$ cannot be reached, and
instead  of a big rip singularity  we get  a `sudden singularity',
despite the brane being filled with phantom matter, as we will next show the brane takes a finite proper time from the present time to reach the singularity.

\begin{figure}[h]
\begin{center}
\includegraphics[width=0.5\columnwidth]{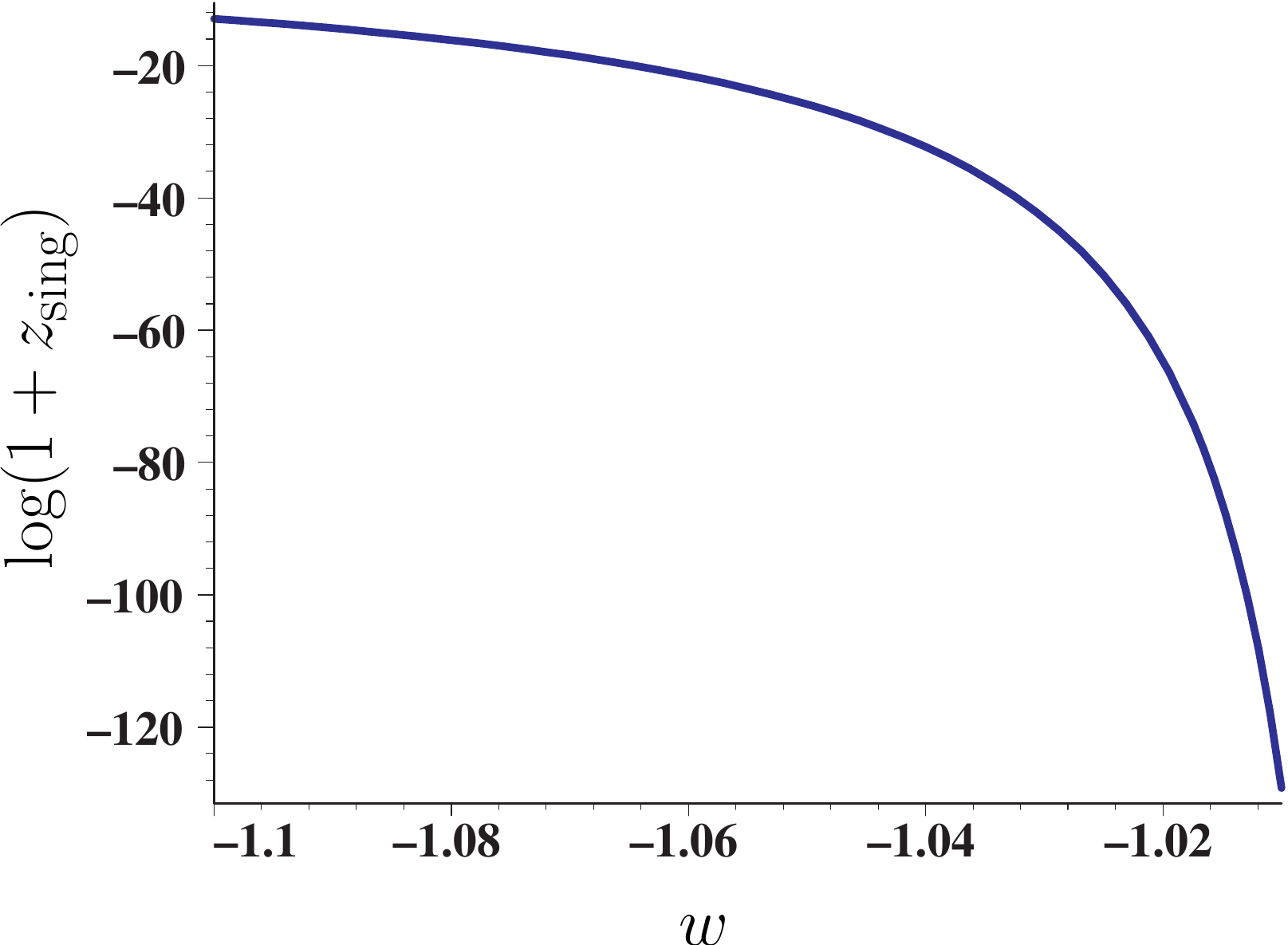}
\end{center}
\caption{Values for the redshift  at the sudden singularity,
$z_{\textrm{sing}}$, for the fixed value  $\Omega_{d}=0.75$, $\Omega_m=0.24$ and $\Omega_{r_{c}}=5.95\times 10^{-6}$ with respect to the constant equation of states.}
\label{zsing}
\end{figure}

\begin{figure}[h]
\begin{center}
\includegraphics[width=0.5\columnwidth]{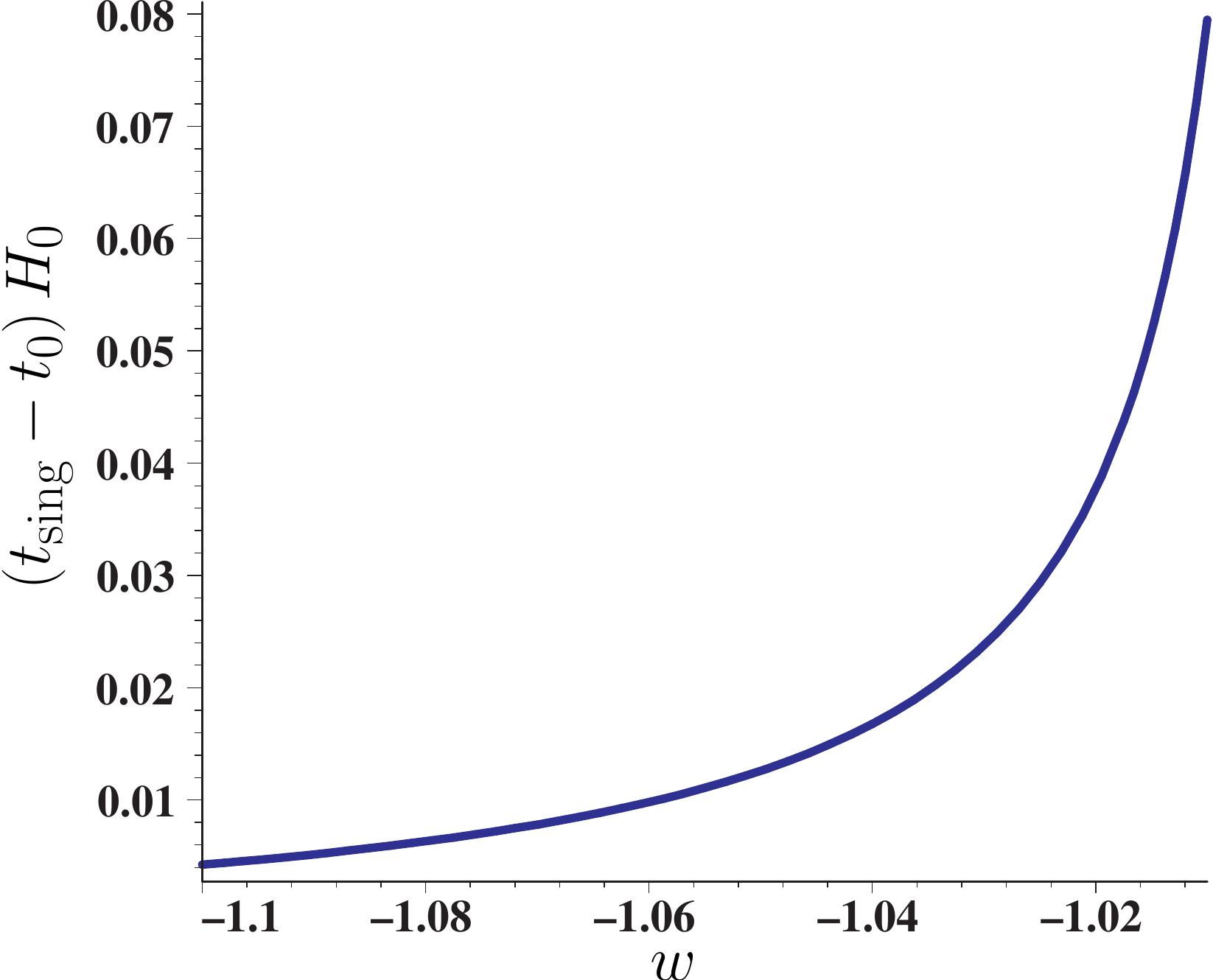}
\end{center}
\caption{Variation of the cosmic time left before the brane hits the sudden singularity,
$(t_{\rm{sing}}-t_0)H_0$ (in dimensionless units), for the values $\Omega_{d}=0.75$, $\Omega_{m}=0.24$
and $\Omega_{r_{c}}=5.95\times 10^{-6}$, with respect to the equation of state
parameter $w$.}
\label{tsing}
\end{figure}

The results above can be further elaborated, by establishing
\emph{when} the sudden singularity  will happen (i.e., at which
redshift and cosmic time values):

\begin{itemize}
\item In order to obtain the redshift $z_{\rm sing}$  where the sudden singularity takes place, we equate
  the dimensionless  total energy density of the brane (\ref{dimensionlessrho}) to its values at the sudden singularity  (cf. Eq.~(\ref{rhoone})). The allowed values for $z_{\rm{sing}}$ are  plotted in Fig.~\ref{zsing} for the
various values of the equation of state parameter $w$.
\item Finally,
using the relation between the scale factor and the redshift
parameter, $a(t)=1/(1+z)$, one can write the Hubble rate as a
function of the redshift parameter and its cosmic time derivative as
follows
\begin{equation}
H=\frac{\dot{a}}{a}=-\frac{\dot{z}}{1+z}~,
\label{eq27}
\end{equation}
and then integrating  this equation,  the cosmic time remaining before the brane hits the  sudden singularity,
reads
\begin{equation} 
(t_{\rm sing} - t_{0})H_0 \  =\   - \int_{z=0}^{z=z_{\rm sing}} \frac{dz}{(1+z)E(z)}~.
\label{t-end}
\end{equation}
In the previous equation $t_{0}$ and $t_{\rm sing}$ indicate the present time and
the time at the sudden singularity, respectively. We can plot
$t_{\rm sing}$ for fixed values of $(\Omega_{m},
\Omega_{d},\Omega_{\alpha})$; see Fig.~\ref{tsing}. As we can notice
from Fig.~\ref{tsing}, the closer is the equation of state of the
phantom matter to that of a cosmological constant,  the further
would be the sudden singularity. The same  plot is quite
enlightening as we can compare the age of the universe, essentially
$H_0^{-1}$, to the time left for such a sudden singularity to take
place on the future of the brane. As we can see, such a sudden
singularity would take place roughly in about 0.1 Gyr.
\end{itemize}

It is of interest to compare this estimate with the time at which a
big rip would take place in a standard four-dimensional universe filled with phantom matter whose equation of state is constant. The time remaining for such a
big rip singularity, previously studied (where $w=-1.5$ and $\Omega_m=0.3$ are used),  was estimated to be about 22 Gyr \cite{O2}.  
In comparison, for the model presented in this section, with $w=-1.5$ and
$\Omega_m=0.3$, $\Omega_d=0.68$, $\Omega_{\alpha}=0.005$, the
singularity emerges in $0.2$ Gyr. On the whole, although the
universe would meet a ``less severe'' singularity in the future, this
would happen much sooner than for the big rip case. 
In addition,  there are a few predictions in other models concerning possible astronomical events
that would indicate the emergence of the singularity herein exposed.
 For example, according to general relativity, by increasing  the  phantom energy,  every gravitationally bound system 
 (e.g., the solar system, the local group, galaxy clusters) will be dissociated at some point in time.
 Thus, the phantom energy rips apart the gravitationally bound objects before the death of the universe in a future doomsday \cite{O2}.
Within the model discussed in this section, some of those events or other of similar impact
could occur rather earlier.

\section{Avoiding the cosmic doomsday in the running vacuum energy scenario}
\label{GRVE}

The cosmological constant term, $\Lambda$, in Einstein’s equations  was first associated to the idea of vacuum energy density;
this  
cannot be a valid theoretical explanation for the accelerated expansion of the universe, and that we necessarily have to `go beyond $\Lambda$'  \cite{Weinberg:1989,Shapiro:2007}. 
The relevance of the cosmological constant problems  has triggered a renewed interest on the dynamical quantum effects on the vacuum energy density and their possible implications in cosmology 
(for a review of cosmological constant and vacuum energy, see Ref.~\cite{RVE-Review}).
The idea of running vacuum energy is a standard way to parameterize the leading quantum effects based on the renormalization group running of cosmological constant,
which provides  an attractive possibility in order to explain certain aspects of the cosmological constant problem \cite{RVE}.  
In this section, we will consider an interesting 
generalization of  this model, namely the GRVE model \cite{GRVE}.
Then, we will study the late time behavior of this cosmological scenario in order to investigate  the nature of the possible future singularity.

\subsection{The  GRVE model}

We assume a spatially flat FLRW universe filled with matter with energy density $\rho$ and pressure $p$, and the GRVE playing the role of dark energy. Then, the evolution of the universe is described by~\cite{GRVE}
\begin{align}
& \frac{\dot{a}^2}{a^2} \ =\ \frac{1}{3M_{\rm{Pl}}^2}\left( \rho+ \rho_{\Lambda}(H,\dot{H})\right), \label{Fried1} \\
& \frac{\ddot{a}}{a} \ =\ -\frac{1}{6M_{\rm{Pl}}^2} \left((\rho+3p) + 2 p_{\Lambda}(H,\dot{H})\right), \label{Fried-o}
\end{align}
where $\rho_{\Lambda}(H,\dot{H})$ is the ``GRVE density"   with $p_{\Lambda}(H,\dot{H})$ being the corresponding pressure:
\begin{align}
\rho_{\Lambda}(H,\dot{H})\ & =\ -p_{\Lambda}(H,\dot{H}) \notag \\ & = \  3M_{\rm{Pl}}^2 \left(C_0+C_H H^2+C_{\dot{H}}\dot{H}\right)  ,
\label{energy-GRVE}
\end{align}
in which the equation of state satisfying $w_{\Lambda}=p_{\Lambda}/\rho_{\Lambda}=-1$ is assumed. The parameters $C_0, C_H$ and $C_{\dot{H}}$ are constants. In addition, the constant\footnote{The coefficient $C_0$ can be estimated by evaluating the Eq.~(\ref{energy-GRVE}) at present. In addition, we assume that $C_H$ can be evaluated as in the standard running vacuum energy, given by $C_H\equiv\nu$, following the approach used in Ref.~\cite{GRVE}. \label{footnote1}} $C_0$ is given by \cite{Sola2,GRVE}
\begin{align}
C_0\  :=\  \frac{\rho_{\Lambda}^0}{3M_{\rm{Pl}}^2}-\nu H_0^2-C_{\dot{H}}\dot{H}_0\  ,
\label{C0}
\end{align}
where
\begin{align}
\nu\  := \ \sum_i\frac{B_i}{48\pi^2}\frac{M_i^2}{M_{\rm{Pl}}^2}\ ,
\end{align}
and $\rho_{\Lambda}^0=\rho_{\Lambda}(H_0)$ is the energy density defined at the present time, $t_0$, or equivalently at the current Hubble rate  $H_0=H(t_0)$. In addition, $M_i$ are the masses of particles contributing in the loops \cite{Sola3}.
The dimensionless parameter $\nu$ provides the main coefficient
of the $\beta$-function for the running of the vacuum energy, and $B_i$ are coefficients computed from the quantum (loop) contributions of fields with masses $M_i$ \cite{Sola3}.
Meanwhile, $C_H$ and $C_{\dot{H}}$ are dimensionless coefficients that can be fitted to the observations. For convenience
we will set henceforth, a new notation as $C_H\equiv \nu$ and $C_{\dot{H}}\equiv \frac{2}{3}\alpha$ (see also footnote \ref{footnote1}), where $\alpha$ and $\nu$ are expected to be small (c.f. Ref.~\cite{GRVE}).
On the other hand, $\rho$ is the remaining matter energy density, given in the standard cosmological case as
\begin{align}
\rho\ =\ \rho_{\rm{m}}+\rho_{\rm{r}}~, \label{menergy}
\end{align}
with $\rho_{\rm{m}}$ being the energy density of the non-relativistic dust-like matter ($p_{\rm{m}}=w_{\rm{m}}~\rho_{\rm{m}}=0$), and $\rho_{\rm{r}}$  the energy density for radiation ($p_{\rm{r}}=w_{\rm{r}}~\rho_{\rm{r}}=\frac{1}{3}~\rho_{\rm{r}}$).

A characteristic of the GRVE model is that while the total energy density is conserved, a given energy density is not conserved.
Therefore, a  local conservation law is employed on the GRVE setup, whereas it does not yield a conservation equation for each component separately. 
More precisely, from the system of Eqs.~(\ref{Fried1})-(\ref{Fried-o}) we get the total conservation equation
\begin{equation}
\dot{\rho}_{\rm m}+\dot{\rho}_{\rm r}+\dot{\rho}_{\Lambda} = -3H\rho_{\rm m} - 4H\rho_{\rm r}~,
\label{cons-GRVE}
\end{equation}
which is indeed a first integral of that system \cite{GRVE}. After some calculations the total conservation equation (\ref{cons-GRVE}) reads
\begin{equation}
\dot{\rho}_{\rm m}+\dot{\rho}_{\rm r} - \alpha \left(\dot{\rho}_{\rm m}+\frac{4}{3}~\dot{\rho}_{\rm r}\right) = -3H(1-\nu) \left(\rho_{\rm m}+\frac{4}{3}~\rho_{\rm r}\right).
\label{conserv-2}
\end{equation}
We note that the model does not yield a conservation equation for each component separately. 
Moreover, the system is not fully defined by Eqs.~(\ref{Fried1})-(\ref{Fried-o}). Indeed, any solution of the following set of equations
\begin{align}
\dot{\rho}_{\rm m} &\ =\ -3H\xi_{\rm m} ~\rho_{\rm m} + Q~,\ \ \  \  \  \  \  \ \ \text{where}\ \ \  \  \  \  \  \ \  \xi_{\rm m}\  := \  \frac{1-\nu}{1-\alpha}\ , \label{energy-m}\\
\dot{\rho}_{\rm r} &\ =\ -4H\xi_{\rm r} ~\rho_{\rm r} -Q ~,\ \  \  \  \  \  \ \ \ \text{where}\ \  \  \  \  \  \  \ \  \xi_{\rm r} \  :=\  \frac{1-\nu}{1-4\alpha/3}\ , \label{energy-r}
\end{align}
for arbitrary function $Q(t)$ will be a solution of Eq.~(\ref{conserv-2}). The decoupling constant $Q(t)$  must vanish during the matter dominated ($\rho_{\rm r}\approx 0$) and radiation dominated ($\rho_{\rm m}\approx 0$) periods where matter and radiation cannot deviate too much  from the standard scaling with the scale factor. This is a plausible condition that can be extended to the whole evolution of the universe \cite{GRVE}.
Therefore, we assume the case $Q=0$, with the solution for the matter components as:
\begin{align}
\rho_{\rm m} &\ =\ \rho_{\rm m}^0 a^{-3\xi_{\rm m}},  \label{energy-m}\\
\rho_{\rm r} &\ =\ \rho_{\rm r}^0 a^{-4\xi_{\rm r}},  \label{energy-r}
\end{align}
where $\rho_{\rm{m}}^0$ and $\rho_{\rm{r}}^0$ are  respectively, the energy density of dust and radiation at the present time.
Note that these decoupled solutions reduce automatically to the behavior of dust-like matter during matter domination and to the radiation component during radiation domination. 
Furthermore, the coefficients $\xi_{\rm m}$ and $\xi_{\rm r}$  take the standard forms,
 $\xi_{\rm m}\approx 1$ and $\xi_{\rm r}\approx 1$, on that case.

\subsection{The late time cosmological scenario}

Substituting now Eqs.~(\ref{energy-m}), (\ref{energy-r}) and (\ref{energy-GRVE}) in Eq.~(\ref{Fried1}) we can rewrite the generalised Friedmann equation in the following form:
\begin{align}
E^2 \ =\ \Omega_{\rm{m}}(1+z)^{3\xi_{\rm{m}}}+\Omega_{\rm{r}}(1+z)^{4\xi_{\rm{r}}} +\Omega_0+\nu E^2+\frac{2\alpha}{3H_0}\dot{E},
\label{Fried2}
\end{align}
with the dimensionless parameters:
\begin{align}
\Omega_{\rm{m}}\ :=\  \frac{\rho_{\rm{m}}^0}{3M_{\rm{Pl}}^2H_0^2}\ ,\ \ \  \  \ \ \Omega_{\rm{r}}\ :=\  \frac{\rho_{\rm{r}}^0}{3M_{\rm{Pl}}^2H_0^2}\ ,\ \ \  \  \  \  \Omega_0\ :=\ \frac{C_0}{H_0^2}\ .
\label{dimless-paraM}
\end{align}
In addition,  the dimensionless parameter for the GRVE density can be written as
\begin{align}
\Omega_{\Lambda} =\frac{\rho_{\Lambda}}{3M_{\rm{Pl}}^2H_0^2}=\Omega_0+\nu E^2+\frac{2\alpha}{3H_0}\dot{E}.
\end{align}

Evaluating the Friedmann equation (\ref{Fried2}) at the present time, $z=0$, gives
a constraint on the cosmological parameters of the model which reads
\begin{align}
1 &\ =\ \Omega_{\rm{m}}+\Omega_{\rm{r}}+\Omega_{\Lambda}^0\ ,
\label{Friedt0}
\end{align}
where $\Omega_{\Lambda}^0$ is defined as
\begin{align}
\Omega_{\Lambda}^0\ :=\ \Omega_{\Lambda}(z=0)=\Omega_0+\nu+\frac{2\alpha}{3H_0}\dot{E}(z=0) .
\label{Friedt0-1}
\end{align}

The quantities $\dot{H}$ and $H^2$ are related through $\dot{H}=-(1+q)H^2$, where $q$ is the deceleration parameter. Therefore, using Eq.~(\ref{Friedt0-1}), we can write down the deceleration parameter at the present time:
\begin{align}
q_0\  =\  -1+\frac{3}{2\alpha}\left(\Omega_0+\nu-\Omega_{\Lambda}^0\right).
\label{Friedt0-2}
\end{align}
Since the universe is currently accelerating, i.e. $q_0<0$,  we obtain the constraint
\begin{align}
\frac{3}{2\alpha}(\Omega_0+\nu-\Omega_{\Lambda}^0)<1\ .
\label{Friedt0-3}
\end{align}
Notice that, a successful cosmological model must be able to produce an accelerated expansion at very low redshifts.

It is convenient to rewrite the generalised Friedmann equation (\ref{Fried2}) by introducing a new variable $x := -\ln(z+1)=\ln (a)$, as follows
\begin{align}
\frac{\dot{E}}{H_0}\ =\ -\frac{3}{2\alpha}\left[\Omega_{\rm{m}}e^{-3\xi_{\rm{m}}x}+\Omega_{\rm{r}}e^{-4\xi_{\rm{r}}x}+(\nu-1)E^2+\Omega_0\right].
\label{Fried3}
\end{align}
Substituting $\dot{E}=H dE/dx$ in Eq.~(\ref{Fried3}) we can further rewrite the Friedmann equation in the following form:
\begin{align}
\frac{dE^2}{dx}\ =\ -\frac{3}{\alpha}\left[\Omega_{\rm{m}}e^{-3\xi_{\rm{m}}x}+\Omega_{\rm{r}}e^{-4\xi_{\rm{r}}x}+(\nu-1)E^2+\Omega_0\right].
\label{Fried4}
\end{align}
Solving the Eq.~(\ref{Fried4}) and rewriting the result in terms of the redshift, we obtain
\begin{align}
E^2(z)  = E_{\rm{m}}(1+z)^{3\xi_{\rm{m}}}+E_{\rm{r}}(1+z)^{4\xi_{\rm{r}}} + E_0 + E_\nu(1+z)^{\frac{3}{\alpha}(\nu-1)},
\label{sol1}
\end{align}
where
\begin{align}
E_0 = \frac{\Omega_0}{(1-\nu)}\ , \ \ \  \ \ \  \ \ \  E_{\rm{m}} = \frac{\Omega_{\rm{m}}}{\xi_{\rm{m}}}\ ,  \ \ \  \ \ \  \ \ \  E_{\rm{r}}  = \frac{\Omega_{\rm{r}}}{\xi_{\rm{r}}}\ ,
\label{sol2}
\end{align}
and $E_\nu$ is an integration constant. Those constants are constrained by
\begin{align}
1  \ =\ E_{\rm{m}}+ E_{\rm{r}} + E_0 + E_\nu \ .
\label{sol2-zerotime}
\end{align}
The first two terms in Eq.~(\ref{sol1}) are related to the energy densities of matter, $\rho_\mathrm{m}$, and radiation, $\rho_\mathrm{r}$, and the last two terms are related to  the GRVE density $\rho_\Lambda$. In the far future, as $z$ decreases, the matter content of the universe is negligible, and therefore the energy density of the universe will be dominated by the GRVE density.
By substituting Eq.~(\ref{sol1}) in the generalised Friedmann equation (\ref{Fried3}), we get the first time derivative for the Hubble parameter:
\begin{align}
\frac{\dot{E}}{H_0}  = -\frac{1}{2} \left[3\Omega_{\rm{m}}(1+z)^{3\xi_{\rm{m}}}+4\Omega_{\rm{r}}(1+z)^{4\xi_{\rm{r}}} + \frac{3}{\alpha}(\nu-1)E_\nu(1+z)^{\frac{3}{\alpha}(\nu-1)}\right].
\label{sol1-2}
\end{align}
Furthermore, using the solution (\ref{sol1}) and its time derivative (\ref{sol1-2}) in Eq.~(\ref{energy-GRVE}), we obtain the GRVE density given as follows
\begin{align}
\rho_\Lambda &\ =\  3H_0^2M_{\rm{P}}^2\left[(\nu-\alpha\xi_{\rm{m}})E_{\rm{m}}(1+z)^{3\xi_{\rm{m}}} + E_0 \right. \notag \\ 
& \  \  \   \  \  \  \  \ \left.+ \left(\nu-\frac{4}{3}\alpha\xi_{\rm{r}}\right)E_{\rm{r}}(1+z)^{4\xi_{\rm{r}}} + E_\nu (1+z)^{\frac{3}{\alpha}(\nu-1)}\right].
\label{venergy-sol}
\end{align}
It can be seen that at present time,  $\rho_\Lambda=M_{\rm{Pl}}^2 \Lambda$ reduces to an effective cosmological constant given by
\begin{align}
\Lambda = 3H_0^2\left[(\nu-\alpha\xi_{\rm{m}})E_{\rm{m}} + \left(\nu-\frac{4}{3}\alpha\xi_{\rm{r}}\right)E_{\rm{r}} + E_0 + E_\nu \right].
\label{venergy-sol-CC}
\end{align}
Using Eq.~(\ref{sol2-zerotime}), we can rewrite $E_{\nu}$  in terms of $E_{\rm{m}}$, $E_{\rm{r}}$ and $E_0$. Finally,
substituting this expression of $E_{\nu}$ in Eq.~(\ref{venergy-sol-CC}) the constraint (\ref{Friedt0}) is recovered,
where $\Lambda/3H_0^2\equiv\Omega_{\Lambda}^0$.

The GRVE density (\ref{venergy-sol}) is obtained herein as a function of the energy density of the matter component and the radiation one, plus the last two terms which play an important role on the fate of the universe at the very low redshift regime in the future. In the following section, we will discuss the fate of the possible singularities in this context.

\subsection{Avoidance of  future singularities}

In the presence of the last term in Eq.~(\ref{venergy-sol}), if\footnote{From now on, we will focus on the case $\frac{3}{\alpha}(\nu-1)<0$, because we are trying to see if this model can avoid the big rip singularity which could appear precisely in this case.}  $\frac{3}{\alpha}(\nu-1)<0$ then
the Hubble rate and its time derivative diverge and hence, the universe undergoes a big rip singularity at $z=-1$. 

On the other hand, the total energy density of the universe, $\rho_{\rm{tot}}=\rho_{\rm{m}}+ \rho_{\rm{r}} +\rho_{\Lambda}$, reads
\begin{align}
\rho_{\rm{tot}}  = 3H_0^2M_{\rm{Pl}}^2  \left[E_{\rm{m}}(1+z)^{3\xi_{\rm{m}}}+ E_{\rm{r}}(1+z)^{4\xi_{\rm{r}}} + E_0 + E_\nu (1+z)^{\frac{3}{\alpha}(\nu-1)}\right].
\label{energy-tot}
\end{align}
The total energy density (\ref{energy-tot}) must satisfy the conservation law
\begin{align}
\dot{\rho}_{\rm{tot}} + 3H (\rho_{\rm{tot}}+p_{\rm{tot}})=0,
\label{cons-law-tot}
\end{align}
with $p_{\rm{tot}}$ being the total pressure of the cosmological system: $p_{\rm{tot}}=p_{\rm{m}}+ p_{\rm{r}} +p_{\Lambda}$, where
$p_{\rm{m}}=0$, $p_{\rm{r}}=\frac{1}{3}~\rho_{\rm{r}}$, and $p_{\Lambda}=-\rho_{\Lambda}$.
Therefore, the total conservation law (\ref{cons-law-tot}) gives
\begin{align}
 (\Omega_{\rm{m}}-\xi_{\rm{m}}E_{\rm{m}})(1+z)^{3\xi_{\rm{m}}} + \frac{4}{3}\left(\Omega_{\rm{r}}-\xi_{\rm{r}}E_{\rm{r}}\right)(1+z)^{4\xi_{\rm{r}}} 
 + \left(\frac{1-\nu}{\alpha}\right)E_\nu(1+z)^{\frac{3}{\alpha}(\nu-1)} =  0.
\label{cons-law-tot2}
\end{align}
By using the definitions (\ref{sol2}) in Eq.~(\ref{cons-law-tot2}), the first two terms vanish spontaneously, but the last term does not. Thus, the conservation equation implies
\begin{align}
\left(\frac{1-\nu}{\alpha}\right)E_\nu(1+z)^{\frac{3}{\alpha}(\nu-1)} = 0.
\label{cons-law-tot3}
\end{align}
The conservation Eq.~(\ref{cons-law-tot}) (or equivalently Eq.~(\ref{cons-law-tot2})) is fulfilled only when $\nu=1$ or $E_\nu=0$ (c.f. Eq.~(\ref{cons-law-tot3})). In the former case ($\nu=1$), the last term in the equality (\ref{venergy-sol}) behaves as a cosmological constant. In the latter case ($E_\nu=0$), the last term in Eq.~(\ref{venergy-sol}) vanishes which corresponds exactly to the case analysed in Ref.~\cite{GRVE}. Indeed, the local conservation law constrains the GRVE density leading to the evolution of  $\rho_{\Lambda}$ \emph{only} in terms of the matter energy component, the radiation component and an effective cosmological constant. Therefore, since the  energy density of matter and radiation vanish in the far future $(z\rightarrow-1)$,  the GRVE density remains finite. Therefore, the  GRVE  scenario is free of future singularities and becomes asymptotically de Sitter in the future.

\section{Late time cosmology with a  holographic dark energy}
\label{holographic}

There is another attractive model of dark energy which was originated from the consideration of the holographic principle; it is the so called holographic dark energy 
\cite{Li:2004rb,Hsu:2004ri}, inspired on applying  the holographic principle to the universe as a whole,
whose energy density is inversely proportional to the square of an appropriate length, $L$,  that characterises the size of the system, in this case the universe, and represents the IR cutoff of it. One of the natural choices of this length, $L$, is the inverse of the Hubble rate. However, this choice does not induce acceleration in a homogeneous and isotropic universe \cite{Hsu:2004ri} (see 
Refs.~\cite{preparation,saridakis:2008a,saridakis:2008b,saridakis:2008c} for an example, where a modification of the model presented in Ref.~\cite{Hsu:2004ri} can explain the current acceleration of the universe).  
Another choice for the length $L$  is given in Ref.~\cite{Holographich} (see also Ref.~\cite{Nojiri:2005pu}), in which the IR cutoff of the  HRDE was taken to be the Ricci scalar curvature, i.e. $L^2\propto 1/{\cal R}$. This model can describe the present accelerating universe.

On this section, we will  analyse the late time behavior of a homogeneous and isotropic universe where the HRDE plays the role of dark energy.
We point out also that in some particular cases the HRDE model, when endowed with a negative cosmological constant, 
can mimic dark matter and explain the late time cosmic acceleration through an asymptotically expanding de Sitter universe.

\subsection{Background dynamics for the HRDE model}

We consider a flat FLRW universe in the presence of non-relativistic matter, radiation and a HRDE component \cite{Holographich} (see also Ref.~\cite{Tavakoli2013c}). The Friedmann equation for this model reads
\begin{align}
\frac{\dot{a}^2}{a^2}\  =\  \frac{1}{3M_{\rm{Pl}}^2}(\rho_{\rm{m}}+\rho_{\rm{r}}+\rho_{\rm{H}}),
\label{fried-H}
\end{align}
where  $\rho_{\rm{H}}$ denotes the energy density of  the HRDE  component.  
The pressureless matter, $\rho_{\rm{m}}$, and radiation, $\rho_{\rm{r}}$, are self-conserved unlike in the GRVE model (discussed in section \ref{GRVE}), that is
\begin{align}
\rho_{\rm{m}} &\ =\ 3 M_{\rm{Pl}}^2H_0^2~\Omega_{\rm{m}}(1+z)^{3}, \notag \\ \rho_{\rm{r}} &\ =\ 3 M_{\rm{Pl}}^2H_0^2~\Omega_{\rm{r}}(1+z)^{4},
\label{energy-m2}
\end{align}
where $\Omega_{\rm{m}}$ and $\Omega_{\rm{r}}$ are the dimensionless energy density parameters defined in Eq.~(\ref{dimless-paraM}). Furthermore, the HRDE density is proportional to the inverse of the Ricci scalar curvature radius ${\cal R}$:
\begin{align}
{\cal R}\ =\ 6(\dot{H}+2H^2)~.
\end{align}
Therefore, the HRDE density is defined as \cite{Holographich}
\begin{align}
\rho_{\rm{H}}\ =\ 3\beta M_{\rm{Pl}}^2\left(\frac{1}{2}\frac{dH^2}{dx}+2H^2\right)~,
\label{energy-H}
\end{align}
where $\beta=c^2$ is a dimensionless parameter that measures the strength of the holographic component. By rewriting Eq.~(\ref{energy-m2}) in terms of $x=-\ln(z+1)$ and substituting it together with Eq.~(\ref{energy-H}) in Eq.~(\ref{fried-H}),  the Friedmann equation can be rewritten as \cite{Holographich}
\begin{align}
E^2 = \Omega_{\rm{m}}e^{-3x}+\Omega_{\rm{r}}e^{-4x}
+ \beta \left(\frac{1}{2}\frac{dE^2}{dx}+2E^2\right).
\label{fried-H2}
\end{align}
Therefore, the dimensionless energy density parameter of the HRDE component can be written as
\begin{align}
\Omega_{\rm{H}}(x)\  =\  \beta \left(\frac{1}{2}\frac{dE^2}{dx}+2E^2\right).
\label{dimless-paraH}
\end{align}

Notice that, the Friedmann equation (\ref{fried-H2}) is pretty much similar to the Friedmann equation (\ref{Fried2}) for the GRVE model. There is a difference which is based on the fact that Eq.~(\ref{Fried2}) contains a phenomenological cosmological constant which is absent in Eq.~(\ref{fried-H2}). For the sake of completeness, we will consider as well a phenomenological cosmological constant, $\tilde{\Omega}_0$, in the model discussed on the present section, therefore Eq.~(\ref{fried-H2}) will be rewritten as
\begin{align}
E^2\ =\ \Omega_{\rm{m}}e^{-3x}+\Omega_{\rm{r}}e^{-4x}+ \tilde{\Omega}_0
+ \beta \left(\frac{1}{2}\frac{dE^2}{dx}+2E^2\right),
\label{fried-H2-2}
\end{align}
and $\tilde\rho_0 := 3M_{\mathrm{Pl}}^2H_0^2 \tilde{\Omega}_0$ is a constant. 
We will compare the model resulting from Eq.~(\ref{fried-H2-2}) with the one of the previous subsection (for the GRVE model) in the presence or absence of the cosmological constant $\tilde{\Omega}_0$.

By evaluating the Friedmann equation (\ref{fried-H2-2}) at the present time, we obtain a constraint on the dimensionless parameters of the model:
\begin{align}
1=\Omega_{\rm{m}}+\Omega_{\rm{r}}+\tilde{\Omega}_0+\Omega_{\rm{H}_0}.
\label{constraint-2b}
\end{align}
We henceforth solve the Friedmann equation  (\ref{fried-H2})  to study the late time evolution of the universe, by assuming different ranges for the holographic parameter $\beta$.

\subsection{Late time cosmology as described by the HRDE model}

After solving the Friedmann equation (\ref{fried-H2-2}), we get
\begin{align}
E^2(z) = \frac{2\Omega_{\rm{m}}}{2-\beta}(1+z)^{3} + \Omega_{\rm{r}}(1+z)^{4} 
+ \Omega_\beta(1+z)^{4-\frac{2}{\beta}} + \frac{\tilde{\Omega}_0}{1-2\beta}\ ,
\label{Hubble-H2}
\end{align}
where $\beta\neq\frac{1}{2}\ , 2$, and $\Omega_\beta$ is an integration constant. Then, by evaluating the solution (\ref{Hubble-H2})
at the present time, we obtain
\begin{align}
1=\frac{2\Omega_{\rm{m}}}{2-\beta} + \Omega_{\rm{r}} + \Omega_\beta + \frac{\tilde{\Omega}_0}{1-2\beta}\ ,
\label{constraint-2c}
\end{align}
which is a complementary constraint to that given in Eq.~(\ref{constraint-2b}).

Substituting $E(z)$ from Eq.~(\ref{Hubble-H2}) in  Eq.~(\ref{energy-H}), we obtain the HRDE density:
\begin{align}
\rho_{\rm{H}}= 3 M_{\rm{Pl}}^2H_0^2 \left[\frac{\beta}{2-\beta}~\Omega_{\rm{m}}(1+z)^{3} 
+ \Omega_\beta(1+z)^{4-\frac{2}{\beta}}+\frac{2\beta}{1-2\beta}~\tilde{\Omega}_0\right].
\label{energy-H-2}
\end{align}
Notice that, in the HRDE model, it is assumed that the energy density of the different components filling the universe is conserved and in particular the one corresponding to the HRDE. So that, by substituting the energy density (\ref{energy-H-2}) in the conservation law $\dot{\rho}_{\rm{H}}+3H(\rho_{\rm{H}}+p_{\rm{H}})=0$, we obtain the  HRDE pressure,  $p_{\rm{H}}$:
\begin{align}
p_{\rm{H}}\ =\ -3 M_{\rm{Pl}}^2H_0^2 \left[\frac{2\beta}{1-2\beta}~ \tilde{\Omega}_0 
+\left( \frac{2}{3\beta}-\frac{1}{3}\right)\Omega_\beta(1+z)^{4-\frac{2}{\beta}}\right].
\label{press-H-2}
\end{align}
Finally, the total energy density reads
\begin{eqnarray}
\rho_{\rm{tot}}\  =\  3 M_{\rm{Pl}}^2H_0^2 \left[\frac{2\Omega_{\rm{m}}}{2-\beta}~(1+z)^{3} + \Omega_{\rm{r}}(1+z)^{4}+ \Omega_\beta(1+z)^{4-\frac{2}{\beta}}+\frac{\tilde{\Omega}_0}{1-2\beta}\right].
\label{energy-H-tot}
\end{eqnarray}
Before continuing, we notice that the term $(1+z)^{4-\frac{2}{\beta}}$ on the previous equation induces acceleration if and only if $0<\beta<1$. We will impose this condition; $0<\beta<1$, to ensure late time acceleration even in the absence of a cosmological constant $\tilde\Omega_0$. In the far future, as $z$ tends to $-1$, the universe would be dominated by the holographic dark energy or the cosmological constant $\tilde\Omega_0$. For a positive cosmological constant ($\tilde{\Omega}_0>0$), if the range of the holographic parameter satisfies $0<\beta<\frac{1}{2}$, then the energy density (\ref{energy-H-tot}) and the Hubble rate (\ref{Hubble-H2}) diverge as well as $\dot H$ and $p_H$; therefore, the universe hits a big rip singularity. 
Notice that for $0<\beta<\frac{1}{2}$, the future singularity is avoided only for vanishing $\Omega_\beta$.
 However,  $\Omega_\beta$ has a crucial role in the acceleration of a  ``holographic" universe, hence, the big rip singularity is unavoidable within the HRDE scenario unless $1>\beta>\frac{1}{2}$. In addition, for the range of the HRDE parameter $1>\beta>\frac{1}{2}$, the last term in Eq.~(\ref{Hubble-H2}) becomes negative, whereas the rest of the terms are positive. Therefore, as the universe evolves, at some redshift $z_\mathrm{b}$ in the future, the positive and negative terms in Eq.~ (\ref{Hubble-H2}) will be cancelled, and hence the Hubble parameter vanishes at that redshift.
Using the relation $\dot{E}/H_0=dE^2/(2dx)$ and  Eq.~(\ref{Hubble-H2}), we get the time derivative of the Hubble rate:
\begin{align}
\frac{\dot{E}}{H_0}\ =\ -\frac{3\Omega_{\rm{m}}}{2-\beta}~(1+z)^{3} - 2\Omega_{\rm{r}}(1+z)^{4}
- \left(\frac{1}{\beta}-2\right)\Omega_\beta(1+z)^{4-\frac{2}{\beta}} \ ,
\label{Hubble-H2bb}
\end{align}
which remains finite at $z_\mathrm{b}$ where the Hubble rate vanishes; i.e. $\dot{E}(z_\mathrm{b})=const.$, when $E(z_\mathrm{b})=0$. Therefore, the universe in this case will bounce in the future and contract afterwards.

For a negative cosmological constant ($\tilde{\Omega}_0<0$), a similar analysis shows that, if $0<\beta<\frac{1}{2}$,  at some redshift $z_b$ the universe will bounce in the future. 
If  $\frac{1}{2}<\beta<1$, the universe is asymptotically de~Sitter, getting therefore, a de~Sitter universe even from a negative cosmological constant.

In order to complete the discussion of this section, we will analyse the cases of the holographic parameter $\beta$, when $\beta=\frac{1}{2}$ and $\beta=2$.

For the case of $\beta=\frac{1}{2}$~, the solution for the Friedmann equation (\ref{fried-H2-2}) in terms of the redshift reads
\begin{align}
E^2(z)\ =\ \frac{4}{3}\Omega_{\rm{m}}(1+z)^{3} + \Omega_{\rm{r}}(1+z)^{4}  + 4\tilde{\Omega}_0\ln(1+z)\ ,
\label{Hubble-H3}
\end{align}
where the time derivative of the Hubble parameter is given by
\begin{align}
\frac{\dot{E}}{H_0} = -2\left[\Omega_{\rm{m}}(1+z)^{3} + \Omega_{\rm{r}}(1+z)^{4}+ \tilde{\Omega}_0 \right].
\label{HubbleDot-H3}
\end{align}
The constraint (\ref{constraint-2b}) at the present time, in this case, can be written as
\begin{align}
1=\frac{4}{3}~\Omega_{\rm{m}} + \Omega_{\rm{r}} .
\label{constraint-2c-1}
\end{align}
Moreover, the holographic energy density is given by
\begin{align}
\rho_{\rm{H}}\ =\ 3 M_{\rm{Pl}}^2H_0^2 \left[\frac{\Omega_{\rm{m}}}{3}(1+z)^{3}+4\tilde{\Omega}_0\ln(1+z)-\tilde{\Omega}_0\right],
\label{energy-H-2-b}
\end{align}
with the holographic pressure reading:
\begin{align}
p_{\rm{H}}\ =\ 3 M_{\rm{Pl}}^2H_0^2 \tilde{\Omega}_0 \left[\frac{7}{3} - 4\ln(1+z) \right].
\label{press-H-2b}
\end{align}
Then, using Eqs.~(\ref{fried-H}) and (\ref{Hubble-H3}), the total energy density filling the universe would be
\begin{align}
\rho_{\rm{tot}}\ =\ 3 M_{\rm{Pl}}^2H_0^2 \left[\frac{4}{3}~\Omega_{\rm{m}}(1+z)^{3} + \Omega_{\rm{r}}(1+z)^{4} + 4\tilde{\Omega}_0\ln(1+z)\right] \ .
\label{Hubble-H3tot}
\end{align}
On the one hand, for a positive cosmological constant ($\tilde{\Omega}_0>0$) the first two terms in Eq.~(\ref{Hubble-H3}) are positive, whereas the last term is negative as the universe evolves at late time. Therefore, there exists a moment in the future, namely at a redshift $z_\mathrm{b}$, at which the Hubble rate vanishes, whereas the time derivative of the Hubble rate remains finite. Therefore, the universe bounces in the future.

On the other hand, for a negative cosmological constant ($\tilde{\Omega}_0<0$), all terms in Eq.~(\ref{Hubble-H3}) are positive in the future. In the far future, the Hubble rate diverges at $z=-1$ while its cosmic time derivative is finite. It can be checked that this event happens at an infinite cosmic time. Therefore,  the universe undergoes a kind of  smooth little rip singularity in the far future.  
This kind of event is the so called ``the little sibling of the big rip singularity"  \cite{Mariam2}.

Finally, in the absence of the cosmological constant ($\tilde{\Omega}_0=0$), the Hubble rate and its time derivative vanish at $z=-1$; the universe becomes Minkowskian in the far future.

It is surprising that in a HRDE model with $\beta=1/2$, the presence of a positive cosmological constant can induce a bounce while the presence of a negative cosmological constant can induce a little sibling of the big rip singularity \cite{Mariam2} (a smoother version of the little rip).

If the holographic parameter is such that $\beta=2$, the solution of the Friedmann equation (\ref{fried-H2-2}) reads
\begin{align}
E^2(z) = \Omega_{\rm{m}}(1+z)^3 \ln(1+z) + \Omega_{\rm{r}}(1+z)^4 + \Omega_2(1+z)^3 - \frac{\tilde{\Omega}_0}{3}\ ,
\label{Hubble-H4}
\end{align}
where $\Omega_2$ is a constant. Furthermore, the constraint equation (\ref{constraint-2b}) for this solution reads
\begin{equation}
1= \Omega_{\rm{r}} + \Omega_2 -  \frac{\tilde{\Omega}_0}{3} \ .
\label{constraint-2c-2}
\end{equation}
The time derivative of the Hubble rate (\ref{Hubble-H4}) is obtained easily as follows,
\begin{align}
\frac{\dot{E}}{H_0} = -\frac{1}{2}\left[\Omega_{\rm{m}}(1+z)^3 + 3\Omega_2(1+z)^3 + 4\Omega_{\rm{r}}(1+z)^{4} + \Omega_{\rm{m}}(1+z)^{3}\ln(1+z) \right].
\label{HubbleDot-H4}
\end{align}
Substituting Eq.~(\ref{Hubble-H4}) in Eq.~(\ref{energy-H}), the holographic energy density $\rho_{\rm{H}}$ reads
\begin{align}
\rho_{\rm{H}}\ =\ 3 M_{\rm{Pl}}^2H_0^2 \left[\Omega_\mathrm{2}(1+z)^3 - \Omega_\mathrm{m}(1+z)^3~\ln(1+z) -\frac{4}{3}~\tilde{\Omega}_0\right].
\label{energy-H-2-c}
\end{align}
Finally, the holographic pressure, $p_{\rm{H}}$,  can be obtained by substituting Eq.~(\ref{energy-H-2-c}) in the conservation equation; then, we get
\begin{align}
p_{\rm{H}}\ =\  M_{\rm{Pl}}^2H_0^2 \left[ \Omega_\mathrm{m}(1+z)^3 + 4\tilde{\Omega}_0 \right].
\label{press-H-2c}
\end{align}
Furthermore, the total energy density of the universe when the holographic parameter fulfils $\beta=2$ reads
\begin{align}
\rho_\mathrm{tot} =  3M_{\rm{Pl}}^2H_0^2\left[\Omega_{\rm{m}}(1+z)^3~ \ln(1+z) +\Omega_{\rm{r}}(1+z)^4 + \Omega_2(1+z)^3 - \frac{\tilde{\Omega}_0}{3}\right]\ .
\label{energy-H4}
\end{align}
This solution shows that, for a given cosmological constant no matter its sign (i.e., $\tilde{\Omega}_0>0$, $\tilde{\Omega}_0<0$, and $\tilde{\Omega}_0=0$),  the Hubble rate (\ref{Hubble-H4}) vanishes at some redshift $z_\mathrm{b}$ in the future, whereas the time derivative of Hubble rate (\ref{HubbleDot-H4}) remains finite; therefore, the universe hits a bounce at $z_\mathrm{b}$ in the future.

On the other hand, if $\Omega_\mathrm{m}=0$, the Hubble rate (\ref{Hubble-H4}) reduces to
\begin{align}
E^2(z) = \Omega_{\rm{r}}(1+z)^4  + \Omega_2(1+z)^3 - \frac{\tilde{\Omega}_0}{3}\ .
\label{Hubble-H4bb}
\end{align}
Notice that, the evolution of the term proportional to $\Omega_2$ in Eq.~(\ref{Hubble-H4bb}) is dust-like indicating that, the HRDE can play the role of dark matter in this case. In addition, for a positive cosmological constant, the Hubble rate would vanish at some $z_b$($\neq -1$) in the future. Furthermore, the time derivative of the Hubble rate, defined as
\begin{align}
\frac{\dot{E}}{H_0} = -\frac{1}{2}\left[ 3\Omega_2(1+z)^3 + 4\Omega_{\rm{r}}(1+z)^{4} \right],
\label{HubbleDot-H4bbb}
\end{align}
remains finite at late time; therefore, the universe hits a bounce within a finite time in the future. In this case there will be no acceleration in the far future of the universe. On the other hand, for a negative cosmological constant, the Hubble rate  (\ref{Hubble-H4bb}) remains finite and non-zero at $z=-1$ while  its time derivative (\ref{HubbleDot-H4bbb}) vanishes at $z=-1$. Therefore, the universe becomes de Sitter in the far future. Consequently, the presence of a negative cosmological constant in this case can explain the late time acceleration of the universe.


\section{Summary}
\label{Summary-chapter4}

In this chapter the study of future cosmological singularities was the subject of
interest and some proposals for their removal or appeasal
have been advanced.   

It was investigated in subsection \ref{DGP}  
whether a composition of specific IR and UV effects could alter a
big rip singularity setting (cf. Ref.~\cite{Mariam2010a}). More concretely, it was employed a simple
model: A  DGP brane model, with phantom matter on the brane and a GB term in the
bulk action. 
The DGP brane configuration has relevant IR effects, whereas
the GB component is important at high energies; phantom matter (with a constant equation of state) in a
standard FLRW model is known to induce the emergence of a big rip
singularity \cite{Caldwell:2003vq}. This analysis indicates that the big rip can be  replaced  by a
smoother singularity, named a  sudden future singularity \cite{Nojiri:2004ip}, through some intertwining between
late time dynamics and high energy effects. Subsequently, it
was determined values of the redshift and the future  cosmic time,  where the brane would
reach the sudden  singularity. These results can be contrasted
with those for the big rip occurrence in a FLRW setting (e.g., see Ref.~\cite{O2}).
Notice  that,  these conclusions are based on a rather
particular result,  that was extracted from a specific model.
Subsequent research work would assist in clarifying some remaining
issues. 
For example,  further studies
of how other singularities can be appeased or removed by means
of the herein combined IR and UV  effects have been done in Ref.~\cite{Mariam:2013}. On the other hand, it might be 
interesting to consider  a modified  Einstein-Hilbert  action on the brane,
which in addition could alleviate the ghost problem present on the
self-accelerating DGP model by self-accelerating the normal DGP
branch \cite{BouhmadiLopez:2009db}, 
and see if some of the dark energy singularities can be removed or at least appeased in this setup.

We further considered in sections  \ref{GRVE}  and  \ref{holographic}  two recently proposed models for dark energy within the context of FLRW cosmology;  the GRVE model (cf. section  \ref{GRVE}  and Ref.~\cite{GRVE}) and the HRDE scenario (cf. section \ref{holographic} and Ref.~\cite{Holographich}). 
Even though the Friedmann equation of both models looks pretty much similar [cf. Eqs.~(\ref{Fried2}) and (\ref{fried-H2})], there is a difference which is based on the fact that Eq.~(\ref{Fried2}) contains a phenomenological cosmological constant which is absent in Eq.~(\ref{fried-H2}). For the sake of completeness and in order to compare both models, we have considered as well a phenomenological cosmological constant on the HRDE scenario.

On the one hand, in the GRVE model, a local conservation law constrains the total energy density of the universe but the energy momentum tensor of the different component filling the universe is not self-conserved. Indeed, this conservation law provides an energy transfer between the matter components and the GRVE density,  leading to a running vacuum energy  dominating the universe at late time. It turns out that, in the far future, the universe is asymptotically de Sitter and free from singularities.

On the other hand, each component of the energy density in the HRDE model is self conserved, leading to  different behaviour of the HRDE energy density depending on the HRDE parameter $\beta$.
In this case, the universe may hit a big rip, a smoother version of the little rip, named recently the little sibling of the big rip singularity \cite{Mariam2}, a bounce or it can even become asymptotically flat or de Sitter. It should be noted that this model becomes asymptotically de Sitter if and only if the HRDE is endowed with a negative cosmological constant\footnote{This is consistent with string theory, since the desire to preserve supersymmetry after compactification to four space-time dimensions favors the compactification on six dimensional Calabi-Yau spaces \cite{Calabi-Yau} on ADS backgrounds,  i.e. on space-times with a negative cosmological constant. This would imply  late time acceleration of the universe  (see also Ref.~\cite{Polchinski}).}. 
In addition, in the particular case where $\beta=2$, the HRDE can mimic dark matter if $\Omega_{\mathbf{m}} =0$.

\chapter{Conclusions and future work}

\lhead{Chapter 5. \emph{Conclusions and future work}} %

\section{Conclusion}

We have studied several types of singularities forming at  the late time evolution of the universe or as the final state of the gravitational collapse.  
In the context of gravitational collapse, we have investigated the status of the singularities forming at the endstate of a collapsing star, where scalar fields are present as collapsing matter source (cf. chapter~\ref{collapse-class}). In particular, we considered a  spherically symmetric space-time for collapse, with a tachyon field and a barotropic fluid, constituting the matter content  \cite{Tavakoli2013a}. Therein, different situations for a black hole and a naked singularity formation were studied.

We investigated  how the loop (quantum) effects could alter the outcome of gravitational collapse. We have shown that, in a semiclassical  collapse with a tachyon field and a barotropic fluid, as matter source (cf. chapter~\ref{QuantumG}),
`inverse triad type' corrections lead to an outward flux of energy at the endstate of the dynamical evolution of the collapse, by avoiding either a naked singularity or a black hole formation (see also Ref.~\cite{Tavakoli2013b}). 
A similar situation can happen when an effective scenario, namely a `holonomy correction', from LQG is employed. 
The corresponding effective Hamiltonian constraint leaded to a quadratic density modification  $H^2\propto \rho(1-\rho/\rho_{\rm crit})$. 
This modification provides an upper limit $\rho_{\rm crit}$ for the energy density $\rho$ of matter, whence  it predicts that the gravitational collapse would include a non-singular bounce at the critical density $\rho=\rho_{\rm crit}$.
We further showed that, classical fixed point solutions (including black hole and naked singularities) provided in section \ref{collapse-tachyon} are no longer present within the loop semiclassical regime \cite{Tach-holonomy:2013}.

Within the cosmological scenarios, we have studied the status of singularities that may appear at the late time evolution of the universe (cf. chapter~\ref{cosmology}). To this aim, we considered the late time cosmology in the context of  recent dark energy models. In particular,  we investigated a flat FLRW universe as a DGP brane world model, in the presence of a GB term on the bulk  (see Ref.~\cite{Mariam2010a}), and a  phantom matter   on the brane. We showed that a composition of specific IR and UV effects could alter a big rip singularity setting, and replace it with a milder singularity named a sudden singularity.
Furthermore, we considered a recently proposed model for dark energy provided by the GRVE scenario, to study the late time behavior of a  FLRW universe. 
The Friedmann equation of this model looks pretty much similar to that of a homogeneous and isotropic universe filled with an HRDE component. 
Despite the analogy between these two models, it turned out that one of them, GRVE, is singularity-free in the future while the other, HRDE, is not. Indeed, a universe filled with an HRDE  component can hit, for example, a big rip singularity. We clarified this issue by solving analytically the Friedmann equation for both models and analysing the role played by the local conservation of the energy density of the different components  filling the universe. In addition, we pointed out that in some particular cases the HRDE, when endowed with a negative cosmological constant and in the absence of an explicit dark matter component, can mimic dark matter and explain the late time cosmic acceleration of the universe through an asymptotically de Sitter universe (see also Ref.~\cite{Tavakoli2013c}).

\section{Outlook}

Singularity problems, as those that would take place at Planck scale, are still an open issue on our understanding about the nature of space-time. 
Gravitational collapse scenario can be used as probes to test the quantum theories of gravity  and to study the nature of such singularities \cite{Goswami2006}.
In particular, there exists some types of singularities (such as naked singularities; cf. Ref. \cite{Chakrabarti1994}) that can be the possible candidates for gamma-ray bursters.

Deformed dispersion relations are a rather natural possibility in quantum gravity \cite{Lorentz2}, requiring a modification of Lorentz symmetry, which is then said to be `broken' by quantum gravity effects. If that is the case, 
Lorentz  invariance is only an approximate symmetry of the low  energy world.  
In the recent paper  \cite{Tavakoli2012} we showed that, in the context of LQC, studying the QFT on a cosmological quantum space-time results in an effective (so-called dressed) space-time on which the Lorentz symmetry can be broken in some approximations. 
Indeed, studying the QFT  on the quantum space-time of gravitational collapse, and the issues of  the Lorentz symmetry breaking may provide an intriguing astrophysical framework for investigating the physics inside the collapsing star at the Planck scale regime.

Another currently active field of research in gravitational physics is numerical relativity which uses numerical methods and algorithms to solve and analyse problems of general relativity, in order to study space-times whose exact form is not known. Numerical relativity is applied to many areas, such as gravitational collapse and black hole physics (e.g. see Refs.~\cite{Numeric1,Numeric2,Numeric3}.
This techniques can be further employed to study the space-time structure at the Planck scale physics, where the quantum gravity effects are important.

For numerical quantum gravity to be useful, methods must be found to make the calculations meaningful by introducing an appropriate definition of quantum observables. 
In the absence of external time, a (massless) scalar field serves the role of internal clock, and physics can be extracted using relational observables in order to describe the dynamics of space-time \cite{A-P-S:2006a}. 
A primary goal  is to study singularity which may arise at the final state of gravitational collapse. To this end, 
numerical techniques to find physical states in the quantum theory were largely developed by
in Ref.~\cite{Singh2012} (see also Refs.~\cite{A-P-S:2006a,A-P-S:2006b}). 
Indeed, numerical methods demonstrated that the big bang singularity is resolved and replaced by a quantum bounce in LQC.
This breakthrough provides a possible solution for finding the desired observables, and is being used for ongoing work into the fully quantum modeling of gravitational collapse and black hole physics on computers. 
Such model  
may provide us a framework to understand better the nature of the black hole singularity.

On the other hand, according to current astrophysical data,  the dark energy equation of state parameter $w_{\rm DE}$ is roughly $-1$ \cite{Planck2013}. 
This has stimulated  studies where  our universe may face future singularities.
If $w_{\rm DE}$ is less than $-1$ then  dark energy (of phantom sort) could drive the universe 
to finite time future singularity, like the big rip or the  sudden singularities among others,  with catastrophic consequences for future civilizations. 
According to various estimations, in this case, the universe will exist for several more billions of years at best, before ending up in a cosmic doomsday (cf. see Ref.~\cite{O2}).

The existence of future singularities in FLRW cosmology reflects the vulnerability of standard Friedmann dynamics whenever the energy density and pressure of the universe become of the order of Planck values.  
Resolution of singularities using WDW quantization has been attempted in Refs.~\cite{Hartle:1984,Kiefer:2011,Kiefer2009,Kiefer:2010}. 
Furthermore, the issues of resolution of  future singularities have been investigated using perturbative corrections such as in string theory models   \cite{Future-sing}.
These analysis indicate that generic resolution of singularities may only be accomplished using non-perturbative corrections. 
 LQC 
has dealt with various  future singularities \cite{Future-sing-LQC}. Nevertheless, in the absence of an analysis which uses non-perturbative quantum gravitational modifications 
to model the dynamics of dark energy, the fate of future singularities is still an open problem. 
Motivated by the above paragraph, the study of dark energy related singularities within the framework of LQC will constitutes the main part of my future work in this context.



\addtocontents{toc}{\vspace{2em}} 

\appendix 


\chapter{Gravitational collapse with a standard scalar field} 
\label{collapse-scalar}

\lhead{Appendix A. Gravitational collapse with a standard scalar field} 

This appendix provides a complementary knowledge  for chapter  \ref{collapse-class}. 
More precisely,  
in this appendix, we will study the gravitational collapse whose matter content is a standard homogeneous  scalar field. 
We describe very briefly the fate of the  collapse by employing a phase space analysis. We consider the interior space-time  to be given by the metric (\ref{metric}).


\section{Interior space-time: Matter Hamiltonian}


 It is convenient to analyse the Hamiltonian formalism of the system. 
From a Hamiltonian perspective, this model may be viewed as arising from the dynamics of the canonically conjugate pairs of the phase space of the system.

The gravitational sector of the phase space,  $\Gamma_{\textrm{grav}}$, 
with the symmetry reduction of the FLRW for the interior space-time (\ref{metric}), is two-dimensional and coordinatized by the scale factor $a$ and its conjugate momentum $\pi_a=-(6/8\pi G)a\dot{a}$, which satisfy the Poisson algebra $\{a, \pi_a\}=1$. 
In terms of these phase space variables, the (interior) gravitational Hamiltonian constraint can be written as \cite{Bojowald2010}
\begin{equation}
C_{\rm grav}\  =\  -\frac{2\pi G}{3} \frac{\pi_a^2}{a}\  .
\label{Ham-grav}
\end{equation}
For a standard scalar field $\phi(t)$ with the potential $V(\phi)$, as interior matter source, 
the matter Lagrangian  reduces to
\begin{equation}
L_{\mathrm{matt}}\  =\  \frac{a^3}{2}  \dot \phi^2 - a^3 V(\phi),
\end{equation}
so that, the conjugate momentum  of the scalar field reads $\pi_\phi=a^3\dot \phi$. Thus, the Hamiltonian constraint of the matter reads
\begin{equation}
C_{\textrm{matt}}\ =\  \frac{\pi_\phi^2}{2a^3} + a^3V(\phi).
\label{matter-const-1}
\end{equation}
Evolution is given by the total Hamiltonian constraint of the system including the scalar field $\phi$ with the conjugate pair  $\pi_\phi$, and the geometrical elements  $(a, \pi_a)$ as:
\begin{equation}
C\  =\  C_{\mathrm{grav}} + C_\mathrm{matt}=-\frac{2\pi G}{3} \frac{\pi_a^2}{a}\ +\  \frac{\pi_\phi^2}{2a^3}+a^3V(\phi)\ .
\label{ham-cont}
\end{equation}
Using the constraint equation (\ref{ham-cont}) we can solve for
$\pi_a$ in terms of $a$ and $\pi_{\phi}$.
It should be noticed that,  the Hamiltonian constraint (\ref{ham-cont})
should be satisfied for any shell $r$, and whence it must be held also on the boundary $r_{\textrm{b}}$, of two regions.

In order to bring the equations in more conventional form in classical setting, we may use $\pi_a=-(6/8\pi G)a\dot a$,  to eliminate $\pi_a$ in Eq. (\ref{ham-cont}) by setting $C(a, \pi_a, \pi_\phi)=0$. In this way we obtain the Friedmann equation
\begin{equation}
\frac{\dot a^2}{a^2}\  =\  \frac{8\pi G}{3} \left(\frac{\pi_\phi^2}{2a^6}+V(\phi)\right)\  .
\label{Friedmann-eq}
\end{equation}
From Eq.~(\ref{Henergy-sf})  the energy density $\rho_\phi$, and the pressure $p_\phi$, of the scalar field are obtained as
\begin{align}
\rho_\phi \  =\   \frac{\pi_\phi^2}{2a^6} +V(\phi), \label{energyPH-1} \\
p_\phi \  =\   \frac{\pi_\phi^2}{2a^6} -V(\phi).
\label{energyPH}
\end{align}

Phenomenologically we assume that the matter  field is described by a perfect (barotropic)
fluid, i.e., satisfying  $p_{\phi}=w_\phi \rho_{\phi}$, where $w_\phi$ is the equation of state for the field $\phi$. So using Eqs. (\ref{energyPH-1}) and (\ref{energyPH}), $w_\phi$ is  given by
\begin{equation}
w_\phi\  =\  \frac{\dot{\phi}^2-2V(\phi)}{\dot{\phi}^2+2V(\phi)}.
\label{state}
\end{equation}
Using the equation (\ref{energyPH}) together with the conservation equation, $\dot{\rho}_\phi+3H(\rho_\phi+p_\phi)=0$,
the Klein-Gordon equation reads,
\begin{equation}
\ddot{\phi}+3\frac{\dot{a}}{a}\dot{\phi}+\frac{\partial V}{\partial\phi}=0.
\label{KleinGordon}
\end{equation}
Furthermore, using the Friedmann equation (\ref{Friedmann-eq}), the dynamical evolution equations of the system for the line element (\ref{metric}) can be written as
\begin{align}
\frac{\dot{a}^2}{a^2} =\frac{8\pi G}{3}\rho_\phi,\  \ \  \  \ \   \ \  \  \ \   \ \  \  \ \  
\frac{\ddot{a}}{a} =-\frac{4\pi G}{3}(\rho_\phi+3p_\phi).
\label{dyneqs}
\end{align}
In this case the continuity equation, $\dot{\rho}_\phi+3(1+w_\phi)\dot{a}/a=0$, can be written in an integrated form:
\begin{equation}
\rho_\phi=\rho_0\exp\left(-\int 3(1+w_{\phi})\frac{da}{a}\right),\label{enerint}
\end{equation}
where $\rho_0$ is an integration constant.
For the collapsing system herein, we will consider the initial condition on the initial hypersurface of the collapsing cloud at $t=0$, given by the initial data $a(0)=a_0,~ \phi(0)=\phi_0$, and $\dot{\phi}(0)=\dot{\phi}(0)$. Then, by starting the collapse at $t=0$, the
scalar field evolves until the stage when the collapsing matter possibly reaches the singular state at $a=0$.

Let us assume the potential of the system to be exponential as\footnote{There are few proposals for the scalar field potential $V(\phi)$ in  cosmology \cite{Copeland1998,Lucchin1985,Kitada1993,Halliwell1987,Burd1988,Coley1997}. In a classical collapsing process, however, it was shown that \cite{Goswami2004} an exponential potential is useful
 to control the divergence of energy density of matter field near the singularity, which in turn governs the development or otherwise of trapped surfaces.
 An alternative has been to use an inverse power law of the field, however, 
it has been argued \cite{Tavakoli2013b} that $V\sim \phi^{-2}$ would not be a convenient choice of the potential for (classical) scalar field collapse. Therefore, motivated by \cite{Goswami2004} we choose an exponential potential for the field $\phi$, for the collapsing system herein. }
\begin{equation}
V(\phi)\ =\  V_0e^{-\lambda_0\kappa\phi}\  .
\label{Pot-SF-1} 
\end{equation}
with $V_0$  being  constants.  Then, the  solution for the scalar field can be given by
\begin{align}
\phi\ =\  \sqrt{\frac{3}{4\pi G}}\ln \left(\frac{a}{a_0}\right)+\phi_0\  . 
\label{scalarF1}
\end{align}
The solution (\ref{scalarF1}) indicates that as collapse evolves,  $\phi$ decreases from its initial condition (at $\phi_0$) and  diverges negatively as $\phi\rightarrow-\infty$ when approaching the center (i.e., $a=0$). 
Substituting  the solution (\ref{scalarF1}) in Eq.~(\ref{Pot-SF-1}), the potential in terms of scale factor reads 
\begin{equation}
V\  \approx \  V_0 \left(\frac{a}{a}\right)^{-m}, 
\label{Pot-fixpoin-a}
\end{equation}
where $m\equiv2\lambda_0\sqrt{12\pi G}$. Eq.~(\ref{Pot-fixpoin-a}) shows that
 the  potential $V(\phi)$ of scalar  field $\phi(a)$ depends on the sign of the constant $\lambda_0$ (or $m$).

From  Eq. (\ref{enerint}), the energy density of the collapsing matter reads 
\begin{equation}
\rho_\phi\  \approx \  \rho_0 a^{-6}. 
\label{energSF1}
\end{equation}
This equation shows that, as $a\rightarrow0$, then the energy density of the collapsing matter diverges, which corresponds to a curvature singularity at $a=0$. 


Let us now consider the situation in which  trapped surfaces can form inside the collapsing matter.
The third case in Eq. (\ref{sp}) characterizes the outermost
boundary of the trapped region, namely, the apparent horizon which
corresponds to the equation $\Theta(t)=0$. 
From Eq. (\ref{THETAboundary}) it is seen that, as collapse proceeds from the initial condition,
the first term in Eq. (\ref{THETAboundary}), which for the model here reads $\dot{a}^2/a^2 \approx \rho_\phi \propto a^{-6}$,  increases faster than
the second term and $\Theta$ becomes positive as $\Theta(t,r)>0$
at any moment $t>t_{i}$ and for any shell $r$, towards the singularity.
Then, as $a\rightarrow0$, $\Theta\rightarrow+\infty$, and hence
the singularity will be covered by apparent horizon. In other words, this means that, as collapse evolves, the mass function $F(R)\approx \rho a^{3}\approx a^{-3}$ increases from its initial condition and diverges at the singularity; this  is accompanied by trapped surfaces forming that cover the final singularity, hence, a black hole forms.

Another solution for the scalar field is
\begin{equation}
\phi\ =\ -\sqrt{\frac{3}{4\pi G}}\ln \left(\frac{a}{a_0}\right) + \phi_0. 
\label{scalarF1-2}
\end{equation}
This equation indicates  that, $\phi(t)$ increases  from the initial configuration of the collapse (at $\phi_0$), and diverges   as $\phi\rightarrow +\infty$ at $a\rightarrow0$. 
Consequently, the scalar field potential (\ref{Pot-SF-1}) changes depending on the choice of $\lambda_0$ as 
\begin{equation}
V\  \approx \  V_0 \left(\frac{a}{a}\right)^{m}. 
\label{Pot-fixpoin-b}
\end{equation}
The energy density of the scalar field in this case is also given by Eq.~(\ref{energSF1}), which increases towards the center and diverges at $a=0$; this corresponds to a curvature singularity. 
The mass function of the collapse in this case also reads $F \approx a^{-3}$, which increases towards the center, and diverges at $a=0$. Thus a black hole forms as collapse end state.

For a scalar field $\phi$ to be a physically relevant  matter content for the  collapse, it must satisfy the WEC;
this amounts to $\rho_\phi=\dot{\phi}^2/2+V(\phi)\geq0$  and $\rho_\phi+p_\phi=\dot{\phi}^2\geq0$, which satisfies the WEC.


\section{Exterior geometry}
\label{exterior-scalar}


To complete the model, the interior geometry must be matched to a suitable exterior space-time as we presented in section \ref{collapse-exterior}.
The procedure for finding the exterior metric function on the boundary uses the Hamiltonian constraint (see Eq. (\ref{ham-cont})), the equation of motion for the scale factor, and the junction conditions.

Substituting $\dot{a}$ by using $\pi_a=-(6/8\pi G)a\dot{a}$ at the boundary
of two regions, $\Sigma$, bearing in mind that $2M(\mathrm{v},r_{\mathrm{v}})=F(R)$ and $F=R\dot{R}^2=r_{\mathrm{b}}^{3}a\dot{a}^2$, we get $M(\mathrm{v},r_{\mathrm{v}})$
in terms of the canonical pairs $a$ and $\pi_a$:
\begin{equation}
2M(\mathrm{v},r_{\mathrm{v}})|_{\Sigma}\ =\  \frac{8\pi G}{3}  r_{\mathrm{b}}^{3} \left(-\frac{2\pi G}{3} \frac{\pi_a^2}{a}\right)= \frac{8\pi G}{3}  r_{\mathrm{b}}^{3}C_{\rm grav}~. 
\label{F/R-1}
\end{equation}
Since the Hamiltonian constraint (\ref{ham-cont}) is satisfied
on any shell $r$, hence $C|_{r_{\mathrm{b}}}=C_{\rm grav}+C_{\rm matt}=0$. So, substituting $C_{\rm grav}$ term in Eq. (\ref{F/R-1}) by the matter Hamiltonian counterpart  (\ref{matter-const-1}), we can rewrite 
$2M(\mathrm{v},r_{\mathrm{v}})$
in terms of $\pi_\phi$ and $a$ at the boundary $\Sigma$ of two regions:
\begin{equation}
2M(\mathrm{v},r_{\mathrm{v}})|_{\Sigma} =   \frac{8\pi G}{3} r_{\mathrm{b}}^{3} C_\mathrm{matt} = \frac{8\pi G}{3} \left(\frac{\pi_{\phi}^{2}}{2a^{3}} + a^3V(\phi)\right) r_{\mathrm{b}}^{3} \  .  
\label{F-phi}
\end{equation}
This equation together with matching conditions enables us to find
the boundary function of the exterior space-time. 
By replacing the potential (\ref{Pot-fixpoin-a}) in Eq.~(\ref{F-phi}) we get,
\begin{equation}
2M(\mathrm{v},r_{\mathrm{v}})|_{\Sigma}\  \approx \    \frac{8\pi G}{3} \left(\frac{\pi_{\phi}^{2}}{2a^{3}} + a^{3-m}\right) r_{\mathrm{b}}^{3} \  .  
\label{F-phi-red}
\end{equation}
For $m<3$, the second term declines by scale factor $a$ while the first term increases towards the center. So that, close to the singularity 
the exterior metric function at the boundary shell is 
\begin{align}
{\cal F}|_{\Sigma}=1-\frac{8\pi G}{6}\frac{\pi_{\phi}^{2}}{a^{4}} r_{\text{b}}^2\  . \label{f-phi}
\end{align}
Notice that, 
near the center, this solution corresponds to a free scalar field for collapsing matter content. 
Then, the matter Hamiltonian constraint (\ref{matter-const-1}) reduces to $C_{\textrm{matt}} = \pi_\phi^2/2a^3$. 
In this case, since the scalar field $\phi$ does not enter in the expression of the constraint $C_\mathrm{matt}$, its momentum $\pi_\phi$ is a constant of motion.
From the exterior metric function given by Eq.~(\ref{f-phi}), we can
get the information regarding the behavior of trapping horizon for
the interior space-time parameters. When the relation \mbox{$2M(\mathrm{v},r_{\mathrm{v}})=r_{\mathrm{v}}$}
is satisfied at the boundary, trapped surfaces will form in the exterior
region close to the matter shells. On classical geometry, ${\cal F}$
becomes negative in the trapped region, and ${\cal F}=0$ on the apparent
horizon. Therefore, the equation for event horizon is given at the
boundary of the collapsing body by ${\cal F}|_{\Sigma}=0$.

The boundary function at vicinity of $a=0$ can be written as
\begin{equation}
{\cal F}=1-\frac{\alpha}{R^{4}}\ ,\label{f-phi2-b}
\end{equation}
 where $\alpha\equiv(8\pi Gr_{\text{b}}^{6}\pi_{\phi}^{2})/6$ is a constant.
The apparent horizon can form when $\Theta$ in Eq.~(\ref{THETAboundary})
vanishes, where $R(t, r_{\mathrm{b}})=\alpha^{1/4}$  and
intersects the matching surface $r=r_{\text{b}}$. Determining the fate of the singularity
depends on existence of a congruence of future-directed non-spacelike
trajectories emerging from a past singularity reaching distant observers.
So that, the classical singularity is covered by the trapping horizon
and a back hole forms classically. Equation (\ref{f-phi2-b}) suggests
the exterior function to be ${\cal F}(r_{\mathrm{v}})|_{\Sigma}=1-\alpha/r_{\mathrm{v}}^{4}$, 
indicating that, the exterior space-time has an exotic black hole geometry
\cite{Hussain2011}. Notice that, in the presence of a non-zero matter
pressure (of the massless scalar field) at the boundary, the (homogeneous)
interior space-time can not be matched with an empty (inhomogeneous)
Schwarzschild exterior \cite{Bojowald2005b}.

A similar analysis  gives the following expression for $M(\mathrm{v},r_{\mathrm{v}})$ at the boundary surface $\Sigma$:
\begin{equation}
2M(\mathrm{v},r_{\mathrm{v}})|_{\Sigma}\  \approx \    \frac{8\pi G}{3} \left(\frac{\pi_{\phi}^{2}}{2a^{3}} + a^{3+m}\right) r_{\mathrm{b}}^{3} \  ,
\label{F-phi-red-pointb}
\end{equation}
where we have substituted the potential of scalar field by Eq.~(\ref{Pot-fixpoin-b}).
For $m> -3$, the first term is dominant, so that, the Eq.~(\ref{F-phi-red-pointb})  reduces to Eq.~(\ref{f-phi}).  Notice that, for $m>0$  and in the region with $a\ll a_*$, this solution corresponds to a free scalar field for the collapsing matter source. Consequently, the exterior space-time for this solution corresponds to a black hole geometry governed by the exterior function ${\cal F}(r_{\mathrm{v}})|_{\Sigma}=1-\alpha/r_{\mathrm{v}}^{4}$.

\chapter{Semiclassical collapse with a  scalar field} 
\label{LQG-scalar}

\lhead{Appendix B. Semiclassical collapse with a standard scalar field} 

%

This appendix provides a complementary knowledge  for chapter  \ref{QuantumG}. 
More precisely,  
in Appendix~\ref{LQG-scalar} we employ the LQG induced effects   in order to study the fate of standard scalar field collapse:
Section~\ref{LQG-scalar1} includes a semiclassical description of scalar field collapse where an inverse triad correction is applied.
In section \ref{effectiveLQC}, we study  the space-time geometry of the collapse within  an effective scenario provided by a holonomy correction of LQG.

\section{Inverse triad corrections}



\label{LQG-scalar1}


For an interior space-time of  collapse with the line element (\ref{metric}), 
and the field $\phi(t)$  as matter source, whose  potential is $V(\phi)$, 
we hence consider the inverse triad modifications 
based on LQG.  
In this case,  modification to the  matter Hamiltonian, Eq. (\ref{matter-const-1}), can be obtained by substituting $|p|^{-3/2}$ from Eq.~(\ref{defD}) with $d_j(a)$ as 
\begin{equation}
C_{\textrm{matt}}^{\rm sc}\  =\   d_{j}(a)\,  \frac{\pi_\phi^2}{2} + a^3  V(\phi) . ~ \label{hamphi}
\end{equation}
Then, 
the vanishing Hamiltonian constraint 
results in the modified Friedmann equation:
\begin{equation} 
\frac{\dot a^2}{a^2}\  =\   \frac{8\pi G}{3} \left(\frac{\dot \phi^2}{2D}  + V(\phi)\right). ~
\label{modfred}
\end{equation}
These also lead to the modified Klein-Gordon equation
\begin{equation}
\ddot \phi + \left(3\frac{\dot a}{a} - \frac{\dot D(q)}{D(q)} \right) 
\, \dot \phi + 
D(q) \, V_{,\phi}(\phi) ~ = 0 ~. \label{kgeq}
\end{equation}
Notice that, for scales $a \lesssim a_*$, the $\dot \phi$ term (which experiences anti-friction in classical regime) acts like a 
frictional term for a collapsing phase.

In semiclassical regime (of inverse triad correction), similar to Eq.~(\ref{hamphi}),  the energy density and pressure of the scalar field is modified due to a replacement of the $|p|^{-3/2}$ term by $d_{j}(a)$, in Eq. (\ref{Henergy-sf}). Therefore, the modified energy density and pressure of the scalar field is obtained as
\begin{equation}
\rho_{\rm sc}\  =\   d_{j}(a) C_{\textrm{matt}}^{\rm sc} = \frac{\dot \phi^2}{2} 
+ D(q) \, V(\phi) \  ,
\label{energyden1}
\end{equation}
and
\begin{equation}
p _{\rm sc}\  =\  \bigg[1 - \frac{2}{3} \, \frac{1}{(\dot a/a)} \, 
\frac{\dot D(q)}{D(q)}
\bigg] \, \frac{\dot \phi^2}{2} - D(q)\, V(\phi) - \frac{\dot D(q)}
{3 (\dot a/a)} \, V(\phi) ~.
\end{equation}
In this  regime, for the limit case $a \ll a_*$, we have $D(a)\approx (a/a_*)^{8}$, so that, the Klein-Gordon equation (\ref{kgeq}) reduces to
\begin{equation}
\ddot \phi - 5 (\dot{a}/a) \dot \phi + 
D(q) \, V_{,\phi}(\phi) ~ = 0 ~. \label{kgeq-2a}
\end{equation}
Furthermore, the modified pressure in this limit becomes
\begin{align}
p _{\rm sc}\ &=\  -\frac{13}{3} \frac{\dot \phi^2}{2} - \frac{11}{3} D(q)\, V(\phi)  \notag \\
& = \  -\frac{13}{3}  \rho_{\rm sc} + \frac{2}{3} D(q)\, V(\phi) ~. 
\end{align}
Therefore, $p_{\rm sc}$ is generically negative for $a \lesssim a_*$ and for $a \ll a_*$ 
it becomes  very strong:  In the limit case $a \sim a_i$, the effective pressure reads $p_{\rm sc} \approx - 4 \rho_{\rm sc}$, which  is super negative, and may result in
an outward energy flux in the  semiclassical regime.

If the scalar field  $\phi$ is a monotonically varying
function of the proper time, then we can present  Eq.~(\ref{modfred}) in a
 Hamilton-Jacobi form \cite{Lidsey:2004}:
\begin{align}
V(\phi) =  \frac{3}{8\pi G}H^2 - \frac{D}{2}H_\phi^2\  ,
\label{potentialS}
\end{align}
where $H_\phi$ is defined as  
\begin{align}
H_\phi\  := \frac{\dot{\phi}}{D} = \frac{1}{r}\frac{dH}{d\phi} \ ,
\label{HJs1}
\end{align}
with $r$ being given by $r :=  \frac{1}{3}(\dot{D}/4HD-3/2)$; notice that in our model for the choice of $D(q)$, $r$ takes the value $r=1/6$. 
This formalism implies that the dynamics of the
semiclassical period  can be determined  once the Hubble parameter
$H(\phi)$ has been specified as a function of the scalar field.

Let us  assume a Hubble parameter of the form 
\begin{equation}
H(\phi) \  :=\  H_1\phi^m~, \label{HubbleS1}
\end{equation}
as in Eq.~(\ref{HubbleS1-tach}). 
The scale factor can be obtained by integrating (\ref{HJs1}).
Then, we get
\begin{equation}
a^4(\phi) = A_0 \phi \ ,  \label{HubbleS2}
\end{equation}
where $A_0 := \left(2/3 |m|\right)^{\frac{1}{2}}a_*^4A^{-\frac{1}{6}}$. 
So that, the potential of the scalar field is given by (\ref{potentialS}):
\begin{equation}
V(\phi)\  \approx\  V_1 \phi^{2m}, \label{HubbleS3}
\end{equation}
where $V_1=const$.
This solution shows that the scalar field remains finite and satisfies the range $0<\phi<\phi_0$ during the collapse, with the initial data at $a(\phi_0)=a_0$. Furthermore, as $a$ decreases, the scalar field decreases towards the center. 
When the scale factor becomes very small and approaches  the Planck scale, the scalar field $\phi$ vanishes very fast. Then, as $\phi\rightarrow0$, the Hubble rate diverges which corresponds to a curvature singularity.
Notice that, the last term in the modified equation of motion (\ref{kgeq-2a}) 
can be approximated as $DV_{,\phi}\sim \phi^{2m+1}$, so that,  for the range of the parameter $m<-1/2$, this term increases towards the singularity and has an important role in dynamics of the system.
In addition, for the case $m=-1$, the solution corresponds to an inverse square potential for the scalar field, i.e.  $V=V_1\phi^{-2}$, which also brings a singular final state for the collapsing model. 
It should be noticed that, this  further suggests that, the results of Ref.~\cite{Goswami2006} (see also section \ref{LQG-scalar}), when taken in view of a general class of potentials, must be discussed with care. 
More precisely,  in the presence of a potential with an inverse power of scalar field, the collapsing model in Ref.~\cite{Goswami2006} may not be regular as long as the semiclassical effects are valid.

For an exponential potential $V(\phi)=V_0e^{\lambda\phi}$ of scalar field, in the regime $a\leq a_{*}$ and $D(q)\ll 1$, the modified dynamics becomes independent of the potential. Thus, 
the last term in Eq.~(\ref{kgeq-2a}) becomes negligible, so that,  the Klein-Gordon equation can be approximated as
$\ddot \phi - 5 (\dot{a}/a) \dot \phi  \approx 0$  \cite{Singh2005b,Goswami2006}.  
This equation  yields $\dot{\phi}\propto a^{5}$, from which we obtain the semiclassical energy density  (\ref{energyden1})  as  $\rho_{\rm sc} \approx a^{10}$; this, instead of blowing up, becomes extremely small and remains finite.
The scalar field now experiences friction leading to decrease of $\dot \phi$.
The slowing down of $\phi$  decreases the rate of collapse and formation of singularity is delayed. 
The classical singularity is thus avoided till the scale factor at which a continuous space-time exists.
Notice that, this result is further valid when we consider a free scalar field for the collapsing matter content \cite{Goswami2006}.

We assume the exterior to be as use the generalized Vaidya geometry, 
so that, the matching 
of interior and exterior space-times remains valid during the semiclassical
evolution. 
The modified mass function of the collapsing cloud can be evaluated by using Eqs. (\ref{einstein2})  and  (\ref{modfred}) as
\begin{equation}
F_{\mathrm{sc}} (r, t)\  =\  \frac{8\pi G}{3} \frac{r^3}{2}  d_{j}^{-1} \dot \phi^2 =   \frac{8\pi G}{3} r^3 C_{\textrm{matt}}^{\rm sc} \ .
\end{equation}
In the regime $a \sim a_i$, the term $d_{j}^{-1} \dot \phi^2$ becomes proportional 
to $a^{5}$,  and thus the mass function becomes vanishingly small at small scale factors.

The phenomena of delay and avoidance of the singularity 
in continuous space-time is accompanied by a burst of matter to the 
exterior. If the mass function at scales $a \gg a_*$ is $F$ and its 
difference with mass of the cloud for $a<a_*$ is $\Delta F = F - F_{\rm sc}$, 
then the mass loss can be computed as
\begin{equation}
\frac{\Delta F}{F}\  =\  \left[1-\frac{\rho_{\rm sc} d_j^{-1}}
{\rho_{\rm cl}a^{3}}\right] ~.
\end{equation}
For $a < a_*$, as the scale factor decreases,  
the energy density and mass in the interior decrease and the negative 
pressure strongly increases; this leads to a  burst
of matter.  The absence of trapped surfaces enables the semiclassical 
gravity induced burst to propagate via the generalized Vaidya exterior (\ref{Vaidya})
to an observer at infinity. 
In the semiclassical regime,  ${\Delta F}/F$ approaches unity 
very rapidly. This feature is independent of the choice of
parameter $j$. 
Thus, non-perturbative semiclassical 
modifications may not allow formation of  singularity as the 
collapsing cloud evaporates away due to super-negative pressures in 
the late regime.

\section{Holonomy correction}
\label{effectiveLQC} 

In this subsection of the appendix, we present a semiclassical description for the gravitational collapse of a  scalar field, by employing a holonomy correction (see section \ref{Holonomy-Tach}).

\begin{figure}
\begin{center}
\includegraphics[width=0.41\textwidth]{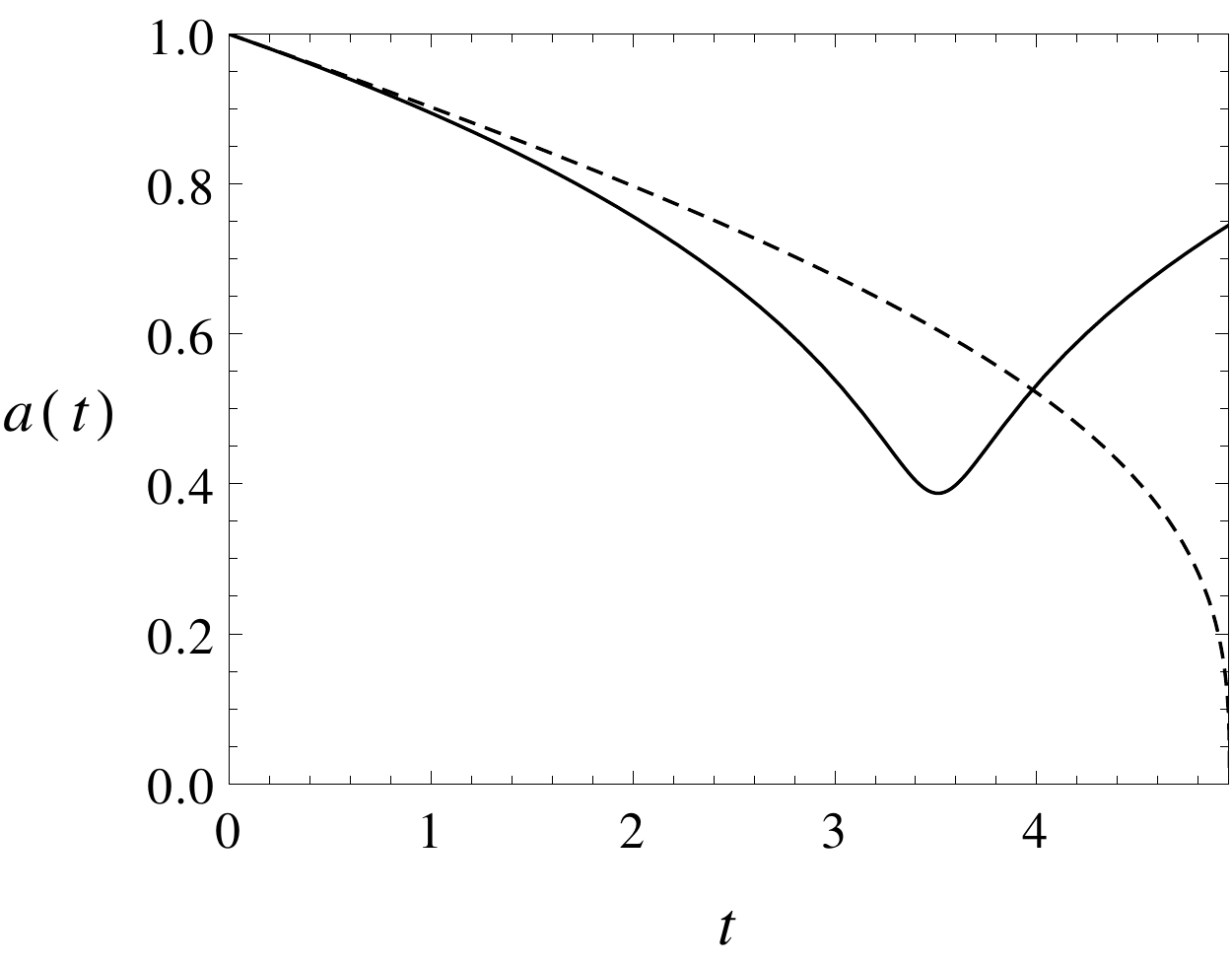}\quad{}\quad{}\quad{}\includegraphics[width=0.415\textwidth]{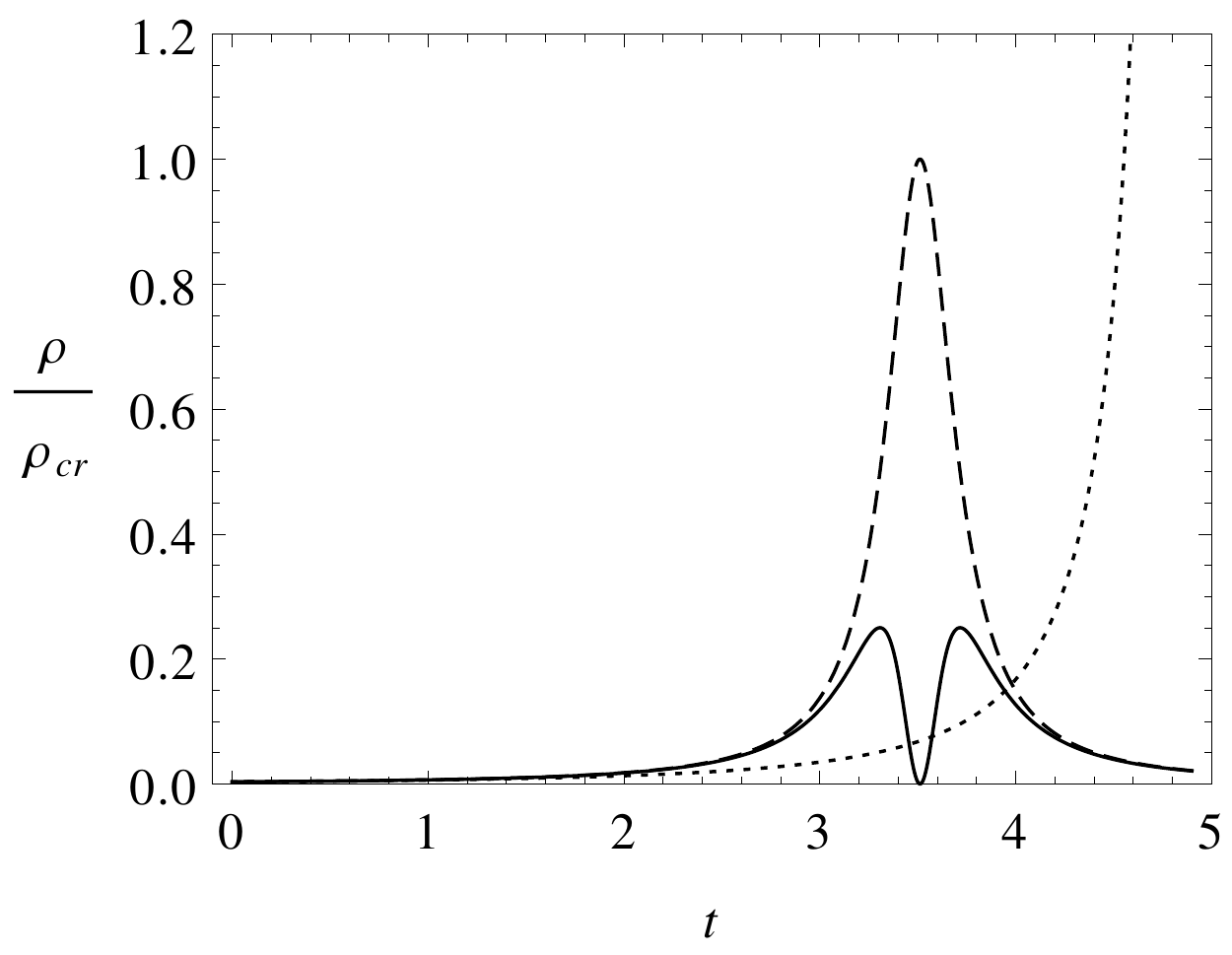}
\end{center}
\caption{The left plot shows the time evolution of scale
factor $a(t)$ in the classical (dotted curve) and semiclassical (solid
curve) regimes. The right plot shows the classical (dotted curve)
and semiclassical (dashed curve) energy densities $\rho$ of scalar
field; the solid curve shows the behaviour of effective energy density
$\rho_{{\rm eff}}$ in the semiclassical regime. We have used the value of parameters
$G=c_{\mathrm{light}}=1$, and $\pi_{\phi}=10~000$.}
\label{F-scalef} 
\end{figure}

The effective Hamiltonian for the system is given by Eq.~(\ref{EFFham-1-tach})   where  the matter is assumed  to be a \emph{massless}  scalar field whose 
Hamiltonian constraint reads  $C_{\mathrm{matt}}=\pi_\phi^2/|p|^{3/2}$. 
The dynamics of the fundamental variables of phase space
is then obtained by solving the system of Hamilton equation  (\ref{HAMeqs-1-tach}) together with Eq.~(\ref{Hamilton-Eq2-tach}).
Then, we obtain 
the modified Friedmann equation  \cite{Taveras2006,A-P-S:2006a}:
\begin{equation}
\left(\frac{\dot a}{a}\right)^{2}\  =\  \frac{8\pi G}{3}\rho_\phi\left(1-\frac{\rho_\phi}{\rho_{\rm crit}}\right),\label{Friedmann-eff-1a}
\end{equation}
where $\rho_\phi=\pi_{\phi}^{2}/(2a^{6})$ with $\pi_\phi$ being a constant. 
Notice that, $\rho_{0}\ll\rho_{\rm crit}$
is the energy density of the star at the initial configuration, $t=0$,
where $\rho_{0}=\pi_{\phi}^{2}/(2a_{0}^{6})$.
Furthermore, in the limit $\rho_\phi\rightarrow\rho_{\rm crit}$,
the Hubble rate vanishes; the classical singularity is thus replaced
by a bounce (cf. figure \ref{F-scalef}). 

\begin{figure}
\begin{center}
\includegraphics[width=0.5\textwidth]{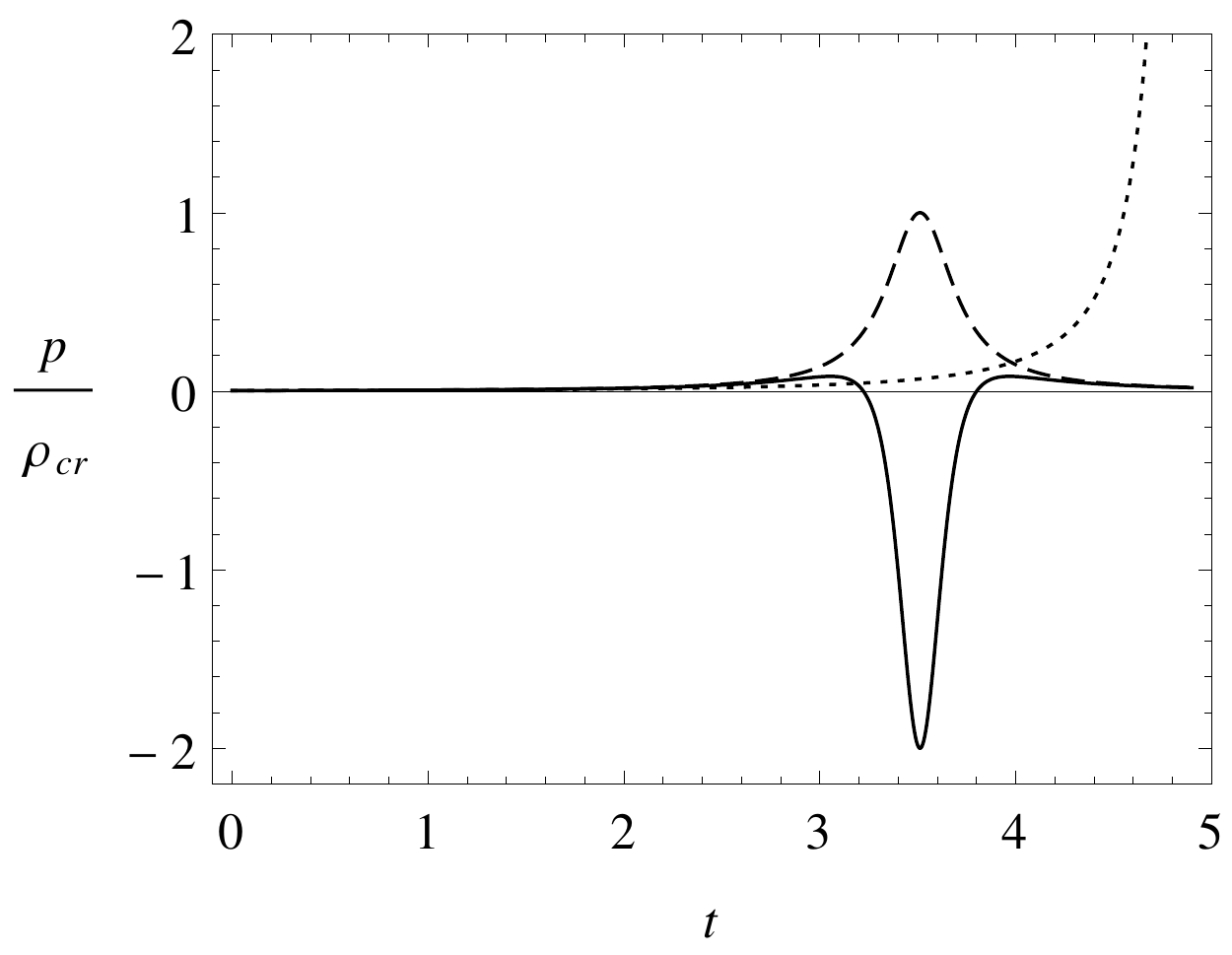}
\end{center}
\caption{This plot represents the behaviour of the pressure in the
classical and semiclassical regimes for the values of parameters $G=c_{\mathrm{light}}=1$,
and $\pi_{\phi}=10~000$. In the classical collapse, the matter pressure
(dotted curve) increases and diverges towards the singularity. In
the semiclassical regime, the matter pressure (dashed curve) increases
and reaches to a maximum $p=\rho_{{\rm crit}}$ at the bounce, wheras
the effective pressure $p_{\textrm{eff}}$ (solid curve) decreases
and takes a minimum super negative value $p_{\textrm{eff}}(\rho_{\textrm{crit}})=-2\rho_{\textrm{crit}}$
at the bounce.}
\label{p-eff-1} 
\end{figure}

From the Raychaudhuri equation (\ref{Raychaudhuri-improv}) we can define the effective pressure (\ref{Peff})
for the massless scalar field as 
\begin{equation}
p_{\mathrm{eff}}\ :=\ \rho_\phi\left(1-3\frac{\rho_\phi}{\rho_{\mathrm{crit}}}\right).\label{press-eff}
\end{equation}
Figure~\ref{p-eff-1} represents the behaviour of the pressures $p_\phi$
and $p_{{\rm eff}}$ in Eq.~(\ref{press-eff}) conveniently scaled
with the critical density $\rho_{\textrm{crit}}$. In the semiclassical
regime, the matter pressure, $p_\phi=\rho_\phi=\pi_{\phi}^{2}/2a^{6}$, increases
during the collapse (see dashed curve in figure~\ref{p-eff-1}),
but remains finite until the bounce where it reaches a maximum at
$p_{{\rm crit}}=\rho_{{\rm crit}}$. The effective pressure (solid curve)
is positive initially, then as energy density increases, $p_{{\rm eff}}$
decreases until it vanishes at $\rho_\phi=\rho_{{\rm crit}}/3$. In the range
$\rho_{{\rm crit}}/3<\rho_\phi<\rho_{\rm crit}$, the effective pressure evolves negatively
until the bounce where it takes the super negative value $p_{\textrm{eff}}(\rho_{\mathrm{crit}})=-2\rho_{\textrm{crit}}$
at the bounce. This indicates that, in the herein homogeneous and
isotropic collapsing model, the singularity resolution is associated
with the violation of (effective) energy conditions (e.g., $\rho_{{\rm eff}}+p_{{\rm eff}}<0$),
which suggests that the quantum gravity effects provide a repulsive
force at the very short distances \cite{Singh:2005c}. This feature
may also result in a strong burst of outward energy flux in the semiclassical
regime.

\vspace*{5mm}
\subsection{Semiclassical dynamics of trapped surfaces}
\label{q-traps}
\vspace*{3mm}

\begin{figure}
\begin{center}
\includegraphics[width=0.5\textwidth]{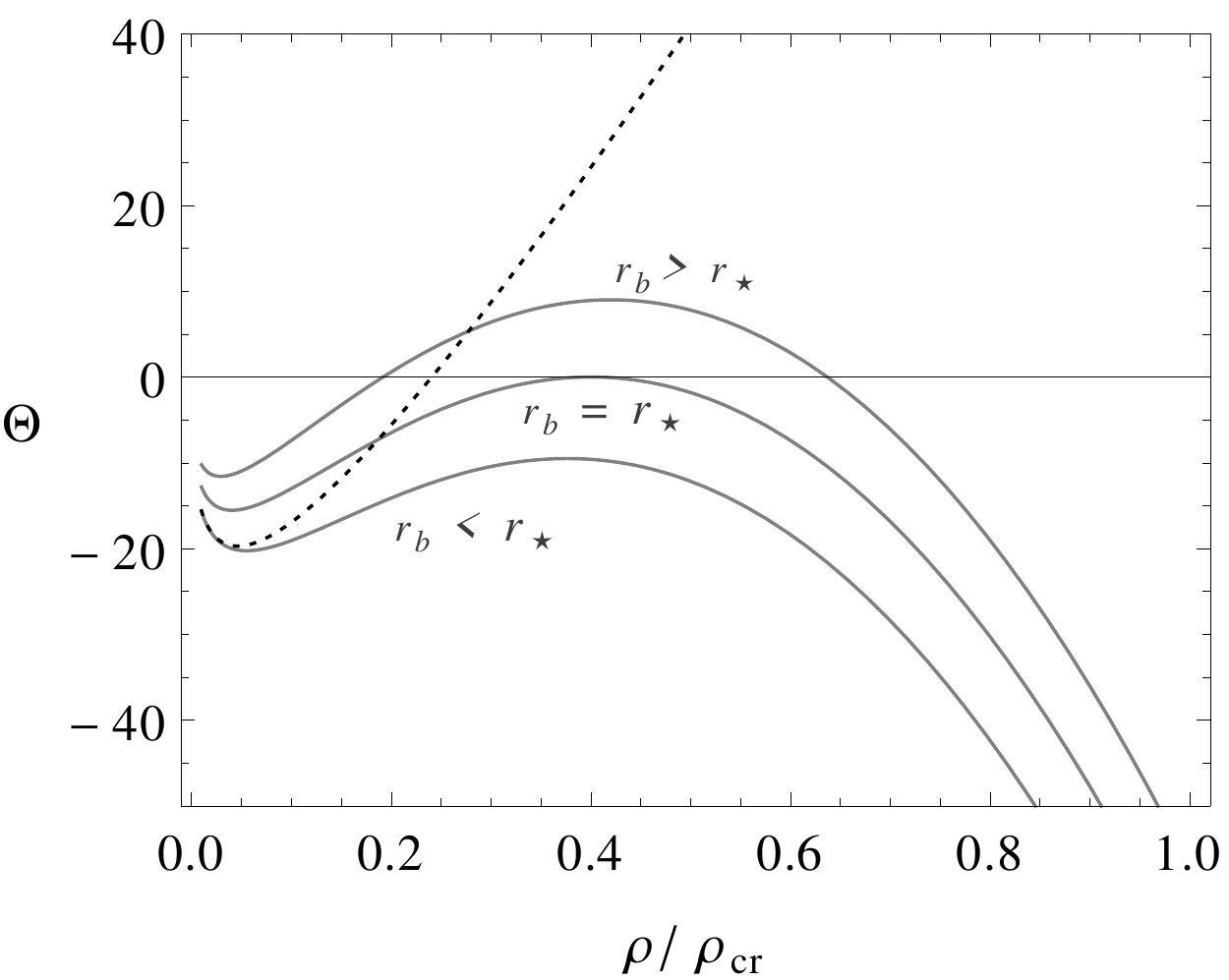}
\end{center}
\caption{Behaviours of $\Theta_{b}(\rho)$ in the classical
(dotted curve), and semiclassical (solid curves) regimes for different
values of $r_{b}$. We have used the value of parameters
$G=c_{\mathrm{light}}=1$, and $\pi_{\phi}=10~000$.}
\label{F-thetaa} 
\end{figure}

To discuss the trapped surfaces dynamics, particular importance is
played by the function $\Theta_{b}$ defined in Eq.~(\ref{THETAboundary}).
Therein, by replacing $\dot{a}/a$ with the effective Friedmann equation
(\ref{Friedmann-eff-1a}) we have 
\begin{equation}
\Theta_{b}\ =\ \frac{64\pi G}{3}\rho_\phi\left(1-\frac{\rho_\phi}{\rho_{\text{crit}}}\right)-\frac{8}{a^{2}r_{b}^{2}}\ .
\label{theta2-q2}
\end{equation}
We will assume that the cloud is initially untrapped, and thus for
$\rho_\phi\ll\rho_{{\rm crit}}$, we have that $\Theta_{b}(t=0)$ is negative.
Now, we can study the behaviour of the effective $\Theta_{b}$ as
a function of the energy density $\rho_\phi$. 
Let us rewrite Eq.~(\ref{theta2-q2})
by setting $X:=\rho_\phi/\rho_{{\rm crit}}$ as 
\begin{equation}
\Theta_{b}(X)\ =\ AX\left(1-X\right)-BX^{1/3}\ ,\label{theta2-q2-X}
\end{equation}
where $A:=(64\pi G/3)\rho_{{\rm crit}}$ and $B:=8(2\rho_{{\rm crit}}/\pi_{\phi}^{2})^{1/3}/r_{b}^{2}$
are constants. The behaviours of $\Theta_{b}$, with respect to $X$,
for the different choices of the initial conditions, are sketched
in figure \ref{F-thetaa}. Therein, the solid curves represent the
trajectories provided by the semiclassical gravitational collapse;
whereas the dotted curve shows the classical trajectories (which coincides
with the semiclassical ones for $X\ll1$). An equation defining the
apparent horizon for the effective geometry can be obtained by equating
(\ref{theta2-q2-X}) to zero. So, we obtain 
\begin{equation}
X^{2}(1-X)^{3}-\left(\frac{B}{A}\right)^{3}=0\ .\label{AH-EQ}
\end{equation}
To solve this last equation, we compute the values
of energy density at which the apparent horizons form. This corresponds
to the intersections of the $\Theta_{b}$ curve with the horizontal
axe in figure \ref{F-thetaa}. Therefore, depending on the initial
conditions, in particular on the choice of the boundary radius $r_{b}$,
three cases can be evaluated, which correspond to \emph{no} apparent
horizon formation, one and two horizons formation. Notice that, denoted
by a dotted curve, only one horizon can form classically. Let us to
be more precise as follows.

In the one hand, the modified Friedmann equation (\ref{Friedmann-eff-1a})
allows to determine the energy density at which speed of the collapse,
$|\dot{a}|$, reaches its maximum. From Eq.~(\ref{Friedmann-eff-1a})
we can present $|\dot{a}|$ as 
\begin{equation}
|\dot{a}|\ =\ \sqrt{A_{0}}X^{1/3}(1-X)^{1/2}\ ,
\end{equation}
where $A_{0}:=(8\pi G/3)(\pi_{\phi}^{2}\rho_{{\rm crit}}^{2}/2)^{\frac{1}{3}}$
is a constant. It follows that for the energy density $\rho_\phi=(2/5)\rho_{{\rm crit}}$,
the speed of the collapse is maximum at 
\begin{equation}
|\dot{a}|_{\text{max}}\ =\ \sqrt{\frac{3}{5}A_{0}}\left(\frac{2}{5}\right)^{\frac{1}{3}}.\label{maxadot}
\end{equation}
The scale factor $a_{\mathrm{max}}$, corresponding to $|\dot{a}|_{\text{max}}$
reads $a_{\mathrm{max}}=(5\pi_{\phi}^{2}/4\rho_{{\rm {crit}}})^{1/6}$.
Notice that this value is independent of $r_{b}$, therefore, it is
the same for any shell. The minimum value of the scale factor, $a_{\text{crit}}$,
is fixed by the requirement that the Hubble rate vanishes, i.e., $\rho_\phi=\rho_{\text{crit}}$,
when the collapse hits a bounce; at this point we have that $a_{\text{crit}}=(\pi_{\phi}^{2}/2\rho_{{\rm {crit}}})^{1/6}=(2/5)^{1/6}a_{\text{max}}$.

On the other hand, by setting $\Theta_{b}=0$ in Eq. (\ref{theta1-q})
we get $\dot{R}^{2}=1$, so that, we can determine the speed of the
collapse, $|\dot{a}|_{\mathrm{AH}}=1/r$, for any shell $r$, at which
one horizon can form; in particular, for the boundary shell, this gives
$|\dot{a}|_{\mathrm{AH}}=1/r_{b}$. When the speed of the collapse,
$|\dot{a}|$, reaches the value $1/r_{b}$, then an apparent horizon
forms. Thus, if the maximum speed $|\dot{a}|_{\text{max}}$ is lower
than the critical speed $|\dot{a}|_{\mathrm{AH}}$, no horizon can
form. Let us introduce a radius $r_{\star}$, as 
\begin{equation}
r_{\star}\ :=\ \frac{1}{|\dot{a}|_{\text{max}}}\ .
\end{equation}
We see that $r_{\star}$ determines a \emph{threshold radius} for
the horizon formation in the scalar field collapse with the momentum
$\pi_{\phi}$; if $r_{b}<r_{\star}$, then \emph{no} horizon can form
at any stage of the collapse. The case $r_{b}=r_{\star}$ corresponds
to the formation of a dynamical horizon at the boundary of the two
spacetime regions \cite{Ashtekar-Horizon:2002,Hayward1994}. Finally, for the case $r_{b}>r_{\star}$
two horizons will form, one inside and the other outside the collapsing
matter \cite{Bojowald2005b}.


\vspace*{5mm}

\subsection{Exterior geometry and collapse endstate}
\label{q-ext}
\vspace*{3mm}

So far, we have analysed the interior collapsing
spacetime in the presence of the quantum gravity effects. This quantum
effects are expected to be carried out to the exterior geometry through
the matching conditions applied on the boundary $r_{b}$ of two regions.
In the following, we will focus on the main physical consequences
that can emerge from this scenario in order to predict the possible
exterior geometry for the collapse.

\begin{figure}
\includegraphics[height=2.1in]{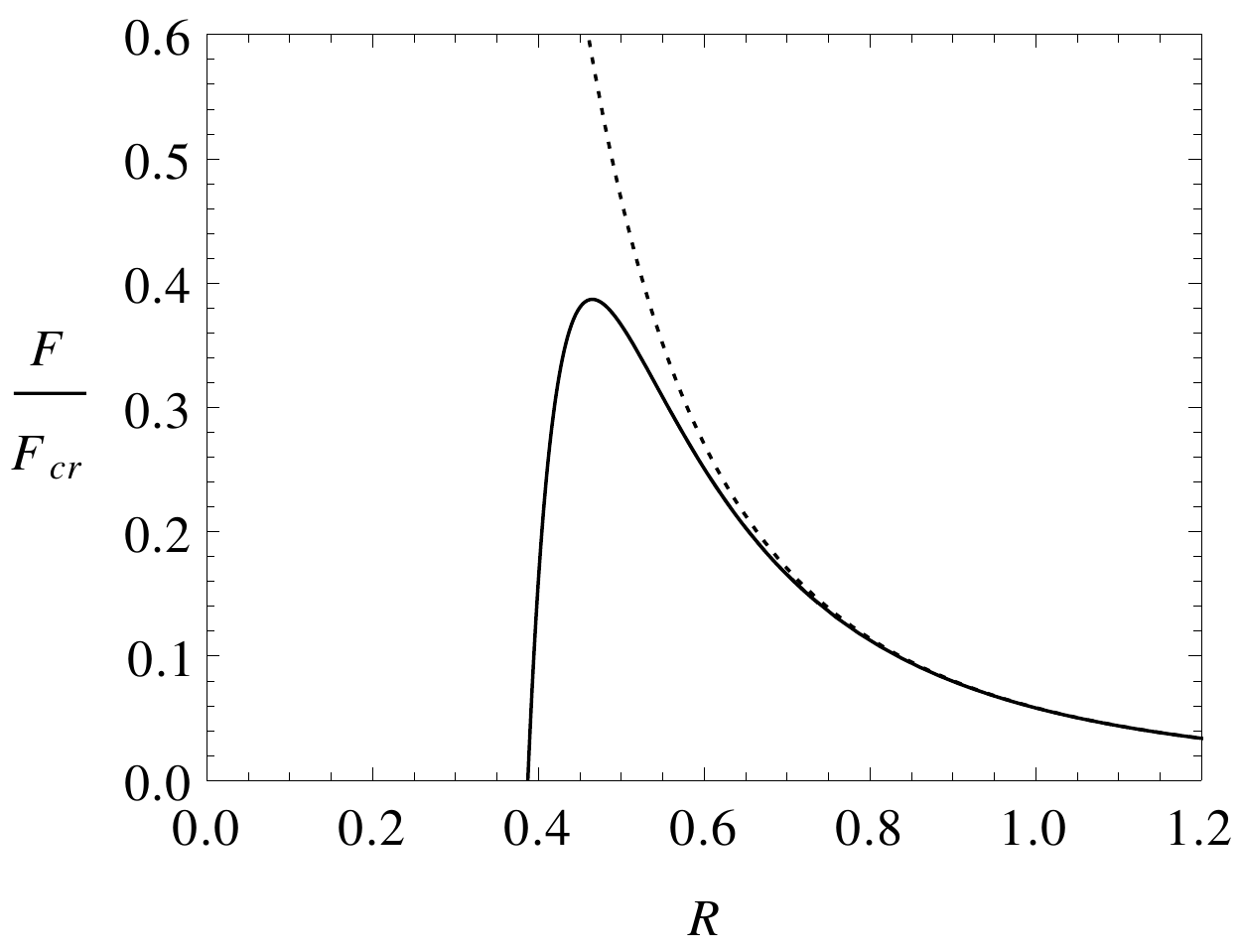} \qquad{} \includegraphics[height=2.1in]{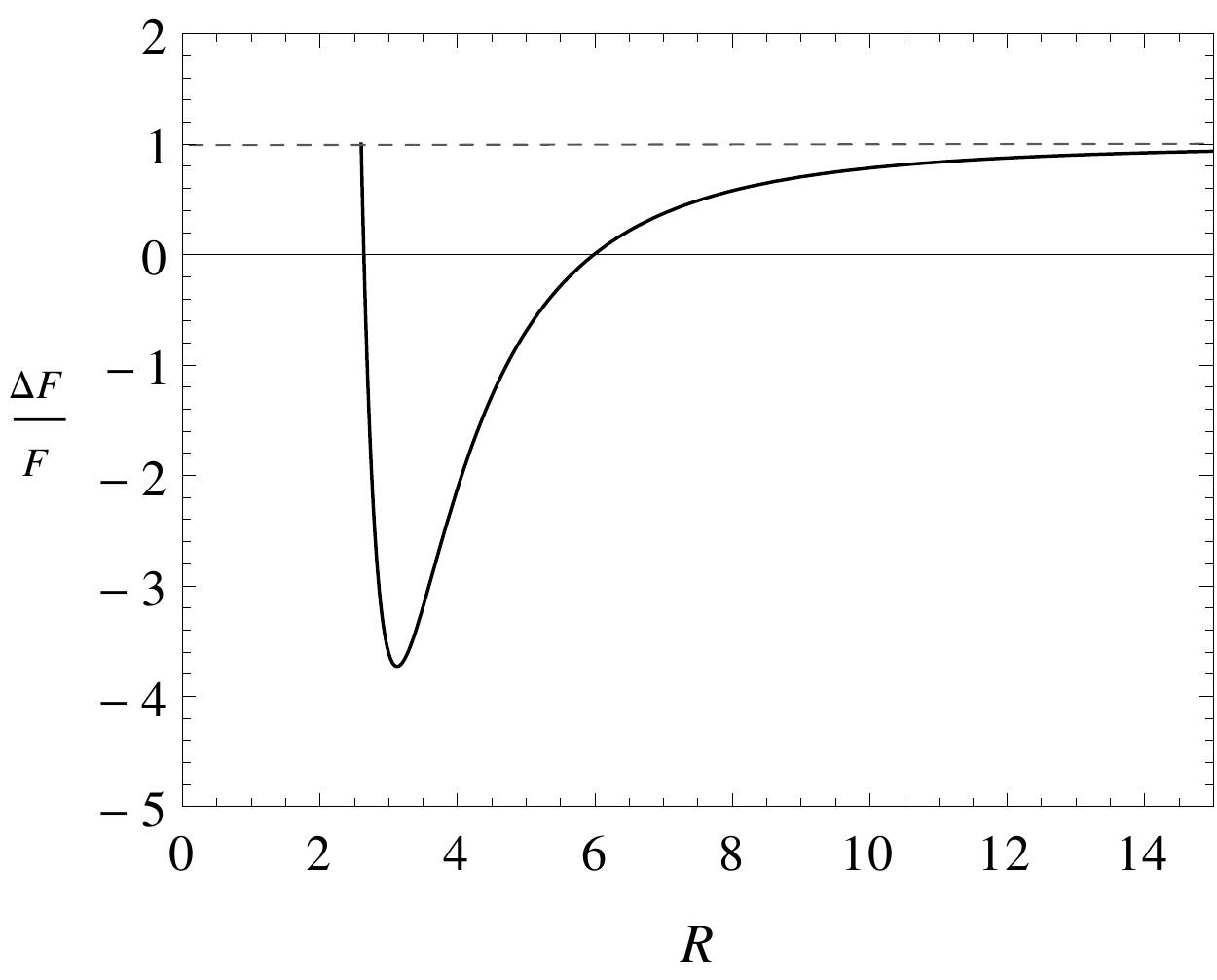}
\caption{The left plot represents the behaviour of classical (dotted
curve) and effective (solid curve) mass function during the collapse.
The right plot shows the behaviour of the mass loss $\Delta F/F$,
as a function of area radius $R$. We have used the value of parameters
$G=c_{\mathrm{light}}=1$, and $\pi_{\phi}=10~000$.}

\label{F-lum1} 
\end{figure}

The classical Friedmann equation corresponds to the last relation
in the classical Einstein's field equation (\ref{einstein-1}), which
can be written in terms of the mass function as $H^{2}=F/R^{3}$.
Consequently, and since in the semiclassical regime the Friedmann
equation is modified to Eq.~(\ref{Friedmann-eff-1a}), this might imply
a modification of the mass function defined by Eq. (\ref{einstein-1}).
In other words, we can introduce an effective mass function $F_{\text{eff}}$
corresponding to the modified Friedmann equation (\ref{Friedmann-eff-1a})
as 
\begin{equation}
F_{\text{eff}}\ =\ \frac{8\pi G}{3}\rho_{\text{eff}}R^{3}\  =\  \frac{8\pi G}{3}\rho_\phi R^{3}\left(1-\frac{\rho_\phi}{\rho_{\rm crit}}\right).\label{mass-eff}
\end{equation}
This describes an effective geometry on which the phase space trajectories
are considered to be classical, whereas the matter content is assumed
to be modified by quantum gravity effects. In the classical limit,
as $\rho_{\text{eff}}\rightarrow\rho_\phi$, the effective mass function
reduces to the classical one, i.e., $F$. In the interior
semiclassical region, since $\rho_{0}<\rho_\phi<\rho_{\text{crit}}$, so
both $F$ and $F_{{\rm eff}}$ remain finite during the collapse.
Using the relations $F=(8\pi G/3)\rho_\phi R^{3}$ and $\rho_\phi/\rho_{\textrm{crit}}=F^{2}/F_{\textrm{crit}}^{2}$  (for a massless scalar field),
it is convenient to rewrite Eq. (\ref{mass-eff}) as 
\begin{equation}
F_{\text{eff}}\ =\ F\left(1-\frac{F^{2}}{F_{\mathrm{crit}}^{2}}\right),\label{mass-effb}
\end{equation}
in which we have defined  
$F_{\mathrm{crit}}:=8\pi G\pi_{\phi}^{2}r_{b}^{3}\sqrt{\rho_{\textrm{crit}}}/3\sqrt{2}$. 
Notice that $F_{\rm crit}$ is a function of the phase space variable $\pi_\phi$; 
since,  for a massless scalar field, $\pi_\phi$ is a constant of motion, fixed by the initial conditions,  
$F_{\mathrm{crit}}$ becomes a constant for any  shell (with a specific choice of  $r_b$) and is determined at the initial configuration of the collapse.
Eq. (\ref{mass-effb}) shows that, the mass function $F$ is allowed
to evolve in the interval $F_{0}<F<F_{\mathrm{crit}}$ along with the
collapse dynamical evolution. Consequently, the effective mass function
$F_{{\rm eff}}$ increases from the initial value
$\sim F_{0}$ (for $\rho_\phi\ll\rho_{\textrm{crit}}$) and reaches a maximum
at $F_{{\rm crit}}/\sqrt{3}$; then, it starts decreasing and vanishes
at $F_{{\rm crit}}$ (cf. see the left plot in figure~\ref{F-lum1}). In
addition, it should be noticed that, classically trapped surfaces
form when $F>R$ at some points during the collapse and $F$ diverges
at the singularity. Nevertheless, in the presence of quantum effects,
this situation is different. For the choice of $r_{b}<r_{\star}$,
the effective mass function remains $F_{{\rm eff}}<R$, so that, no
trapped surface forms; if $r_{b}\geq r_{\star}$, then $F_{{\rm eff}}\geq R$
and trapped surfaces form during the collapse.
In the following of this section, we will study these two cases with more details.

\subsubsection*{(i) Outward flux of energy}

For a collapsing star whose initial boundary radius $r_{b}$ is less
than $r_{\star}$, we study the resulting mass loss due to the semiclassical
modified interior geometry. Let us designate the initial mass function
at scales $\rho_\phi\ll\rho_{{\rm crit}}$, i.e, in the classical regime,
as $F_{0}=(8\pi G/3)\rho_{0}R_{0}^{3}$, where $\rho_{0}=\pi_{\phi}^{2}/2a_{0}^{6}$,
and for $\rho_\phi\lesssim\rho_{{\rm crit}}$ (in the semiclassical regime)
we have $F_{\text{eff}}$ given by Eq. (\ref{mass-effb}). Then, the
(quantum geometrical) mass loss, $\Delta F/F_{0}$ (where $\Delta F=F_{0}-F_{\text{eff}}$),
for any shell (similar to Eq.~(\ref{mass-loss-eff})) is provided by the following expression: 
\begin{align}
\frac{\triangle F}{F(a_{0})}\   =\ 1-\frac{F_{\text{eff}}}{F_{0}}\ =\ 1-\sqrt{\frac{\rho_\phi}{\rho_{0}}}\left(1-\frac{\rho_\phi}{\rho_{{\rm crit}}}\right)\ .
\label{mass-loss-eff-SF}
\end{align}
As $\rho_\phi$ increases the mass loss decreases positively until it vanishes
at a point. Then, $\Delta F/F$ continues decreasing (negatively)
until it reaches to a minimum at $\rho_\phi=\rho_{{\rm crit}}/3$. Henceforth,
in the energy interval $\rho_{{\rm crit}}/3<\rho_\phi<\rho_{{\rm crit}}$,
the mass loss increases until the bouncing point at
$\rho_\phi\rightarrow\rho_{\text{crit}}$, where $\Delta F/F\rightarrow1$;
this means that the quantum gravity corrections, applied
to the interior region, give rise to an outward flux of energy near
the bounce in the semiclassical regime (see the right plot in figure \ref{F-lum1}).

It is worthy to mention that, when an inverse triad correction is
applied to the collapsing system (with a scalar field, cf. see  appendix \ref{LQG-scalar}, or
a tachyon field, cf. section~\ref{LQG-tach}, as matter sources), the (quantum) modified
energy density decreases as collapse evolves. Whence, as the collapsing
cloud approaches the center (with a vanishing scale factor, where
the classical singularity is located) the energy density reaches its
minimum value, whereas the mass loss tends to one. In the holonomy
corrected semiclassical collapse herein, the energy density increases
and reaches to a maximum value $\rho_{\text{crit}}$ at the bounce (with
a finite non-zero volume). Nevertheless, the dynamics of the collapse
is governed by an effective energy density which decreases close to
the bounce and vanishes at $a=a_{\textrm{crit}}$. Consequently,
the effective mass function also decreases and vanishes at the bounce,
which happens at $t_{\textrm{crit}}<t_{{\rm sing}}$ (with $t_{{\rm sing}}$
being the time when the classical singularity is reached).

\subsubsection*{(ii) Non singular black hole formation}

If the initial condition for the collapsing star is such that $r_{b}\geq r_{\star}$,
then a black hole will form at the collapse final state. We will now
analyse a possible prediction for the exterior geometry of the collapsing
system in this case. The total mass measured by an asymptotic observer
is given by $m_{\mathrm{ext}}=m_{M}+m_{\phi}$, where $m_{M}$ is
the total mass in the generalized Vaidya region, and $m_{\phi}=\int\rho dV$
is the interior mass related to the scalar field $\phi$. Since the
matter related to $m_{M}$ is not specified in the
exterior Vaidya geometry in our model, we just focus on a qualitative
analysis of behaviour of the horizon close to the matter shells.

From the matching conditions (\ref{V2})-(\ref{V4-a}), we can
get the information regarding the behaviour of trapping horizons in
the exterior region. Indeed, when the relation $2M(\mathrm{v},r_{\mathrm{v}})G=r_{\mathrm{v}}$
is satisfied at the boundary, trapped surfaces will form in the exterior
region close to the matter shells. On the classical geometry,
the boundary function, ${\cal F}=(1-2M(\mathrm{v})G/r_{\mathrm{v}})$,
becomes negative for the trapped region and vanishes at the apparent
horizon. Therefore, the equation for event horizon is given at the
boundary of the collapsing body by ${\cal F}|_{\Sigma}=0$. Nevertheless,
in the semiclassical regime, the boundary function is expected to
be modified by employing the matching conditions due to the fact that
the interior spacetime was modified by the quantum gravity effects.
Using the conditions (\ref{V1-a}) and (\ref{V2}), we have
that $2M(\mathrm{v},r_{\mathrm{v}})G/r_{\mathrm{v}}=F(t)/R(t)$ at
the boundary surface $\Sigma$ with $r=r_{b}$. Since the mass function
is modified as in the Eq. (\ref{mass-eff}) in the semiclassical regime,
therefore, the mass $M(\mathrm{v})$ is also modified as $\tilde{M}(\mathrm{v})=F_{{\rm eff}}/2G$
at $\Sigma$: 
\begin{equation}
\tilde{M}(\mathrm{v})\ =\ M-\frac{M^{3}}{M_{\mathrm{crit}}^{2}}\ ,\label{bounF-eff}
\end{equation}
where $2GM_{{\rm crit}}:=F_{{\rm crit}}=const$. Eq. (\ref{bounF-eff})
shows that the quantum gravity induced effects leads to a modification
of the boundary function by a cubic term $M^{3}$. By substituting
the classical mass function with $F=(8\pi G/3)\rho_\phi R^{3}$, we can
rewrite the Eq. (\ref{bounF-eff}) as 
\begin{equation}
\tilde{M}(R)\ =\ \frac{C}{R^{3}}-\frac{D}{R^{9}}\ ,\label{bounF-eff2}
\end{equation}
where $C:=(2\pi/3)\pi_{\phi}^{2}r_{b}^{6}$, and $D:=C\pi_{\phi}^{2}r_{b}^{6}/(2\rho_{\mathrm{cr}})$
are constants. Eq. (\ref{bounF-eff2}) represents a non singular,
exotic black hole geometry. Notice that, the effective exterior function
${\cal F}_{{\rm eff}}=(1-2G\tilde{M}/R)$ in the classical limit,
where $\rho_{\mathrm{crit}}\rightarrow\infty$, tends to ${\cal F}=1-2GC/R^{4}$,
which represents a classical singular black hole geometry (cf. see Eq.~(\ref{f-phi2-b})).
In addition, as we expected, in the presence of a nonzero matter pressure
(of the massless scalar field) at the boundary, the (homogeneous)
interior spacetime is not matched with an empty (inhomogeneous) Schwarzschild
exterior.

\addtocontents{toc}{\vspace{2em}} 

\backmatter


\newpage
\parskip 0mm
\bibliographystyle{Classes/jmb}
\renewcommand{\bibname}{References} 

\addcontentsline{toc}{chapter}{References} 





\end{document}